\documentclass[aps,prx,twocolumn,superscriptaddress,superscriptaddress,longbibliography]{revtex4-2}
\usepackage{braket}
\usepackage{graphicx}
\usepackage{amsmath,amssymb}    
\usepackage{amsfonts}
\usepackage[hidelinks]{hyperref}
\usepackage{lipsum}
\usepackage{xcolor}
\usepackage{quantikz}
\usepackage{bbold}
\usepackage{adjustbox}
\usepackage{multirow}
\usepackage{rotating}
\usepackage{array,mathtools,amssymb,booktabs}
\usepackage[T1]{fontenc}
\usepackage[tiny, center, uppercase]{titlesec}
\usepackage[caption=false]{subfig}
\usepackage[version=3]{mhchem} 

\newcommand{\select}{\mathsf{SELECT}}
\newcommand{\prepare}{\mathsf{PREPARE}}

\begin{document}

\title{Fault-tolerant quantum algorithm \\
for symmetry-adapted perturbation theory }

\newcommand{\BI}{\affiliation{
Quantum Lab, Boehringer Ingelheim, 55218 Ingelheim am Rhein, Germany}}

\newcommand{\QCW}{\affiliation{QC Ware Corp, Palo Alto, CA 94306, USA}}

\newcommand{\PSIQ}{\affiliation{PsiQuantum, 700 Hansen Way, Palo Alto, CA 94304, USA}}

\newcommand{\BIMedChem}{\affiliation{
Boehringer Ingelheim Pharma GmbH \& Co KG, Birkendorfer Strasse 65, 88397 Biberach, Germany}}

\newcommand{\UIBK}{\affiliation{Department of General, Inorganic and Theoretical Chemistry,
University of Innsbruck, 6020 Innsbruck, Austria}}

\author{Cristian L. Cortes}
\email{cris.cortes@qcware.com}
\QCW
\author{Matthias Loipersberger}
\QCW
\author{Robert M.~Parrish}
\QCW

\author{Sam Morley-Short}
\PSIQ

\author{William Pol}
\email{wpol@psiquantum.com}
\PSIQ

\author{Sukin Sim}
\PSIQ

\author{Mark Steudtner}
\PSIQ

\author{Christofer S.~Tautermann}
\BIMedChem
\UIBK

\author{Matthias Degroote}
\BI
\author{Nikolaj Moll} 
\BI
\author{Raffaele Santagati}
\email{raffaele.santagati@boehringer-ingelheim.com}
\BI
\author{Michael Streif}
\BI

\begin{abstract}
The efficient computation of observables beyond the total energy is a key challenge and opportunity for fault-tolerant quantum computing approaches in quantum chemistry. Here we consider the symmetry-adapted perturbation theory (SAPT) components of the interaction energy as a prototypical example of such an observable. We provide a guide for calculating this observable on a fault-tolerant quantum computer while optimizing the required computational resources. Specifically, we present a quantum algorithm that estimates interaction energies at the first-order SAPT level with a Heisenberg-limited scaling. To this end, we exploit a high-order tensor factorization and block encoding technique that efficiently represents each SAPT observable. To quantify the computational cost of our methodology, we provide resource estimates in terms of the required number of logical qubits and Toffoli gates to execute our algorithm for a range of benchmark molecules, also taking into account the cost of the eigenstate preparation and the cost of block encoding the SAPT observables. Finally, we perform the resource estimation for a heme and artemisinin complex as a representative large-scale system encountered in drug design, highlighting our algorithm's performance in this new benchmark study and discussing possible bottlenecks that may be improved in future work.
\end{abstract}

\maketitle

The computation of expectation values of observables other than the total molecular energy is a foundational task in quantum chemistry, for which fault-tolerant quantum computers (FTQC) are expected to provide speed-ups for those systems where classical computers cannot find an accurate solution~\cite{cao2019,mcardle2020,bauer2020,Liu2022}. Some of the most important observable properties, such as the Born-Oppenheimer potential energy landscape, the adiabatic excitation energy landscape, the total intermolecular interaction energies, polarizabilities, and various spectroscopical properties, can all be written in terms of the  total molecular energy or its derivatives ~\cite{huggins2021nearly,o2021efficient}. For these cases, many recent efforts have focused on optimizing the computational cost, bringing several orders of magnitude improvements~\cite{Reiher2017,Lee2021,poulin2018quantum,o2021efficient}. However, many other properties of molecular systems cannot be optimally defined as a linear combination of total energies and require a specific quantum algorithm to calculate their expectation value~\cite{brassard2002quantum,knill2007optimal}. For example, the one-particle density, total kinetic energy, and multipole moments of a molecular wavefunction do not allow an efficient expression in terms of the total energy~\cite{jensen2017introduction}. The components of the symmetry-adapted perturbation theory (SAPT) represent an additional example of such kind of observables ~\cite{bukowski2013sapt2012, smith2020psi4}.  This type of observables plays a pivotal role in the characterization and featurization of intermolecular interactions between two weakly interacting sub-systems, with practical applications in molecules and materials design for polymers, catalysts, batteries, and drugs~\cite{VonBurg2021,santagati2023drug}.

\begin{figure}[tb]
    \centering
    \includegraphics[width=\columnwidth]{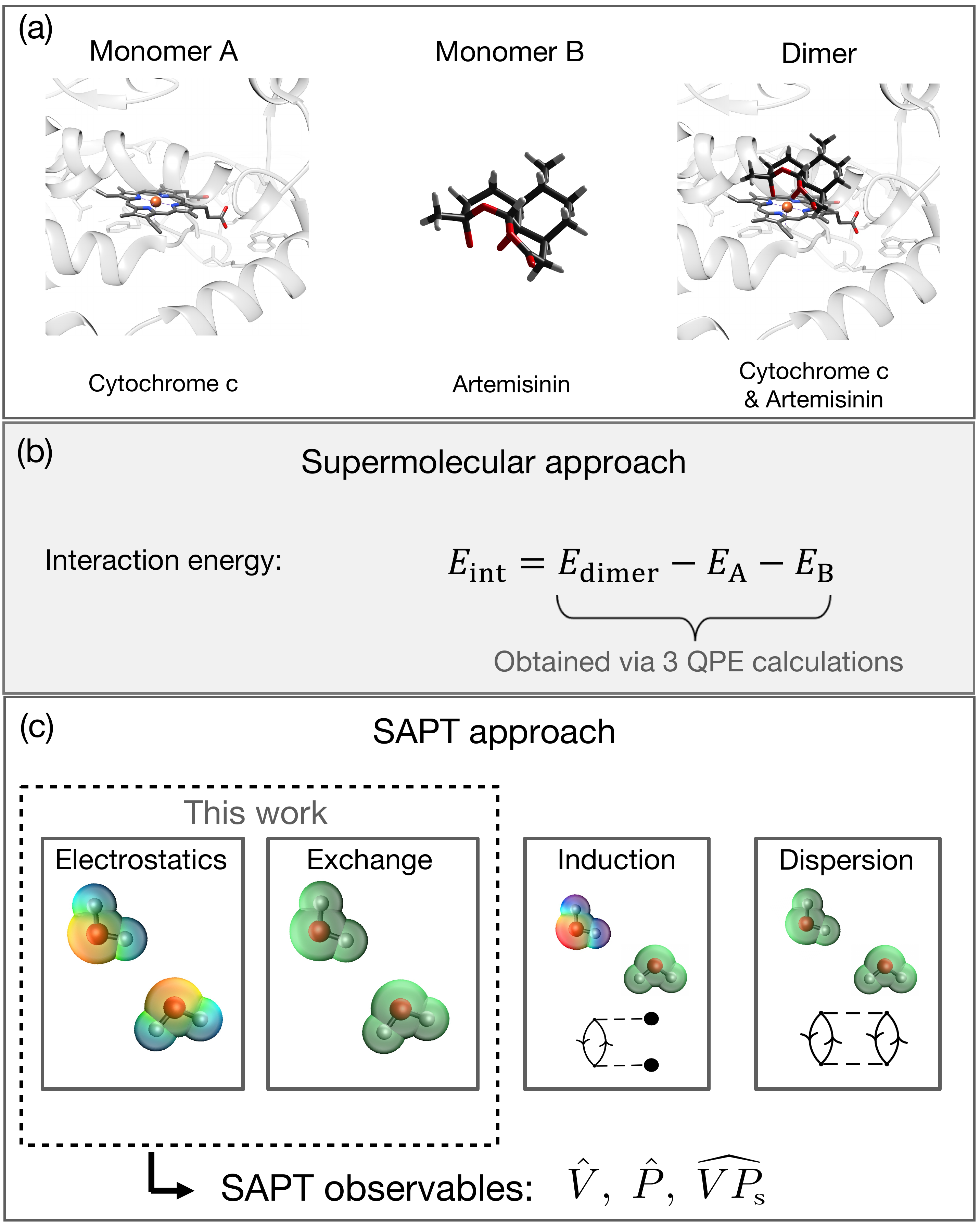}
    \caption{Overview of an interaction energy problem encountered in drug design. (a) Schematics of monomer A, monomer B, and dimer. (b) Supermolecular and (c) SAPT approaches to calculate the interaction energy. First-order SAPT energy terms are determined from expectation values of electrostatic $\hat{V}$, exchange $\hat{P}$ and electrostatic-exchange  $\widehat{VP}_\mathrm{s}$ operators. }
    \label{fig:overview}
\end{figure}

SAPT is a variant of Rayleigh-Schr\"odinger perturbation theory specifically designed to describe fermionic systems. It restores the Pauli exclusion principle for anti-symmetric wavefunctions at every order of perturbation theory. In classical computing, SAPT is a state-of-the-art method that directly calculates the interaction energy, $E_\mathrm{int} = E_{AB} - E_A - E_B$, defined as the energy difference between the weakly interacting systems (monomers A and B), and the system in which the monomers interact, referred to as the dimer (AB), see Refs.~\cite{jeziorski1993sapt, szalewicz2012symmetry, hohenstein2011large}. While the so-called supermolecular approach~\cite{gutowski1986basis, gutowski1987proper} calculates the interaction energy by combining the results of three separate energy calculations, SAPT computes $E_\mathrm{int}$ directly as an observable estimation task by decomposing the interaction energy in terms of physically interpretable quantities such as electrostatic, exchange, dispersion and induction energy contributions (see Fig.~\ref{fig:overview}).

Previous studies have focused on noisy intermediate-scale quantum (NISQ) computations of these observables~\cite{Malone2022,Loipersberger2023}, where the uncertainty scaling is far from the optimal Heisenberg-limited scaling that can be achieved by  fault-tolerant quantum algorithms~\cite{knill2007optimal}. Furthermore, the scientific literature has been missing an in-depth study of the computational cost of calculating the SAPT components on a fault-tolerant quantum computer.

Here, we present a Heisenberg-limited methodology for calculating the first-order SAPT terms on a fault-tolerant quantum computer. We take advantage of a recently proposed Heisenberg-limited  method for expectation values estimation (called QSP-EVE)~\cite{steudtner2023fault}, exploiting the latest techniques in quantum simulation, such as block encodings and quantum signal processing. We tailor the QSP-EVE algorithm to calculate the expectation value $\braket{\Psi_A\Psi_B|\hat{F}|\Psi_A\Psi_B}$ defined with respect to the two ground states of the two sub-systems $\ket{\Psi_A}$ and $\ket{\Psi_B}$, to compute first-order symmetry-adapted perturbation theory observables. 

In this work, we introduce a framework for implementing SAPT on a fault-tolerant quantum computer with Heisenberg scaling of the uncertainty and we determine the corresponding algorithmic resource costs. We accomplish this goal through three major contributions we outline in the following sections: (i) derivation of first-order SAPT operators using a second quantization picture (Sec.~\ref{sec:SAPT_theory}). The operators were derived in both the full and active space pictures and are useful in both the monomer and dimer centered bases, (ii) tailoring of the QSP-EVE algorithm for the estimation of expectation values $\braket{\Psi_A\Psi_B|\hat{F}|\Psi_A\Psi_B}$ of observables $\hat{F}$ (Sec.~\ref{sec:SAPT-EVE algorithm}) and (iii) the development of tensor factorization and block encoding arithmetic techniques specifically designed for SAPT operators that reduce the total cost of the $\mathsf{PREPARE}$ and $\mathsf{SELECT}$ operators associated with the block encodings, as well as the $\ell_1$ norm of the block encoded observables (Sec.~\ref{sec:SAPT_operator_encoding}). To evaluate the performance of our framework, we compare the cost of the proposed tensor-factorized SAPT-EVE algorithm with a set of benchmark molecules against an equivalent \emph{sparse} SAPT-EVE algorithm that encodes the SAPT operators through the conventional Jordan-Wigner mapping (Sec.~\ref{sec:small_mol}). Finally, we present a new benchmark study relevant to drug design, which may be of independent interest to the quantum computing community, involving the interaction of a heme with an artemisinin drug molecule (Sec.~\ref{sec:heme_art}). We finalize our work by discussing the complete resource cost of our algorithm and outlining possible avenues of improvement for future work (Sec.~\ref{sec:conclusion}).

\section{Theoretical Results}

\subsection{Symmetry-adapted perturbation theory}
\label{sec:SAPT_theory}
Ranking ligands according to their binding strength with a substrate is a fundamental task in computational drug design~\cite{cournia2017}. SAPT aims to calculate the interaction energy $E_\mathrm{int}$ of the full system (the dimer) as a sum of physically interpretable contributions from the substrate and ligand (monomer A and monomer B), $E_\mathrm{int} = E_\mathrm{pol}^{(1)} + E_\mathrm{exch}^{(1)} + E_\mathrm{pol}^{(2)} + E_\mathrm{exch}^{(2)} \cdots$, through the use of a symmetry-adapted Rayleigh-Schr\"odinger perturbative expansion. In this manuscript, we derive the second quantized operators based on a methodology first presented in the paper by Moszynski {\em et al}.~\cite{moszynski1994many}, which presented the SAPT formulation using a density matrix formalism. The advantage of this approach is that it will work both on the dimer-centered basis as well as on the monomer-centered basis. In Appendix~\ref{sec:Appendix_SAPT}, we provide a full derivation of the first-order SAPT operators highlighting the permutational symmetries inherent to all operators. Here, we summarize the results for the first-order interaction energy under the $S^2$ approximation (see Appendix~\ref{sec:Appendix_SAPT}), which may be decomposed in terms of electrostatic polarization energy $E^{(1)}_\mathrm{pol}$ and exchange energy $E^{(1)}_\mathrm{exch}$ contributions,
\begin{equation}
    E^{(1)}_\mathrm{int} = E^{(1)}_\mathrm{pol} + E^{(1)}_\mathrm{exch} = \braket{\hat{V}} + \left( \braket{\widehat{VP}_\mathrm{s}} - \braket{\hat{V}}\braket{\hat{P}}\right) ,
    \label{Interaction_Energy}
\end{equation}
and is fully captured by defining an electrostatic $\hat{V}$, exchange $\hat{P}$, and symmetric electrostatic-exchange $\widehat{VP}_\mathrm{s}$ operator with excitation operators, $\hat{E}_{\textsc{p}\textsc{p}'} = \hat{a}^\dagger_{\textsc{p}}\hat{a}_{\textsc{p}'}$ and $\hat{E}_{\textsc{q}\textsc{q}'} = \hat{b}^\dagger_{\textsc{q}}\hat{b}_{\textsc{q}'}$, defined for each monomer. In the SAPT framework, two independent sets of fermionic operators $\hat{a}^\dagger_{\textsc{p}}/\hat{a}_{\textsc{p}}$ and $\hat{b}^\dagger_{\textsc{q}}/\hat{b}_{\textsc{q}}$ are used which obey the conventional fermionic commutation relations but fully commute with one another, $[\hat{a}^\dagger_{\textsc{p}},\hat{b}_{\textsc{q}}] = [\hat{a}_{\textsc{p}},\hat{b}^\dagger_{\textsc{q}}] = [\hat{a}^\dagger_{\textsc{p}},\hat{b}^\dagger_{\textsc{q}}] = [\hat{a}_{\textsc{p}},\hat{b}_{\textsc{q}}] = 0$. Explicitly, the SAPT operators are written as:
\begin{align}
   \hat{V} &= \tfrac{1}{2}\sum_{\textbf{\textsc{p}},\textbf{\textsc{q}}} \left( v^{\textsc{p}_1\textsc{p}_2}_{\textsc{q}_1\textsc{q}_2}\hat{E}_{\textsc{p}_1\textsc{p}_2} \hat{E}_{\textsc{q}_1\textsc{q}_2} + \text{h.c.} \right)  , \label{eq:V}\\
   \hat{P} &= -\tfrac{1}{2}\sum_{\substack{\textbf{\textsc{p}},\textbf{\textsc{q}}}} \left(S^{\textsc{p}_1}_{\textsc{q}_2}S^{\textsc{p}_2}_{\textsc{q}_1} \hat{E}_{\textsc{p}_1\textsc{p}_2} \hat{E}_{\textsc{q}_1\textsc{q}_2}  + \text{h.c.} \right) , \, \label{eq:P} \\
    \widehat{VP}_\mathrm{s} &= -\tfrac{1}{2}\sum_{\substack{\textbf{\textsc{p}},\textbf{\textsc{q}}}} \left( \nu^{\textsc{p}_1\textsc{q}_2}_{\textsc{q}_1\textsc{p}_2}\hat{E}_{\textsc{p}_1\textsc{p}_2}\hat{E}_{\textsc{q}_1\textsc{q}_2} + \text{h.c.} \right)    \label{eq:VP} \\
    &-\tfrac{1}{2}\sum_{\substack{\textbf{\textsc{p}},\textbf{\textsc{q}}}} \left(\nu^{\textsc{p}_1\textsc{p}_2}_{\textsc{q}_1\textsc{p}_4} S^{\textsc{p}_3}_{\textsc{q}_2} \hat{E}_{\textsc{p}_1\textsc{p}_2}\hat{E}_{\textsc{p}_3\textsc{p}_4}\hat{E}_{\textsc{q}_1\textsc{q}_2} + \text{h.c.} \right) \nonumber \\
    &-\tfrac{1}{2}\sum_{\substack{\textbf{\textsc{p}},\textbf{\textsc{q}}}} \left(\nu_{\textsc{q}_1\textsc{q}_2}^{\textsc{p}_1\textsc{q}_4}S_{\textsc{p}_2}^{\textsc{q}_3} 
    \hat{E}_{\textsc{p}_1\textsc{p}_2} \hat{E}_{\textsc{q}_1\textsc{q}_2}\hat{E}_{\textsc{q}_3\textsc{q}_4} + \text{h.c.} \right) \nonumber \\
    & -\tfrac{1}{2}\sum_{\substack{\textbf{\textsc{p}},\textbf{\textsc{q}}}} \left( v^{\textsc{p}_1\textsc{p}_2}_{\textsc{q}_1\textsc{q}_2} S^{\textsc{p}_3}_{\textsc{q}_4}S_{\textsc{p}_4}^{\textsc{q}_3} \hat{E}_{\textsc{p}_1\textsc{p}_2}\hat{E}_{\textsc{p}_3\textsc{p}_4} \hat{E}_{\textsc{q}_1\textsc{q}_2}\hat{E}_{\textsc{q}_3\textsc{q}_4} + \text{h.c.} \right) \, , \nonumber   
\end{align}
The indices $\textsc{p}/\textsc{q}$ label the molecular spin-orbitals of monomers A/B respectively, while h.c. abbreviates the Hermitian conjugate. The electrostatic $v^{\textsc{p}_1\textsc{p}_2}_{\textsc{q}_1\textsc{q}_2}$ and exchange $S^\textsc{p}_{\textsc{q}}$ tensors are defined with respect to molecular spin-orbitals $\phi_\textsc{p}(\mathbf{\mathbf{x}})$ and $\phi_\textsc{q}(\mathbf{\mathbf{x}})$ as:
\begin{align}
    v^{\textsc{p}_1\textsc{p}_2}_{\textsc{q}_1\textsc{q}_2} &= \int\! \mathrm{d}\mathbf{x}_i \mathrm{d}\mathbf{x}_j\, \phi_{\textsc{p}_1}^*(\mathbf{x}_i)\phi^{}_{\textsc{p}_2}(\mathbf{x}_i)v_{ij}\phi_{\textsc{q}_1}^*(\mathbf{x}_j)\phi^{}_{\textsc{q}_2}(\mathbf{x}_j) \, , 
    \label{eq:vppqq} \\
    S^\textsc{p}_{\textsc{q}} &= \int\!\! \mathrm{d}\mathbf{\mathbf{x}}\, \phi_\textsc{p}^*(\mathbf{\mathbf{x}})\phi_\textsc{q}(\mathbf{\mathbf{x}}) \, ,
    \label{eq:Spq}
\end{align}
where the coordinate $\mathbf{\mathbf{x}}$ collectively represents the spatial coordinate $\mathbf{r}$ and the spin coordinate $\sigma$ of the electron and,
\begin{equation}
    v_{ij} = \frac{1}{r_{ij}} - \sum_J \frac{Z_{J}}{\eta_B} \frac{1}{r_{iJ}} - \sum_I \frac{Z_{I}}{\eta_A} \frac{1}{r_{Ij}} + \sum_{IJ} \frac{Z_{I} Z_{J}}{\eta_A \eta_B} \frac{1}{r_{IJ}},
\end{equation}
is equal to the intermolecular interaction operator as a function of distances $r$ between the electronic (lower case index) and nuclear (upper case index) degrees of freedom. The total number of electrons is given by $\eta_A/\eta_B$, and $Z_I/Z_J$ represents the charges of nuclei $I$/$J$ in monomers A/B, respectively. The tensor components of the electrostatic-exchange operator $\widehat{VP}_\mathrm{s}$ are given by,  
\begin{align}
\nu^{\textsc{p}_1\textsc{q}_2}_{\textsc{q}_1\textsc{p}_2} &= \,v^{\textsc{p}_1\textsc{q}_2}_{\textsc{q}_1\textsc{p}_2}  + \sum_{\textsc{p}_3\textsc{q}_3} v^{\textsc{p}_1\textsc{p}_3}_{\textsc{q}_1\textsc{q}_3} S^{\textsc{p}_3}_{\textsc{q}_2}S_{\textsc{p}_2}^{\textsc{q}_3} \label{eq:vpqqp_dressed} \\
&- \sum_{\textsc{q}_3} v_{\textsc{q}_1\textsc{q}_3}^{\textsc{p}_1\textsc{q}_2}S_{\textsc{p}_2}^{\textsc{q}_3} - \sum_{\textsc{p}_3} v^{\textsc{p}_1\textsc{p}_3}_{\textsc{q}_1\textsc{p}_2} S^{\textsc{p}_3}_{\textsc{q}_2}, \nonumber \\   
\nu^{\textsc{p}_1\textsc{p}_2}_{\textsc{q}_1\textsc{p}_4} &= v^{\textsc{p}_1\textsc{p}_2}_{\textsc{q}_1\textsc{p}_4} - \sum_{\textsc{q}_3}v^{\textsc{p}_1\textsc{p}_2}_{\textsc{q}_1\textsc{q}_3} S^{\textsc{p}_4}_{\textsc{q}_3} \, , \\
\nu_{\textsc{q}_1\textsc{q}_2}^{\textsc{p}_1\textsc{q}_4} &= v_{\textsc{q}_1\textsc{q}_2}^{\textsc{p}_1\textsc{q}_4} - \sum_{\textsc{p}_3}v^{\textsc{p}_1\textsc{p}_3}_{\textsc{q}_1\textsc{q}_2} S^{\textsc{p}_3}_{\textsc{q}_4} \, .
\label{eq:vpqqq_dressed}
\end{align}
In this form, all of the SAPT operators are manifestly Hermitian, as required for the QSP-EVE algorithm. It is worth noting that the SAPT operators in Eqs.~\eqref{eq:V}-\eqref{eq:VP} are applicable to both complex and real molecular spin-orbitals. For the rest of the paper, however, we will assume all orbitals are real to provide the most compact representation. We have also derived the corresponding active space SAPT operators, which may be found in Appendix~\ref{sec:Appendix_Active_Space}. The SAPT formulation presented in the current manuscript is based on a definition of the electrostatic-exchange operator $\widehat{VP}_\mathrm{s}$ which consists of four terms including the symmetric product $\tfrac{1}{2}(\hat{V}\hat{P} + \hat{P}\hat{V})$. The origin of the four terms is due to the use of finite basis sets that preclude certain completeness relations from occurring. Details of this observation are provided in Appendix~\ref{sec:Complete_basis_set_limit}. We emphasize that our methodology is applicable in the finite basis set limit and is directly comparable to classical SAPT numerical results. 

\subsection{SAPT-EVE algorithm}
\label{sec:SAPT-EVE algorithm}
Calculating the separate SAPT contributions comes down to the estimation of the expectation value of three operators in the most resource efficient way. This is a motive that has been studied before, both in NISQ and FTQC. In the NISQ-era, observable estimation through the direct measurement of the density matrix has been considered extensively and is used for energy and gradient estimation tasks encountered in variational quantum algorithms~\cite{cerezo2021variational,bharti2022noisy}. However, these methods rely on shot-noise-limited sampling that scales as $\mathcal{O}(1/\epsilon^2)$ for a single observable and $\mathcal{O}(M/\epsilon^2)$ for $M$ non-commuting observables with respect to the target precision $\epsilon$. Shadow tomography reduces the cost for large $M$ resulting in a logarithmic scaling, $\mathcal{O}(\log(M)/\epsilon^4)$. The method of classical shadows reduces the precision dependence even further, $\mathcal{O}(\log(M)/\epsilon^2)$, through the use of randomized Clifford or Pauli measurements \cite{aaronson2018shadow,huang2020predicting}.  

In contrast, a Heisenberg-limited observable estimation has a runtime that scales as $\mathcal{O}(1/\epsilon)$, providing a quadratic speed-up with respect to the precision $\epsilon$ while also saturating the fundamental limit in precision scaling imposed by nature. However, this benefit comes at the expense of longer circuits making these methods more suitable for fault-tolerant quantum computing. The first proposal for Heisenberg-limited observable estimation was given by \citet{Knill2007} based on an amplitude estimation with a Szegedy walk operator. The work by Rall~\cite{rall2020quantum} generalized this result to the case of block encoded observables. A subset of the current authors also developed a similar framework for estimating molecular forces and energy gradients~\cite{o2021efficient}. Recently, we proposed the quantum-signal-processing-based expectation value estimation (QSP-EVE) algorithm~\cite{steudtner2023fault}.

\begin{figure}[tb]
    \centering
    \includegraphics[width=\linewidth]{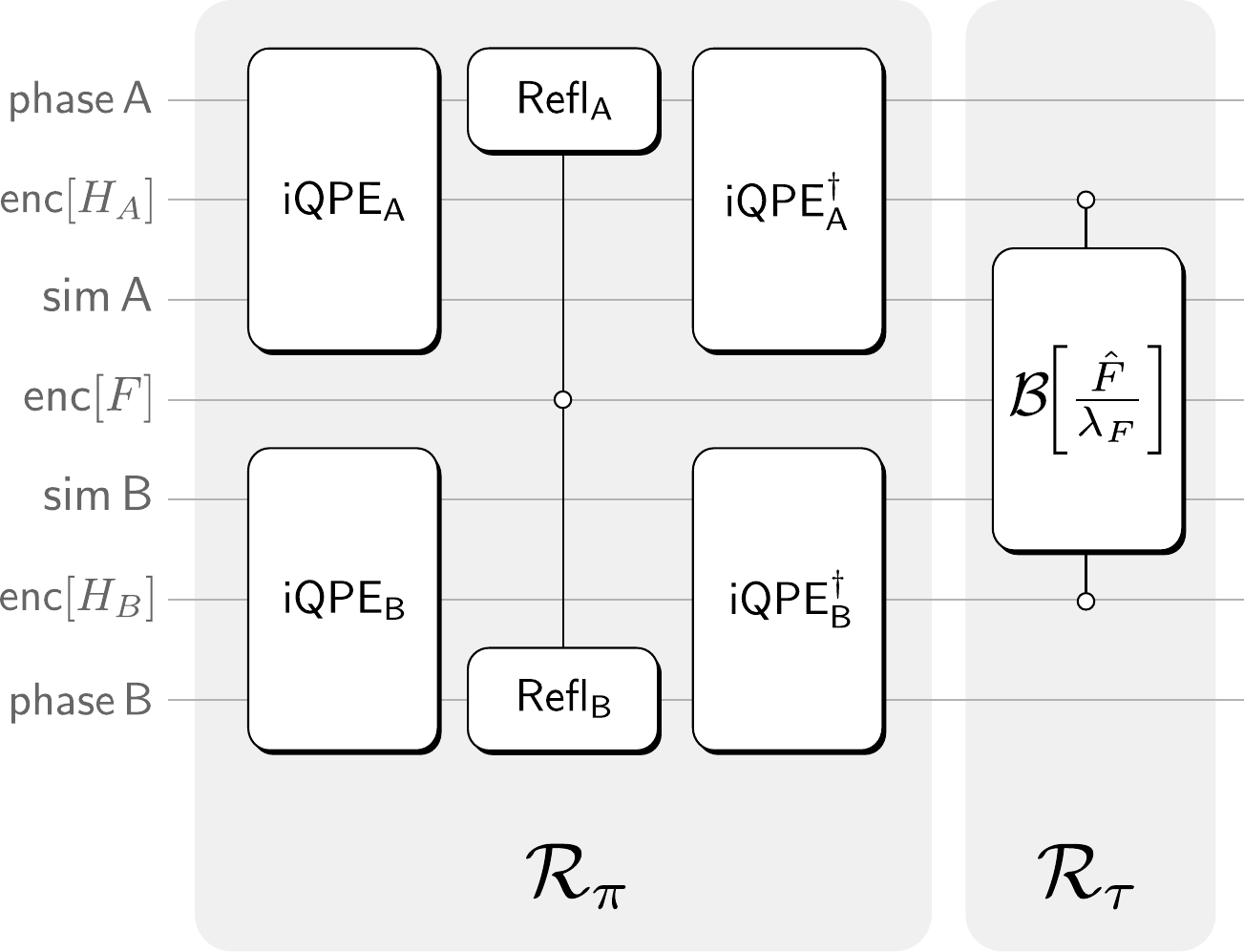}
    \caption{Iterate $\mathcal{U}_F$ used in SAPT-EVE algorithm to determine the expectation values of operators, $\hat{F} = \{\hat{V},\hat{P},\widehat{VP}_\mathrm{s}\}$. The iterate consists of two reflection oracles, $\mathcal{R}_\pi$ and $\mathcal{R}_\tau$, which act on 7 quantum registers in total. $\mathsf{enc} [A] $ / $\mathsf{enc} [B]$ are auxiliary registers used for the block encodings of the monomer Hamiltonians $H_A$ and $H_B$. $\mathsf{Phase\, A}$ / $\mathsf{Phase\, B}$ are phase registers that hold expressions of eigenphases for monomer Hamiltonians $H_A$ and $H_B$.  $\mathsf{sim\, A}$ / $\mathsf{sim\, B}$ are quantum registers representing the wavefunctions of monomers A and B. $\mathsf{enc}[F]$ is the auxiliary register for the block encoding of the observable. Additional details of the iterate may be found in the main text.
    }
    \label{fig:sapt_circuit}
\end{figure}

In the following, we present the SAPT expectation-value-estimation (SAPT-EVE) algorithm, which allows for calculating SAPT observables. Our proposal is based on the QSP-EVE algorithm, but it is tailored to account for the two uncoupled monomer wavefunctions required for the SAPT calculation. The QSP-EVE algorithm's correctness, probability of success, and other details can be found in Ref.~\cite{steudtner2023fault}. 

The SAPT-EVE algorithm is comprised of three different quantum phase estimation procedures. Each phase estimation calculation features a special iterate $\mathcal{U}_F$ for each observable $\hat{F} = \lbrace \hat{V}, \hat{P}, \widehat{VP}_\mathrm{s} \rbrace$. The iterate $\mathcal{U}_F$ is deliberately constructed such that for some eigenstates $|F{+}\rangle$ and $|F{-}\rangle$, we have 
\begin{align}
\label{eq:SEVE0}
   \mathcal{U}_F |F\pm\rangle = \exp\left(\pm i 2 \arccos \sqrt{\frac{1 - \langle \hat{F}/\lambda_F\rangle}{8}}\right)|F{\pm}\rangle \,
\end{align}
where $\langle \hat{F}/\lambda_F \rangle$ is the desired expectation value of the respective observable (relative to its $\ell_1$ norm $\lambda_F$) on the monomer reference states. Phase estimation with the operator $\mathcal{U}_F$ provides an estimate of the eigenphase of Eq.~\eqref{eq:SEVE0}, allowing us to extract $\langle \hat{F} \rangle$. 

The circuit for $\mathcal{U}_F = \mathcal{R}_\tau\mathcal{R}_\pi$ is visualized in Fig.~\ref{fig:sapt_circuit}. The first reflection circuit, $\mathcal{R}_\pi$, is used to flag the monomer ground-states, $\ket{\Psi_A,\Psi_B}$, non-destructively allowing them to be used in the rest of the algorithm coherently, thereby preserving Heisenberg scaling. $\mathcal{R}_\pi$ features inner phase estimation circuits $\mathsf{iQPE_A}$ and $\mathsf{iQPE_B}$ applied on the registers associated with  monomers A and B. Within these circuits, we employ QSP techniques as in \cite{steudtner2023fault} to implement the rounding of the eigenphases. Quantum signal processing requires a block encoding of the Hamiltonian for each monomer, which will bring an $\ell_1$ norm dependence of both Hamiltonians to the total complexity of the algorithm, as shown below. The necessary analysis of the expectation value error that sets the degree of the QSP polynomials is similar to Ref.~\cite{steudtner2023fault} and is summarized for the SAPT-EVE algorithm in {Appendix~\ref{sec:Appendix_SAPT_EVE}}. An open-controlled, coupled-reflection circuit using $\mathsf{Refl_A}$ and $\mathsf{Refl_B}$ is also used to reflect the binary eigenphases (known to $p$ bits of precision) associated with the monomer ground states.  

The second reflection, $\mathcal{R}_\tau$, in the circuit of Fig.~\ref{fig:sapt_circuit} is a controlled version of the observable block encoding, $\mathcal{B}[\hat{F}/\lambda_F]$, discussed in more detail in Refs.~\cite{berry2018improved,Berry_2019,babbush2018encoding,Lee2021}. The asymptotic Toffoli gate cost,  $C_\mathrm{Toff}$, of the SAPT-EVE algorithm is given by,
 \begin{align}
     C_\mathrm{Toff} = \mathcal{O}\left(\frac{\lambda_F}{\varepsilon_F} \left[\left( \frac{\lambda_A}{\Delta_A}C_{A}+\frac{\lambda_B}{\Delta_B}C_{B}\right)\log \frac{\lambda_F}{\varepsilon_F} + C_{F}\right] \right) \, ,
     \label{eq:scaling}
 \end{align}
 where $\lambda_A$, $\lambda_B$ and $\lambda_F$ denote the $\ell_1$ norms of $\hat{H}_A$, $\hat{H}_B$ and $\hat{F}$, respectively. $\Delta_A$ and $\Delta_B$ are the respective spectral gaps between the ground state and first excited state of Hamiltonians $\hat{H}_{A/B}$, $\varepsilon_F$ is the specific precision allocated to the respective observable $\hat{F}$, and $C_A$, $C_B$, and $C_F$ correspond to the cost of the block encoded operators for the Hamiltonian operators of monomers $A$ and $B$ as well as the observable $\hat{F}$.

\subsection{SAPT operator encoding}
\label{sec:SAPT_operator_encoding}
In this section, we present the \emph{sparse} and \emph{tensor factorization} schemes, which can be used for block encoding all of the SAPT operators in the SAPT-EVE algorithm. To unify both approaches, we use the Majorana representation, which expresses Eqs.~\eqref{eq:V}-\eqref{eq:VP} in terms of Hermitian and self-inverse operators, $\hat{\gamma}_{\textsc{p},0}/\hat{\gamma}_{\textsc{p},1}$ and $\hat{\gamma}_{\textsc{q},0}/\hat{\gamma}_{\textsc{q},1}$, defined for each monomer separately (see Appendix \ref{sec:Appendix_Operator_Encoding} for details). The electrostatic and exchange operators are written as,
\begin{align}
    \hat{V} &=  \tfrac{1}{4}\sum_{\textsc{p}\textsc{q}}v^{\textsc{p}\textsc{p}}_{\textsc{q}\textsc{q}} + \tfrac{i}{4}\sum_{\substack{\textsc{p}_1\textsc{p}_2 }}  f^{(A)}_{\textsc{p}_1\textsc{p}_2}\hat{\gamma}_{\textsc{p}_1,0}\hat{\gamma}_{\textsc{p}_2,1} \label{eq:V_majorana1} \\
    &+ \tfrac{i}{4}\sum_{\substack{\textsc{q}_1\textsc{q}_2 } } f^{(B)}_{\textsc{p}_1\textsc{p}_2} \hat{\gamma}_{\textsc{q}_1,0}\hat{\gamma}_{\textsc{q}_2,1} \nonumber \\
    &- \tfrac{1}{4}\sum_{\substack{\textbf{\textsc{p}}\textbf{\textsc{q}}  }} \text{sym}( v^{\textsc{p}_1\textsc{p}_2}_{\textsc{q}_1\textsc{q}_2})\hat{\gamma}_{\textsc{p}_1,0}\hat{\gamma}_{\textsc{p}_2,1} \hat{\gamma}_{\textsc{q}_1,0}\hat{\gamma}_{\textsc{q}_2,1}, \nonumber 
\end{align}
and
\begin{align}
    \hat{P} &= -\tfrac{1}{4}\sum_{\textsc{p}\textsc{q}}S^{\textsc{p}}_{\textsc{q}}S^{\textsc{p}}_{\textsc{q}} - \tfrac{i}{4}\sum_{\substack{\textsc{p}_1\textsc{p}_2 \\ \textsc{q}}} S^{\textsc{p}_1}_{\textsc{q}}S^{\textsc{p}_2}_{\textsc{q}} \hat{\gamma}_{\textsc{p}_1,0}\hat{\gamma}_{\textsc{p}_2,1} \label{eq:P_majorana1} \\
    &- \tfrac{i}{4}\sum_{\substack{\textsc{q}_1\textsc{q}_2 \\ \textsc{p}} }  S^{\textsc{p}}_{\textsc{q}_2}S^{\textsc{p}}_{\textsc{q}_1} \hat{\gamma}_{\textsc{q}_1,0}\hat{\gamma}_{\textsc{q}_2,1} \nonumber \\
    &+ \tfrac{1}{4}\sum_{\substack{\textbf{\textsc{p}}\textbf{\textsc{q}}  }} \text{sym}( S^{\textsc{p}_1}_{\textsc{q}_2}S^{\textsc{p}_2}_{\textsc{q}_1}) \hat{\gamma}_{\textsc{p}_1,0}\hat{\gamma}_{\textsc{p}_2,1} \hat{\gamma}_{\textsc{q}_1,0}\hat{\gamma}_{\textsc{q}_2,1},  \nonumber 
\end{align}
where we have defined $f^{(A)}_{\textsc{p}_1\textsc{p}_2} = \sum_\textsc{q} v^{\textsc{p}_1\textsc{p}_2}_{\textsc{q}\textsc{q}}$ and $f^{(B)}_{\textsc{p}_1\textsc{p}_2} = \sum_\textsc{p} v^{\textsc{p}\textsc{p}}_{\textsc{q}_1\textsc{q}_2}$. Here, the operator, $\text{sym}(\cdot)$, symmetrizes all of the tensors with respect to the monomer indices $\textsc{p}/\textsc{q}$ independently (see Appendix~\ref{sec:Appendix_Spatial_Orbitals} for details). Furthermore, the electrostatic-exchange operator is found to be composed of seven terms (ignoring the additive constant),
\begin{equation}
    \widehat{VP}_\mathrm{s} = \widehat{VP}_A + \widehat{VP}_B + \widehat{VP}_{1\mathrm{m}} + \widehat{VP}_{1\ell} + \widehat{VP}_2 + \widehat{VP}_3 + \widehat{VP}_4
    \label{eq:VP_majorana1}
\end{equation}
where
\begin{align}
    \widehat{VP}_A &= -\tfrac{i}{4}\sum_{ \textbf{\textsc{p}} } \text{sym}(\kappa_{\textsc{p}_1\textsc{p}_2}^{(A)})  \hat{\gamma}_{\textsc{p}_1,0}\hat{\gamma}_{\textsc{p}_2,1} \label{eq:VPA}\\
    &+\tfrac{1}{8} \sum_{ \substack{\textbf{\textsc{p}}  }} \text{sym}(\Lambda^{\textsc{p}_1\textsc{p}_2}_{\textsc{p}_3\textsc{p}_4}  )\hat{\gamma}_{\textsc{p}_1,0}\hat{\gamma}_{\textsc{p}_2,1} \hat{\gamma}_{\textsc{p}_3,0}\hat{\gamma}_{\textsc{p}_4,1} ,  \nonumber \\
    \widehat{VP}_B &=  -\tfrac{i}{4}\sum_{ \textbf{\textsc{q}} } \text{sym}(\kappa_{\textsc{q}_1\textsc{q}_2}^{(B)})  \hat{\gamma}_{\textsc{q}_1,0}\hat{\gamma}_{\textsc{q}_2,1} \label{eq:VPB} \\
    &+\tfrac{1}{8} \sum_{ \substack{\textbf{\textsc{q}} }} \text{sym}(\Lambda^{\textsc{q}_1\textsc{q}_2}_{\textsc{q}_3\textsc{q}_4} ) \hat{\gamma}_{\textsc{q}_1,0}\hat{\gamma}_{\textsc{q}_2,1} \hat{\gamma}_{\textsc{q}_3,0}\hat{\gamma}_{\textsc{q}_4,1}  
    ,\nonumber  \\
    \widehat{VP}_{1\mathrm{m}} &= \tfrac{1}{8}\sum_{ \substack{ \textbf{\textsc{p}},\textbf{\textsc{q}} } } \text{sym}(\Lambda^{\textsc{p}_1\textsc{p}_2}_{\textsc{q}_1\textsc{q}_2})\; \hat{\gamma}_{\textsc{p}_1,0}\hat{\gamma}_{\textsc{p}_2,1} \hat{\gamma}_{\textsc{q}_1,0}\hat{\gamma}_{\textsc{q}_2,1} 
    ,\label{eq:VP1m} \\
    \widehat{VP}_{1\ell} &= \tfrac{1}{4}\sum_{ \substack{ \textbf{\textsc{p}},\textbf{\textsc{q}} } } \text{sym}(\Lambda^{\textsc{p}_1\textsc{q}_2}_{\textsc{q}_1\textsc{p}_2})\hat{\gamma}_{\textsc{p}_1,0}\hat{\gamma}_{\textsc{p}_2,1} \hat{\gamma}_{\textsc{q}_1,0}\hat{\gamma}_{\textsc{q}_2,1}
    ,\label{eq:VP1l} \\
    \widehat{VP}_{2} &= \tfrac{i}{8}\sum_{ \substack{ \textbf{\textsc{p}},\textbf{\textsc{q}} } } \text{sym}(\Lambda^{\textsc{p}_1\textsc{p}_2}_{\textsc{q}_1\textsc{p}_4} S^{\textsc{p}_3}_{\textsc{q}_2} )
    \Big[ \hat{\gamma}_{\textsc{p}_1,0}\hat{\gamma}_{\textsc{p}_2,1} \hat{\gamma}_{\textsc{p}_3,0}\hat{\gamma}_{\textsc{p}_4,1} \nonumber \\
    &\;\;\;\;\; \hat{\gamma}_{\textsc{q}_1,0}\hat{\gamma}_{\textsc{q}_2,1}\Big]   ,\label{eq:VP2} \\
    \widehat{VP}_{3} &= \tfrac{i}{8}\sum_{ \substack{ \textbf{\textsc{p}},\textbf{\textsc{q}} } } \text{sym}(\Lambda^{\textsc{p}_1\textsc{q}_4}_{\textsc{q}_1\textsc{q}_2} S^{\textsc{q}_3}_{\textsc{p}_2})\Big[ \hat{\gamma}_{\textsc{p}_1,0}\hat{\gamma}_{\textsc{p}_2,1} \nonumber  \\
    &\;\;\;\;\;\hat{\gamma}_{\textsc{q}_1,0}\hat{\gamma}_{\textsc{q}_2,1} \hat{\gamma}_{\textsc{q}_3,0}\hat{\gamma}_{\textsc{q}_4,1}\Big] 
    ,\label{eq:VP3} \\
    \widehat{VP}_4&= \tfrac{1}{2}(\hat{V}\hat{P} + \hat{P}\hat{V}).
    \label{eq:VP4}
\end{align}
In this representation, we find two independent monomer operators given by $\widehat{VP}_A$ and $\widehat{VP}_B$ consisting of both one-body and two-body terms, as well as five intermonomer operators: $\widehat{VP}_{1\mathrm{m}}$, $\widehat{VP}_{1\ell}$, $\widehat{VP}_2$, $\widehat{VP}_3$, and $\widehat{VP}_4$. Here, $\widehat{VP}_{1\mathrm{m}}$ and $\widehat{VP}_{1\ell}$ correspond to spin-mixed and spin-locked contributions of the same order. Additionally, we have defined $\kappa_{\textsc{p}_1\textsc{p}_2}^{(A)}$/$\kappa_{\textsc{q}_1\textsc{q}_2}^{(B)}$ as two-index tensors that appear within the one-body terms of monomers A/B as well as the general four-index tensor, $\Lambda^{\textsc{j}_1\textsc{j}_2}_{\textsc{j}_3\textsc{j}_4}$ where $\textsc{j}\in \{\textsc{p},\textsc{q}\}$, which is assumed to be symmetrized in these equations and represents a dressed version of the intermolecular tensor appearing in the electrostatic-exchange operator, Eq.~\eqref{eq:VP}. Due to the verbosity of the equations, all of the tensor elements are defined explicitly in Appendix \ref{sec:Appendix_Operator_Encoding}. It is important to note that the form of the SAPT operators in Eqs.~\eqref{eq:V_majorana1}-\eqref{eq:VP_majorana1} is the same in both the full space and active space pictures apart from the definition of tensor elements themselves. This property emerges naturally in the Majorana representation, but it is not clearly evident otherwise. 

To block encode the SAPT operators given by Eqs.~\eqref{eq:V_majorana1}-\eqref{eq:VP_majorana1} in the \emph{sparse} representation, we use the Jordan-Wigner transformation on each of the two monomers written as,
\begin{align}
    \hat{\gamma}_{\textsc{p},0} = \hat{X}_{\textsc{p},}\hat{Z}_{\textsc{p}-1,}\cdots \hat{Z}_{0,}, \\
    \hat{\gamma}_{\textsc{p},1} = \hat{Y}_{\textsc{p},}\hat{Z}_{\textsc{p}-1,}\cdots \hat{Z}_{0,},
\end{align}
for monomer A and,
\begin{align}
    \hat{\gamma}_{\textsc{q},0} = \hat{X}_{\textsc{q},}\hat{Z}_{\textsc{q}-1,}\cdots \hat{Z}_{0,}, \\
    \hat{\gamma}_{\textsc{q},1} = \hat{Y}_{\textsc{q},}\hat{Z}_{\textsc{q}-1,}\cdots \hat{Z}_{0,},
\end{align}
for monomer B. The $\mathsf{SELECT}$ circuits required to implement these Pauli strings are presented explicitly in Ref.~\cite{babbush2018encoding}. The \emph{sparse} SAPT-EVE algorithm uses the Jordan-Wigner mapping of the second-quantized operators above and proceeds to load the non-zero tensor coefficients found in Eqs.~\eqref{eq:V_majorana1}-\eqref{eq:VP_majorana1}, thereby applying the so-called sparse block encoding method introduced by \citet{Berry_2019}. We use the sparse method to benchmark the tensor factorization procedure in the Numerical Results section, Sec.~\ref{sec:Numerical_results}. In Appendix \ref{sec:Appendix_Operator_Encoding}, we present the full expressions for the $\ell_1$ norms in the \emph{sparse} representation accounting for additional permutational symmetries of certain tensors which help reduce the $\ell_1$ norm further, as pointed out in Ref.~\cite{koridon2021orbital}. 

We now consider the tensor factorization procedure required for decomposing the four-index tensors appearing in Eq.~\eqref{eq:V_majorana1} as well as Eqs.~\eqref{eq:VPA}-\eqref{eq:VP3}. In the following, we will work in the spatial orbital basis with lower-case indices $p/q$ denoting the spatial molecular orbitals of monomer A/B respectively. Our approach is analogous to the low-rank double factorization technique outlined in previous works \cite{VonBurg2021, motta2021low}, but contains small differences to account for the unique permutational symmetries of each tensor. The tensor factorization procedure consists of a two-step process. In the first step, the general SAPT tensor, $\Lambda^{j_1j_2}_{j_3j_4}$, is factorized as,
\begin{equation}
    \Lambda^{j_1j_2}_{j_3j_4} = 
    \begin{cases}
        \sum_t s_{t}^{(z)} [\mathbf{u}^{(z)}_t]_{j_1j_2} [\mathbf{u}^{(z)}_t]_{j_3j_4} & \text{if}\;\Lambda^{j_1j_2}_{j_3j_4} = \Lambda_{j_1j_2}^{j_3j_4}, \\
        \sum_t s_{t}^{(z)} [\mathbf{u}^{(z)}_t]_{j_1j_2} [\mathbf{w}^{(z)}_t]_{j_3j_4} & \text{otherwise}, 
    \end{cases}
    \label{first_factorization}
\end{equation}
where $s^{(z)}_t$ correspond to eigenvalues/singular values and $\mathbf{u}^{(z)}_t/\mathbf{w}^{(z)}_t$ correspond to eigenvectors or singular vectors depending on whether the symmetry condition is satisfied. Here, we use the label $z$ to represent all of the unique tensor sub-blocks that appear in first-order SAPT theory found in Eqs.~\eqref{eq:V_majorana1} and \eqref{eq:VPA}-\eqref{eq:VP3}. The factorization procedure is based on grouping the $j_1j_2/j_3j_4$ indices into two indices $j/j'$ where $j \in \{p,q\}$. This implies that the four-index tensor $\Lambda^{j_1j_2}_{j_3j_4}$ is mapped to a matrix with indices $j$ and $j'$. As a result, the condition, $\Lambda^{j_1j_2}_{j_3j_4} = \Lambda_{j_1j_2}^{j_3j_4}$, in the first line of Eq.~\eqref{first_factorization} applies when the matrix is symmetric. The tensors appearing in Eqs.~\eqref{eq:VPA}, \eqref{eq:VPB} and \eqref{eq:VP1l} will use a standard eigendecomposition in the first factorization step, while those in Eqs.~\eqref{eq:V_majorana1}, \eqref{eq:VP1m}, \eqref{eq:VP2}, \eqref{eq:VP3} will use a singular value decomposition (SVD). All seven tensors will have a different rank, which we label by $N_1^{(z)}$.

In the second step, each column vector $\mathbf{u}^{(z)}_t/\mathbf{w}^{(z)}_t$ is reshaped into a  matrix and the following factorization procedure is performed:
\begin{equation}
    [\mathbf{u}^{(z)}_t]_{j_1j_2}\! = \!\!
    \begin{cases}
        \!\sum\limits_{k} \alpha_{kt}^{(z)} U^{(z)}_{t,kj_1} U^{(z)}_{t,kj_2} & \!\!\text{if}\; [\mathbf{u}^{\!(z)}_{t}]_{j_1j_2} \!\!=\! [\mathbf{u}^{\!(z)}_{t}]_{j_2j_1},\\
        \!\sum\limits_{k} \beta_{kt}^{(z)} U^{(z)}_{t,kj_1} V^{(z)}_{t,kj_2} & \!\!\text{otherwise}.
    \end{cases}
    \label{second_factorization}
\end{equation}
The second factorization procedure is performed for each index $t$ from the first factorization step and corresponds to an eigendecomposition/SVD with eigenvalues/singular values given by, $\alpha_{kt}^{(z)}/\beta^{(z)}_{kt}$, labeled by the index, $k$. For a particular $t$, the rank of the second factorization will be given by $N_2^{t}$. Here, $U^{(z)}_{t,kj_1}$ and $V^{(z)}_{t,kj_2}$ may be interpreted as three-index tensors where we use the convention that $U$ corresponds to monomer A and $V$ is associated with monomer B. For a particular $t$, the two-index tensors $[U^{(z)}_{t}]_{k,j_1}$ and $[V^{(z)}_{t}]_{k,j_2}$ are equal to unitary matrices that rotate the molecular orbitals of each monomer independently. Eqs.~\eqref{first_factorization}-\eqref{second_factorization} completely define the tensor factorization process that is required for all of the SAPT tensors. We emphasize that the tensor factorization procedure outlined above is not unique. However, we found that it provided the smallest $\ell_1$ norms consistently for all benchmark cases. 

In addition to the four-index tensor decomposition outlined above, we also performed a singular value decomposition of the rectangular overlap matrix $S^p_q$,
\begin{equation}
    S^p_q = \sum_{n}^{N_S} s_n U^{(s)}_{pn}V^{(s)}_{qn} \, ,
    \label{eq:S}
\end{equation}
where $s_n$ denotes the singular values, while $U^{(s)}_{pn}$ and $V^{(s)}_{qn}$ denote the unitary orbital transformation matrices for monomer A and B, respectively. The rank $N_S$ of this decomposition will be upper bounded by the number of spatial orbitals of the smaller monomer, $N_S \leq \text{min}(N_A,N_B)$. This dependence provides several advantages when one of the monomers is much smaller than the other. 

To finalize the tensor factorization representation for all of the SAPT operators, a final decomposition is required for the tensors appearing in the one-body terms in Eqs.~\eqref{eq:V_majorana1} and \eqref{eq:VPA}-\eqref{eq:VPB}. In all cases, the two-index tensors $f^{(A)}_{p_1p_2}$, $f^{(B)}_{q_1q_2}$, $\kappa^{(A)}_{p_1p_2}$ and $\kappa^{(B)}_{q_1q_2}$, will be symmetric therefore a standard eigendecomposition can be performed,
\begin{align}
    f^{(A)}_{p_1p_2} &= \sum_k s_k^{(A)} U^{(f)}_{p_1k} U^{(f)}_{p_2k}, \\
    f^{(B)}_{q_1q_2} &= \sum_l s_l^{(B)} V^{(f)}_{q_1l} V^{(f)}_{q_2l}, \\
    \kappa^{(A)}_{p_1p_2} &= \sum_k \tilde{s}_k^{(A)} {U}^{(\kappa)}_{p_1k} {U}^{(\kappa)}_{p_2k}, \\
    \kappa^{(B)}_{q_1q_2} &= \sum_l \tilde{s}_l^{(B)} {V}^{(\kappa)}_{q_1l} {V}^{(\kappa)}_{q_2l},
    \label{one_body_factorization}
\end{align}
where $s_k^{(A)}$, $s_l^{(B)}$, $\tilde{s}_k^{(A)}$, $\tilde{s}_l^{(B)}$ denote the eigenvalues while $U^{(f)}_{p_ik}/{U}^{(\kappa)}_{p_ik}$ denote the eigenvectors for monomer A, and $V^{(f)}_{q_il}/{V}^{(\kappa)}_{q_il}$ denote the eigenvectors of monomer B. In matrix form, the eigenvectors correspond to the unitary orbital transformation matrices required for monomer A and B respectively. 

Based on the tensor factorization procedure outlined in Eqs.~\eqref{first_factorization}-\eqref{one_body_factorization}, the appropriate block encoded operators can be constructed by carefully loading the eigenvalues $s_k^{(A)}$, $s_l^{(B)}$, $\tilde{s}_k^{(A)}$, $\tilde{s}_l^{(B)}$ as well as the singular coefficients $s^{(z)}_t$, $\alpha_{kt}^{(z)}/\beta_{kt}^{(z)}$ and $s_n$ using  the $\mathsf{PREPARE}$ and $\mathsf{SELECT}$ circuits outlined in {Appendix~\ref{sec:SAPT_block_encoding}}. After some additional manipulations, we found the following expressions for the $\ell_1$ norms of all of the SAPT operators in the tensor factorization (tf) representation,
 \begin{align}
    \lambda_V^{(\mathrm{tf})} &= \!\sum_{k} |s_k^{(A)}| + \!\sum_l |s_l^{(B)}| + \!\sum_{tkl} |s_t^{(v)} \alpha_{kt}^{(A_{v})}\alpha_{lt}^{(B_{v})}|, \\
    \lambda_{P}^{(\mathrm{tf})} &= \frac{1}{2}\lambda_s^2 + \sum_n |s_n|^2, \\
    \lambda_{\mathrm{VP}_s}^{(\mathrm{tf})} &=  \tfrac{1}{2} \sum_{k} |\tilde{s}_k^{(A)}| + \tfrac{1}{2}\sum_l |\tilde{s}_l^{(B)}| \\
    &+ \tfrac{1}{4}\sum_{tkl} ( |s_t^{(A_2)} \alpha_{kt}^{(A_2)}\alpha_{lt}^{(A_2)}| + |s_t^{(B_2)} \alpha_{kt}^{(B_2)}\alpha_{lt}^{(B_2)}|) \nonumber\\
    &+ \tfrac{1}{2}\sum_{tkl} (|s_t^{(1\mathrm{m})} \alpha_{kt}^{(A_{1\mathrm{m}})}\alpha_{lt}^{(B_{1\mathrm{m}})}| + |s_t^{(1\ell)} \beta_{kt}^{(1\ell)}\beta_{lt}^{(1\ell)}|) \nonumber\\
    &+ \tfrac{\lambda_s}{2}\sum_{tkl} ( |s_t^{(2)} \alpha_{kt}^{(2)}\beta_{lt}^{(2)}| + |s_t^{(3)} \alpha_{kt}^{(3)}\beta_{lt}^{(3)}| ) \nonumber \\
    &+ \lambda_P^{(\mathrm{tf})} \sum_{tkl} |s_t^{(v)} \alpha_{kt}^{(A_{v})}\alpha_{lt}^{(B_{v})}|  , \nonumber
    \label{eq:lam_VP}
\end{align}
where we have defined $\lambda_s = \sum_n |s_n|$ for notational clarity. We have also derived the corresponding $\ell_1$ norms for the active space picture in {Appendix~\ref{sec:Appendix_Operator_Encoding}}. The complete block encoding circuits for $\hat{V}$ and $\hat{P}$ can be found in Appendix~\ref{sec:SAPT_block_encoding}. For illustration purposes, we have also derived the block encoding circuits for the intermonomer operators $\widehat{VP}_{1\mathrm{m}}$, $\widehat{VP}_{1\ell}$ and $\widehat{VP}_{4}$ of the electrostatic-exchange operator, $\widehat{VP}_\mathrm{s}$, in Eq.~\eqref{eq:VP_majorana1}. The block encoding circuits for $\widehat{VP}_A$ and $\widehat{VP}_B$ are equivalent to the conventional second-quantized, double-factorized Hamiltonian and may be found in Refs.~\cite{VonBurg2021,Lee2021}. It is important to emphasize that $\widehat{VP}_{4} $ represents the dominant contribution to the total block encoding compilation cost of the electrostatic-exchange operator. In Appendix~\ref{subsec:PBEO}, we exploit the fact that $\widehat{VP}_4$ may be written as the product of two operators, $\hat{V}$ and $\hat{P}$, in order to reduce the overall compilation cost. In this regard, while we found that all of the terms in the electrostatic-exchange operator have an upper bound asymptotic Toffoli gate scaling of $\mathcal{O}(N_A^3)$ with respect to the number of spatial orbitals $N_A$ (assuming $N_A \approx N_B $), $\widehat{VP}_4$ is accompanied by an additive $\mathcal{O}(N_A^2)$ contribution that makes it quadratically larger than any other term. Our resource estimates for the electrostatic-exchange operator in Section \ref{sec:Numerical_results} take into account $\widehat{VP}_A$, $\widehat{VP}_B$, $\widehat{VP}_{1\mathrm{m}}$,  $\widehat{VP}_{1\ell}$ and $\widehat{VP}_4$ (additional details can be found in Appendix~\ref{sec:Appendix_Operator_Encoding}) which allows for an adequate assessment of the total resource cost for the SAPT-EVE algorithm.

\begin{figure}[tb]
    \centering
    \includegraphics[width=0.95\linewidth]{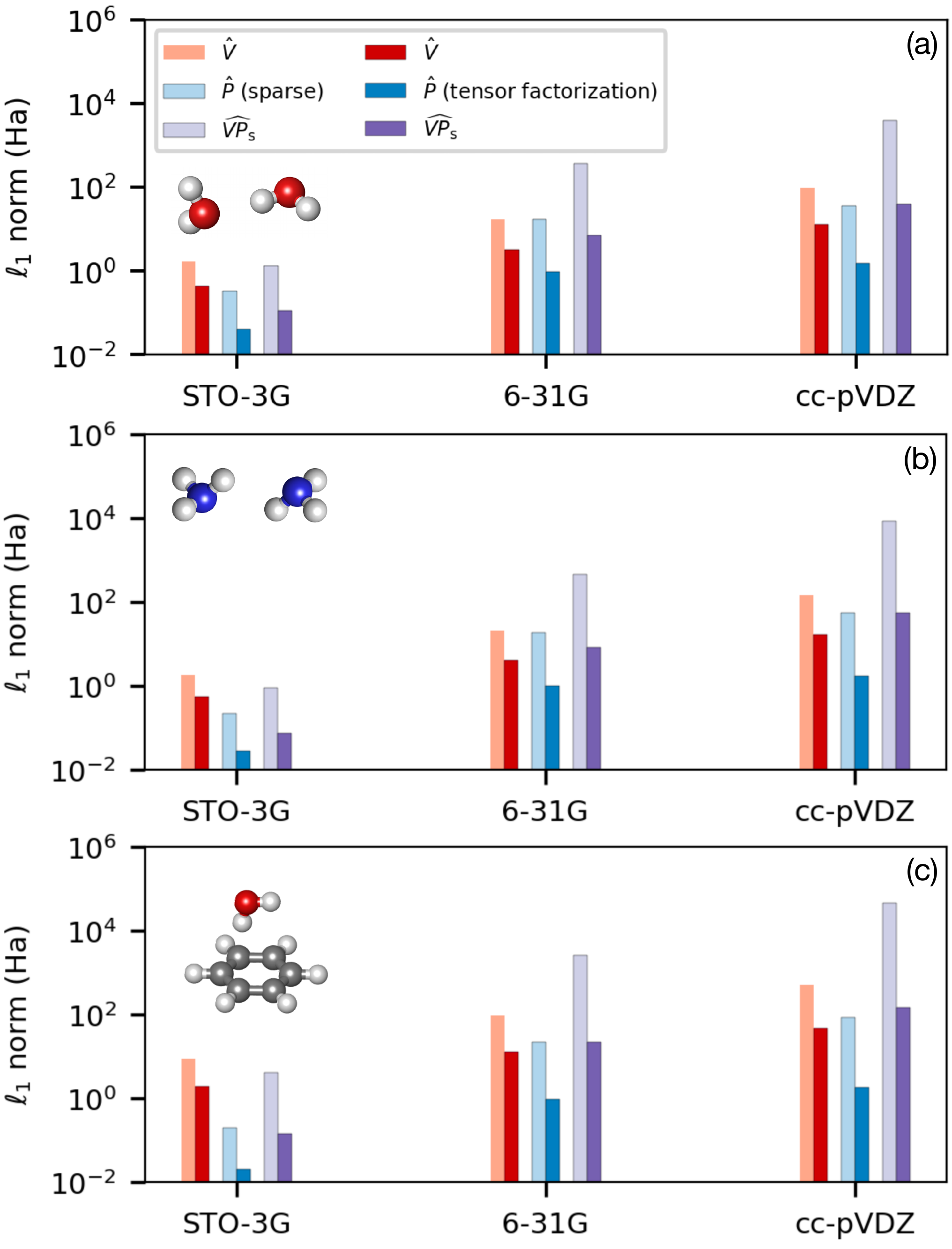}
    \caption{Basis set dependence of the $\ell_1$ norm for the sparse and tensor factorized SAPT observables for three different molecular systems (top: water dimer, middle: ammonia dimer and bottom: benzene-water dimer).}
    \label{fig:basis_set_dependance}
\end{figure}

\begin{figure*}
\subfloat[\label{fig:sub1}]{%
  \includegraphics[width=.49\linewidth]{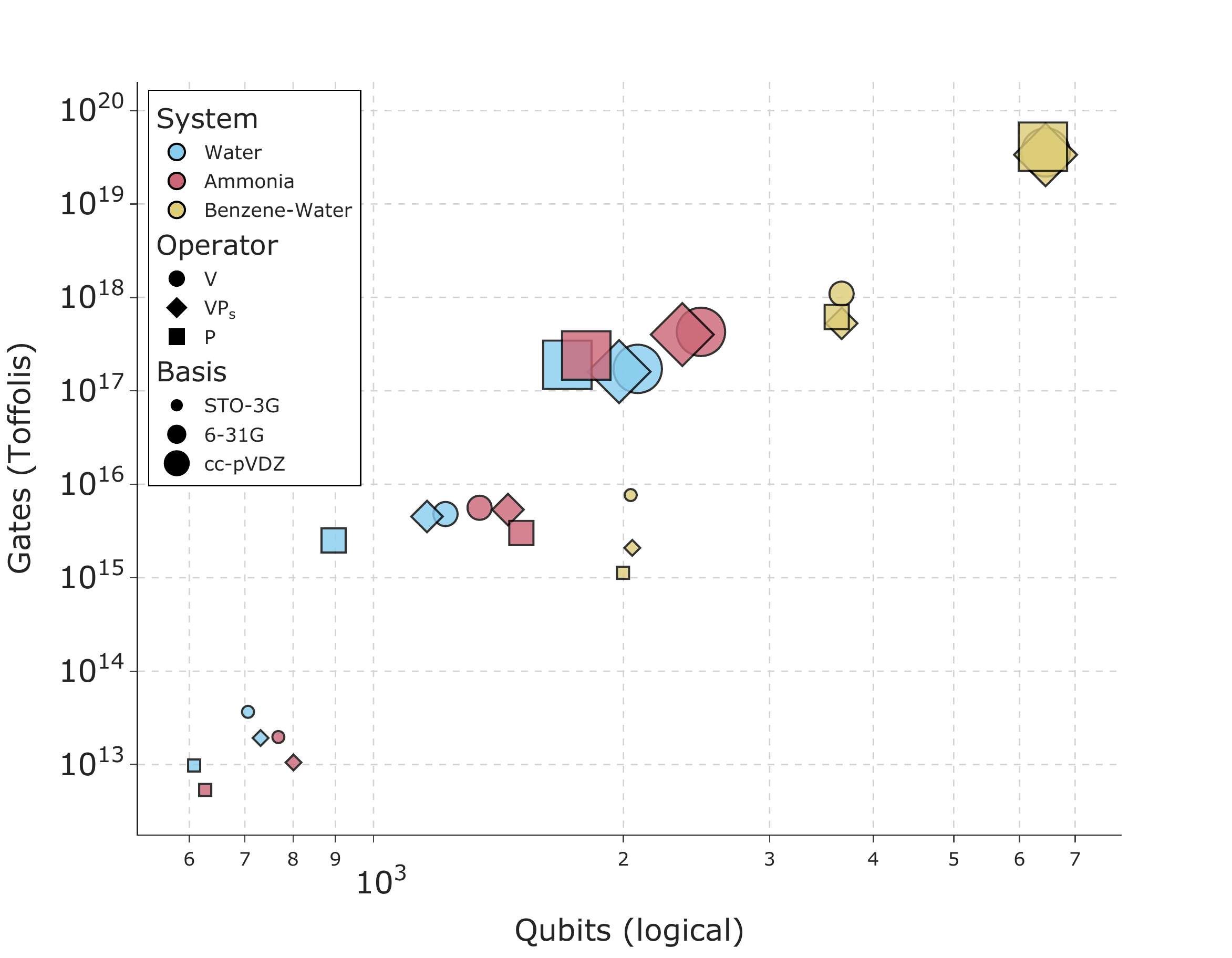}%
}\hfill
\subfloat[\label{fig:sub2}]{%
  \includegraphics[width=.49\linewidth]{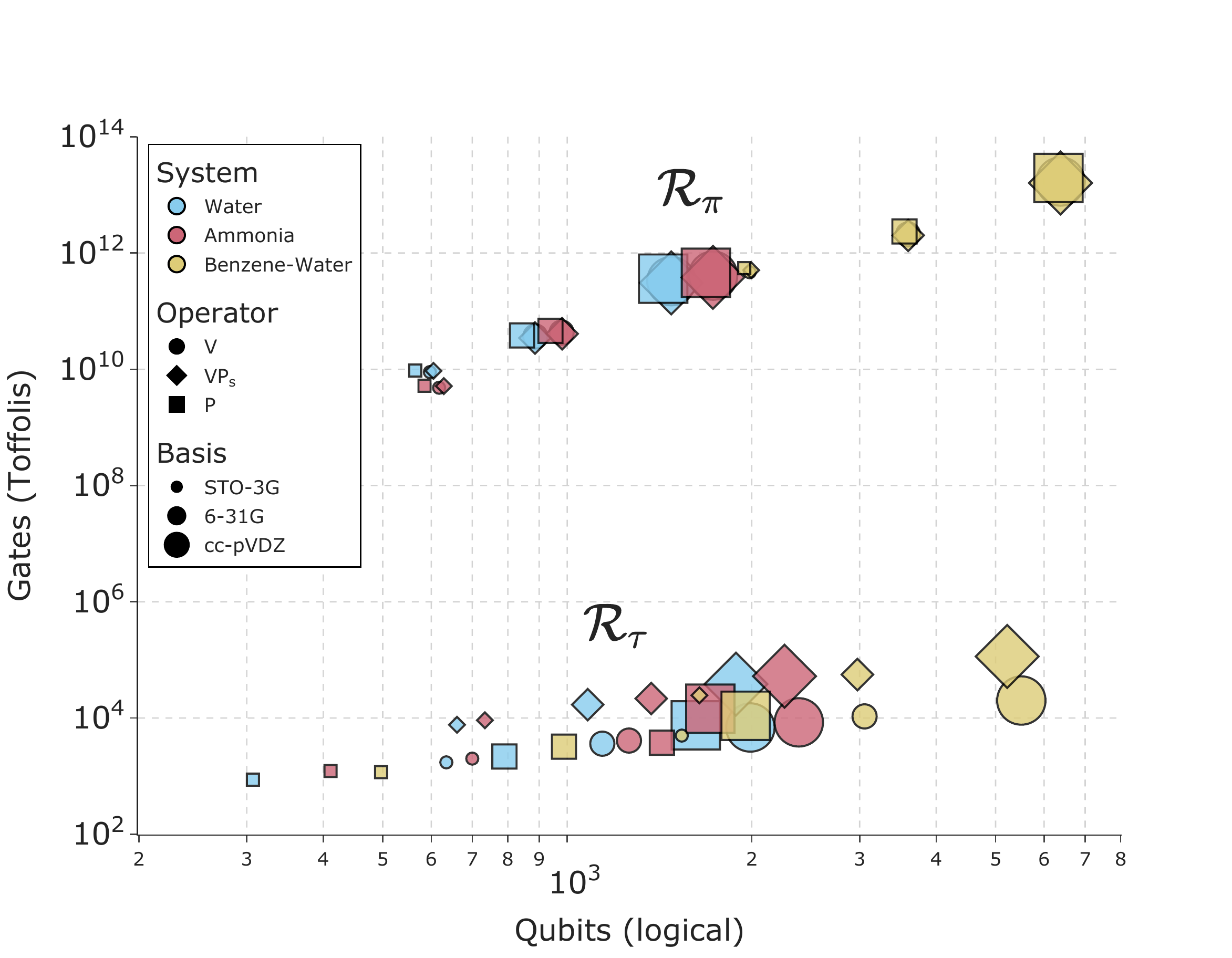}%
}
\caption{Resource estimation for benchmark molecules (separated by the color of the symbols) in the all-electron (no active space) picture. Different basis sets ($\{$STO-3G, 6-31G, cc-pVDZ $\}$) are encoded by the hue of the symbols while the different operators ($\{\hat{V},\hat{P},\widehat{VP}_\mathrm{s}\}$) are encoded by the symbol shape. (a) Toffoli and logical qubit cost for total SAPT-EVE algorithm for our molecule test set. The block encoding cost included in the total cost of estimating the electrostatic-exchange operator, $\widehat{VP}_\mathrm{s}$, include the $\widehat{VP}_A$, $\widehat{VP}_B$, $\widehat{VP}_{1\mathrm{m}}$, $\widehat{VP}_{1\ell}$, and $\widehat{VP}_4$ contributions as discussed in the main text. (b) Toffoli and qubit cost comparison between the eigenstate reflection oracle, $\mathcal{R}_\pi$, and the observable block encoding oracle, $\mathcal{R}_\tau$. Even though the cost of both reflections scales is similar in terms of number of qubits, it is clear from the plot that the Toffoli gate cost of $\mathcal{R}_\pi$ is orders of magnitude larger than $\mathcal{R}_\tau$ and will dominate the total cost of the algorithm. }
\label{fig:resource_count_estimation}
\end{figure*}

\section{Numerical Results}
\label{sec:Numerical_results}

\subsection{Benchmark results for small molecular systems}
\label{sec:small_mol}
We first investigate the advantages of the tensor factorization procedure over the standard sparse method. The overall performance of any scheme is primarily determined by the $\ell_1$ norm $\lambda_F$ of the observable $F$. In Fig.~\ref{fig:basis_set_dependance}, we compare the $\ell_1$ norms of the sparse and tensor factorization schemes as a function of basis set size for three different molecular systems: an ammonia dimer, a water dimer, and a benzene-water complex (see Appendix~\ref{sec:Appendix_Small_Benchmark} for details of the molecular systems). 

We observe that the $\ell_1$ norm of the exchange operator is often the smallest while $\lambda_\mathrm{VP}$ is often the largest in the limit of large basis sets, regardless of the encoding scheme. The difference between sparse encoding and double factorization $\ell_1$ norm is small at smaller basis set sizes. However, for large basis sets, such as cc-pVDZ, the tensor factorization encoding scheme greatly reduces the $\ell_1$ norm ranging from two to three orders of magnitude. 

While the $\ell_1$ norm provides a multiplicative factor that is fundamental to the total runtime, the total cost of the algorithm is also dependent on the eigenstate reflection circuit $\mathcal{R}_\pi$ as well as the block encoding cost of each observable. The eigenstate reflection cost, in turn, depends on the $\ell_1$ norm of each monomer Hamiltonian required for block encoding and is inversely proportional to the spectral gap between the ground and first excited state in each of the monomers, as detailed in Eq.~\eqref{eq:scaling}. In Fig.~\ref{fig:resource_count_estimation}, we present the overall logical qubit and Toffoli gate costs for the three molecular systems considered above. The left panel of Fig.~\ref{fig:resource_count_estimation} presents the total cost of the SAPT-EVE algorithm for the three molecular systems. The right panel of Fig.~\ref{fig:resource_count_estimation} presents a breakdown of the costs of the eigenstate preparation reflection oracle, $\mathcal{R}_{\pi}$, as well as SAPT block encoding oracle, $\mathcal{R}_\tau$ respectively. 

In addition to using the tensor-factorization techniques outlined above, our resource estimates used various optimization schemes for reducing the Toffoli count using efficient quantum adders and data-loading QROM implementation techniques outlined in previous works \cite{babbush2018encoding, low2018trading}. In particular, we optimize the total runtime of the algorithm in order to determine the total first-order interaction energy, Eq.\eqref{Interaction_Energy}, to chemical accuracy ($\epsilon = 0.0016$ Hartree). This optimization was based on $\ell_1$ norm upper bounds for the expectation values resulting in conservative and pessimistic estimates, which could be improved with alternative approaches as discussed in Appendix~\ref{sec:Appendix_SAPT_EVE}.

Based on the data from Fig.~\ref{fig:resource_count_estimation}, we find that the resource cost of each dimer system is highly dependent on the atomic composition and corresponding basis set size, which, in turn, affects the total number of qubits and $\ell_1$ norm of the Hamiltonian and observable. One of the main conclusions of this work is readily seen by comparing the Toffoli gate count for the total algorithm and the eigenstate reflection, $\mathcal{R}_\pi$. We observe that the dominant cost of the total algorithm is due to $\mathcal{R}_\pi$, highlighting one of the key bottlenecks for observable estimation and emphasizing the subroutine that requires further improvement in future work. The data from this figure also suggests that brute-force basis set extrapolation, which amounts to increasing the basis set size and number of spin-orbitals on the quantum computer, does not scale well even for small molecular benchmark systems. For instance, the benzene-water system in the cc-pVDZ basis will require over 6000 logical qubits and Toffoli gate counts on the order of $10^{19}$. Both of these requirements are substantial and highlight the need for alternative strategies. For this reason, one of the major contributions of our work is the derivation of the first-order active space SAPT operators, which drastically reduces resource costs. In Appendix~\ref{sec:Appendix_Small_Benchmark}, we present the active space resource costs for the small molecule benchmark set. The following section presents the active space resource cost analysis for a benchmark test system relevant to drug design. 

\subsection{Benchmark result for drug design: heme-artemisinin}
\label{sec:heme_art}
\begin{figure}[t]
    \includegraphics[width=0.4\textwidth]{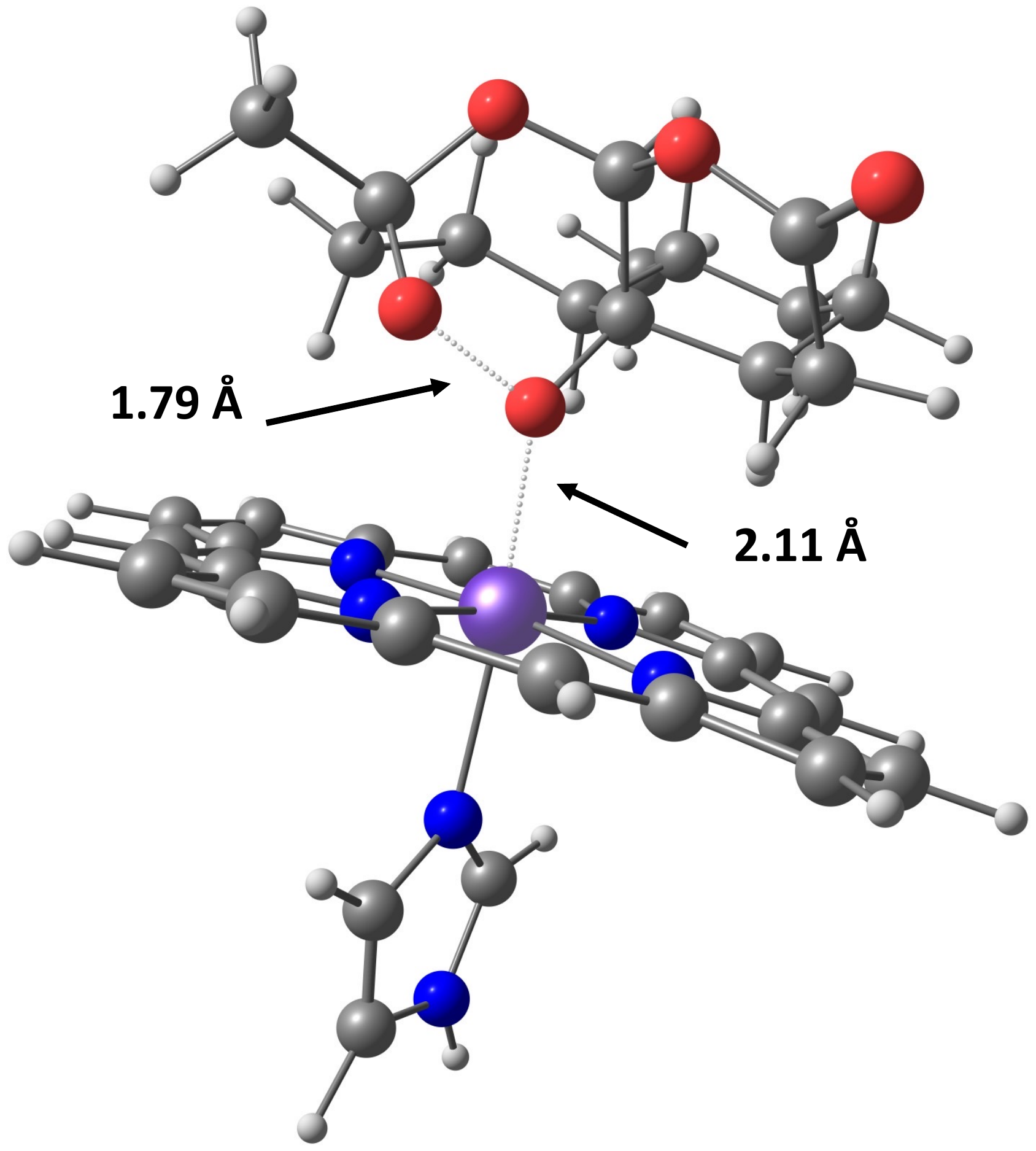}
    \makeatletter\long\def\@ifdim#1#2#3{#2}\makeatother
    \caption{Transition state structure of the binding of artemisinin to the Fe at the center of the heme complex; the process is concerted with an electron transfer and cleavage of the peroxo bond (using the $\omega$B97X-D functional see computational methods in Appendix~\ref{sec:Appendix_Benchmark_Artemisinin} for details). }
    \label{fig:ts_structure}
\end{figure}

We now consider a more interesting application for the quantum computer that approaches the limit of current classical computing consisting of a heme interacting with an artemisinin molecule at a separation distance of 2.11~\AA, as shown in Fig.~\ref{fig:ts_structure}. In the following, we discuss why this benchmark system is interesting for drug design. 

Artemisinin is a popular plant-derived anti-malaria drug. While the exact mechanisms of action of the artemisinin drug are not completely understood~\cite{posner2004knowledge,o2010molecular}, the interaction with the iron center of heme is known to play an integral role~\cite{posner2004knowledge}. An established mechanism of action suggests that artemisinin gets reduced by the Fe(II) center upon heme binding and is concerted with the cleavage of the peroxide bond resulting in an oxygen-centered radical. It is postulated that a rearrangement yields carbon-centered radicals (see~\cite{mercer2011role} for a detailed mechanism), which have been observed experimentally by spin trapping \cite{wu1998unified}. To further optimize synthetic analogs to this drug, a better understanding of the mechanism of action would be key. To this end, the activation of artemisinin by O-O bond cleavage has been studied using density functional theory (DFT) ~\cite{taranto2006dft, moles2006modeling, moles2008theoretical}. However, none of these studies included a heme model complex explicitly. Several other studies have investigated the reactivity of heme complexes using DFT~\cite{hirao2014applications, relt2017electronic, derrick2022templating}, as well as more advanced quantum computational methods~\cite{altun2019local, lee2020utilizing, tarrago2021experimental, li2018understanding}, including a resource estimation cost for fault-tolerant quantum computing~\cite{goings}. Several studies~\cite{goings, li2018understanding} have recommended using a large active space with more than 40 orbitals to include the key Fe-$d$ and heme-$\pi$ orbitals~\cite{goings}. To study its interaction with artemisinin, the ideal next step would be to simulate the interaction of both systems together. Unfortunately, this requires even larger active spaces (see below), making it intractable for many classical methods. As a result, this system is an appealing target for fault-tolerant quantum computing in a pharmaceutical context. Here, we focus on the first step of the decomposition mechanism, the binding, which is accompanied by an electron transfer and the cleavage of the peroxo bond, see Fig.~\ref{fig:art_chemdraw} in Appendix \ref{sec:Appendix_Benchmark_Artemisinin}. 

In this work, we located the transition state for this reaction step and the corresponding molecular geometry is the core example in this study (see Fig.~\ref{fig:ts_structure}). The heme is treated as a neutral-charge open-shell system with two unpaired electrons. In contrast, the artemisinin system is treated as a closed-shell system. However, we emphasize that our SAPT derivation remains applicable in the most general case of two open-shell monomers. In Table~\ref{tbl:heme-artemisin-benchmark}, we estimate the FTQC resources for the three SAPT observables in the (42e, 43o) active space for heme and (48e, 40o) active space for artemisinin. The orbital-optimized active spaces were determined via DMRG calculations (additional details can be found in Appendix~\ref{sec:Appendix_Benchmark_Artemisinin}). For comparison, the all-electron (full space) picture would require (222e, 529o) for heme and (144e, 366o) for artemisinin in the cc-pVDZ basis. 

In Tab.~\ref{tbl:heme-artemisin-benchmark}, we observe larger Toffoli gate count estimates in the total SAPT-EVE algorithm for the heme-artemisinin benchmark compared to the small molecule benchmark set ($10^{20}$ compared to $10^{19}$), which we found was largely due to the smaller spectral gaps observed in heme and artemisinin. As pointed out in Ref.~\cite{koridon2021orbital}, the $\ell_1$ norm of the Hamiltonian operators (and SAPT observables) also seem to acquire a slightly larger power dependence with respect to the number of active space orbitals, which also contributes to the large resource count estimate presented in the table. This is also observed in Tab.~\ref{fig:Resource_estimate_data} and Fig.~\ref{fig:trend_line} in Appendix~\ref{sec:data_code}. While these estimates show that the resource cost is substantially reduced compared to the non-active space methodology, the eigenstate reflection remains a fundamental bottleneck for the total algorithm. This may become detrimental for systems with small energy gaps and further work is required to determine whether alternative approaches for the eigenstate reflection subroutine could be used. For more details on the algorithm's performance, we also present call graphs in Figs.~\ref{fig:call_graph1}-\ref{fig:call_graph3} in Appendix ~\ref{sec:data_code}. Tab.~\ref{fig:Resource_estimate_data} also provides a system parameter and resource cost breakdown of every dimer system considered in this work for completeness.

To finalize this section, we also discuss the cost of the supermolecular approach which is used to estimate the interaction energy, $E_\mathrm{int}$. For the quantum resource estimation, we used the double-factorized representation of the monomer and dimer Hamiltonians presented in Refs.~\cite{VonBurg2021,Lee2021}. In the supermolecular approach, three separate quantum phase estimation runs are performed. The full resource cost is provided in Tab.~\ref{tab:supermolecular} in Appendix \ref{sec:supermolecular_data}, where we found that the total number of Toffoli gates required for each run was typically on the order of $10^{10}$ with a total qubit cost on the order of $10^3$. While the supermolecular approach provides an accurate estimation of the interaction energy, it does not provide a decomposition with respect to the electrostatic or exchange energy contributions which can be used to improve the understanding of binding mechanisms in drug molecules. Nevertheless, this highlights the chasm between the two methodologies as well as the broader difference between eigenvalue/energy estimation and observable estimation in the context of fault-tolerant quantum computing.

\begin{table*}
\caption{Resource estimates for SAPT-EVE algorithm for heme-artemisinin active space system. $N_A/N_B$ denote the number of spatial orbitals of monomers A/B. $\lambda_A/\lambda_B$ denote $\ell_1$ norms for the double factorized Hamiltonians,  $H^{(\mathrm{df})}_A/H^{(\mathrm{df})}_B$, in units of Hartree. $\Delta_A/\Delta_B$ denotes the spectral gap for monomers $A/B$ determined through DMRG calculations in units of Hartree. $\lambda_F$ denotes the $\ell_1$ norm for observable, $\hat{F}$. $\lambda_V/\lambda_{VP}$ are in units of Hartree, while $\lambda_P$ is unitless.  $\epsilon_F$ is the allocated precision for observable $\hat{F}$. $\Lambda_F = \lambda_F/\epsilon_F$. The final cost is given in terms of the total number of Toffoli gates and qubits required to implement the SAPT-EVE algorithm to reach chemical accuracy in the first-order interaction energy, Eq.~\ref{Interaction_Energy}. The block encoding cost for the electrostatic-exchange operator, $\widehat{VP}_\mathrm{s}$, include the $\widehat{VP}_A$, $\widehat{VP}_B$, $\widehat{VP}_{1\mathrm{m}}$, $\widehat{VP}_{1\ell}$, and $\widehat{VP}_4$ contributions as discussed in the main text. }
\label{tbl:heme-artemisin-benchmark}
\begin{tabular}{rrrrrr|@{\hskip 1mm}crcc@{\hskip 4mm}ccc@{\hskip 4mm}rrr}
\hline\hline
 \multicolumn{6}{c}{System Parameters} & \multicolumn{4}{c}{Observable Parameters} & \multicolumn{3}{c}{Gates} & \multicolumn{3}{c}{Qubits} \\
 $N_A$ & $N_B$ & $\lambda_{A}$ & $\lambda_{B}$ & $\Delta_{A}$ & $\Delta_{B}$ & $\hat{F}$ & $\lambda_F$ & $\!\!\!\!\varepsilon_F$ & $\Lambda_F$ & Total & $\mathcal{R}_\pi$ & $\mathcal{R}_\tau$ & Total & $\mathcal{R}_\pi$ & $\mathcal{R}_\tau$ \\
\midrule
\multirow[c]{3}{*}{43} & \multirow[c]{3}{*}{40} & \multirow[c]{3}{*}{232.2} & \multirow[c]{3}{*}{361.8} & \multirow[c]{3}{*}{0.0069} & \multirow[c]{3}{*}{0.1212}& $\hat{V}$ & 65.5 & $7.29 \times 10^{-5}$ & $8.99 \times 10^{5}$ & $9.74 \times 10^{19}$ & $1.16 \times 10^{13}$ & $1.34 \times 10^{4}$ & 3724 & 2615 & 3617 \\
 & & & & & & $\widehat{VP}_\mathrm{s}$ & 537.3 & $5.66 \times 10^{-4}$ & $9.49 \times 10^{5}$ & $8.63 \times 10^{19}$ & $1.03 \times 10^{13}$ & $1.10 \times 10^{5}$ & 3449 & 2611 & 3342 \\
 & & & & & & $\hat{P}$ & 6.3 & $7.60 \times 10^{-6}$ & $8.35 \times 10^{5}$ & $1.10 \times 10^{20}$ & $1.31 \times 10^{13}$ & $5.13 \times 10^{4}$ & 2631 & 2562 & 1981 \\
\hline\hline
\end{tabular}
\end{table*}

\section{Discussion and conclusion}
\label{sec:conclusion}
In this paper, we have investigated the calculation of binding energies through the first-order SAPT formalism on a fault-tolerant device. We see this as a paradigmatic task in drug discovery that is well understood classically but where no quantum primitive yet exists. Translating this subroutine from classical to quantum computation requires considerable work on the algorithm selection (see previous work in Ref.~\cite{steudtner2023fault}) and adjusting all elements of this algorithm to the specific use case, as highlighted in this paper. Blind application of QSP-EVE without splitting into monomers, block encoding the SAPT operator without taking into account the product of operators structure, or input of an all-electron molecular Hamiltonian all would lead to much higher costs. This work has systematically compared several implementations of observable estimation and showed that observable-specific tensor factorizations and block encoding methodologies can dramatically reduce the algorithm's total runtime by many orders of magnitude. We presented an end-to-end resource estimation of the algorithm considering the ground-state preparation for each of the two monomers and the cost of block encodings tailored to the specific SAPT observables working both in the full space and active space pictures. While this work represents a solid first step in developing Heisenberg-limited fault-tolerant quantum algorithms beyond energy estimation, there remains much room for improvement in future work. For instance, developing the second-order SAPT energy contributions is still required, and explorations of alternative techniques that go beyond the $S^2$ approximation should also be considered. 

Ultimately, we have observed two major bottlenecks for this algorithm. The first bottleneck consists of the eigenstate reflection subroutine. Our work suggests that alternative approaches for this reflection are needed to make the cost of observable estimation more practical. The second bottleneck consists of the $\ell_1$ norm of the observable, $F$. In this regard, however, there are possible paths forward that we envision could help reduce the $\ell_1$ norm. For instance, methods such as tensor hypercontraction \cite{Lee2021} and regularized double factorization methods \cite{rubin2022compressing, oumarou2022accelerating} should help reduce the $\ell_1$ norm of the SAPT observables and hence improve the performance of the current SAPT-EVE algorithm. In conjunction, quantile estimation methods have also been proposed  to replace the $\ell_1$ norm dependence with the standard deviation ($\sigma_F$) of the observable that could reduce the overall runtime~\cite{cornelissen2022near}. While our work represents a key step in developing SAPT for fault-tolerant quantum computing, future methods that integrate all of the mentioned techniques will have the ability to make the SAPT-EVE algorithm much more competitive and a useful alternative to the supermolecular approach. 

Looking at this work through a broader lens, we wonder if this amount of effort is roughly universal in mapping non-total-energy-observables to efficient Heisenberg-limited FTQC methodologies. In this work, we encountered two main barriers to progress: (1) verbosity of mapping and (2) extensive overhead over standard QPE-type methodology to obtain the desired observable estimation values, even after extensive optimization. Overall, the first point seems palatable, as this effort is roughly similar in scope to the efforts needed to write an efficient block-encoded variant of the original QPE-type total energy method and is done once during the algorithm design stage. The second point is more concerning - it still seems that eigenvalue-based observables are considerably more efficient than general observables, even after extensive optimization. This deserves more research, and the mapping of even more non-total-energy observable properties to the FTQC environment will likely help to resolve this story and the question at the top of this paragraph. In any case, the present work may be taken as just one of many forays into this difficult but rewarding effort to map general observables to FTQC. 
\vspace{0.5cm}
\begin{acknowledgments}
\vspace{-0.1cm}
MS, WP, SS, and SMS thank all our colleagues at PsiQuantum for useful discussions. In particular, we thank Owen Williams for implementing the callgraph visualization engine and Sam Pallister for discussing some details of the factorized operator block encodings. The authors thank Clemens Utschig-Utschig for his comments on the manuscript and support during the project. 
\end{acknowledgments}

\bibliography{bibfile.bib}
\bibliographystyle{apsrev4-2}

\onecolumngrid
\newpage

\appendix

\section{Overview of Appendices}

\subsection{Organization}
Appendix A presents the notation and relevant definitions that will be used throughout this document. Appendix B presents the derivation of the first order SAPT operators. Appendix C presents the SAPT operators in the complete basis set limit. Appendix D derives the active space formulation of SAPT. Appendix E presents relevant details to the SAPT-EVE algorithm. Appendix F presents the SAPT operator encoding in the Majorana, \emph{sparse}, and \emph{tensor factorization} representations. Appendix G presents the SAPT block encoding compilation techniques. Appendix H presents details regarding the benchmark set for small molecules. Appendix I presents details regarding the heme and Artemisinin benchmark system. 

\subsection{Notation}
Throughout all of the appendices, we will use the following notation:
\begin{itemize}
    \item $\textsc{p}/\textsc{q}$ - orthogonal molecular spin orbital basis indices.
    \item $p/q$ - orthogonal molecular spatial orbital basis indices.
    \item $i/j$ - orthogonal occupied spatial orbital basis indices.
    \item $t/u$ - orthogonal active spatial orbital basis indices.
    \item $\sigma/\tau$ - orthogonal spin function indices.
\end{itemize}
Conventional SAPT theory considers two separate monomer calculations that are effectively independent from one another apart from the atomic orbital choice. Each monomer wavefunction will be solved so that is fully anti-symmetric with itself but not with respect to the total dimer system. Formally, this requires two sets of fermionic operators for each of the two monomers obeying the conventional fermionic commutation relations,
\begin{equation}
    \{\hat{a}_{\textsc{p}_1},\hat{a}_{\textsc{p}_2}^\dagger \} = \delta_{\textsc{p}_1\textsc{p}_2} \;\;\;\text{and}\;\;\; \{\hat{b}_{\textsc{q}_1},\hat{b}_{\textsc{q}_2}^\dagger \} = \delta_{\textsc{q}_1\textsc{q}_2}.
\end{equation}
The monomer's molecular orbitals are orthonormal to themselves but not each other. Furthermore, it is assumed that the monomer operators fully commute with one another:
\begin{equation}
    [\hat{a}_\textsc{p},b_\textsc{q}] = [\hat{a}_\textsc{p},\hat{b}_\textsc{q}^\dagger] = [\hat{a}_\textsc{p}^\dagger,\hat{b}_\textsc{q}] = [\hat{a}_\textsc{p}^\dagger,\hat{b}_\textsc{q}^\dagger] = 0.
\end{equation}
Throughout all of the appendices, we use the single-excitation operator defined in the spin-orbital basis, 
\begin{align}
    \hat{E}_{\textsc{p}_1\textsc{p}_2} = \hat{a}^\dagger_{\textsc{p}_1} \hat{a}_{\textsc{p}_2} \;\;\text{and}\;\;\hat{E}_{\textsc{q}_1\textsc{q}_2} = \hat{b}^\dagger_{\textsc{q}_1} \hat{b}_{\textsc{q}_2}
\end{align}
as well as the double-excitation operator, 
\begin{align}
    \hat{e}_{\textbf{\textsc{p}}} &\equiv \hat{e}_{\textsc{p}_1\textsc{p}_2\textsc{p}_3\textsc{p}_4} =  \hat{a}^\dagger_{\textsc{p}_1}\hat{a}^\dagger_{\textsc{p}_3}\hat{a}_{\textsc{p}_4}\hat{a}_{\textsc{p}_2}= \hat{E}_{\textsc{p}_1\textsc{p}_2}\hat{E}_{\textsc{p}_3\textsc{p}_4} - \delta_{\textsc{p}_2\textsc{p}_3}\hat{E}_{\textsc{p}_1\textsc{p}_4} \\
    \hat{e}_{\textbf{\textsc{q}}} &\equiv \hat{e}_{\textsc{q}_1\textsc{q}_2\textsc{q}_3\textsc{q}_4} = \hat{a}^\dagger_{\textsc{q}_1}\hat{a}^\dagger_{\textsc{q}_3}\hat{a}_{\textsc{q}_4}\hat{a}_{\textsc{q}_2}  = \hat{E}_{\textsc{q}_1\textsc{q}_2}\hat{E}_{\textsc{q}_3\textsc{q}_4} - \delta_{\textsc{q}_1\textsc{q}_4}\hat{E}_{\textsc{q}_1\textsc{q}_4}.
\end{align} 
We also define the corresponding operators in the spatial orbital basis,
\begin{align}
    \hat{E}^\sigma_{p_1p_2} = \hat{a}^\dagger_{p_1\sigma} \hat{a}_{p_2\sigma} \;\;\text{and}\;\;\hat{E}^\sigma_{q_1q_2} = \hat{b}^\dagger_{q_1\sigma} \hat{b}_{q_2\sigma}
\end{align}
as well as the spin-summed excitation operators, 
\begin{align}
    \hat{E}^+_{p_1p_2} = \hat{a}^\dagger_{p_1\alpha} \hat{a}_{p_2\alpha} + \hat{a}^\dagger_{p_1\beta} \hat{a}_{p_2\beta} \;\;\text{and}\;\;\hat{E}^+_{q_1q_2} = \hat{b}^\dagger_{q_1\alpha} \hat{b}_{q_2\alpha} + \hat{b}^\dagger_{q_1\beta} \hat{b}_{q_2\beta}.
\end{align}
We will generally use the convention that $p$ indices belong to monomer A and $q$ indices belong to monomer B unless it is explicitly stated otherwise.

\section{Symmetry-adapted perturbation theory}\label{sec:Appendix_SAPT}

\subsection{Brief overview of SAPT}

Symmetry-adapted perturbation theory is a state-of-the-art method used for the calculation of interaction energies $E_\text{int}$ in large molecular dimer systems. Within the context of SAPT, the interaction energy $E_\text{int}$ is computed directly as a sum of well-defined, physically interpretable polarization and exchange contributions,
\begin{equation}
    E_\text{int} = E_\text{pol}^{(1)} + E_\text{exch}^{(1)} + E_\text{pol}^{(2)} + E_\text{exch}^{(2)} + \cdots
\end{equation}
The interaction terms are calculated by applying a symmetry-adapted Rayleigh-Schrodinger (RS) perturbative expansion with respect to the interaction operator, $V = H - H_o$, where $H$ corresponds to the total dimer Hamiltonian and $H_o$ corresponds to the unperturbed operator, $H_o = H_A + H_B$, of the uncoupled monomers. The first-order polarization energy $E_\text{pol}^{(1)}$ describes the effect of the classical electrostatic interaction of the unperturbed charge distribution in the monomers. In contrast, the second-order polarization energy is the sum of induction and dispersion energies. The exchange corrections, $E_\text{exch}^{(n)}$, are specific to SAPT and provide the short-range repulsion which distinguishes this approach from conventional Rayleigh-Schrodinger perturbation theory, making it applicable to correlated systems. The first-order perturbative energy correction is written as:
\begin{equation}
    E^{(1)} = E_\text{pol}^{(1)} +E_\text{exch}^{(1)} = \frac{\braket{\Psi_A\Psi_B|\mathcal{A}\hat{V}\mathcal{A}|\Psi_A\Psi_B} }{\braket{\Psi_A\Psi_B|\mathcal{A}|\Psi_A\Psi_B}}
    \label{first_order_energy}
\end{equation}
where $\hat{V}$ is the intermolecular interaction operator and $\mathcal{A}$ is a symmetry-projection idempotent operator satisfying the property ($\mathcal{A}^2=\mathcal{A})$, known as the anti-symmetrizer. In conventional SAPT theory, each of the monomers is treated independently of one another, resulting in a dimer wavefunction that is not fully anti-symmetric. A typical approximation that is made within SAPT theory is commonly referred to as the $S^2$ approximation, where the anti-symmetrizer is approximated as, 
 $\mathcal{A} \approx 1 + \hat{P}$. The operator $\hat{P}$ is the exchange operator, $\hat{P} = -\sum_{i\in A}\sum_{j\in B} \hat{P}_{i_1j_1}$, describing the interchange between the spin and spatial coordinates of the electrons between the two monomers. The polarization energy is then defined as:
\begin{equation}
    E_\text{pol}^{(1)} = \braket{\Psi_A\Psi_B|\hat{V}|\Psi_A\Psi_B}
\end{equation}
while the exchange energy is written as:
\begin{equation}
    E_\text{exch}^{(1)} = E^{(1)} - E_\text{pol}^{(1)} = \braket{\Psi_A\Psi_B|\widehat{VP}_\mathrm{s}|\Psi_A\Psi_B} - E_\text{pol}^{(1)}\braket{\Psi_A\Psi_B|\hat{P}|\Psi_A\Psi_B},
\end{equation}
where we have kept first order terms in $\hat{P}$ after expanding the denominator in Eq.~\eqref{first_order_energy}. Note that we have defined $\widehat{VP}_\mathrm{s}$ rather than $\hat{V}\hat{P}$ due to a convention in the SAPT literature where all SAPT energy contributions are derived from a first-quantized formulation, which are then projected into a finite basis of orbitals. In order to validate our results with numerical results from classical SAPT calculations, we follow the same path. In order to estimate the total first-order SAPT energy on a quantum computer, we will require the estimation of the three observables: $\hat{F}=\{\hat{V},\hat{P},\widehat{VP}_\mathrm{s} \}$.

\subsection{Derivation of SAPT operators}
\subsubsection{Intermolecular operator}

In first quantization, the intermolecular interaction operator between monomers A and B is defined as:
\begin{equation}
    V_c = \sum_{i\in A}^{\eta_A} \sum_{j\in B}^{\eta_B} v_{ij}
    \label{V_c}
\end{equation}
where 
\begin{equation}
    v_{ij} = \frac{1}{r_{ij}} - \sum_J \frac{Z_{J}}{\eta_B} \frac{1}{r_{iJ}} - \sum_I \frac{Z_{I}}{\eta_A} \frac{1}{r_{Ij}} + \sum_{IJ} \frac{Z_{I} Z_{J}}{\eta_A \eta_B} \frac{1}{r_{IJ}}
    \label{eq:intermolecular_tensor}
\end{equation}
Monomer A/B consists of a number of $\eta_A/\eta_B$ electrons respectively. The first term describes the electron-electron interaction between the electrons in monomer A and monomer B. The second term describes the attractive electron-nuclear interaction between the electrons in monomer A with the nuclear atoms in monomer B; vice versa for the third term. The fourth term describes the repulsion interaction between the nuclear atoms of the two monomers. Interestingly, we found that keeping all of the terms combined in a single four-index tensor helps reduce the $\ell_1$ norm of the operator required for block encoding.

\subsubsection{Single-exchange operator}
In addition, conventional SAPT under the $S^2$ approximation uses the first-quantized form of the single-exchange operator,
\begin{equation}
    P_c = -  \sum_{i\in A}^{\eta_A} \sum_{j\in B}^{\eta_B} {P}_{ij} \label{single-exchange},
\end{equation}
where ${P}_{ij}$ exchanges the spin and space coordinates of the $i$th and $j$th electron of monomer A and monomer B respectively.

\subsubsection{Reduced density matrices}
In first quantization, the one and two-body density matrices are defined as:
\begin{align}
    \rho_X(\mathbf{x}_1|\mathbf{x}_1') &= \eta_X \int \mathrm{d}\mathbf{x}_{2,\eta_X} \Psi_X^*(\mathbf{x}_1,\mathbf{x}_2,\cdots,\mathbf{x}_{\eta_X}) \Psi_X(\mathbf{x}_1',\mathbf{x}_2,\cdots,\mathbf{x}_{\eta_X})  \label{1RDM}, \\
    \Gamma_X(\mathbf{x}_1,\mathbf{x}_2|\mathbf{x}_1',\mathbf{x}_2') &= \eta_X(\eta_X-1) \int \mathrm{d}\mathbf{x}_{3,\eta_X} \Psi_X^*(\mathbf{x}_1,\mathbf{x}_2,\mathbf{x}_3,\cdots,\mathbf{x}_{\eta_X}) \Psi_X(\mathbf{x}_1',\mathbf{x}_2',\mathbf{x}_3,\cdots,\mathbf{x}_{\eta_X}) \label{2RDM},
\end{align}
where $\mathrm{d}\mathbf{x}_{i,\eta_X} = \mathrm{d}\mathbf{x}_i\mathrm{d}\mathbf{x}_{i+1}\cdots \mathrm{d}\mathbf{x}_{\eta_X}$ and each monomer is assumed to have $\eta_X$ total electrons. The integral over the variable $\mathbf{x}_i$ is interpreted as an integral over both the spatial and spin components. Notably, these density matrices obey the permutational symmetries, 
\begin{align}
    \rho_X(\mathbf{x}_1|\mathbf{x}_1') &= \rho_X^*(\mathbf{x}_1'|\mathbf{x}_1), \\
    \Gamma_X(\mathbf{x}_1,\mathbf{x}_2|\mathbf{x}_1',\mathbf{x}_2') &= \Gamma_X^*(\mathbf{x}_1',\mathbf{x}_2'|\mathbf{x}_1,\mathbf{x}_2)=\Gamma_X(\mathbf{x}_2,\mathbf{x}_1|\mathbf{x}_2',\mathbf{x}_1')=\Gamma^*_X(\mathbf{x}_2',\mathbf{x}_1'|\mathbf{x}_2,\mathbf{x}_1).
\end{align}
In second quantization, the density matrices are written in terms of molecular spin-orbitals $\phi_{\textsc{p}}(\mathbf{\mathbf{x}})$:
\begin{align}
    \rho_X(\mathbf{x}_1|\mathbf{x}_1') &= \sum_{\textsc{p}\textsc{q}} D^X_{\textsc{p}\textsc{q}}\, \phi^*_{\textsc{p}}({\mathbf{\mathbf{x}}}_1)\phi_{\textsc{q}}({\mathbf{\mathbf{x}}}_1'), \\
    \Gamma_X(\mathbf{x}_1,\mathbf{x}_2|\mathbf{x}_1',\mathbf{x}_2') &= \sum_{\textsc{p}\textsc{qrs}} d^X_{\textsc{p}\textsc{qrs}}\, \phi^*_{\textsc{p}}(\mathbf{x}_1)\phi_{\textsc{q}}(\mathbf{x}_1')\phi^*_{\textsc{r}}(\mathbf{x}_2)\phi_{\textsc{s}}(\mathbf{x}_2'),
\end{align}
with the reduced density matrix (RDM) coefficients given by:
\begin{align}
    D^X_{\textsc{p}\textsc{q}} &= \braket{\Psi_X|\hat{c}^\dagger_{\textsc{p}}\hat{c}_{\textsc{q}}|\Psi_X},  \\
    d^X_{\textsc{pqrs}} &= \braket{\Psi_X|\hat{c}^\dagger_{\textsc{p}}\hat{c}^\dagger_{\textsc{r}}\hat{c}_{\textsc{s}}\hat{c}_{\textsc{q}}|\Psi_X}, \label{2RDM_2}
\end{align}
where $\textsc{p},\textsc{q},\textsc{r},\textsc{s}$ denote the indices of the molecular spin-orbitals for monomer $X$, and $\hat{c}^\dagger/\hat{c}$ denote the creation/annihilation operators for either monomer A or B. The permutational symmetries associated with the RDMs are:
\begin{align}
    D_{\textsc{p}\textsc{q}} &= D^*_{\textsc{q}\textsc{p}}, \label{permutational_symmetries_1RDM} \\
    d_{\textsc{pqrs}} &= d^*_{\textsc{qpsr}} = d_{\textsc{rspq}}  = d^*_{\textsc{srqp}}.
    \label{permutational_symmetries_2RDM}
\end{align}

\subsubsection{Electrostatic Operator: $\hat{V}$}
In first quantization, the expectation value with the intermolecular interaction operator is given by:
\begin{equation}
    E_\mathrm{pol}^{(1)} \equiv  \braket{\Psi_A\Psi_B|V_c|\Psi_A\Psi_B}  = \int \mathrm{d}\mathbf{x}_{1,\eta_A}^A \mathrm{d}\mathbf{x}_{1,\eta_B}^B \Psi_A(\mathbf{x}_{1,\eta_A}^A)\Psi_B(\mathbf{x}_{1,\eta_B}^B)V_c\Psi_A^*(\mathbf{x}_{1,\eta_A}^A)\Psi_B^*(\mathbf{x}_{1,\eta_B}^B),
\end{equation}
where $\mathbf{x}_{i,\eta_X} = \mathbf{x}_i\mathbf{x}_{i+1}\cdots \mathbf{x}_{\eta_X}$. Using the density matrix formalism defined above, it is possible to derive the second quantized operator of interest. Using equations \eqref{1RDM}-\eqref{2RDM_2}, we find:
\begin{align}
    E_\mathrm{pol}^{(1)}  &= \int \mathrm{d}\mathbf{x}_i \mathrm{d}\mathbf{x}_j\, \rho_A(\mathbf{x}_i|\mathbf{x}_i) v_{ij} \rho_B(\mathbf{x}_j|\mathbf{x}_j) , \nonumber \\
     &= \sum_{\textsc{p}_1\textsc{p}_2}\sum_{\textsc{q}_1\textsc{q}_2} D^A_{\textsc{p}_1\textsc{p}_2} D^B_{\textsc{q}_1\textsc{q}_2} \int \mathrm{d}\mathbf{x}_i \mathrm{d}\mathbf{x}_j  v_{ij} \phi_{\textsc{p}_1}^*(\mathbf{x}_i)\phi^{}_{\textsc{p}_2}(\mathbf{x}_i)\phi_{\textsc{q}_1}^*(\mathbf{x}_j)\phi^{}_{\textsc{q}_2}(\mathbf{x}_j)  
\end{align}
where we have defined the intermolecular four-index tensor, ${v}^{\textsc{p}_1\textsc{p}_2}_{\textsc{q}_1\textsc{q}_2}$, as:
\begin{align}
    {v}^{\textsc{p}_1\textsc{p}_2}_{\textsc{q}_1\textsc{q}_2} &= \int \mathrm{d}\mathbf{x}_i \mathrm{d}\mathbf{x}_j\;   \phi_{\textsc{p}_1}^*(\mathbf{x}_i)\phi_{\textsc{p}_2}(\mathbf{x}_i) v_{ij}\phi_{\textsc{q}_1}^*(\mathbf{x}_j)\phi_{\textsc{q}_2}(\mathbf{x}_j).
\label{electrostatic_tei}
\end{align}
The molecular overlap matrix is also defined with respect to molecular spin-orbitals $\phi_{\textsc{p}}(\mathbf{\mathbf{x}})$ and $\phi_\textsc{q}(\mathbf{\mathbf{x}})$ as,
\begin{equation}
    S^\textsc{p}_\textsc{q} = \int\! \mathrm{d}\mathbf{\mathbf{x}}\; \phi_{\textsc{p}}^*(\mathbf{\mathbf{x}})\phi_{\textsc{q}}(\mathbf{\mathbf{x}}).
\end{equation}
Since the operators from monomer A commute with those in monomer B, we find the final expression for the electrostatic operator $\hat{V}$:
\begin{equation}
    \hat{V} = \sum_{\textsc{p}_1\textsc{p}_2}\sum_{\textsc{q}_1\textsc{q}_2} {v}^{\textsc{p}_1\textsc{p}_2}_{\textsc{q}_1\textsc{q}_2} \hat{E}_{\textsc{p}_1\textsc{p}_2}\hat{E}_{\textsc{q}_1\textsc{q}_2}.
    \label{electrostatics_unified}
\end{equation}

\subsubsection{Exchange Operator: $\hat{P}$}
The exchange operator is found by considering the expectation value of $\hat{P}$ with respect to the two-monomer wavefunction in first quantization:
\begin{equation}
    \braket{\hat{P}} = \int \mathrm{d}\mathbf{x}_{1,\eta_A}^A \mathrm{d}\mathbf{x}_{1,\eta_B}^B \Psi_A(\mathbf{x}_{1,\eta_A}^A)\Psi_B(\mathbf{x}_{1,\eta_B}^B)P_c\Psi_A^*(\mathbf{x}_{1,\eta_A}^A)\Psi_B^*(\mathbf{x}_{1,\eta_B}^B).
\end{equation}
Performing this exchange and using the density matrix formalism as before, we find:
\begin{align}
    \braket{\hat{P}}  &= -\int \mathrm{d}\mathbf{x}_i \mathrm{d}\mathbf{x}_j\,\rho_A(\mathbf{x}_j|\mathbf{x}_i) \rho_B(\mathbf{x}_i|\mathbf{x}_j) , \nonumber \\
     &= -\sum_{\textsc{p}_1\textsc{p}_2}\sum_{\textsc{q}_1\textsc{q}_2} D^A_{\textsc{p}_1\textsc{p}_2} D^B_{\textsc{q}_1\textsc{q}_2} \int \mathrm{d}\mathbf{x}_i \mathrm{d}\mathbf{x}_j  \phi_{\textsc{p}_1}^*(\mathbf{x}_j)\phi_{\textsc{p}_2}(\mathbf{x}_i)\phi_{\textsc{q}_1}^*(\mathbf{x}_i)\phi_{\textsc{q}_2}(\mathbf{x}_j). 
\end{align}
As a result,  we find the final expression for the exchange operator:
\begin{equation}
    \hat{P} = -\sum_{\textsc{p}_1\textsc{p}_2}\sum_{\textsc{q}_1\textsc{q}_2} S^{\textsc{p}_1}_{\textsc{q}_2}S^{\textsc{q}_1}_{\textsc{p}_2}\hat{E}_{\textsc{p}_1\textsc{p}_2}\hat{E}_{\textsc{q}_1\textsc{q}_2}.
    \label{full_exchange}
\end{equation}
Note that this operator can also be derived with the interaction density matrix approach that we discuss below.

\subsubsection{Electrostatic-Exchange Operator: $\widehat{VP}_\mathrm{s}$}
The electrostatic-exchange operator can be found in a similar fashion by using the following first quantized expectation value expression:
\begin{equation}
    \braket{\widehat{VP}_\text{s}} = \braket{\Psi_A\Psi_B|\tfrac{1}{2}(V_cP_c + P_cV_c)|\Psi_A\Psi_B}.
    \label{symmetric_VP_expectation_value}
\end{equation}
It is important to note that we have used the symmetric expression, $\tfrac{1}{2}(V_cP_c + P_cV_c)$, rather than $V_cP_c$ which is normally the convention in the classical SAPT literature. This ensures that the electrostatic-exchange operator will remain strictly Hermitian as required for the SAPT-EVE algorithm. In the following, we will derive the resulting equations based on the first term, $V_cP_c$, but will include both contributions in the final result. The expectation value of the first term is written as:
\begin{equation}
    \braket{V_cP_c} = \int \mathrm{d}\mathbf{x}_i \mathrm{d}\mathbf{x}_j  v_{ij}  \rho_{\mathrm{int}}(\mathbf{x}_i,\mathbf{x}_j) \label{Eexch_DM},
\end{equation}
where we have written this expression in terms of the interaction density matrix $\rho_{\mathrm{int}}(\mathbf{x}_i,\mathbf{x}_j)$ \cite{moszynski1994many},
\begin{equation}
    \rho_{\mathrm{int}}(\mathbf{x}_i,\mathbf{x}_j) = \eta_A \eta_B \int \mathrm{d}\mathbf{x}_{2,\eta_A}^A \mathrm{d}\mathbf{x}_{2,\eta_B}^B \Psi_A^*(\mathbf{x}_i^A,\mathbf{x}_{2,\eta_A}^A) \Psi_B^*(\mathbf{x}_j^B,\mathbf{x}_{2,\eta_B}^B) P_c \Psi_A (\mathbf{x}_i^A,\mathbf{x}_{2,\eta_A}^A) \Psi_B (\mathbf{x}_j^B,\mathbf{x}_{2,\eta_B}^B) . \label{intRDM}
\end{equation}
If we apply (\ref{single-exchange}) to the interaction density matrix (\ref{intRDM}), we obtain
\begin{align}
    \rho_{\mathrm{int}}(\mathbf{x}_i,\mathbf{x}_j) = &- \rho_A(\mathbf{x}_i|\mathbf{x}_j)\rho_B(\mathbf{x}_j|\mathbf{x}_i) - \int \mathrm{d}\mathbf{x}_k  \,\Gamma_A(\mathbf{x}_i,\mathbf{x}_k|\mathbf{x}_i,\mathbf{x}_j)\rho_B(\mathbf{x}_j|\mathbf{x}_k) \nonumber \\
    &- \int \mathrm{d}\mathbf{x}_l \,\rho_A(\mathbf{x}_i|\mathbf{x}_l)\Gamma_B(\mathbf{x}_j,\mathbf{x}_l|\mathbf{x}_j,\mathbf{x}_i) - \int\!\mathrm{d}\mathbf{x}_k \mathrm{d}\mathbf{x}_l\,\Gamma_A(\mathbf{x}_i,\mathbf{x}_k|\mathbf{x}_i\mathbf{x}_l)\Gamma_B(\mathbf{x}_j,\mathbf{x}_l|\mathbf{x}_j\mathbf{x}_k) .
\end{align}
The electrostatic-exchange observable can then be expressed as a sum of four terms:
\begin{align*}
    \braket{\widehat{VP}}= \int \mathrm{d}\mathbf{x}_i \mathrm{d}\mathbf{x}_j {v}_{ij}\rho_{\mathrm{int}}(\mathbf{x}_i,\mathbf{x}_j)   &= \braket{\widehat{VP}_1} + \braket{\widehat{VP}_2} + \braket{\widehat{VP}_3} + \braket{\widehat{VP}_4}
\end{align*}
where
\begin{align}
    \braket{\widehat{VP}_1} &= - \int \mathrm{d}\mathbf{x}_i \mathrm{d}\mathbf{x}_j \,{v}_{ij}\rho_A(\mathbf{x}_i|\mathbf{x}_j)\rho_B(\mathbf{x}_j|\mathbf{x}_i)\\
    \braket{\widehat{VP}_2} &= - \int \mathrm{d}\mathbf{x}_k \mathrm{d}\mathbf{x}_i \mathrm{d}\mathbf{x}_j\,{v}_{ij} \Gamma_A(\mathbf{x}_i,\mathbf{x}_k|\mathbf{x}_i,\mathbf{x}_j)\rho_B(\mathbf{x}_j|\mathbf{x}_k)  \\
    \braket{\widehat{VP}_3} &= - \int \mathrm{d}\mathbf{x}_l\mathrm{d}\mathbf{x}_i \mathrm{d}\mathbf{x}_j\,{v}_{ij} \rho_A(\mathbf{x}_i|\mathbf{x}_l)\Gamma_B(\mathbf{x}_j,\mathbf{x}_l|\mathbf{x}_j,\mathbf{x}_i) \\
    \braket{\widehat{VP}_4} &= - \int \mathrm{d}\mathbf{x}_k \mathrm{d}\mathbf{x}_l\mathrm{d}\mathbf{x}_i \mathrm{d}\mathbf{x}_j\,{v}_{ij} \Gamma_A(\mathbf{x}_i,\mathbf{x}_k|\mathbf{x}_i\mathbf{x}_l)\Gamma_B(\mathbf{x}_j,\mathbf{x}_l|\mathbf{x}_j\mathbf{x}_k) .
\end{align}
Following the same steps as the electrostatics case, we now substitute the 1- and 2-RDMs to obtain the electrostatic-exchange operators. The first term is given by:
\begin{align}
    \braket{\widehat{VP}_1} &=  -\sum_{\textbf{\textsc{p}}\textbf{\textsc{q}}} D^A_{\textsc{p}_1\textsc{p}_2} D^B_{\textsc{q}_1\textsc{q}_2} \int \mathrm{d}\mathbf{x}_i \mathrm{d}\mathbf{x}_j {v}_{ij} \phi_{\textsc{p}_1}^*(\mathbf{x}_j)\phi_{\textsc{p}_2}(\mathbf{x}_i)\phi_{\textsc{q}_1}^*(\mathbf{x}_i)\phi_{\textsc{q}_2}(\mathbf{x}_j) 
\end{align}
which leads to the operator,
\begin{equation}
    \widehat{VP}_1 = -\sum_{\textbf{\textsc{p}}\textbf{\textsc{q}}}{v}^{\textsc{p}_1\textsc{q}_2}_{\textsc{q}_1\textsc{p}_2} \hat{E}_{\textsc{p}_1\textsc{p}_2} \hat{E}_{\textsc{q}_1\textsc{q}_2}.
    \label{VP1_initial}
\end{equation}
The second term is then evaluated as:
\begin{align}
    \braket{\widehat{VP}_2} &= - \int  \mathrm{d}\mathbf{x}_k \mathrm{d}\mathbf{x}_i \mathrm{d}\mathbf{x}_j \, {v}_{ij} \Gamma_A(\mathbf{x}_i,\mathbf{x}_k|\mathbf{x}_i,\mathbf{x}_j)\rho_B(\mathbf{x}_j|\mathbf{x}_k) \\
    &= -\sum_{\textbf{\textsc{p}}\textbf{\textsc{q}}}\!\! \int\! \mathrm{d}\mathbf{x}_k \mathrm{d}\mathbf{x}_i \mathrm{d}\mathbf{x}_j {v}_{ij} d^A_{\textsc{p}_1\textsc{p}_2\textsc{p}_3\textsc{p}_4}  D^B_{\textsc{q}_1\textsc{q}_2} \phi_{\textsc{p}_1}^*(\mathbf{x}_i)\phi_{\textsc{p}_2}(\mathbf{x}_i)\phi_{\textsc{p}_3}^*(\mathbf{x}_k)\phi_{\textsc{p}_4}(\mathbf{x}_j) \phi_{\textsc{q}_1}^*(\mathbf{x}_j) \phi_{\textsc{q}_2}(\mathbf{x}_k) 
\end{align}
The effective operator is then written as:
\begin{align}
    \widehat{VP}_2 &= - \sum_{\textbf{\textsc{p}}\textbf{\textsc{q}}}  {v}^{\textsc{p}_1\textsc{p}_2}_{\textsc{q}_1\textsc{p}_4} S^{\textsc{p}_3}_{\textsc{q}_2} \, \hat{a}^\dagger_{\textsc{p}_1}\hat{a}^\dagger_{\textsc{p}_3}\hat{a}_{\textsc{p}_4}\hat{a}_{\textsc{p}_2}
    \hat{b}_{\textsc{q}_1}^\dagger\hat{b}_{\textsc{q}_2} \\
    &=  -  \sum_{\textbf{\textsc{p}}\textbf{\textsc{q}}} {v}^{\textsc{p}_1\textsc{p}_2}_{\textsc{q}_1\textsc{p}_4} S^{\textsc{p}_3}_{\textsc{q}_2}\hat{e}_{\textsc{p}_1\textsc{p}_2\textsc{p}_3\textsc{p}_4}\hat{E}_{\textsc{q}_1\textsc{q}_2}.
    \label{VP2_initial}
\end{align}
The evaluation of the third term follows in the same fashion,
\begin{align}
    \braket{\widehat{VP}_3} &= - \int \mathrm{d}\mathbf{x}_l\mathrm{d}\mathbf{x}_i \mathrm{d}\mathbf{x}_j\,{v}_{ij} \rho_A(\mathbf{x}_i|\mathbf{x}_l)\Gamma_B(\mathbf{x}_j,\mathbf{x}_l|\mathbf{x}_j,\mathbf{x}_i)  \\
    &= -\sum_{\textbf{\textsc{p}}\textbf{\textsc{q}}} 
    \int \mathrm{d}\mathbf{x}_l \mathrm{d}\mathbf{x}_i \mathrm{d}\mathbf{x}_j {v}_{ij} D^A_{\textsc{p}_1\textsc{p}_2} d^B_{\textsc{q}_1\textsc{q}_2\textsc{q}_3\textsc{q}_4} \phi^*_{\textsc{p}_1}(\mathbf{x}_i)\phi_{\textsc{p}_2}(\mathbf{x}_l)  \phi_{\textsc{q}_1}^*(\mathbf{x}_j)\phi_{\textsc{q}_2}(\mathbf{x}_j)\phi^*_{\textsc{q}_3}(\mathbf{x}_l)\phi_{\textsc{q}_4}(\mathbf{x}_i)  \\ 
    &= - \sum_{\textbf{\textsc{p}}\textbf{\textsc{q}}} S^{\textsc{q}_3}_{\textsc{p}_2} {v}^{\textsc{p}_1\textsc{q}_4}_{\textsc{q}_1\textsc{q}_2} D^A_{\textsc{p}_1\textsc{p}_2} d^B_{\textsc{q}_1\textsc{q}_2\textsc{q}_3\textsc{q}_4} 
\end{align}
The effective operator is written as:
\begin{align}
    \widehat{VP}_3 &= -\sum_{\textbf{\textsc{p}}\textbf{\textsc{q}}} S^{\textsc{q}_3}_{\textsc{p}_2} {v}^{\textsc{p}_1\textsc{q}_4}_{\textsc{q}_1\textsc{q}_2} \hat{a}_{\textsc{p}_1}^\dagger\hat{a}_{\textsc{p}_2} \hat{b}^\dagger_{\textsc{q}_1}\hat{b}^\dagger_{\textsc{q}_3}\hat{b}_{\textsc{q}_4}\hat{b}_{\textsc{q}_2}  \\
    &= -\sum_{\textbf{\textsc{p}}\textbf{\textsc{q}}} S^{\textsc{q}_3}_{\textsc{p}_2} {v}^{\textsc{p}_1\textsc{q}_4}_{\textsc{q}_1\textsc{q}_2} \hat{E}_{\textsc{p}_1\textsc{p}_2}\hat{e}_{\textsc{q}_1\textsc{q}_2\textsc{q}_3\textsc{q}_4}.
    \label{VP3_initial}
\end{align}
 Finally, the evaluation of the last term is given by:
\begin{align}
    \braket{\widehat{VP}_4} &= - \int \mathrm{d}\mathbf{x}_k \mathrm{d}\mathbf{x}_l\mathrm{d}\mathbf{x}_i \mathrm{d}\mathbf{x}_j\, {v}_{ij} \Gamma_A(\mathbf{x}_i,\mathbf{x}_k|\mathbf{x}_i\mathbf{x}_l)\Gamma_B(\mathbf{x}_j,\mathbf{x}_l|\mathbf{x}_j\mathbf{x}_k)   \\
    &= - \sum_{\textbf{\textsc{p}}\textbf{\textsc{q}}} 
    \int \mathrm{d}\mathbf{x}_k \mathrm{d}\mathbf{x}_l\mathrm{d}\mathbf{x}_i \mathrm{d}\mathbf{x}_j {v}_{ij}  d^A_{\textsc{p}_1\textsc{p}_2\textsc{p}_3\textsc{p}_4}  d^B_{\textsc{q}_1\textsc{q}_2\textsc{q}_3\textsc{q}_4}\times \\ &\phi_{\textsc{p}_1}^*(\mathbf{x}_i)\phi_{\textsc{p}_2}(\mathbf{x}_i)\phi^*_{\textsc{p}_3}(\mathbf{x}_k)\phi_{\textsc{p}_4}(\mathbf{x}_l) \phi_{\textsc{q}_1}^*(\mathbf{x}_j)\phi_{\textsc{q}_2}(\mathbf{x}_j)\phi^*_{\textsc{q}_3}(\mathbf{x}_l)\phi_{\textsc{q}_4}(\mathbf{x}_k) \nonumber \\
    &= - \sum_{\textbf{\textsc{p}}\textbf{\textsc{q}}} S^{\textsc{q}_3}_{\textsc{p}_4} S^{\textsc{p}_3}_{\textsc{q}_4} {v}^{\textsc{p}_1\textsc{p}_2}_{\textsc{q}_1\textsc{q}_2} \;d^A_{\textsc{p}_1\textsc{p}_2\textsc{p}_3\textsc{p}_4}  d^B_{\textsc{q}_1\textsc{q}_2\textsc{q}_3\textsc{q}_4} \\
    &= - \sum_{\textbf{\textsc{p}}\textbf{\textsc{q}}} S^{\textsc{q}_3}_{\textsc{p}_4} S^{\textsc{p}_3}_{\textsc{q}_4} {v}^{\textsc{p}_1\textsc{p}_2}_{\textsc{q}_1\textsc{q}_2} d^A_{\textsc{p}_1\textsc{p}_2\textsc{p}_3\textsc{p}_4}  d^B_{\textsc{q}_1\textsc{q}_2\textsc{q}_3\textsc{q}_4}
\end{align}
resulting in the operator,
\begin{align}
    \widehat{VP}_4 &= - \sum_{\textbf{\textsc{p}}\textbf{\textsc{q}}} {v}^{\textsc{p}_1\textsc{p}_2}_{\textsc{q}_1\textsc{q}_2} S^{\textsc{q}_3}_{\textsc{p}_4} S^{\textsc{p}_3}_{\textsc{q}_4}  \hat{a}^\dagger_{\textsc{p}_1}\hat{a}^\dagger_{\textsc{p}_3}\hat{a}_{\textsc{p}_4}\hat{a}_{\textsc{p}_2}
    \hat{b}^\dagger_{\textsc{q}_1}\hat{b}^\dagger_{\textsc{q}_3}\hat{b}_{\textsc{q}_4}\hat{b}_{\textsc{q}_2} \\
    &= - \sum_{\textbf{\textsc{p}}\textbf{\textsc{q}}} {v}^{\textsc{p}_1\textsc{p}_2}_{\textsc{q}_1\textsc{q}_2} S^{\textsc{q}_3}_{\textsc{p}_4} S^{\textsc{p}_3}_{\textsc{q}_4} \,\hat{e}_{\textsc{p}_1\textsc{p}_2\textsc{p}_3\textsc{p}_4}\hat{e}_{\textsc{q}_1\textsc{q}_2\textsc{q}_3\textsc{q}_4}.
    \label{VP4_initial}
\end{align}
Combining all of the results from above, we find the total electrostatic-exchange operator,
\begin{align}
    \widehat{VP} = &-\sum_{\substack{\textbf{\textsc{p}}\textbf{\textsc{q}}}} v^{\textsc{p}_1\textsc{q}_2}_{\textsc{q}_1\textsc{p}_2} \hat{E}_{\textsc{p}_1\textsc{p}_2} \hat{E}_{\textsc{q}_1\textsc{q}_2}  \nonumber \\ 
    &- \sum_{\textbf{\textsc{p}}\textbf{\textsc{q}}} S^{\textsc{p}_3}_{\textsc{q}_2} {v}^{\textsc{p}_1\textsc{p}_2}_{\textsc{q}_1\textsc{p}_4}\hat{e}_{\textsc{p}_1\textsc{p}_2\textsc{p}_3\textsc{p}_4}\hat{E}_{\textsc{q}_1\textsc{q}_2}  \nonumber \\
    &-\sum_{\textbf{\textsc{p}}\textbf{\textsc{q}}} S^{\textsc{q}_3}_{\textsc{p}_2} {v}^{\textsc{p}_1\textsc{q}_4}_{\textsc{q}_1\textsc{q}_2} \hat{E}_{\textsc{p}_1\textsc{p}_2}\hat{e}_{\textsc{q}_1\textsc{q}_2\textsc{q}_3\textsc{q}_4}  \nonumber \\
    &- \sum_{\textbf{\textsc{p}}\textbf{\textsc{q}}} {v}^{\textsc{p}_1\textsc{p}_2}_{\textsc{q}_1\textsc{q}_2} S^{\textsc{q}_3}_{\textsc{p}_4} S^{\textsc{p}_3}_{\textsc{q}_4} \hat{e}_{\textsc{p}_1\textsc{p}_2\textsc{p}_3\textsc{p}_4}\hat{e}_{\textsc{q}_1\textsc{q}_2\textsc{q}_3\textsc{q}_4}.
    \label{VP_initial}
\end{align}
The contribution of the second term, $P_cV_c$, in Eq.~\eqref{symmetric_VP_expectation_value} will be given by:
\begin{align}
    \widehat{PV} = &-\sum_{\substack{\textbf{\textsc{p}}\textbf{\textsc{q}}}} v^{\textsc{q}_1\textsc{p}_2}_{\textsc{p}_1\textsc{q}_2} \hat{E}_{\textsc{p}_1\textsc{p}_2} \hat{E}_{\textsc{q}_1\textsc{q}_2}   \nonumber\\ 
    &- \sum_{\textbf{\textsc{p}}\textbf{\textsc{q}}} S^{\textsc{q}_1}_{\textsc{p}_4} {v}^{\textsc{p}_1\textsc{p}_2}_{p_3\textsc{q}_2}\hat{e}_{\textsc{p}_1\textsc{p}_2\textsc{p}_3\textsc{p}_4}\hat{E}_{\textsc{q}_1\textsc{q}_2}  \nonumber \\
    &-\sum_{\textbf{\textsc{p}}\textbf{\textsc{q}}} S^{\textsc{p}_1}_{\textsc{q}_4} {v}^{\textsc{q}_3\textsc{p}_2}_{\textsc{q}_1\textsc{q}_2} \hat{E}_{\textsc{p}_1\textsc{p}_2}\hat{e}_{\textsc{q}_1\textsc{q}_2\textsc{q}_3\textsc{q}_4}  \nonumber \\
    &- \sum_{\textbf{\textsc{p}}\textbf{\textsc{q}}} {v}^{\textsc{p}_1\textsc{p}_2}_{\textsc{q}_1\textsc{q}_2} S^{\textsc{q}_3}_{\textsc{p}_4} S^{\textsc{p}_3}_{\textsc{q}_4} \hat{e}_{\textsc{p}_1\textsc{p}_2\textsc{p}_3\textsc{p}_4}\hat{e}_{\textsc{q}_1\textsc{q}_2\textsc{q}_3\textsc{q}_4}.
\end{align}
Resulting in the total electrostatic-exchange operator given by:
\begin{align}
    \widehat{VP}_{\text{s}} = &\;\;\;\;\tfrac{1}{2}(\widehat{VP} + \widehat{PV})  \\
    = &-\tfrac{1}{2}\sum_{\textbf{\textsc{p}}\textbf{\textsc{q}}} \left( v^{\textsc{p}_1\textsc{q}_2}_{\textsc{q}_1\textsc{p}_2}  \hat{E}_{\textsc{p}_1\textsc{p}_2} \hat{E}_{\textsc{q}_1\textsc{q}_2} + v^{\textsc{q}_2\textsc{p}_1}_{\textsc{p}_2\textsc{q}_1} \hat{E}_{\textsc{p}_2\textsc{p}_1} \hat{E}_{\textsc{q}_2\textsc{q}_1} \right)  \nonumber \\
    &-\tfrac{1}{2} \sum_{\textbf{\textsc{p}}\textbf{\textsc{q}}} \left(  {v}^{\textsc{p}_1\textsc{p}_2}_{\textsc{q}_1\textsc{p}_4}S^{\textsc{p}_3}_{\textsc{q}_2} \hat{e}_{\textsc{p}_1\textsc{p}_2\textsc{p}_3\textsc{p}_4}\hat{E}_{\textsc{q}_1\textsc{q}_2} +
    {v}^{\textsc{p}_2\textsc{p}_1}_{\textsc{p}_4\textsc{q}_1}S_{\textsc{p}_3}^{\textsc{q}_2} \hat{e}_{\textsc{p}_2\textsc{p}_1\textsc{p}_4\textsc{p}_3} \hat{E}_{\textsc{q}_2\textsc{q}_1} \right)  \nonumber\\
    &-\tfrac{1}{2} \sum_{\textbf{\textsc{p}}\textbf{\textsc{q}}} \left(   {v}^{\textsc{p}_1\textsc{q}_4}_{\textsc{q}_1\textsc{q}_2}S^{\textsc{q}_3}_{\textsc{p}_2} \, \hat{E}_{\textsc{p}_1\textsc{p}_2}\hat{e}_{\textsc{q}_1\textsc{q}_2\textsc{q}_3\textsc{q}_4} +
    {v}^{\textsc{q}_4\textsc{p}_1}_{\textsc{q}_2\textsc{q}_1}S_{\textsc{q}_3}^{\textsc{p}_2} \, \hat{E}_{\textsc{p}_2\textsc{p}_1}\hat{e}_{\textsc{q}_2\textsc{q}_1\textsc{q}_4\textsc{q}_3}\right)  \nonumber \\
    &-\tfrac{1}{2} \sum_{\textbf{\textsc{p}}\textbf{\textsc{q}}} \left( {v}^{\textsc{p}_1\textsc{p}_2}_{\textsc{q}_1\textsc{q}_2} S^{\textsc{q}_3}_{\textsc{p}_4} S^{\textsc{p}_3}_{\textsc{q}_4}\hat{e}_{\textsc{p}_1\textsc{p}_2\textsc{p}_3\textsc{p}_4}\hat{e}_{\textsc{q}_1\textsc{q}_2\textsc{q}_3\textsc{q}_4} +
    {v}^{\textsc{p}_2\textsc{p}_1}_{\textsc{q}_2\textsc{q}_1} S_{\textsc{q}_3}^{\textsc{p}_4} S_{\textsc{p}_3}^{\textsc{q}_4}\hat{e}_{\textsc{p}_2\textsc{p}_1\textsc{p}_4\textsc{p}_3}\hat{e}_{\textsc{q}_2\textsc{q}_1\textsc{q}_4\textsc{q}_3}\right).
\end{align}
This recovers the electrostatic-exchange operator defined in the main text. Note that compared to the electrostatic and exchange operators, $\hat{V}$ and $\hat{P}$, the electrostatic-exchange operator contains products of high-order tensors. It is possible to gain more insight by rewriting this operator in the so-called chemist notation where single-excitation and spin-summed operators are combined,
\begin{align}
    \widehat{VP}_{\text{s}} 
    = &-\tfrac{1}{2}\sum_{\textbf{\textsc{p}}\textbf{\textsc{q}}} \left( \nu^{\textsc{p}_1\textsc{q}_2}_{\textsc{q}_1\textsc{p}_2}  \hat{E}_{\textsc{p}_1\textsc{p}_2} \hat{E}_{\textsc{q}_1\textsc{q}_2} + \nu^{\textsc{q}_2\textsc{p}_1}_{\textsc{p}_2\textsc{q}_1} \hat{E}_{\textsc{p}_2\textsc{p}_1} \hat{E}_{\textsc{q}_2\textsc{q}_1} \right)  \nonumber\\
    &- \tfrac{1}{2} \sum_{\textbf{\textsc{p}}\textbf{\textsc{q}}} \left(  \nu^{\textsc{p}_1\textsc{p}_2}_{\textsc{q}_1\textsc{p}_4}S^{\textsc{p}_3}_{\textsc{q}_2} \hat{E}_{\textsc{p}_1\textsc{p}_2} \hat{E}_{\textsc{p}_3\textsc{p}_4}\hat{E}_{\textsc{q}_1\textsc{q}_2} +
    \nu^{\textsc{p}_2\textsc{p}_1}_{\textsc{p}_4\textsc{q}_1}S_{\textsc{p}_3}^{\textsc{q}_2} \hat{E}_{\textsc{p}_2\textsc{p}_1} \hat{E}_{\textsc{p}_4\textsc{p}_3} \hat{E}_{\textsc{q}_2\textsc{q}_1} \right) 
      \nonumber\\
    &- \tfrac{1}{2} \sum_{\textbf{\textsc{p}}\textbf{\textsc{q}}} \left(   \nu^{\textsc{p}_1\textsc{q}_4}_{\textsc{q}_1\textsc{q}_2}S^{\textsc{q}_3}_{\textsc{p}_2} \, \hat{E}_{\textsc{p}_1\textsc{p}_2}\hat{E}_{\textsc{q}_1\textsc{q}_2} \hat{E}_{\textsc{q}_3\textsc{q}_4} +
    \nu^{\textsc{q}_4\textsc{p}_1}_{\textsc{q}_2\textsc{q}_1}S_{\textsc{q}_3}^{\textsc{p}_2} \, \hat{E}_{\textsc{p}_2\textsc{p}_1}\hat{E}_{\textsc{q}_2\textsc{q}_1} \hat{E}_{\textsc{q}_4\textsc{q}_3}\right) 
      \nonumber \\
    &- \tfrac{1}{2} \sum_{\textbf{\textsc{p}}\textbf{\textsc{q}}} \left( v^{\textsc{p}_1\textsc{p}_2}_{\textsc{q}_1\textsc{q}_2} S^{\textsc{q}_3}_{\textsc{p}_4} S^{\textsc{p}_3}_{\textsc{q}_4}\hat{E}_{\textsc{p}_1\textsc{p}_2} \hat{E}_{\textsc{p}_3\textsc{p}_4} 
    \hat{E}_{\textsc{q}_1\textsc{q}_2} \hat{E}_{\textsc{q}_3\textsc{q}_4}  +
    v^{\textsc{p}_2\textsc{p}_1}_{\textsc{q}_2\textsc{q}_1} S_{\textsc{q}_3}^{\textsc{p}_4} S_{\textsc{p}_3}^{\textsc{q}_4}\hat{E}_{\textsc{p}_2\textsc{p}_1} \hat{E}_{\textsc{p}_4\textsc{p}_3}\hat{E}_{\textsc{q}_2\textsc{q}_1} \hat{E}_{\textsc{q}_4\textsc{q}_3}\right),
    \label{symmetric_VP}
\end{align}
where the renormalized tensor coefficients used in the chemist notation are defined as:
\begin{align}
    \nu^{\textsc{p}_1\textsc{q}_2}_{\textsc{q}_1\textsc{p}_2} &= v^{\textsc{p}_1\textsc{q}_2}_{\textsc{q}_1\textsc{p}_2} - \sum_{\textsc{q}_3} S^{\textsc{q}_3}_{\textsc{p}_2} {v}^{\textsc{p}_1\textsc{q}_2}_{\textsc{q}_1\textsc{q}_4}  - \sum_{\textsc{p}_3} v^{\textsc{p}_1\textsc{p}_3}_{\textsc{q}_1\textsc{p}_2} S^{\textsc{p}_3}_{\textsc{q}_2} + \sum_{\textsc{p}_3\textsc{q}_3} v^{\textsc{p}_1\textsc{p}_3}_{\textsc{q}_1\textsc{q}_4} S^{\textsc{p}_3}_{\textsc{q}_2}S^{\textsc{p}_2}_{\textsc{q}_3}, \\
    \nu^{\textsc{p}_1\textsc{p}_2}_{\textsc{q}_1\textsc{p}_4} &= v^{\textsc{p}_1\textsc{p}_2}_{\textsc{q}_1\textsc{p}_4}  - \sum_{\textsc{q}_3} v^{\textsc{p}_1\textsc{p}_2}_{\textsc{q}_1\textsc{q}_4} S^{\textsc{p}_4}_{\textsc{q}_3}, \\
    \nu^{\textsc{p}_1\textsc{q}_4}_{\textsc{q}_1\textsc{q}_2} &= v_{\textsc{q}_1\textsc{q}_2}^{\textsc{p}_1\textsc{q}_4} - \sum_{\textsc{p}_3} v^{\textsc{p}_1\textsc{p}_3}_{\textsc{q}_1\textsc{q}_2} S^{\textsc{p}_3}_{\textsc{q}_4}.
\end{align}
The renormalized tensor coefficients for the complex conjugate terms are also defined appropriately. In summary, these equations recover the final expressions in the main paper using the spin-orbital notation. In this form, it is possible to see that the first term in the fourth line of Eq.~\eqref{symmetric_VP} exactly equals the product of the electrostatic and exchange operators, $\hat{V}\hat{P}$. Using the symmetry, $\hat{e}_{\textsc{p}_1\textsc{p}_2\textsc{p}_3\textsc{p}_4} = \hat{e}_{\textsc{p}_3\textsc{p}_4\textsc{p}_1\textsc{p}_2}$ and $\hat{e}_{\textsc{q}_1\textsc{q}_2\textsc{q}_3\textsc{q}_4} = \hat{e}_{\textsc{q}_3\textsc{q}_4\textsc{q}_1\textsc{q}_2}$, it is possible to see that the second term equals the product $\hat{P}\hat{V}$. The first three lines are now more clearly contributions that arise from the truncated basis set. In Appendix B, we show how these terms cancel exactly in the complete basis set limit. 
\newpage
\subsection{SAPT operator symmetries}
In the following, we discuss the permutational symmetries inherent to the SAPT operators defined above. For complex orbitals, the four-index intermolecular tensor obeys the Hermitian symmetry,
\begin{equation}
    v^{\textsc{p}_1\textsc{p}_2}_{\textsc{q}_1\textsc{q}_2} = [v^{\textsc{p}_2\textsc{p}_1}_{\textsc{q}_2\textsc{q}_1}]^*,
\end{equation}
while the intermolecular overlap matrix obeys,
\begin{equation}
    S^{\textsc{p}}_\textsc{q} = [S^\textsc{q}_\textsc{p}]^*.
\end{equation}
For real orbitals, the intermolecular tensor obeys the four-fold symmetry,
\begin{equation}
    v^{\textsc{p}_1\textsc{p}_2}_{\textsc{q}_1\textsc{q}_2} = v^{\textsc{p}_2\textsc{p}_1}_{\textsc{q}_1\textsc{q}_2} = v^{\textsc{p}_1\textsc{p}_2}_{\textsc{q}_2\textsc{q}_1} = v^{\textsc{p}_2\textsc{p}_1}_{\textsc{q}_2\textsc{q}_1},
\end{equation}
while the intermolecular overlap matrix symmetry becomes,
\begin{equation}
    S^{\textsc{p}}_\textsc{q} = S^{\textsc{q}}_\textsc{p}. 
\end{equation}
It is important to note that while the two-electron integral, $(\textsc{p}_1\textsc{p}_2|\textsc{q}_1\textsc{q}_2) = \int \mathrm{d}\mathbf{r}_i \mathrm{d}\mathbf{r}_j\;   \phi_{\textsc{p}_1}^*(\mathbf{r}_i)\phi_{\textsc{p}_2}(\mathbf{r}_i) r_{ij}^{-1}\phi_{\textsc{q}_1}^*(\mathbf{r}_j)\phi_{\textsc{q}_2}(\mathbf{r}_j)$, obeys the symmetry, $(\textsc{p}_1\textsc{p}_2|\textsc{q}_1\textsc{q}_2) = (\textsc{q}_1\textsc{q}_2|\textsc{p}_1\textsc{p}_2)$, the full intermolecular tensor in Eq.~\eqref{eq:intermolecular_tensor} does not obey this type of symmetry. The intermolecular tensor only has a four-fold symmetry as opposed to the typical eight-fold symmetry encountered in the conventional quantum chemistry Hamiltonian. These considerations will be important for the development of the SAPT-EVE algorithm and the block encoding methodologies encountered shortly. For the rest of the appendices as well as the main manuscript, we will assume real orbitals for all of the calculations and resource estimates. As a result, we will also assume that the prepared ground-state wavefunctions will be fully real resulting in the following permutational symmetries for the reduced density matrix elements of monomer A:
\begin{align}
    D^A_{\textsc{p}_1\textsc{p}_2} &= D^A_{\textsc{p}_2\textsc{p}_1}, \label{permutational_symmetries_1RDM_real_A}\\
    d^A_{\textsc{p}_1\textsc{p}_2\textsc{p}_3\textsc{p}_4} &= d^A_{\textsc{p}_2\textsc{p}_1\textsc{p}_4\textsc{p}_3} = d^A_{\textsc{p}_3\textsc{p}_4\textsc{p}_1\textsc{p}_2}  = d^A_{\textsc{p}_4\textsc{p}_3\textsc{p}_2\textsc{p}_1},
    \label{permutational_symmetries_2RDM_real_A}
\end{align}
as well as monomer B:
\begin{align}
    D^B_{\textsc{q}_1\textsc{q}_2} &= D^B_{\textsc{q}_2\textsc{q}_1}, \label{permutational_symmetries_1RDM_real_B}\\
    d^B_{\textsc{q}_1\textsc{q}_2\textsc{q}_3\textsc{q}_4} &= d^B_{\textsc{q}_2\textsc{q}_1\textsc{q}_4\textsc{q}_3} = d^B_{\textsc{q}_3\textsc{q}_4\textsc{q}_1\textsc{q}_2}  = d^B_{\textsc{q}_4\textsc{q}_3\textsc{q}_2\textsc{q}_1}.
    \label{permutational_symmetries_2RDM_real_B}
\end{align}
Taking into account all of these symmetries, we re-write all of the SAPT operators as:
\begin{align}
   \hat{V} = &\;\;\;\sum_{\textbf{\textsc{p}},\textbf{\textsc{q}}} \text{sym}(v^{\textsc{p}_1\textsc{p}_2}_{\textsc{q}_1\textsc{q}_2})\hat{E}_{\textsc{p}_1\textsc{p}_2} \hat{E}_{\textsc{q}_1\textsc{q}_2} \, , \label{eq:V_real_sym}\\
   \hat{P} = &-\sum_{\substack{\textbf{\textsc{p}},\textbf{\textsc{q}}}} \text{sym}(S^{\textsc{p}_1}_{\textsc{q}_2}S^{\textsc{p}_2}_{\textsc{q}_1} )\hat{E}_{\textsc{p}_1\textsc{p}_2} \hat{E}_{\textsc{q}_1\textsc{q}_2}, \, \label{eq:P_real_sym} \\
    \widehat{VP}_\mathrm{s} = &-\sum_{\substack{\textbf{\textsc{p}},\textbf{\textsc{q}}}} \text{sym}(\bar{\nu}^{\textsc{p}_1\textsc{q}_2}_{\textsc{q}_1\textsc{p}_2})\hat{E}_{\textsc{p}_1\textsc{p}_2}\hat{E}_{\textsc{q}_1\textsc{q}_2}     \label{eq:VP_real_sym} \\ 
    &- \sum_{\substack{\textbf{\textsc{p}},\textbf{\textsc{q}}}} \text{sym}( \nu^{\textsc{p}_1\textsc{p}_2}_{\textsc{q}_1\textsc{p}_4} S^{\textsc{p}_3}_{\textsc{q}_2} )\hat{E}_{\textsc{p}_1\textsc{p}_2}\hat{E}_{\textsc{p}_3\textsc{p}_4}\hat{E}_{\textsc{q}_1\textsc{q}_2} \nonumber \\
    &- \sum_{\substack{\textbf{\textsc{p}},\textbf{\textsc{q}}}} \text{sym}(\nu_{\textsc{q}_1\textsc{q}_2}^{\textsc{p}_1\textsc{q}_4}S^{\textsc{p}_2}_{\textsc{q}_3} )
    \hat{E}_{\textsc{p}_1\textsc{p}_2} \hat{E}_{\textsc{q}_1\textsc{q}_2}\hat{E}_{\textsc{q}_3\textsc{q}_4} \nonumber \\
    &- \sum_{\substack{\textbf{\textsc{p}},\textbf{\textsc{q}}}} \text{sym}(v^{\textsc{p}_1\textsc{p}_2}_{\textsc{q}_1\textsc{q}_2} S^{\textsc{p}_3}_{\textsc{q}_4}S^{\textsc{p}_4}_{\textsc{q}_3} )\hat{E}_{\textsc{p}_1\textsc{p}_2}\hat{E}_{\textsc{p}_3\textsc{p}_4} \hat{E}_{\textsc{q}_1\textsc{q}_2}\hat{E}_{\textsc{q}_3\textsc{q}_4} \, , \nonumber      
\end{align}
where 
\begin{align}
\bar{\nu}^{\textsc{p}_1\textsc{q}_2}_{\textsc{q}_1\textsc{p}_2} &=  v^{\textsc{p}_1\textsc{q}_2}_{\textsc{q}_1\textsc{p}_2}  + \tfrac{1}{2}\Big[\sum_{\textsc{p}_3\textsc{q}_3} (v^{\textsc{p}_1\textsc{p}_3}_{\textsc{q}_1\textsc{q}_3} S^{\textsc{p}_3}_{\textsc{q}_2}S^{\textsc{p}_2}_{\textsc{q}_3} + 
v^{\textsc{p}_1\textsc{p}_3}_{\textsc{q}_1\textsc{q}_3} S^{\textsc{p}_3}_{\textsc{q}_3}S^{\textsc{p}_2}_{\textsc{q}_2}) - \sum_{\textsc{q}_3}(v_{\textsc{q}_1\textsc{q}_3}^{\textsc{p}_1\textsc{q}_2}S^{\textsc{p}_2}_{\textsc{q}_3} + v_{\textsc{q}_1\textsc{q}_3}^{\textsc{p}_1\textsc{q}_3}S^{\textsc{p}_2}_{\textsc{q}_2}) -  \sum_{\textsc{p}_3} (v^{\textsc{p}_1\textsc{p}_3}_{\textsc{q}_1\textsc{p}_2} S^{\textsc{p}_3}_{\textsc{q}_2} + v^{\textsc{p}_1\textsc{p}_3}_{\textsc{q}_1\textsc{p}_3} S^{\textsc{p}_2}_{\textsc{q}_2}) \Big] \label{eq:vpqqp_dressed} \\   
\nu^{\textsc{p}_1\textsc{p}_2}_{\textsc{q}_1\textsc{p}_4} &= v^{\textsc{p}_1\textsc{p}_2}_{\textsc{q}_1\textsc{p}_4} - \sum_{\textsc{q}_3}v^{\textsc{p}_1\textsc{p}_2}_{\textsc{q}_1\textsc{q}_3} S^{\textsc{p}_4}_{\textsc{q}_3} \, , \\
\nu_{\textsc{q}_1\textsc{q}_2}^{\textsc{p}_1\textsc{q}_4} &= v_{\textsc{q}_1\textsc{q}_2}^{\textsc{p}_1\textsc{q}_4} - \sum_{\textsc{p}_3}v^{\textsc{p}_1\textsc{p}_3}_{\textsc{q}_1\textsc{q}_2} S^{\textsc{p}_3}_{\textsc{q}_4},
\end{align}
where we emphasize that $\bar{\nu}^{\textsc{p}_1\textsc{q}_2}_{\textsc{q}_1\textsc{p}_2}$ is different from $\nu^{\textsc{p}_1\textsc{q}_2}_{\textsc{q}_1\textsc{p}_2}$ in the main text. The operator, $\text{sym}(\cdot)$, symmetrizes all of the tensors with respect to the monomer indices $\textsc{p}/\textsc{q}$ independently using the permutational symmetries associated with the 1- and 2- RDMS in Eqs.~\eqref{permutational_symmetries_1RDM_real_A}-\eqref{permutational_symmetries_2RDM_real_B}. The permutational-invariant SAPT tensors are written explicitly as:
\begin{align}
    \text{sym}(S^{\textsc{p}_1}_{{\textsc{q}_2}}S^{{\textsc{p}_2}}_{{\textsc{q}_1}}) &= \tfrac{1}{4}(S^{\textsc{p}_1}_{{\textsc{q}_2}}S^{{\textsc{p}_2}}_{{\textsc{q}_1}} + S^{{\textsc{p}_2}}_{{\textsc{q}_2}}S^{{\textsc{p}_1}}_{{\textsc{q}_1}} + S^{\textsc{p}_1}_{{\textsc{q}_1}}S^{{\textsc{p}_2}}_{{\textsc{q}_2}} + S^{{\textsc{p}_2}}_{{\textsc{q}_1}}S^{{\textsc{p}_1}}_{{\textsc{q}_2}}) = \tfrac{1}{2}(S^{\textsc{p}_1}_{{\textsc{q}_2}}S^{{\textsc{p}_2}}_{{\textsc{q}_1}} + S^{{\textsc{p}_2}}_{{\textsc{q}_2}}S^{{\textsc{p}_1}}_{{\textsc{q}_1}} ), \\
    \text{sym}(\nu^{{\textsc{p}_1}{\textsc{p}_2}}_{{\textsc{q}_1}{\textsc{q}_2}}) &= \tfrac{1}{4}( \nu^{{\textsc{p}_1}{\textsc{p}_2}}_{{\textsc{q}_1}{\textsc{q}_2}} + \nu^{{\textsc{p}_2}{\textsc{p}_1}}_{{\textsc{q}_1}{\textsc{q}_2}} + \nu^{{\textsc{p}_1}{\textsc{p}_2}}_{{\textsc{q}_2}{\textsc{q}_1}} + \nu^{{\textsc{p}_2}{\textsc{p}_1}}_{{\textsc{q}_2}{\textsc{q}_1}} ),  \\
     \text{sym}(\bar{\nu}^{\textsc{p}_1\textsc{q}_2}_{\textsc{q}_1\textsc{p}_2}) &= \tfrac{1}{4}( \bar{\nu}^{\textsc{p}_1\textsc{q}_2}_{\textsc{q}_1\textsc{p}_2} + \bar{\nu}^{\textsc{p}_2\textsc{q}_2}_{\textsc{q}_1\textsc{p}_1} + \bar{\nu}^{\textsc{p}_1\textsc{q}_1}_{\textsc{q}_2\textsc{p}_2} + \bar{\nu}^{\textsc{p}_2\textsc{q}_1}_{\textsc{q}_2\textsc{p}_1}), \\
     \text{sym}(\nu^{\textsc{p}_1\textsc{p}_2}_{\textsc{q}_1\textsc{p}_4} S^{\textsc{p}_3}_{\textsc{q}_2}) &= \tfrac{1}{8}( 
     \nu^{\textsc{p}_1\textsc{p}_2}_{\textsc{q}_1\textsc{p}_4} S^{\textsc{p}_3}_{\textsc{q}_2} + \nu^{\textsc{p}_2\textsc{p}_1}_{\textsc{q}_1\textsc{p}_3} S^{\textsc{p}_4}_{\textsc{q}_2} + \nu^{\textsc{p}_3\textsc{p}_4}_{\textsc{q}_1\textsc{p}_2} S^{\textsc{p}_1}_{\textsc{q}_2} + \nu^{\textsc{p}_4\textsc{p}_3}_{\textsc{q}_1\textsc{p}_1} S^{\textsc{p}_2}_{\textsc{q}_2} \\
     &+\nu^{\textsc{p}_1\textsc{p}_2}_{\textsc{q}_2\textsc{p}_4} S^{\textsc{p}_3}_{\textsc{q}_1} + \nu^{\textsc{p}_2\textsc{p}_1}_{\textsc{q}_2\textsc{p}_3} S^{\textsc{p}_4}_{\textsc{q}_1} + \nu^{\textsc{p}_3\textsc{p}_4}_{\textsc{q}_2\textsc{p}_2} S^{\textsc{p}_1}_{\textsc{q}_1} + \nu^{\textsc{p}_4\textsc{p}_3}_{\textsc{q}_2\textsc{p}_1} S^{\textsc{p}_2}_{\textsc{q}_1}), \nonumber\\
     \text{sym}(\nu^{\textsc{p}_1\textsc{q}_4}_{\textsc{q}_1\textsc{q}_2} S^{\textsc{p}_2}_{\textsc{q}_3}) &= \tfrac{1}{8}( 
     \nu^{\textsc{p}_1\textsc{q}_4}_{\textsc{q}_1\textsc{q}_2} S^{\textsc{p}_2}_{\textsc{q}_3} + \nu^{\textsc{p}_1\textsc{q}_3}_{\textsc{q}_2\textsc{q}_1} S^{\textsc{p}_2}_{\textsc{q}_4} + \nu^{\textsc{p}_1\textsc{q}_2}_{\textsc{q}_3\textsc{q}_4} S^{\textsc{p}_2}_{\textsc{q}_1} + \nu^{\textsc{p}_1\textsc{q}_1}_{\textsc{q}_4\textsc{q}_3} S^{\textsc{p}_2}_{\textsc{q}_2} \\
     &+\nu^{\textsc{p}_2\textsc{q}_4}_{\textsc{q}_1\textsc{q}_2} S^{\textsc{p}_1}_{\textsc{q}_3} + \nu^{\textsc{p}_2\textsc{q}_3}_{\textsc{q}_2\textsc{q}_1} S^{\textsc{p}_1}_{\textsc{q}_4} + \nu^{\textsc{p}_2\textsc{q}_2}_{\textsc{q}_3\textsc{q}_4} S^{\textsc{p}_1}_{\textsc{q}_1} + \nu^{\textsc{p}_2\textsc{q}_1}_{\textsc{q}_4\textsc{q}_3} S^{\textsc{p}_1}_{\textsc{q}_2}), \nonumber \\
     \text{sym}(v^{\textsc{p}_1\textsc{p}_2}_{\textsc{q}_1\textsc{q}_2} S^{\textsc{p}_3}_{\textsc{q}_4}S^{\textsc{p}_4}_{\textsc{q}_3} ) &= \tfrac{1}{16}\Big[ 
        v^{\textsc{p}_1\textsc{p}_2}_{\textsc{q}_1\textsc{q}_2} S^{\textsc{p}_3}_{\textsc{q}_4}S^{\textsc{p}_4}_{\textsc{q}_3} + v^{\textsc{p}_1\textsc{p}_2}_{\textsc{q}_2\textsc{q}_1} S^{\textsc{p}_3}_{\textsc{q}_3}S^{\textsc{p}_4}_{\textsc{q}_4} + v^{\textsc{p}_1\textsc{p}_2}_{\textsc{q}_3\textsc{q}_4} S^{\textsc{p}_3}_{\textsc{q}_1}S^{\textsc{p}_4}_{\textsc{q}_2} +  v^{\textsc{p}_1\textsc{p}_2}_{\textsc{q}_4\textsc{q}_3} S^{\textsc{p}_3}_{\textsc{q}_2}S^{\textsc{p}_4}_{\textsc{q}_1}  \\
     &\;\;\;\;\;\; + v^{\textsc{p}_2\textsc{p}_1}_{\textsc{q}_1\textsc{q}_2} S^{\textsc{p}_4}_{\textsc{q}_4}S^{\textsc{p}_3}_{\textsc{q}_3} + v^{\textsc{p}_2\textsc{p}_1}_{\textsc{q}_2\textsc{q}_1} S^{\textsc{p}_4}_{\textsc{q}_3}S^{\textsc{p}_3}_{\textsc{q}_4} + v^{\textsc{p}_2\textsc{p}_1}_{\textsc{q}_3\textsc{q}_4} S^{\textsc{p}_4}_{\textsc{q}_1}S^{\textsc{p}_3}_{\textsc{q}_2} +  v^{\textsc{p}_2\textsc{p}_1}_{\textsc{q}_4\textsc{q}_3} S^{\textsc{p}_4}_{\textsc{q}_2}S^{\textsc{p}_3}_{\textsc{q}_1}  \nonumber \\
     &\;\;\;\;\;\; + v^{\textsc{p}_3\textsc{p}_4}_{\textsc{q}_1\textsc{q}_2} S^{\textsc{p}_1}_{\textsc{q}_4}S^{\textsc{p}_2}_{\textsc{q}_3} + v^{\textsc{p}_3\textsc{p}_4}_{\textsc{q}_2\textsc{q}_1} S^{\textsc{p}_1}_{\textsc{q}_3}S^{\textsc{p}_2}_{\textsc{q}_4} + v^{\textsc{p}_3\textsc{p}_4}_{\textsc{q}_3\textsc{q}_4} S^{\textsc{p}_1}_{\textsc{q}_1}S^{\textsc{p}_2}_{\textsc{q}_2} +  v^{\textsc{p}_3\textsc{p}_4}_{\textsc{q}_4\textsc{q}_3} S^{\textsc{p}_1}_{\textsc{q}_2}S^{\textsc{p}_2}_{\textsc{q}_1} \nonumber \\
     &\;\;\;\;\;\; + v^{\textsc{p}_4\textsc{p}_3}_{\textsc{q}_1\textsc{q}_2} S^{\textsc{p}_2}_{\textsc{q}_4}S^{\textsc{p}_1}_{\textsc{q}_3} + v^{\textsc{p}_4\textsc{p}_3}_{\textsc{q}_2\textsc{q}_1} S^{\textsc{p}_2}_{\textsc{q}_3}S^{\textsc{p}_1}_{\textsc{q}_4} + v^{\textsc{p}_4\textsc{p}_3}_{\textsc{q}_3\textsc{q}_4} S^{\textsc{p}_2}_{\textsc{q}_1}S^{\textsc{p}_1}_{\textsc{q}_2} +  v^{\textsc{p}_4\textsc{p}_3}_{\textsc{q}_4\textsc{q}_3} S^{\textsc{p}_2}_{\textsc{q}_2}S^{\textsc{p}_1}_{\textsc{q}_1} \Big]. \nonumber
\end{align}

\subsection{Spatial orbital basis}
\label{sec:Appendix_Spatial_Orbitals}
The intermolecular interaction and overlap tensors in the spin-orbital basis are related to the corresponding spatial orbital tensors through the expressions,
\begin{align}
    {v}^{\textsc{p}_1\textsc{p}_2}_{\textsc{q}_1\textsc{q}_2} &= {v}^{p_1p_2}_{q_1q_2}\delta_{\sigma_1\sigma_2}\delta_{\tau_1\tau_2},
\end{align}
and
\begin{equation}
    S^\textsc{p}_\textsc{q} = S^p_q \delta_{\sigma\tau}.
\end{equation}
The spatial orbital basis reduces the memory storage requirements on the classical computer while also speeding up the factorization procedures (i.e. eigendecomposition and/or singular value decomposition) required for the tensor factorization techniques discussed in the main manuscript. As we show below, the spatial orbital basis also helps reduce the total number of terms required for the data-loading oracle on the quantum computer by exploiting the spin-projection symmetries for each of the two monomers defined as:
\begin{align}
    \hat{S}_z^{(A)} &= \tfrac{1}{2}\sum_p( \hat{a}^\dagger_{p\alpha}\hat{a}_{p\alpha} - \hat{a}^\dagger_{p\beta}\hat{a}_{p\beta}),
\end{align}
and
\begin{align}
    \hat{S}_z^{(B)} &= \tfrac{1}{2}\sum_q( \hat{b}^\dagger_{q\alpha}\hat{b}_{q\alpha} - \hat{b}^\dagger_{q\beta}\hat{b}_{q\beta}).
\end{align}
Converting into the spatial orbital basis and taking into account the spin-projection symmetries for each monomer, we find that the electrostatic and exchange operators may be written as:
\begin{align}
   \hat{V} = &\;\;\;\,\sum_{\substack{\mathbf{p}\mathbf{q}\\\sigma\tau}} \text{sym}(v^{p_1p_2}_{q_1q_2})\hat{E}_{p_1p_2}^\sigma \hat{E}_{q_1q_2}^\tau \, , \label{eq:V_symm} \\
   \hat{P} = &-\sum_{\substack{\textbf{p}\textbf{q}\\ \sigma}} \text{sym}(S^{p_1}_{q_2}S^{p_2}_{q_1} )\hat{E}_{p_1p_2}^\sigma \hat{E}_{q_1q_2}^\sigma,  \label{eq:P_symm} 
\end{align}
while the electrostatic-exchange operator arising from $V_cP_c$ in Eq.~\eqref{VP_initial} is written as:
\begin{align}
    \widehat{VP} = &-\sum_{\substack{\textbf{p}\textbf{q} \\ \sigma }} \left({\nu}^{p_1q_2}_{q_1p_2} \hat{E}^\sigma_{p_1p_2}\hat{E}^\sigma_{q_1q_2} 
    + \nu^{p_1p_2}_{q_1p_4} S^{p_3}_{q_2} \hat{E}^+_{p_1p_2}\hat{E}^\sigma_{p_3p_4}\hat{E}^\sigma_{q_1q_2} 
    +  \nu_{q_1q_2}^{p_1q_4}S^{p_2}_{q_3} 
    \hat{E}^\sigma_{p_1p_2} \hat{E}^+_{q_1q_2}\hat{E}^\sigma_{q_3q_4} \right) 
    + \hat{V}\hat{P}   .   
    \label{VP_spatial_orbital}
\end{align}

\section{Complete basis set limit}
\label{sec:Complete_basis_set_limit}
In the following, we show that the equality, $\widehat{VP}=\hat{V}\hat{P}$, holds in the complete basis limit for the electrostatic-exchange operator defined in the spatial orbital basis defined by Eq.~\eqref{VP_spatial_orbital}. The proof is based on the identity,
\begin{equation}
    \delta(\mathbf{r}-\mathbf{r}') = \sum_n \phi^*_n(\mathbf{r})\phi_n(\mathbf{r}'),
\end{equation}
which only holds for complete basis sets. This identity leads to the following formula in the spatial orbital basis,
\begin{align}
    \sum_{p_3} S^{p_3}_{q_4}v^{p_1 p_3}_{q_1 q_2} &= \sum_{p_3} \iint\! d\mathbf{r} d\mathbf{r}_i d\mathbf{r}_j\, \phi^*_{p_3}(\mathbf{r})\phi_{q_4}(\mathbf{r})\phi^*_{p_1}(\mathbf{r}_i)\phi_{p_3}(\mathbf{r}_i)v_{ij}\phi^*_{q_1}(\mathbf{r}_j)\phi_{q_2}(\mathbf{r}_j) \\
    &= \iint\! d\mathbf{r} d\mathbf{r}_i d\mathbf{r}_j\,\delta(\mathbf{r}-\mathbf{r}_i)\phi_{q_4}(\mathbf{r})\phi^*_{p_1}(\mathbf{r}_i)v_{ij}\phi^*_{q_1}(\mathbf{r}_j)\phi_{q_2}(\mathbf{r}_j) \\
    &= v^{p_1q_4}_{q_1 q_2},
\end{align}
which is used throughout the calculation. Starting with the third term of Eq.~\eqref{VP_spatial_orbital}, we have:
\begin{align}
    \sum_{\substack{\textbf{p}\textbf{q} \\ \sigma }}\nu_{q_1q_2}^{p_1q_4}S_{p_2}^{q_3} 
    \hat{E}^\sigma_{p_1p_2} \hat{E}^+_{q_1q_2}\hat{E}^\sigma_{q_3q_4} &= \sum_{\substack{\textbf{p}\textbf{q} \\ \sigma }} \Big(v_{q_1q_2}^{p_1q_4}S_{p_2}^{q_3} - \sum_{p} v^{p_1p}_{q_1q_2} S^{p}_{q_4}S_{p_2}^{q_3}\Big) E_{p_1p_2}^{\sigma} \hat{E}^{+}_{q_1q_2} \hat{E}^{\sigma}_{q_3 q_4} \\
    &= \sum_{\substack{\textbf{p}\textbf{q} \\ \sigma }} \Big( v_{q_1q_2}^{p_1q_4}S_{p_2}^{q_3} - v^{p_1q_4}_{q_1q_2} S_{p_2}^{q_3}\Big) E_{p_1p_2}^{\sigma} \hat{E}^{+}_{q_1q_2} \hat{E}^{\sigma}_{q_3 q_4} \\
    &= 0.
\end{align}
The second term in Eq.~\eqref{VP_spatial_orbital} is evaluated as:
\begin{align}
    \sum_{\substack{\textbf{p}\textbf{q} \\ \sigma }} \nu^{p_1p_2}_{q_1p_4} S^{p_3}_{q_2} \hat{E}^+_{p_1p_2}\hat{E}^\sigma_{p_3p_4}\hat{E}^\sigma_{q_1q_2} &= \sum_{\substack{\textbf{p}\textbf{q} \\ \sigma }}\Big(v^{p_1p_2}_{q_1p_4} S^{p_3}_{q_2} - \sum_{q} v^{p_1p_2}_{q_1q} S^{p_3}_{q_2}S_{p_4}^{q} \Big) E_{p_1p_2}^{+}E_{p_3p_4}^{\sigma} \hat{E}^{\sigma}_{q_1q_2} \\
    &= \sum_{\substack{\textbf{p}\textbf{q} \\ \sigma }} \Big(v^{p_1p_2}_{q_1p_4} S^{p_3}_{ q_2} - v^{p_1p_2}_{q_1p_4} S^{p_3}_{q_2}\Big) E_{p_1p_2}^{+}E_{p_3p_4}^{\sigma} \hat{E}^{\sigma}_{q_1q_2} \\
    &= 0,
\end{align}
while the first term is given by:
\begin{align}
     \sum_{\substack{\textbf{p}\textbf{q} \\ \sigma }} {\nu}^{p_1q_2}_{q_1p_2} \hat{E}^\sigma_{p_1p_2}\hat{E}^\sigma_{q_1q_2}  &= \sum_{\substack{\textbf{p}\textbf{q} \\ \sigma }} \Big(v^{p_1q_2}_{q_1p_2} - \sum_{q} v_{q_1q}^{p_1q_2}S^{q}_{p_2} - \sum_{p} v^{p_1p}_{q_1p_2} S^{p}_{ q_2} + \sum_{pq} v^{p_1p}_{q_1q} S^{p}_{q_2}S_{p_2}^{q} \Big) E_{p_1p_2}^{\sigma} E_{q_1q_2}^{\sigma} \\
     &= \sum_{\substack{\textbf{p}\textbf{q} \\ \sigma }} \Big(v^{p_1q_2}_{q_1p_2} - v_{q_1p_2}^{p_1q_2} - v^{p_1q_2}_{q_1p_2} + v^{p_1q_2}_{q_1p_2} \Big) E_{p_1p_2}^{\sigma} E_{q_1q_2}^{\sigma} \\
     &= 0.
\end{align}
As a result, we are only left with the fourth term such that, $\widehat{VP} = \hat{V}\hat{P}$. Ultimately, this explains the form of $\widehat{VP}$ consisting of four terms compared the simpler $\hat{V}\hat{P}$ term that one might have expected from the beginning. It is important to note that this equality does not hold exactly for the electrostatic-exchange operator defined by Eq. \eqref{eq:VP_real_sym}. Further studies of these properties and their relevance to the SAPT-EVE algorithm and the $\ell_1$ norm is left for future work.

\section{Active Space Formulation}\label{sec:Appendix_Active_Space}

In the following, we develop an active space formulation of SAPT where the orbitals are decomposed into three parts: (1) core orbitals which are assumed to be fully occupied, (2) active space orbitals where we expect the main physics/chemistry to occur, and (3) virtual orbitals which have higher energy than the active space orbitals but provide a negligible contribution to the ground-state energy. Active space methods are important for the description of large-scale systems. In the following, we use a notation where $i/t$ and $j/u$ indices correspond to the core/active orbitals of monomers A and B respectively. The active space decomposition is illustrated for the electrostatic (Sec. \ref{subsec:V_active_space}), exchange (Sec. \ref{subsec:P_active_space}), and electrostatic-exchange (Sec. \ref{subsec:VP_active_space}) operators below. Throughout this appendix, we will use Einstein notation exclusively with the core orbital indices ($i/j$) to reduce the verbosity of the equations. We will also ignore the symmetrization of the tensor elements but will discuss the symmetrized version of the active space expressions near the end. Real orbitals are also assumed throughout.

\subsection{Electrostatics}\label{subsec:V_active_space}
\noindent
We first consider the full-space electrostatic operator,    
\begin{align}
    \hat{V} &= \sum_{\textbf{\textsc{p}}\textbf{\textsc{q}}} {v}^{\textsc{p}_1\textsc{p}_2}_{\textsc{q}_1\textsc{q}_2} \hat{E}_{\textsc{p}_1\textsc{p}_2}\hat{E}_{\textsc{q}_1\textsc{q}_2} = \sum_{\substack{\textbf{{p}}\textbf{{q}} \\ \sigma\tau} } {v}^{p_1p_2}_{q_1q_2} \hat{E}^\sigma_{p_1p_2}\hat{E}^\tau_{q_1q_2}
\end{align}
The active space operator is derived by tracing out the inactive degrees of freedom with a Hartree-Fock-like wavefunction in the inactive space. As a result, we require decomposing the summation of each of the indices in terms of core $i/j$ and active space $t/u$ indices, 
\begin{align}
     \sum_{p_1p_2\in A}\sum_{q_1q_2 \in B} &= \sum_{i_1i_2\in A}\sum_{j_1j_2 \in B} + \sum_{t_1t_2\in A}\sum_{j_1j_2 \in B} + \sum_{i_1i_2\in A}\sum_{u_1u_2 \in B} + \sum_{t_1t_2\in A}\sum_{u_1u_2 \in B}.
\end{align}
The virtual orbitals are assumed to remain unoccupied therefore they will not contribute to the active space renormalization procedure. By tracing out the core orbitals, the active space electrostatic operator may be written as:
\begin{align}
    \hat{V}_{\mathrm{active}}
    &= v^{i_1i_2}_{j_1j_2} D^A_{i_1i_2}D^B_{j_1j_2} + \sum_{ t_1t_2\in A} \!\!v^{t_1t_2}_{j_1j_2}\hat{E}^+_{t_1t_2} D^B_{j_1j_2} + \sum_{ u_1u_2 \in B} \!\!v^{i_1i_2}_{u_1u_2} D^A_{i_1i_2} \hat{E}^+_{u_1u_2} + \sum_{ \substack{ t_1t_2\in A \\ u_1u_2 \in B}} \!\! v^{t_1t_2}_{u_1u_2} \hat{E}^+_{t_1t_2}\hat{E}^+_{u_1u_2}.
\end{align}
defined with respect to the core-orbital-based reduced density matrix, $D_{i_1i_2}^{(A)}$/$D_{j_1j_2}^{B}$, for monomers A/B. Assuming that the core orbitals are doubly occupied in a closed-shell configuration, $D^X_{nn'} = 2\delta_{nn'}$, we obtain the following active space electrostatic operator:
\begin{equation}
    \hat{V}_{\mathrm{active}} = 4 v^{ii}_{jj} + 2\sum_{\mathbf{t}} v^{t_1t_2}_{jj} \hat{E}^+_{t_1t_2} + 2\sum_{\mathbf{u}}  v^{ii}_{u_1u_2} \hat{E}^+_{u_1u_2} + \sum_{\mathbf{t}\mathbf{u}}v^{t_1t_2}_{u_1u_2} \hat{E}^+_{t_1t_2}\hat{E}^+_{u_1u_2},
\end{equation}
Note that it is possible to combine all of the zero-body and one-body terms into a spin-independent four-index tensor:
\begin{align}
    \tilde{v}^{t_1t_2}_{u_1u_2} = v^{t_1t_2}_{u_1u_2} + 2v^{t_1t_2}_{jj}\delta_{u_1u_2}/\tilde{\eta}_{B} + 2v^{ii}_{u_1u_2}\delta_{t_1t_2}/\tilde{\eta}_{A} + 4 v^{ii}_{jj}\delta_{t_1t_2}\delta_{u_1u_2}/(\tilde{\eta}_{A}\tilde{\eta}_{B}),
\end{align}
where $\tilde{\eta}_{A}$ and $\tilde{\eta}_{B}$ correspond to the total number of electrons in the active spaces of monomer A and B respectively. Using this formulation, we obtain the final form of the active-space electrostatic operator:
\begin{equation}
    \hat{V}_{\mathrm{active}} = \sum_{\mathbf{t}\mathbf{u}}\tilde{v}^{t_1t_2}_{u_1u_2} \hat{E}^+_{t_1t_2}\hat{E}^+_{u_1u_2}.
\end{equation}

\subsection{Exchange}
\label{subsec:P_active_space}
The full space exchange term is given by Eq.~\eqref{full_exchange} which, following the same steps as above, results in the following expression for the active space exchange operator:
\begin{equation}
    \hat{P}_{\mathrm{active}} = -2S^{i}_{j}S_{i}^{j} - \sum_{\mathbf{t}} S^{t_1}_j S_{t_2}^{j} \hat{E}^+_{t_1t_2} - \sum_{\mathbf{u}} S_{i}^{u_1} S^{i}_{u_2} \hat{E}^+_{u_1u_2} - \sum_{\mathbf{t}\mathbf{u}} S_{u_2}^{t_1}S_{t_2}^{u_1} \hat{E}^\sigma_{t_1t_2}\hat{E}^\sigma_{u_1u_2}.
\end{equation}

\subsection{Electrostatic-Exchange}
\label{subsec:VP_active_space}
The derivation of the active space operators for the electrostatic-exchange term is much more involved. We proceed by deriving the active space operators for each of the four terms separately. For notational clarily, we also define 
$\hat{e}_{\mathbf{t}}^{\sigma_1\sigma_2}=(\hat{E}^{\sigma_1}_{t_1t_2}\hat{E}^{\sigma_2}_{t_3t_4} - \delta_{t_2t_3}^{\sigma_1\sigma_2}\hat{E}^{\sigma_2}_{t_1t_4} )$ and $\hat{e}_{\mathbf{u}}^{\sigma_1\sigma_2}=(\hat{E}^{\sigma_1}_{u_1u_2}\hat{E}^{\sigma_2}_{u_3u_4} - \delta_{u_2u_3}^{\sigma_1\sigma_2}\hat{E}^{\sigma_2}_{u_1u_4} )$.
\\
\subsubsection{$\widehat{VP}_1$} 
The first term of the full space electrostatic-exchange operator is given by Eq.~\eqref{VP_initial} which, following the same steps as above, results in the following expression for the active space operator:
\begin{align}
     \widehat{VP}_\mathrm{1,active} &= - 2v^{ij}_{ji} - \sum_{\mathbf{u}}v^{iu_2}_{u_1i}\hat{E}_{u_1u_2}^+ - \sum_{\mathbf{t}}v^{t_1j}_{jt_2}\hat{E}_{t_1t_2}^+ - \sum_{\substack{\mathbf{t}\mathbf{u}\\\sigma}}v^{t_1u_2}_{u_1t_2}\hat{E}^\sigma_{t_1t_2}\hat{E}^\sigma_{u_1u_2}.
\end{align}

\subsubsection{$\widehat{VP}_2$} 
For the second term, Eq.~\eqref{VP2_initial}, we find the following active space expression:
\begin{align}
     \widehat{VP}_\mathrm{2,active} = &- (4 {v}^{i_1i_1}_{j_1i_2}S^{i_2}_{j_1} - 2{v}^{i_2i_1}_{j_1i_2} S^{i_1}_{j_1} ) - \sum_{\mathbf{u}}(2 {v}^{i_1i_1}_{u_1i_2}S^{i_2}_{u_2} - {v}^{i_2i_1}_{u_1 i_2}S^{i_1}_{u_2} )\hat{E}_{u_1u_2}^+ \nonumber \\
    &-\sum_{\mathbf{t}}(2 {v}^{t_1t_2}_{j_1i_1}S^{i_1}_{j_1} + 2 {v}^{i_1i_1}_{j_1t_2}S^{t_1}_{j_1} - {v}^{t_1 i}_{j t_2}S^{i_1}_{j_1} - {v}^{i_1t_2}_{j_1i_1}S^{t_1}_{j_1} )\hat{E}_{t_1t_2}^+ - \sum_{\mathbf{t}}{v}^{t_1t_2}_{j_1t_4} S^{t_3}_{j_1}\hat{E}^+_{t_1t_2} \nonumber \\
    &- \sum_{\mathbf{t}\mathbf{u}} v^{t_1t_2}_{u_1i_1}S^{i_1}_{u_2}\hat{E}^+_{t_1t_2}\hat{E}^+_{u_1u_2} - \sum_{\substack{\mathbf{t}\mathbf{u}\\ \sigma}}(2v^{i_1i_1}_{u_1t_2}S^{t_1}_{u_2} - v^{t_1i_1}_{u_1t_2}S^{i_1}_{u_2} - v^{i_1t_2}_{u_1i_1}S^{t_1}_{u_2})\hat{E}^\sigma_{t_1t_2}\hat{E}^\sigma_{u_1u_2} \nonumber \\ 
    &- \sum_{\substack{\mathbf{t}\mathbf{u}\\\sigma_1\sigma_2}}{v}^{t_1t_2}_{u_1t_4}S^{t_3}_{u_2} \hat{e}^{\sigma_1\sigma_2}_{\mathbf{t}}\hat{E}^{\sigma_2}_{u_1u_2}.
\end{align}
\subsubsection{$\widehat{VP}_3$} 
Next, we find the following expression for the active space version of the third term, Eq.~\eqref{VP3_initial}:
\begin{align}
     \widehat{VP}_\mathrm{3,active} = &-(4 v^{i_1j_2}_{j_1j_2}S_i^{j_2} - 2 v^{i_1j_2}_{j_2j}S_{i_1}^{j_1}) - \sum_{\mathbf{t}}( v^{t_1j_2}_{j_1j_2}S_{t_2}^{j_2} - v^{t_1j_2}_{j_2j}S_{t_2}^{j_1})\hat{E}^+_{t_1t_2} \nonumber \\
    &- \sum_{\mathbf{u}}(2v^{i_1j_1}_{u_1u_2}S_{i_1}^{j_1} + 2v^{i_1u_2}_{j_1j_2}S_{i_1}^{u_1} - v^{i_1u_2}_{u_1j_1}S_{i_1}^{j_1} - v^{i_1j_1}_{ju_1}S_{i_1}^{u_2}) \hat{E}^+_{u_1u_2} - \sum_{\mathbf{u}}v^{i_1u_4}_{u_1u_2}S_{i_1}^{u_3}\hat{E}^+_{u_1u_2} \nonumber \\
    &- \sum_{\mathbf{t}\mathbf{u}}v^{t_1j_1}_{u_1u_2}S_{t_2}^{j_1} \hat{E}^+_{t_1t_2}\hat{E}^+_{u_1u_2} - \sum_{ \substack{\mathbf{t}\mathbf{u}\\\sigma}}(2 v^{t_1u_2}_{j_1j_2}S_{t_2}^{u_1} - v^{t_1u_2}_{u_1j_1}S_{t_2}^{j_1} - v^{t_1j_1}_{ju_1}S_{t_2}^{u_2} )\hat{E}^\sigma_{t_1t_2}\hat{E}^\sigma_{u_1u_2} \nonumber \\
    &- \sum_{\substack{\mathbf{t}\mathbf{u}\\\sigma_1\sigma_2}}v^{t_1u_4}_{u_1u_2} S_{t_2}^{u_3} \hat{E}^{\sigma_1}_{t_1t_2}\hat{e}^{\sigma_2\sigma_1}_{\mathbf{u}}.
\end{align}
\subsubsection{$\widehat{VP}_4$} 
Finally, we find the following active space operator for the fourth term, Eq.~\eqref{VP4_initial}: 
\begin{align}
    \widehat{VP}_\mathrm{4,active} =  
     &-(8 v^{i_1i_1}_{j_1j_2}S^{i_2}_{j_2}S_{i_2}^{j_2} - 4 v^{i_1i_2}_{j_1j_2}S^{i_2}_{j_2}S^{i_1}_{j_1} - 4 v^{i_1i_1}_{j_1j_2} S^{i_2}_{j_1} S^{i_2}_{j_2} + 2 v^{i_1i_2}_{j_1j_2} S^{i_2}_{j_1}S^{i_1}_{j_2}) \nonumber \\
     &-\sum_{\mathbf{u}}(4 v^{i_1i_1}_{ u_1u_2} S^{i_2}_{j_1}S^{i_2}_{j_1} - 2v^{i_1i_2}_{ u_1u_2} S^{i_2}_{j_1}S^{i_1}_{j_1} 
    + 4 v^{i_1i_1}_{ j_1j_2} S^{i_2}_{u_2}S^{i_2}_{u_1} - 2v^{i_1i_2}_{ j_1j_2} S^{i_2}_{u_2}S^{i_1}_{u_1} \nonumber \\
    &- 2 v^{i_1i_1}_{ u_1j_1} S^{i_2}_{u_2}S^{i_2}_{j_1} + v^{i_1i_2}_{ u_1j_1} S^{i_2}_{u_2}S^{i_1}_{j_1} 
    - 2 v^{i_1i_1}_{j_1u_1} S^{i_2}_{j_1}S^{i_2}_{u_2} + v^{i_1i_2}_{j_1u_1} S^{i_2}_{j_1}S^{i_1}_{u_2})\hat{E}^+_{u_1u_2} \nonumber \\
    &- \sum_{\substack{\mathbf{u}\\\sigma_1\sigma_2}}(2 v^{i_1i_1}_{ u_1u_2} S^{i_2}_{u_4}S^{i_2}_{u_3} - v^{i_1i_2}_{ u_1u_2} S^{i_2}_{u_4}S^{i_1}_{u_3})\hat{e}^+_{u_1u_2} \nonumber \\
    &- \sum_{\mathbf{t}}(4 v^{t_1t_2}_{j_1j_2} S^{i_1}_{j_2}S^{i_1}_{j_2} - 2v^{t_1t_2}_{ j_1j_2} S^{i_1}_{j_1}S^{i_1}_{j_2} 
    + 4 v^{i_1i_1}_{ j_1j_2} S^{t_1}_{j_2}S^{t_2}_{j_2} - 2v^{i_1i_1}_{ j_1j_2} S^{t_1}_{j_1}S^{t_2}_{j_2} \nonumber \\
    &- 2v^{t_1i_1}_{ j_1j_2} S^{i_1}_{j_2}S^{t_2}_{j_2} + v^{t_1i_1}_{ j_1j_2} S^{i_1}_{j_1}S^{t_2}_{j_2} 
    - 2 v^{it_1}_{ j_1j_2} S^{t_2}_{j_2}S^{i_1}_{j_2} + v^{it_1}_{ j_1j_2} S^{t_2}_{j_1}S^{i_1}_{j_2})\hat{E}^+_{t_1t_2} \nonumber \\
    &- \sum_{\substack{\mathbf{t}\\\sigma_1\sigma_2}}(2 v^{t_1t_2}_{ j_1j_2} S_{t_3j_2}S_{t_4j_2} - v^{t_1t_2}_{ j_1j_2} S_{t_3j}S_{t_4j_2})\hat{e}^{\sigma_1\sigma_2}_{t_1t_2} \nonumber \\
    &- \sum_{ \substack{\mathbf{t}\mathbf{u}}}( 2v^{t_1t_2}_{ u_1u_2} S^{i_1}_{j_1}S^{i_1}_{j_1} + 2v^{t_1t_2}_{ j_1j_2} S^{i_1}_{u_1}S^{i_1}_{u_2} 
    - v^{t_1t_2}_{ u_1j_1} S^{i_1}_{u_2}S^{i_1}_{j_1} - v^{t_1t_2}_{j_1u_1} S^{i_1}_{j_1}S^{i_1}_{u_2} + 2v^{i_1i_1}_{ u_1u_2} S^{t_1}_{j_1}S^{t_2}_{j_1} \nonumber 
    \\ 
    &- v^{t_1i_1}_{ u_1u_2} S^{i_1}_{j_1}S^{t_2}_{j_1} - v^{i_1t_2}_{ u_1u_2} S^{t_1}_{j_1}S^{i_1}_{j_1} )\hat{E}^+_{t_1t_2}\hat{E}^+_{u_1u_2} \nonumber \\
    &- \sum_{ \substack{\mathbf{t}\mathbf{u}\\\sigma}}( 4v^{i_1i_1}_{ j_1j_2} S^{t_1}_{u_2}S^{t_2}_{u_1} - 2v^{i_1i_1}_{ u_1j_1 } S^{t_1}_{u_2}S^{t_2}_{j } - 2v^{i_1i_1}_{ j u_2} S^{t_1}_{j }S^{t_2}_{u_1} 
    - 2v^{t_1i_1}_{ j_1j_2} S^{i_1}_{u_2}S^{t_2}_{u_1} - 2v^{t_1i_1}_{ j_1j_2} S^{i_1}_{u_2}S^{t_2}_{u_1} \nonumber \\
    &+ v^{t_1i_1}_{ u_1j_1} S^{i_1}_{u_2}S^{t_2}_{j_1} + v^{t_1i_1}_{j_1u_1} S^{i_1}_{j_1}S^{t_2}_{u_2} + v^{i_1t_2}_{ u_1j_1} S^{t_1}_{u_2}S^{i_1}_{j_1} + v^{i_1t_2}_{j_1u_2} S^{t_1}_{j_1}S^{i_1}_{u_1}   )\hat{E}^\sigma_\mathbf{t}\hat{E}^\sigma_\mathbf{u} \nonumber \\ 
    &- \sum_{ \substack{\mathbf{t}\mathbf{u}\\\sigma_1\sigma_2}} (2v^{i_1i_1}_{ u_1u_2} S^{t_1}_{u_4}S^{t_2}_{u_3} - v^{t_1i_1}_{ u_1u_2} S^{i_1}_{u_4}S^{t_2}_{u_3} - v^{it_1 }_{u_1u_2} S^{t_2}_{u_4}S^{i_1}_{u_3})\hat{E}^{\sigma_2}_{\mathbf{t}}\hat{e}^{\sigma_1\sigma_2}_{\mathbf{u}} \nonumber \\
    &- \sum_{ \substack{\mathbf{t}\mathbf{u}\\\sigma_1\sigma_2\sigma_3}} v^{t_1t_2}_{ u_1u_2} S^{i_1}_{u_4}S^{i_2}_{u_3} 
    \hat{E}^{\sigma_1}_{t_1t_2}\hat{e}^{\sigma_2\sigma_3}_{u_1u_2} - \sum_{ \substack{\mathbf{t}\mathbf{u}\\\sigma_1\sigma_2\sigma_3}} v^{t_1t_2 }_{u_1u_2} S^{t_3}_{j_1}S^{t_4}_{j_1}  
    \hat{e}^{\sigma_1\sigma_2}_{t_1t_2}\hat{E}^{\sigma_3}_{u_1u_2} \nonumber \\  
    &- \sum_{ \substack{\mathbf{t}\mathbf{u}\\\sigma_1\sigma_2}}( 2v^{t_1t_2}_{ j_1j_2} S^{t_3}_{u_2}S^{t_4}_{u_1} - v^{t_1t_2}_{ u_1j_1} S^{t_3}_{u_2}S^{t_4}_{j_1} - v^{t_1t_2}_{j_1u_1} S^{t_3}_{j_1}S^{t_4}_{u_2} ) \hat{e}^{\sigma_1\sigma_2}_{\mathbf{t}} \hat{E}^{\sigma_2}_{\mathbf{u}} \nonumber \\
    &- \sum_{ \substack{\mathbf{t}\mathbf{u}\\\sigma_1\sigma_2\sigma_3}} v^{t_1t_2}_{u_1u_2} S^{t_3}_{u_4}S^{t_4}_{u_3}   \hat{e}^{\sigma_1\sigma_3}_{\mathbf{t}} \hat{e}^{\sigma_2\sigma_3}_{\mathbf{u}}
\end{align}
Combining all four terms, the active space electrostatic-exchange operator may be written as:
\begin{align}
    \widehat{VP}_{\mathrm{active}} &= VP_{0,\mathrm{active}} + \widehat{VP}_{A,\mathrm{active}} + \widehat{VP}_{B,\mathrm{active}} + \widehat{VP}_{1m,\mathrm{active}} \nonumber \\
    &+ \widehat{VP}_{1\ell,\mathrm{active}} + \widehat{VP}_{2,\mathrm{active}} + \widehat{VP}_{3,\mathrm{active}} + \widehat{VP}_{4,\mathrm{active}}
    \label{eq:VP_majorona}
\end{align}
where
\begin{align}
    \widehat{VP}_{A,\mathrm{active}} &= \sum_{ \substack{\mathbf{t}  \\ \sigma}} \tilde{f}_{t_1t_2}^{(A)}  \hat{E}^\sigma_{t_1t_2} 
    + \sum_{ \substack{\mathbf{t}  \\ \sigma\tau}} \tilde{v}^{t_1t_2}_{t_3t_4}\, \hat{e}_{\mathbf{t}}^{\sigma\tau}  ,\\
    \widehat{VP}_{B,\mathrm{active}} &=  \sum_{ \substack{\mathbf{u}  \\ \sigma}} \tilde{f}_{u_1u_2}^{(B)} \hat{E}^\sigma_{u_1u_2} 
    + \sum_{ \substack{\mathbf{u}  \\ \sigma\tau}} \tilde{v}^{u_1u_2}_{u_3u_4} \hat{e}_{\mathbf{u}}^{\sigma\tau} 
    ,\\
    \widehat{VP}_{1m,\mathrm{active}} &= \sum_{ \substack{\mathbf{t}\mathbf{u}\\\sigma } }  \tilde{v}^{t_1t_2}_{u_1u_2} \hat{E}^+_{t_1t_2}\hat{E}^+_{u_1u_2} 
    ,\\ 
    \widehat{VP}_{1\ell,\mathrm{active}} &= \sum_{ \substack{\mathbf{t}\mathbf{u} } }  \tilde{v}^{t_1u_2}_{u_1t_2} \hat{E}^\sigma_{t_1t_2}\hat{E}^\sigma_{u_1u_2}
    ,\\
    \widehat{VP}_{2,\mathrm{active}} &= \sum_{ \substack{ \mathbf{t}\mathbf{u} \\ \sigma\tau} } \tilde{v}^{t_1t_2}_{u_1t_4}S^{t_3}_{u_2}\,   \hat{e}^{\sigma_1\sigma_2}_{\mathbf{t}}\hat{E}^{\sigma_2}_{\mathbf{u}}   
    ,\\
    \widehat{VP}_{3,\mathrm{active}} &= \sum_{ \substack{ \mathbf{t}\mathbf{u} \\ \sigma_1\sigma_2} } \tilde{v}^{t_1u_4}_{u_1u_2}S^{t_2}_{u_3} \hat{E}^{\sigma_2}_{\mathbf{t}}  \hat{e}^{\sigma_1\sigma_2}_{\mathbf{u}}
    ,\\
    \widehat{VP}_{4,\mathrm{active}} &= \sum_{\substack{ \mathbf{t}\mathbf{u} \\ \sigma_1\sigma_2 \sigma_3}} \tilde{v}^{t_1t_2}_{u_1u_2} \tilde{p}^{t_3t_4}_{u_3u_4} \; \hat{e}^{\sigma_1\sigma_2}_{\mathbf{t}}\hat{e}^{\sigma_3\sigma_2}_{\mathbf{u}},
\end{align}
with renormalized active space tensor coefficients defined as:
\begin{align}
    VP_{0,\mathrm{active}} = &- 2v^{i_1j_1}_{j_1i_1} 
    - (4 {v}^{i_1i_1}_{j_1i_2}S^{i_2}_{j_1} - 2{v}^{i_2i_1}_{j_1i_2} S^{i_1}_{j_1} ) 
    -(4 v^{i_1j_2}_{j_1j_2}S^{i_1}_{j_2} - 2 v^{i_1j_2}_{j_2j}S^{i_1}_{j_1}) \\ 
    &-(8 v^{i_1i_1}_{j_1j_2}S^{i_2}_{j_2}S^{i_2}_{j_2} - 4 v^{i_1i_2}_{j_1j_2}S^{i_2}_{j_2}S^{i_1}_{j_2} - 4 v^{i_1i_1}_{j_1j_2} S^{i_2}_{j_1} S^{i_2}_{j_2} + 2 v^{i_1i_2}_{j_1j_2} S^{i_2}_{j_1}S^{i_1}_{j_2}) \\ 
    \tilde{f}_{t_1t_2}^{(A)} = &- v^{j_1t_2}_{t_1j_1} 
    -(2v^{t_1t_2}_{j_1i_1}S^{i_1}_{j_1} + 2 {v}^{i_1i_1}_{j_1t_2}S^{t_1}_{j_1} - {v}^{t_1 i}_{j t_2}S^{i_1}_{j_1} - {v}^{i_1t_2}_{j_1i_1}S^{t_1}_{j_1} )  
    - ( v^{t_1j_2}_{j_1j_2}S^{t_2}_{j_2} - v^{t_1j_2}_{j_2j}S^{t_2}_{j_1}) \\
    &- (4 v^{t_1t_2}_{j_1j_2} S^{i_1}_{j_2}S^{i_1}_{j_2} - 2v^{t_1t_2}_{ j_1j_2} S^{i_1}_{j_1}S^{i_1}_{j_2} 
     - 2v^{i_1i_1}_{ j_1j_2} S^{t_1}_{j_1}S^{t_2}_{j_2}  
    - 2v^{t_1i_1}_{ j_1j_2} S^{i_1}_{j_2}S^{t_2}_{j_2} + v^{t_1i_1}_{ j_1j_2} S^{i_1}_{j_1}S^{t_2}_{j_2} 
    - 2 v^{it_1}_{ j_1j_2} S^{t_2}_{j_2}S^{i_1}_{j_2} + v^{it_1}_{ j_1j_2} S^{t_2}_{j_1}S^{i_1}_{j_2}) \nonumber \\
    \tilde{f}_{u_1u_2}^{(B)} = &- v^{i_1u_2}_{u_1i_1} 
    - (2 {v}^{i_1i_1}_{u_1i_2}S^{i_2}_{u_2} - {v}^{i_2i_1}_{u_1 i_2}S^{i_1}_{u_2} ) 
    - (2v^{i_1j_1}_{u_1u_2}S^{i_1}_{j_1} + 2v^{i_1u_2}_{j_1j_2}S^{i_1}_{u_1} - v^{i_1u_2}_{u_1j_1}S^{i_1}_{j_1} - v^{i_1j_1}_{ju_1}S^{i_1}_{u_2}) \\ 
    &-(4 v^{i_1i_1}_{ u_1u_2} S^{i_2}_{j_1}S^{i_2}_{j_1} - 2v^{i_1i_2}_{ u_1u_2} S^{i_2}_{j_1}S^{i_1}_{j_1} 
     - 2v^{i_1i_2}_{ j_1j_2} S^{i_2}_{u_2}S^{i_1}_{u_1} 
    - 2 v^{i_1i_1}_{ u_1j_1} S^{i_2}_{u_2}S^{i_2}_{j_1} + v^{i_1i_2}_{ u_1j_1} S^{i_2}_{u_2}S^{i_1}_{j_1} 
    - 2 v^{i_1i_1}_{j_1u_1} S^{i_2}_{j_1}S^{i_2}_{u_2} + v^{i_1i_2}_{j_1u_1} S^{i_2}_{j_1}S^{i_1}_{u_2}) \nonumber \\
    \tilde{v}^{t_1t_2}_{t_3t_4} = &- {v}^{t_1t_2}_{j_1t_4} S^{t_3}_{j_1} 
    + v^{t_1t_2}_{ j_1j_2} S^{t_3}_{j_1}S^{t_4}_{j_2} \\
     \tilde{v}^{u_1u_2}_{u_3u_4} = &- v^{i_1u_4}_{u_1u_2}S^{i_1}_{u_3} 
     + v^{i_1i_2}_{ u_1u_2} S^{i_2}_{u_4}S^{i_1}_{u_3} \\
    \tilde{v}^{t_1u_2}_{u_1t_2} = &- v^{t_1u_2}_{u_1t_2} 
    - (2v^{i_1i_1}_{u_1t_2}S^{t_1}_{u_2} - v^{t_1i_1}_{u_1t_2}S^{i_1}_{u_2} - v^{i_1t_2}_{u_1i_1}S^{t_1}_{u_2}) 
    - (2 v^{t_1u_2}_{j_1j_2}S^{t_2}_{u_1} - v^{t_1u_2}_{u_1j_1}S^{t_2}_{j_1} - v^{t_1j_1}_{ju_1}S^{t_2}_{u_2} ) 
    \\ 
    &+ ( 4v^{i_1i_1}_{ u_1j_1 } S^{t_1}_{u_2}S^{t_2}_{j }
    + 4v^{t_1i_1}_{ j_1j_2} S^{i_1}_{u_2}S^{t_2}_{u_1}   \nonumber 
    - 2v^{t_1i_1}_{ u_1j_1} S^{i_1}_{u_2}S^{t_2}_{j_1} - 2v^{t_1i_1}_{j_1u_1} S^{i_1}_{j_1}S^{t_2}_{u_2}   ) \\
    \tilde{v}^{t_1t_2}_{u_1u_2} = &- v^{t_1t_2}_{u_1i_1}S^{i_1}_{u_2} 
    - v^{t_1j_1}_{u_1u_2}S^{t_2}_{j_1} 
     - ( 2v^{t_1t_2}_{ u_1u_2} S^{i_1}_{j_1}S^{i_1}_{j_1} 
    - 2v^{t_1t_2}_{ u_1j_1} S^{i_1}_{u_2}S^{i_1}_{j_1}  - 2v^{t_1i_1}_{ u_1u_2} S^{i_1}_{j_1}S^{t_2}_{j_1} ) \\
    \tilde{v}^{t_1t_2}_{u_1t_4} = &-(v^{t_1t_2}_{u_1t_4}  - 2v^{t_1t_2}_{u_1j_1}S^{t_4}_{j_1} )\\
    \tilde{v}^{t_1u_4}_{u_1u_2} = &-({v}^{t_1u_4}_{u_1u_2} - 2 v^{t_1i_1}_{u_1u_2} S^{i_1}_{u_4} )\\
    \tilde{v}^{t_1t_2}_{u_1u_2} = & + v^{t_1t_2}_{u_1u_2} + 2v^{t_1t_2}_{j_1j_2}\delta_{u_3u_4}/\tilde{\eta}_{B} + 2 v^{i_1i_1}_{u_1u_2}\delta_{t_3t_4}/\tilde{\eta}_{A} + 4v^{i_1i_1}_{ j_1j_2}\delta_{t_3t_4}\delta_{u_3u_4}/(\tilde{\eta}_A\tilde{\eta}_{B})  \\
    \tilde{p}^{t_3t_4}_{u_3u_4} = &-S^{t_3}_{u_4}S^{t_4}_{u_3} - S^{t_3}_{j_1}S^{t_4}_{j_1}\delta_{u_3u_4}/\tilde{\eta}_{B\sigma} - S^{i_1}_{u_3}S^{i_1}_{u_4}\delta_{t_3t_4}/\tilde{\eta}_{A\sigma} .
\end{align}
Converting to chemist notation, we obtain:
\begin{align}
    \widehat{VP}_{A,\mathrm{active}} &= \sum_{ \substack{\mathbf{t}  \\ \sigma}} (\tilde{f}_{t_1t_2}^{(A)} - \sum_{t} \tilde{v}^{t_1t}_{tt_2}) \hat{E}^\sigma_{t_1t_2} 
    + \sum_{ \substack{\mathbf{t}  \\ \sigma\tau}} \tilde{\nu}^{t_1t_2}_{t_3t_4}\,  \hat{E}^\sigma_{t_1t_2} \hat{E}^\tau_{t_3t_4}  , \label{eq:VPA_active} \\
    \widehat{VP}_{B,\mathrm{active}} &=  \sum_{ \substack{\mathbf{u}  \\ \sigma}} (\tilde{f}_{u_1u_2}^{(B)} - \sum_{u} \tilde{v}^{u_1u}_{uu_2}) \hat{E}^\sigma_{u_1u_2} 
    + \sum_{ \substack{\mathbf{u}  \\ \sigma\tau}} \tilde{\nu}^{u_1u_2}_{u_3u_4} \hat{E}^\sigma_{u_1u_2}\hat{E}^\tau_{u_3u_4} 
    ,\label{eq:VPB_active} \\
    \widehat{VP}_{1m,\mathrm{active}} &= \sum_{ \substack{\mathbf{t}\mathbf{u} } }  \tilde{\nu}^{t_1t_2}_{u_1u_2} \hat{E}^+_{t_1t_2}\hat{E}^+_{u_1u_2}  
    ,\label{eq:VP1m_active} \\
    \widehat{VP}_{1\ell,\mathrm{active}} &= \sum_{ \substack{\mathbf{t}\mathbf{u} } }  \tilde{\nu}^{t_1u_2}_{u_1t_2}\hat{E}^\sigma_{t_1t_2}\hat{E}^\sigma_{u_1u_2} 
    ,\label{eq:VP1l_active} \\
    \widehat{VP}_{2,\mathrm{active}} &= \sum_{ \substack{ \mathbf{t}\mathbf{u} \\ \sigma_1\sigma_2} } \tilde{\nu}^{t_1t_2}_{u_1t_4}S^{t_3}_{u_2}\,   \hat{E}^{\sigma_1}_{t_1t_2} \hat{E}^{\sigma_2}_{t_3t_4}\hat{E}^{\sigma_2}_{u_1u_2}   
    ,\label{eq:VP2_active} \\
    \widehat{VP}_{3,\mathrm{active}} &= \sum_{ \substack{ \mathbf{t}\mathbf{u} \\ \sigma_1\sigma_2} } \tilde{\nu}^{t_1u_4}_{u_1u_2}S^{t_2}_{u_3} \hat{E}^{\sigma_2}_{t_1t_2}  \hat{E}^{\sigma_1}_{u_1u_2}\hat{E}^{\sigma_2}_{u_3u_4} 
    ,\label{eq:VP3_active} \\
    \widehat{VP}_{4,\mathrm{active}} &= \sum_{\substack{ \mathbf{t}\mathbf{u} \\ \sigma_1\sigma_2 \sigma_3}} \tilde{v}^{t_1t_2}_{u_1u_2} \tilde{p}^{t_3t_4}_{u_3u_4} \;  \hat{E}^{\sigma_1}_{t_1t_2} \hat{E}^{\sigma_2}_{t_3t_4}\hat{E}^{\sigma_3}_{u_1u_2}\hat{E}^{\sigma_2}_{u_3u_4},
    \label{eq:VP4_active}
\end{align}
where the active space tensor coefficients in the chemist notation are given by,
\begin{align}
    \tilde{\nu}^{t_1t_2}_{u_1u_2} &= \tilde{v}^{t_1t_2}_{u_1u_2} - \sum_{t_3} \tilde{v}^{t_1t_3}_{u_1u_2}S^{t_3}_{j_1}S^{t_2}_{j_1} - \sum_{u_3} \tilde{v}^{t_1t_2}_{u_1u_3}S^{i_1}_{u_3}S^{i_1}_{u_2},\\
    \label{nu_1M_active}
    \tilde{\nu}^{t_1u_2}_{u_1t_2} &= \tilde{v}^{t_1u_2}_{u_1t_2}  + \sum_{t_3u_3} \tilde{v}^{t_1t_3}_{u_1u_3} S^{t_3}_{u_2}S^{t_2}_{u_2} - \sum_{u_3}\tilde{v}_{u_1u_3}^{t_1u_2}S^{t_2}_{u_3} - \sum_{t_3} \tilde{v}^{t_1t_3}_{u_1t_2} S^{t_3}_{u_2}, \\
    \tilde{\nu}^{t_1t_2}_{u_1t_4} &= \tilde{v}^{t_1t_2}_{u_1t_4} - \sum_{u_3}\tilde{v}^{t_1t_2}_{u_1u_3} S^{t_4}_{u_3} \, , \\
    \tilde{\nu}_{u_1u_2}^{t_1u_4} &= 
    \tilde{v}_{u_1u_2}^{t_1u_4} - \sum_{t_3}\tilde{v}^{t_1t_3}_{u_1u_2} S^{t_3}_{u_4} \,.
\end{align}
The intra-monomer tensors under the chemist notation remain equivalent, i.e $\tilde{\nu}^{t_1t_2}_{t_3t_4} = \tilde{v}^{t_1t_2}_{t_3t_4}$ and $\tilde{\nu}^{u_1u_2}_{u_3u_4} = \tilde{v}^{u_1u_2}_{u_3u_4}$. 

\section{SAPT-EVE algorithm}\label{sec:Appendix_SAPT_EVE}
To ensure a first order interaction energy estimate to chemical accuracy, the target accuracy of the three observables $\hat{F}=\{\hat{V},\hat{P},\widehat{VP}\}$ must be optimally chosen to simultaneously satisfy the desired accuracy and minimize computational costs. That is, we must find a target precision $\varepsilon_F$ for every observable $\hat{F}$, such that they satisfy one over all precision $\varepsilon_{\mathrm{targ}}$ while minimizing the cost of the three prime phase estimation algorithms. The resource cost minimization can be formulated as the following optimization problem,
\begin{align}
\min_{\varepsilon_V, \varepsilon_{VP}, \varepsilon_{P}} \left( \frac{\lambda_V}{\varepsilon_{V}} + \frac{\lambda_{VP}}{\varepsilon_{VP}} + \frac{\lambda_P}{\varepsilon_{P}}\right) \\ 
\text{s.t.}\quad (1 + \lambda_P) \varepsilon_V + \varepsilon_{VP} + \lambda_V \varepsilon_{P} = \varepsilon_{\mathrm{targ}}\, .
\label{constraint}
\end{align}
which can be solved with the Lagrange method yielding
 \begin{align}
 \left\lbrace
 \vphantom{\frac{\sum_1}{\sum_1}}
\varepsilon_V, \, \varepsilon_{V\!P}, \, \varepsilon_{P} \right\rbrace = 
\frac{\varepsilon_{\mathrm{targ}}}{ \sqrt{(1 + \lambda_P) \lambda_V} + \sqrt{\lambda_{V\!P}} + \sqrt{\lambda_V \lambda_{P}}} 
\left\lbrace \vphantom{\frac{\sum_1}{\sum_1}}
\sqrt{\frac{ \lambda_V}{1 + \lambda_P }}, \,  
\sqrt{ \lambda_{V\!P}}, \,
\sqrt{\frac{ \lambda_{P}}{\lambda_{V}}} \right\rbrace\, .
 \end{align}
This solution is only optimal if given no further information about the expected values of $\langle \hat{V}\rangle $, $\langle \hat{P}\rangle $ and $\langle \widehat{VP}\rangle$. We have also upper bounded expectation values $\langle F\rangle $ with their respective $\ell_1$ norms $\lambda_F$  in Eq.~\eqref{constraint}, which is a very loose bound.  In practice, one will certainly be able to relax the target accuracies by bootstrapping Eq.~\eqref{constraint} with low accuracy estimates  $\langle F\rangle_\text{low}$, replacing $\lambda_F \mapsto \langle F\rangle_\text{low}$.

 For every observable $\hat{F}$ we have to adjust the polynomial degrees of SAPT-EVE's inner phase estimation routines $\mathsf{iQPE_A}$ and $\mathsf{iQPE_B}$ according to the target errors $\varepsilon_F$. While we have demonstrated this procedure in Ref.~\cite{steudtner2023fault} for QSP-EVE, it is not immediately clear how the discretization errors of two inner phase estimation routines would be dealt with in SAPT-EVE. Let us quickly summarize the estimation procedure behind QSP-EVE and SAPT-EVE. The iterate of Figure~\ref{fig:sapt_circuit} is the product of two reflections $\mathcal{R}_{\pi}$ and $\mathcal{R}_{\tau}$, each of which has a structure $\mathcal{R}_{\mathfrak{X}} = 1 - 2 \hat{\mathfrak{X}}$ with some arbitrary-rank projectors $\hat{\mathfrak{X}} = \hat{\pi}$, $\hat{\tau}$. The product $\mathcal{R}_{\tau}\mathcal{R}_{\pi}$ now has the eigenvalues $\exp(\pm i2 \arccos w_k )$, where $w_k$ are the singular values of the product of projectors $\hat{\pi} \cdot\hat{\tau}$. The idea behind SAPT-EVE is now that one of the singular values is a function of $\langle\hat{F} \rangle$. In a world without discretization errors, the circuits we employ would fix the reference states exactly:
 \begin{align}
    \label{eq:plan}
     \mathsf{iQPE}^{\dagger}_q \, \mathsf{Refl}^{\phantom{\dagger}}_q \, \mathsf{iQPE}^{\phantom{\dagger}}_q \approx 1 - 2 |Q_q\rangle\!\langle Q_q|_{\mathsf{sim} \,q,\mathsf{enc}[H_q] } \otimes |\boldsymbol{0}\rangle\!\langle \boldsymbol{0}|_{\mathsf{phase}\, q} \, - \dots
 \end{align}
  for $q=\mathsf{A}, \mathsf{B}$ labeling routines associated with the two subsystems, and $|Q_{\mathsf{A}}\rangle$, $|Q_{\mathsf{B}}\rangle$ are their respective qubitization ground states. Following the state on the right-hand side of Eq.~\eqref{eq:plan} are other terms orthogonal to the all zero state $|\boldsymbol{0}\rangle$ on the $\mathsf{phase}$ register of the corresponding monomer $q=\mathsf{A}, \mathsf{B}$. Considering the leading order contribution of the discretization errors, we introduce the qubitization states $|\mathcal{E}_q\rangle$ associated with the first excited states in the respective subsystem $q$. Note that $| Q_{\mathsf{A}}\rangle$ and $| \mathcal{E}_{\mathsf{A}}\rangle$ are associated with Hamiltonian $H_A$, they are 
 generally formed with respect to different eigenenergies than $|Q_{\mathsf{B}}\rangle$ and $| \mathcal{E}_{\mathsf{B}}\rangle$, which are associated with $H_B$. For both subsystems we introduce the projectors $\varrho^{\mathsf{A}}(E)_{\mathsf{phase\, A}}$ and $\varrho^{\mathsf{B}}(E)_{\mathsf{phase\, B}}$ associated with either excited state $E=\mathcal{E}$
 or ground state $E=Q$. We find
 \begin{align}
     &\hat{\pi} = \sum_{E = Q, \mathcal{E}} \left(\bigotimes_{q=\mathsf{A}, \mathsf{B}} |E_q\rangle\!\langle E_q|_{\mathsf{sim} \,q,\mathsf{enc}[H_q] } \otimes \varrho^{q}(E)_{\mathsf{phase}\, q} \right) \otimes |\boldsymbol{0}\rangle \!\langle\boldsymbol{0}|_{\mathsf{enc}[F]}\, , \\
     &\hat{\tau} = \frac{1}{2}\left(1 - \mathcal{B}[\hat{F}]\right)_{\mathsf{sim\, A, \, sim\,B, \, enc}[F]} \otimes \bigotimes_{q=\mathsf{A}, \mathsf{B}}|\boldsymbol{0}\rangle\!\langle \boldsymbol{0}|_{\mathsf{enc}[H_q]}\, .
 \end{align} 
To investigate the product $\hat{\pi}\cdot \hat{\tau}$, we consider the singular value decomposition
 \begin{align}
 \label{singvaldecompagain}
 \varrho^{q}(Q) \cdot \varrho^{q}(\mathcal{E}) = \sum_{j=1}^r \Omega^{(q)} |\xi_{Q,q,j}\rangle\!\langle \xi_{\mathcal{E},q,j}|\,
 \end{align}
where $\Omega^{(q)}$ are singular values, $|\xi_{Q,q,j}\rangle$ and $|\xi_{\mathcal{E},q,j}\rangle$ are singular vectors and where the sum over $j$ is in the range from $1$ to $r$, the rank of $ \varrho^{q}(Q) \cdot \varrho^{q}(\mathcal{E})$. We will now learn the singular values $w_k$ of $\hat{\pi} \cdot \hat{\tau}$ by computing the eigenvalues $(w_k)^2$ of $\hat{\pi} \cdot \hat{\tau} \cdot \hat{\pi}$, which has a block-diagonal form due to the singular value decomposition in Eq.~\eqref{singvaldecompagain}:
\begin{align}
\label{eq:EVEmtx}
\hat{\pi} \cdot \hat{\tau} \cdot \hat{\pi} = \bigoplus_{i,j} \frac{1}{8}\left[\begin{matrix}
 1 - F_{00} & - F_{01} \Omega^{(\mathsf{B})}_j & - F_{02} \Omega^{(\mathsf{A})}_i & - F_{03} \Omega^{(\mathsf{A})}_i \Omega^{(\mathsf{B})}_j \\[4pt]
 - F_{10} \Omega^{(\mathsf{B})}_j & 1- F_{11} & - F_{12}\Omega^{(\mathsf{A})}_i \Omega^{(\mathsf{B})}_j & -F_{13}\Omega^{(\mathsf{A})}_i \\[4pt]
 - F_{20} \Omega^{(\mathsf{A})}_i & - F_{21}\Omega^{(\mathsf{A})}_i\Omega^{(\mathsf{B})}_j & 1- F_{22} & - F_{23} \Omega^{(\mathsf{B})}_j\\[4pt]
  - F_{30} \Omega^{(\mathsf{A})}_i \Omega^{(\mathsf{B})}_j & - F_{31} \Omega^{(\mathsf{A})}_i & - F_{32} \Omega^{(\mathsf{B})}_j& 1-F_{33} 
\end{matrix}
\right]_{i_1j_1}, 
\end{align}
where $F_{lm}$ are some matrix elements of the observable $\hat{F}$, and the block matrices $[\cdots]_{i_1j_1}$ denote operators with respect to the bases states
\begin{align}
&|Q_{\mathsf{A}}\rangle_{\mathsf{sim\, A, \, enc}[H_A]} \otimes | \xi_{Q,\mathsf{A},i}\rangle_{\mathsf{phase\, A}} \otimes |Q_{\mathsf{B}}\rangle_{\mathsf{sim\, B, \, enc}[H_B]} \otimes | \xi_{Q,\mathsf{B},j}\rangle_{\mathsf{phase\, B}} \otimes |\boldsymbol{0}\rangle_{\mathsf{enc}[F]}, \\
&|Q_{\mathsf{A}}\rangle_{\mathsf{sim\, A, \, enc}[H_A]} \otimes | \xi_{Q,\mathsf{A},i}\rangle_{\mathsf{phase\, A}} \otimes |\mathcal{E}_{\mathsf{B}}\rangle_{\mathsf{sim\, B, \, enc}[H_B]} \otimes | \xi_{\mathcal{E},\mathsf{B},j}\rangle_{\mathsf{phase\, B}} \otimes |\boldsymbol{0}\rangle_{\mathsf{enc}[F]}, \\
&|\mathcal{E}_{\mathsf{A}}\rangle_{\mathsf{sim\, A, \, enc}[H_A]} \otimes | \xi_{\mathcal{E},\mathsf{A},i}\rangle_{\mathsf{phase\, A}} \otimes |Q_{\mathsf{B}}\rangle_{\mathsf{sim\, B, \, enc}[H_B]} \otimes | \xi_{Q,\mathsf{B},j}\rangle_{\mathsf{phase\, B}} \otimes |\boldsymbol{0}\rangle_{\mathsf{enc}[F]}, \\
&|\mathcal{E}_{\mathsf{A}}\rangle_{\mathsf{sim\, A, \, enc}[H_A]} \otimes | \xi_{\mathcal{E},\mathsf{A},i}\rangle_{\mathsf{phase\, A}} \otimes |\mathcal{E}_{\mathsf{B}}\rangle_{\mathsf{sim\, B, \, enc}[H_B]} \otimes | \xi_{\mathcal{E},\mathsf{B},j}\rangle_{\mathsf{phase\, B}} \otimes |\boldsymbol{0}\rangle_{\mathsf{enc}[F]}\, .
\end{align}
For vanishing singular values, $\Omega_{j}^{(q)} = 0$, we find that there is one solution of Eq.~\eqref{eq:EVEmtx} with eigenvalue $w_{\hat{k}}^2 = (1-F_{00})/8$. Since $F_{00}$ is the matrix element associated with the ground states of both subsystems, $w_{\hat{k}}^2$ is the solution with the eigenphase $\pm 2 \arccos \sqrt{(1 - \langle F\rangle)/8}$. For at least one $\Omega_{j}^{(q)} \neq 0$, the block matrices $[\cdots]_{i_1j_1}$ must be diagonalized. The good solution is the one whose eigenvalue $w_{\hat{k}}^2$ is closest to $(1- F_{00})/8$. Assuming that the contamination of this solution with contributions of excited states will increase, we replace individual $\Omega_{j}^{(\mathsf{A})}$ and $\Omega_{j}^{(\mathsf{B})}$ in Eq.~\eqref{eq:EVEmtx} with one 
\begin{align}
    \Omega = \max_{j}\max_{q= \mathsf{A, B}} \Omega_{j}^{(q)} \, .
\end{align}
Perturbation theory in $\Omega$ informs us that the deviation of the solution from $(1-F_{00})/8$ is $O(\Omega)$, only if the diagonal elements of $F_{j_1j_2}$ are all equal. Assuming that the structure of $F_{i_1j_1}$ is as malicious as possible, we set all $F_{j_1j_2}$ to be equal. Neglecting quadratic contributions $\Omega^2$ due to their size, we get a deviation of $w_{\hat{k}}^2$ from $(1-F_{00})/8$ that is entirely linear in $\Omega$. Note that for $\Omega=1$, we would bound $0<w_{\hat{k}}^2<1/4$ due to $||\hat{F}||\leq 1$. This means that we can bound the error of estimated observable $F_{\text{est}}$ and actual observable $\langle \hat{F}\rangle$ by 
\begin{align}
\left| F_{\text{est}} - \langle \hat{F}\rangle\right| \leq 2 \Omega\, .
\end{align}
We therefore need to fortify both $\mathsf{iQPE}_q$ routines equally well against errors using quantum signal processing (QSP). Given the QSP routines, the values of $\Omega^{(\mathsf{A})}_i$ and $\Omega^{(\mathsf{B})}_j$ can be obtained with a numerical procedure outlined in \cite{steudtner2023fault}. The success probability of this routine, if one prepares an initial state of 
\begin{align}
    \left(\bigotimes_{q=\mathsf{A,B}} |Q_{q}\rangle_{\mathsf{sim}\, q, \, \mathsf{enc}[H_q]} \otimes |\boldsymbol{0}\rangle_{\mathsf{phase}\, q}\right) \otimes |\boldsymbol{0}\rangle_{\mathsf{enc}[F]}
\end{align}
is between $\langle\boldsymbol{0}|\varrho^{\mathsf{A}}(Q)\otimes \varrho^{\mathsf{B}}(Q)|\boldsymbol{0}\rangle/4 $ and $\langle\boldsymbol{0}|\varrho^{\mathsf{A}}(Q)\otimes \varrho^{\mathsf{B}}(Q)|\boldsymbol{0}\rangle$.

\section{SAPT operator encoding}\label{sec:Appendix_Operator_Encoding}
In the following, we present the \emph{sparse} and \emph{tensor factorization} encoding methods that allow us to design a full algorithm for SAPT observable estimation. To this end, we first re-write all of the SAPT operators in terms of Majorana operators as they provide a clear and direct decomposition in terms of self-inverse operators required for block encoding.

\subsection{Majorana representation}
\noindent
In the Majorana representation, the fermionic operators $\hat{\gamma}_{\textsc{p},0}$ and $\hat{\gamma}_{\textsc{p},1}$ for monomer A are defined as:
\begin{align}
    \hat{\gamma}_{\textsc{p},0} = \hat{a}_{\textsc{p}} + \hat{a}^\dagger_{\textsc{p}} \;\;,\;\;
     \hat{\gamma}_{\textsc{p},1} = -i(\hat{a}_{\textsc{p}} - \hat{a}^\dagger_{\textsc{p}}),
\end{align}
satisfying the properties,
\begin{align}
    \{\hat{\gamma}_{\textsc{p},i}\;,\hat{\gamma}_{\textsc{p}',j} \} = 2\delta_{\textsc{p}\textsc{p}'}\delta_{ij}\mathbb{1}\;\;,\;\;
    \hat{\gamma}_{\textsc{p},i}^\dagger = \hat{\gamma}_{\textsc{p},i} \;\;,\;\;
    \hat{\gamma}_{\textsc{p},i}^2 = \mathbb{1}.
\end{align}
For monomer B, the Majorana operators $\hat{\gamma}_{\textsc{q},0}$ and $\hat{\gamma}_{\textsc{q},1}$ are defined as:
\begin{align}
    \hat{\gamma}_{\textsc{q},0} = \hat{b}_{\textsc{q}} + \hat{b}^\dagger_{\textsc{q}} \;\;,\;\;
     \hat{\gamma}_{\textsc{q},1} = -i(\hat{b}_{\textsc{q}} - \hat{b}^\dagger_{\textsc{q}}),
\end{align}
while also satisfying the properties,
\begin{align}
    \{\hat{\gamma}_{\textsc{q},i}\;,\hat{\gamma}_{\textsc{q}',j} \} = 2\delta_{\textsc{q}\textsc{q}'}\delta_{ij}\mathbb{1}\;\;,\;\;
    \hat{\gamma}_{\textsc{q},i}^\dagger = \hat{\gamma}_{\textsc{q},i} \;\;,\;\;
    \hat{\gamma}_{\textsc{q},i}^2 = \mathbb{1}.
\end{align}
To find the representation of the SAPT operators in terms of the Majorana operators defined above, one may simply use the inverse relations, $\hat{a}^\dagger_{\textsc{p}} = \tfrac{1}{2}(\hat{\gamma}_{\textsc{p},0} - i \hat{\gamma}_{\textsc{p},1})$, $\hat{a}_{\textsc{p}} = \tfrac{1}{2}(\hat{\gamma}_{\textsc{p},0} + i \hat{\gamma}_{\textsc{p},1})$ and $\hat{b}^\dagger_{\textsc{q}} = \tfrac{1}{2}(\hat{\gamma}_{\textsc{q},0} - i \hat{\gamma}_{\textsc{q},1})$, $\hat{b}_{\textsc{q}} = \tfrac{1}{2}(\hat{\gamma}_{\textsc{q},0} + i \hat{\gamma}_{\textsc{q},1})$. In the following, we provide the final result in both the full space and active space pictures after simplification. Throughout this section, we will use the symmetrized SAPT operators defined in Eqs.~\eqref{eq:V_real_sym}-\eqref{eq:VP_real_sym}.

\subsubsection{Full space picture}
The all-electron (full space) electrostatic and exchange operators in the Majorana representation are defined as:
\begin{align}
    \hat{V} &=  \tfrac{1}{4}\sum_{\textsc{p}\textsc{q}}v^{\textsc{p}\textsc{p}}_{\textsc{q}\textsc{q}} + \tfrac{i}{4}\sum_{\substack{\textsc{p}_1\textsc{p}_2 }}  f^{(A)}_{\textsc{p}_1\textsc{p}_2}\hat{\gamma}_{\textsc{p}_1,0}\hat{\gamma}_{\textsc{p}_2,1}  
    + \tfrac{i}{4}\sum_{\substack{\textsc{q}_1\textsc{q}_2 } } f^{(B)}_{\textsc{p}_1\textsc{p}_2} \hat{\gamma}_{\textsc{q}_1,0}\hat{\gamma}_{\textsc{q}_2,1} \label{eq:V_majorana2} \\
    &- \tfrac{1}{4}\sum_{\substack{\textbf{\textsc{p}}\textbf{\textsc{q}}  }} \text{sym}( v^{\textsc{p}_1\textsc{p}_2}_{\textsc{q}_1\textsc{q}_2})\hat{\gamma}_{\textsc{p}_1,0}\hat{\gamma}_{\textsc{p}_2,1} \hat{\gamma}_{\textsc{q}_1,0}\hat{\gamma}_{\textsc{q}_2,1}, \nonumber \\
    \hat{P} &= -\tfrac{1}{4}\sum_{\textsc{p}\textsc{q}}S^{\textsc{p}}_{\textsc{q}}S^{\textsc{p}}_{\textsc{q}} - \tfrac{i}{4}\sum_{\substack{\textsc{p}_1\textsc{p}_2 \\ \textsc{q}}} p^{(A)}_{\textsc{p}_1\textsc{p}_2} \hat{\gamma}_{\textsc{p}_1,0}\hat{\gamma}_{\textsc{p}_2,1} 
    - \tfrac{i}{4}\sum_{\substack{\textsc{q}_1\textsc{q}_2 \\ \textsc{p}} }  p^{(B)}_{\textsc{q}_1\textsc{q}_2} \hat{\gamma}_{\textsc{q}_1,0}\hat{\gamma}_{\textsc{q}_2,1} \label{eq:P_majorana2} \\
    &+ \tfrac{1}{4}\sum_{\substack{\textbf{\textsc{p}}\textbf{\textsc{q}}  }} \text{sym}( S^{\textsc{p}_1}_{\textsc{q}_2}S^{\textsc{p}_2}_{\textsc{q}_1}) \hat{\gamma}_{\textsc{p}_1,0}\hat{\gamma}_{\textsc{p}_2,1} \hat{\gamma}_{\textsc{q}_1,0}\hat{\gamma}_{\textsc{q}_2,1}, \nonumber 
\end{align}
where we have defined $f^{(A)}_{\textsc{p}_1\textsc{p}_2} = \sum_\textsc{q} v^{\textsc{p}_1\textsc{p}_2}_{\textsc{q}\textsc{q}}$, $f^{(B)}_{\textsc{p}_1\textsc{p}_2} = \sum_\textsc{p} v^{\textsc{p}\textsc{p}}_{\textsc{q}_1\textsc{q}_2}$ and $    p^{(A)}_{\textsc{p}_1\textsc{p}_2} =\sum_{\textsc{q}} S^{\textsc{p}_1}_{\textsc{q}}S^{\textsc{p}_2}_{\textsc{q}}$, $p^{(B)}_{\textsc{q}_1\textsc{q}_2} = \sum_{\textsc{p}} S^{\textsc{p}}_{\textsc{q}_2}S^{\textsc{p}}_{\textsc{q}_1}$, as the single-body coefficients of the electrostatic and exchange operators in the Majorana representation. The electrostatic-exchange operator is written as:
\begin{equation}
    \widehat{VP}^{(m)}_\mathrm{s} = \widehat{VP}_A^{(m)} + \widehat{VP}_B^{(m)} + \widehat{VP}_{1\mathrm{m}}^{(m)} + \widehat{VP}_{1\ell}^{(m)} + \widehat{VP}_2^{(m)} + \widehat{VP}_3^{(m)} + \widehat{VP}_4^{(m)},
    \label{VP_majorana2}
\end{equation}
where
\begin{align} 
    \widehat{VP}_A^{(m)} &= -\tfrac{i}{4}\sum_{ \textbf{\textsc{p}} } \text{sym}(\kappa_{\textsc{p}_1\textsc{p}_2}^{(A)})  \hat{\gamma}_{\textsc{p}_1,0}\hat{\gamma}_{\textsc{p}_2,1} 
    +\tfrac{1}{8} \sum_{ \substack{\textbf{\textsc{p}}  }} \text{sym}(\Lambda^{\textsc{p}_1\textsc{p}_2}_{\textsc{p}_3\textsc{p}_4}  )\hat{\gamma}_{\textsc{p}_1,0}\hat{\gamma}_{\textsc{p}_2,1} \hat{\gamma}_{\textsc{p}_3,0}\hat{\gamma}_{\textsc{p}_4,1} , \label{eq:VPA_majorana} \\
    \widehat{VP}_B^{(m)} &=  -\tfrac{i}{4}\sum_{ \textbf{\textsc{q}} } \text{sym}(\kappa_{\textsc{q}_1\textsc{q}_2}^{(B)})  \hat{\gamma}_{\textsc{q}_1,0}\hat{\gamma}_{\textsc{q}_2,1} 
    +\tfrac{1}{8} \sum_{ \substack{\textbf{\textsc{q}} }} \text{sym}(\Lambda^{\textsc{q}_1\textsc{q}_2}_{\textsc{q}_3\textsc{q}_4} ) \hat{\gamma}_{\textsc{q}_1,0}\hat{\gamma}_{\textsc{q}_2,1} \hat{\gamma}_{\textsc{q}_3,0}\hat{\gamma}_{\textsc{q}_4,1}  
    ,\label{eq:VPB_majorana} \\
    \widehat{VP}_{1\mathrm{m}}^{(m)} &= \tfrac{1}{8}\sum_{ \substack{ \textbf{\textsc{p}},\textbf{\textsc{q}} } } \text{sym}(\Lambda^{\textsc{p}_1\textsc{p}_2}_{\textsc{q}_1\textsc{q}_2})\; \hat{\gamma}_{\textsc{p}_1,0}\hat{\gamma}_{\textsc{p}_2,1} \hat{\gamma}_{\textsc{q}_1,0}\hat{\gamma}_{\textsc{q}_2,1} 
    ,\label{eq:VP1m_majorana} \\
    \widehat{VP}_{1\ell}^{(m)} &= \tfrac{1}{4}\sum_{ \substack{ \textbf{\textsc{p}},\textbf{\textsc{q}} } } \text{sym}(\Lambda^{\textsc{p}_1\textsc{q}_2}_{\textsc{q}_1\textsc{p}_2})\hat{\gamma}_{\textsc{p}_1,0}\hat{\gamma}_{\textsc{p}_2,1} \hat{\gamma}_{\textsc{q}_1,0}\hat{\gamma}_{\textsc{q}_2,1}
    ,\label{eq:VP1l_majorana} \\
    \widehat{VP}_{2}^{(m)} &= \tfrac{i}{8}\sum_{ \substack{ \textbf{\textsc{p}},\textbf{\textsc{q}} } } \text{sym}(\Lambda^{\textsc{p}_1\textsc{p}_2}_{\textsc{q}_1\textsc{p}_4} S^{\textsc{p}_3}_{\textsc{q}_2} )
    \hat{\gamma}_{\textsc{p}_1,0}\hat{\gamma}_{\textsc{p}_2,1} \hat{\gamma}_{\textsc{p}_3,0}\hat{\gamma}_{\textsc{p}_4,1} 
    \hat{\gamma}_{\textsc{q}_1,0}\hat{\gamma}_{\textsc{q}_2,1}   ,\label{eq:VP2_majorana} \\
    \widehat{VP}_{3}^{(m)} &=\tfrac{i}{8}\sum_{ \substack{ \textbf{\textsc{p}},\textbf{\textsc{q}} } } \text{sym}(\Lambda^{\textsc{p}_1\textsc{q}_4}_{\textsc{q}_1\textsc{q}_2} S^{\textsc{q}_3}_{\textsc{p}_2}) \hat{\gamma}_{\textsc{p}_1,0}\hat{\gamma}_{\textsc{p}_2,1} \hat{\gamma}_{\textsc{q}_1,0}\hat{\gamma}_{\textsc{q}_2,1} \hat{\gamma}_{\textsc{q}_3,0}\hat{\gamma}_{\textsc{q}_4,1} 
    ,\label{eq:VP3_majorana} \\
    \widehat{VP}_4^{(m)} &= \tfrac{1}{2}(\hat{V}'\hat{P}' + \hat{P}'\hat{V}').
    \label{eq:VP4_majorana}
\end{align}
Here, we have defined the modified electrostatic and exchange operators $\hat{V}'$ and $\hat{P}'$,
\begin{align}
    \hat{V}' &=  - \tfrac{1}{4}\sum_{\textbf{\textsc{p}}\textbf{\textsc{q}}}v^{\textsc{p}_1\textsc{p}_2}_{\textsc{q}_1\textsc{q}_2} \hat{\gamma}_{\textsc{p}_1,0}\hat{\gamma}_{\textsc{p}_2,1} \hat{\gamma}_{\textsc{q}_1,0}\hat{\gamma}_{\textsc{q}_2,1}, \label{modified_V}\\
    \hat{P}' &= - \tfrac{i}{4}\sum_{\substack{\textsc{p}_1\textsc{p}_2} } p^{(A)}_{\textsc{p}_1\textsc{p}_2} \hat{\gamma}_{\textsc{p}_1,0}\hat{\gamma}_{\textsc{p}_2,1} - \tfrac{i}{4}\sum_{\substack{\textsc{q}_1\textsc{q}_2} }  p^{(B)}_{\textsc{q}_1\textsc{q}_2} \hat{\gamma}_{\textsc{q}_1,0}\hat{\gamma}_{\textsc{q}_2,1} + \tfrac{1}{4}\sum_{\substack{\textbf{\textsc{p}}\textbf{\textsc{q}}\\   }} \text{sym}(S^{\textsc{p}_1}_{\textsc{q}_2}S^{\textsc{p}_2}_{\textsc{q}_1})\hat{\gamma}_{\textsc{p}_1,0}\hat{\gamma}_{\textsc{p}_2,1} \hat{\gamma}_{\textsc{q}_1,0}\hat{\gamma}_{\textsc{q}_2,1},
    \label{modified_P}
\end{align}
in order to reduce the $\ell_1$ norm contribution of the total electrostatic-exchange operator. The renormalized tensor coefficients in the Majorana representation are given by,
\begin{align}
    {VP}_0^{(m)} &= -\tfrac{1}{4} \sum_{\textsc{p}\textsc{q}} \text{sym}(\bar{\nu}^{\textsc{p}\textsc{q}}_{\textsc{q}\textsc{p}}) -  \tfrac{1}{8} \sum_{\textsc{p}\textsc{q}\textsc{q}'} \text{sym}(\nu^{\textsc{p}\textsc{q}'}_{\textsc{q}\textsc{q}} S^\textsc{p}_{\textsc{q}'})  -   \tfrac{1}{8} \sum_{\textsc{p}\textsc{p}'\textsc{q}} \text{sym}(\nu^{\textsc{p}\textsc{p}}_{\textsc{q}\textsc{p}'} S^{\textsc{p}'}_{\textsc{q}} ) -   \tfrac{1}{16} \sum_{\substack{\textsc{p}\textsc{p}'\\\textsc{q}\textsc{q}'}}\text{sym}(v^{\textsc{p}\textsc{p}}_{\textsc{q}\textsc{q}}S^{\textsc{p}'}_{\textsc{q}'}S^{\textsc{p}'}_{\textsc{q}'}) ,\\
    \kappa^{(A)}_{\textsc{p}_1\textsc{p}_2} &=  \sum_{\textsc{q}} \bar{\nu}^{\textsc{p}_1\textsc{q}}_{\textsc{q}\textsc{p}_2} + \tfrac{1}{2}\sum_{\textsc{q}\textsc{q}'}  \nu^{\textsc{p}_1\textsc{q}'}_{\textsc{q}\textsc{q}}S^{\textsc{p}_2}_{\textsc{q}'} + \tfrac{1}{2}\sum_{\textsc{p}\textsc{q}} \nu^{\textsc{p}\textsc{p}}_{\textsc{q}\textsc{p}_2}S^{\textsc{p}_1}_{\textsc{q}} + \tfrac{1}{2}\sum_{\textsc{p}\textsc{q}} \nu^{\textsc{p}_1\textsc{p}_2}_{\textsc{q}\textsc{p}}S^\textsc{p}_\textsc{q} +  \tfrac{1}{4}\sum_{\textsc{p}\textsc{q}\textsc{q}'}v^{\textsc{p}_1\textsc{p}_2}_{\textsc{q}\textsc{q}}S^{\textsc{p}'}_{\textsc{q}'}S^{\textsc{p}'}_{\textsc{q}'} + \tfrac{1}{4}\sum_{\textsc{p}\textsc{q}\textsc{q}'}v^{\textsc{p}\textsc{p}}_{\textsc{q}\textsc{q}}S^{\textsc{p}_1}_{\textsc{q}'}S^{\textsc{p}_2}_{\textsc{q}'} ,\\
    \kappa^{(B)}_{\textsc{q}_1\textsc{q}_2} &= \sum_{\textsc{p}} \bar{\nu}^{\textsc{p}\textsc{q}_2}_{\textsc{q}_1\textsc{p}} +  \tfrac{1}{2}\sum_{\textsc{p}\textsc{p}'} \nu^{\textsc{p}\textsc{p}}_{\textsc{q}_1\textsc{p}'}S^{\textsc{p}'}_{\textsc{q}_2} + \tfrac{1}{2}\sum_{\textsc{p}\textsc{q}}\nu^{\textsc{p}\textsc{q}_2}_{\textsc{q}\textsc{q}}S^{\textsc{p}}_{\textsc{q}_1} + \tfrac{1}{2}\sum_{\textsc{p}\textsc{q}}\nu^{\textsc{p}\textsc{q}}_{\textsc{q}_1\textsc{q}_2}S^{\textsc{p}}_{\textsc{q}} +  \tfrac{1}{4}\sum_{\textsc{p}\textsc{p}'\textsc{q}} v^{\textsc{p}\textsc{p}}_{\textsc{q}_1\textsc{q}_2}S^{\textsc{p}'}_{\textsc{q}}S^{\textsc{p}'}_{\textsc{q}} + \tfrac{1}{4}\sum_{\textsc{p}\textsc{p}'\textsc{q}} v^{\textsc{p}\textsc{p}}_{\textsc{q}\textsc{q}}S^{\textsc{p}'}_{\textsc{q}_1}S^{\textsc{p}'}_{\textsc{q}_2}   ,\\    
    \Lambda^{\textsc{p}_1\textsc{p}_2}_{\textsc{p}_3\textsc{p}_4} &= \sum_{\textsc{q}} \nu^{\textsc{p}_1\textsc{p}_2}_{\textsc{q}\textsc{p}_4}S^{\textsc{p}_3}_\textsc{q} + \tfrac{1}{2}\sum_{\textsc{q}\textsc{q}'} v^{\textsc{p}_1\textsc{p}_2}_{\textsc{q}\textsc{q}}S^{\textsc{p}_3}_{\textsc{q}'}S^{\textsc{p}_4}_{\textsc{q}'}  ,\\
    \Lambda^{\textsc{q}_1\textsc{q}_2}_{\textsc{q}_3\textsc{q}_4} &= \sum_{\textsc{p}} \nu^{\textsc{p}\textsc{q}_4}_{\textsc{q}_1\textsc{q}_2}S^{\textsc{p}}_{\textsc{q}_3} +  \tfrac{1}{2}\sum_{\textsc{p}\textsc{p}'}v^{\textsc{p}\textsc{p}}_{\textsc{q}_1\textsc{q}_2}S^{\textsc{p}'}_{\textsc{q}_3}S^{\textsc{p}'}_{\textsc{q}_4} ,\\
    \Lambda^{\textsc{p}_1\textsc{p}_2}_{\textsc{q}_1\textsc{q}_2} &=   \tfrac{1}{2}\sum_{\textsc{p}\textsc{q}} (S^{\textsc{p}}_{\textsc{q}}S^{\textsc{p}}_{\textsc{q}} v^{\textsc{p}_1\textsc{p}_2}_{\textsc{q}_1\textsc{q}_2} +  v^{\textsc{p}_1\textsc{p}_2}_{\textsc{q}\textsc{q}} S^\textsc{p}_{\textsc{q}_1}S^\textsc{p}_{\textsc{q}_2} + v^{\textsc{p}\textsc{p}}_{\textsc{q}_1\textsc{q}_2}S^{\textsc{p}_1}_\textsc{q} S^{\textsc{p}_2}_\textsc{q} ) + \sum_\textsc{p}\nu^{\textsc{p}_1\textsc{p}_2}_{\textsc{q}_1\textsc{p}}S^\textsc{p}_{\textsc{q}_2} + \sum_\textsc{q} \nu^{\textsc{p}_1\textsc{q}}_{\textsc{q}_1\textsc{q}_2}S^{\textsc{p}_2}_{\textsc{q}},\\
    \Lambda^{\textsc{p}_1\textsc{q}_2}_{\textsc{q}_1\textsc{p}_2} &=  \bar{\nu}^{\textsc{p}_1\textsc{q}_2}_{\textsc{q}_1\textsc{p}_2} + \tfrac{1}{2}\sum_{\textsc{p}}\nu^{\textsc{p}\textsc{p}}_{\textsc{q}_1\textsc{p}_2}S^{\textsc{p}_1}_{\textsc{q}_2} + \tfrac{1}{2}\sum_{\textsc{q}}\nu^{\textsc{p}_1\textsc{q}_2}_{\textsc{q}\textsc{q}}S^{\textsc{p}_2}_{\textsc{q}_1} +  \tfrac{1}{4}\sum_{\textsc{p}\textsc{q}}v^{\textsc{p}\textsc{p}}_{\textsc{q}\textsc{q}}S^{\textsc{p}_1}_{\textsc{q}_2}S^{\textsc{p}_2}_{\textsc{q}_1}   ,\\
    \Lambda^{\textsc{p}_1\textsc{p}_2}_{\textsc{q}_1\textsc{p}_4} &= \nu^{\textsc{p}_1\textsc{p}_2}_{\textsc{q}_1\textsc{p}_4} + \tfrac{1}{2}\sum_{\textsc{q}} v^{\textsc{p}_1\textsc{p}_2}_{\textsc{q}\textsc{q}}S^{\textsc{p}_4}_{\textsc{q}_1}  ,\\
    \Lambda^{\textsc{p}_1\textsc{q}_4}_{\textsc{q}_1\textsc{q}_2} &= \nu^{\textsc{p}_1\textsc{q}_4}_{\textsc{q}_1\textsc{q}_2} +  \tfrac{1}{2}\sum_\textsc{p} v^{\textsc{p}\textsc{p}}_{\textsc{q}_1\textsc{q}_2}S^{\textsc{p}_1}_{\textsc{q}_4}.
\end{align}

\subsubsection{Active space picture}
The active space SAPT operators in the Majorana represetnation are exactly equivalent apart from the definitions of the two-index and four-index tensors, which are mapped as: $v\rightarrow\tilde{v}$, $\Lambda\rightarrow \tilde{\Lambda}$, and so forth. In the spatial orbital basis, the renormalized active space tensor coefficients for the electrostatic-exchange operator are given by:
\begin{align}
    VP_{0,\mathrm{active}}^{(m)} &= VP_{0,\mathrm{active}} - \sum_t \tilde{f}^{(A)}_{tt} + \sum_{tt'}(\tilde{v}^{tt'}_{t't} - \tilde{v}^{tt}_{t't'}) 
    - \sum_u \tilde{f}^{(A)}_{uu} + \sum_{uu'}( \tilde{v}^{uu'}_{u'u} - \tilde{v}^{uu}_{u'u'})  - \sum_{tu} \tilde{\nu}^{tt}_{uu} \\
    &-    \tfrac{1}{2} \sum_{tuu'}\tilde{\nu}^{tu'}_{uu} S^t_{u'}  - \tfrac{1}{2}  \sum_{tt'u}\tilde{\nu}^{tt}_{ut'} S^{t'}_{u}  - \tfrac{1}{2} \sum_{tu} \tilde{\nu}^{tu}_{ut}  - \tfrac{1}{2} \sum_{ \substack{tt'\\uu'}}\tilde{v}^{tt}_{uu}S^{t'}_{u'}S^{t'}_{u'},  \nonumber \\
    \tilde{\kappa}^{(A)}_{t_1t_2} &= 2\tilde{f}^{(A)}_{t_1t_2} - 2\sum_t(\tilde{\nu}^{t_1t}_{tt_2} - \tilde{\nu}^{t_1t_2}_{tt} - \tilde{\nu}^{tt}_{t_1t_2})  +  2\sum_u \tilde{\nu}^{t_1t_2}_{uu} + \sum_{uu'} \tilde{\nu}^{t_1u'}_{uu}S^{t_2}_{u'} \\
    &+ \sum_{tu}(\tilde{\nu}^{tt}_{ut_2}S^{t_1}_{u} + \tilde{\nu}^{t_1t_2}_{ut}S^t_u) + \sum_u \tilde{\nu}^{t_1u}_{ut_2}
    + \sum_{tuu'}\tilde{v}^{t_1t_2}_{uu}S^{t}_{u'}S^{t'}_{u'} + \sum_{tuu'}\tilde{v}^{tt}_{uu}S^{t_1}_{u'}S^{t_2}_{u'} \nonumber \\
    \tilde{\kappa}^{(A)}_{u_1u_2} &= 2\tilde{f}^{(A)}_{u_1u_2} \!- \!2\sum_u( \tilde{\nu}^{u_1u}_{uu_2} \!- \! \tilde{\nu}^{u_1u_2}_{uu} - \!\tilde{\nu}^{uu}_{u_1u_2}) \!  +  2\!\sum_t\tilde{\nu}^{tt}_{u_1u_2} + \!\sum_{tt'} \tilde{\nu}^{tt}_{u_1t'}S^{t'}_{u_2} \\
    &+\sum_{tu}(\tilde{\nu}^{tu_2}_{uu}S^{t}_{u_2}\! + \tilde{\nu}^{tu}_{u_1u_2}S^{t}_{u}) + \!\sum_t \tilde{\nu}^{tu_2}_{u_1t}
    + \sum_{tt'u} \tilde{v}^{tt}_{u_1u_2}S^{t'}_{u}S^{t'}_{u} + \sum_{tt'u} \tilde{v}^{tt}_{uu}S^{t'}_{u_1}S^{t'}_{u_2} , \nonumber \\  
    \tilde{\Lambda}^{t_1t_2}_{t_3t_4} &= 2\tilde{\nu}^{t_1t_2}_{t_3t_4} + \sum_u
    \tilde{\nu}^{t_1t_2}_{ut_4}S^{t_3}_u + \sum_{uu'} \tilde{v}^{t_1t_2}_{uu}S^{t_3}_{u'}S^{t_4}_{u'}, \\
    \tilde{\Lambda}^{u_1u_2}_{u_3u_4} &= 2\tilde{\nu}^{u_1u_2}_{u_3u_4} + \sum_t
    \tilde{\nu}^{tu_4}_{u_1u_2}S^{t}_{u_3} +  \sum_{tt'} \tilde{v}^{tt}_{u_1u_2}S^{t'}_{u_3}S^{t'}_{u_4}, \\
    \tilde{\Lambda}^{t_1t_2}_{u_1u_2} &= 2\tilde{\nu}^{t_1t_2}_{u_1u_2}  + \sum_{tu} (
     S^{t}_{u}S^{t}_{u} \tilde{v}^{t_1t_2}_{u_1u_2}  + \tilde{v}^{t_1t_2}_{uu} S^t_{u_1}S^t_{u_2} + \tilde{v}^{tt}_{u_1u_2}S^{t_1}_u S^{t_2}_u) + \sum_t \tilde{\nu}^{t_1t_2}_{u_1t}S^t_{u_2} + \sum_u \tilde{\nu}^{t_1u}_{u_1u_2}S^{t_2}_{u}, \\
    \tilde{\Lambda}^{t_1u_2}_{u_1t_2} &=  \tilde{\nu}^{t_1u_2}_{u_1t_2} + \sum_t\tilde{\nu}^{tt}_{u_1t_2}S^{t_1}_{u_2} + \sum_u\tilde{\nu}^{t_1u_2}_{uu}S^{t_2}_{u_1} +  \sum_{tu}\tilde{v}^{tt}_{uu}S^{t_1}_{u_2}S^{t_2}_{u_1},      
\end{align}
\begin{align}
    \tilde{\Lambda}^{t_1t_2}_{u_1t_4} &=  \tilde{\nu}^{t_1t_2}_{u_1t_4} + \sum_u \tilde{\nu}^{t_1t_2}_{uu}S^{t_4}_{u_1},  \\
    \tilde{\Lambda}^{t_1u_4}_{u_1u_2} &= \tilde{\nu}^{t_1u_4}_{u_1u_2} + \sum_t \tilde{\nu}^{tt}_{u_1u_2}S^{t_1}_{u_4}.
\end{align}
Eqs.~\eqref{modified_V} and \eqref{modified_P} in the active space picture are given by,
\begin{align}
    \hat{V}_\mathrm{active}' &=  - \tfrac{1}{4}\sum_{\substack{\mathbf{t}\mathbf{u}\\ \sigma \tau }} \tilde{v}^{t_1t_2}_{u_1u_2} \hat{\gamma}_{t_1\sigma,0}\hat{\gamma}_{t_2\sigma,1} \hat{\gamma}_{u_1\tau,0}\hat{\gamma}_{u_2\tau,1}, \\
    \hat{P}_\mathrm{active}' &= - \tfrac{i}{4}\sum_{\substack{t_1t_2 \\\sigma}} \tilde{p}_{t_1t_2}^{(A)} \hat{\gamma}_{t_1\sigma,0}\hat{\gamma}_{t_2\sigma,1} - \tfrac{i}{4}\sum_{\substack{u_1u_2 \\ \sigma} }  \tilde{p}_{u_1u_2}^{(B)} \hat{\gamma}_{u_1\sigma,0}\hat{\gamma}_{u_2\sigma,1} + \tfrac{1}{4}\sum_{\substack{\mathbf{t}\mathbf{u}\\ \sigma  }} \text{sym}(S^{t_1}_{u_2}S^{t_2}_{u_1})\hat{\gamma}_{t_1\sigma,0}\hat{\gamma}_{t_2\sigma,1} \hat{\gamma}_{u_1\sigma,0}\hat{\gamma}_{u_2\sigma,1}.
\end{align}
with
\begin{align}
    \tilde{p}^{(A)}_{t_1t_2} &=\sum_{u} S^{t_1}_{u}S^{t_2}_{u}  + 2\sum_{j} S^{t_1}_{j} S^{t_2}_{j}, \\
    \tilde{p}^{(B)}_{u_1u_2} &= \sum_{t} S^{t}_{u_2}S^{t}_{u_1} + 2\sum_{i} S^{i}_{u_1}S^{i}_{u_2}.
\end{align}
In the following two sections, we present the \emph{sparse} and \emph{tensor factorization} encoding schemes which can apply to both the all-electron (full-space) and active space pictures.

\subsection{Sparse representation}
In the \emph{sparse} encoding scheme \cite{berry2018improved} scheme, a data-loading oracle loads the non-zero entries of the tensor coefficients reducing the overall cost compared to an equivalent \emph{dense} scheme. The Jordan-Wigner mapping from the Majorana operators for monomer A is given by,
\begin{align}
    \hat{\gamma}_{p\sigma,0} = \hat{X}_{p,\sigma}\hat{Z}_{p-1,\sigma}\cdots \hat{Z}_{0,\sigma}, \\
    \hat{\gamma}_{p\sigma,1} = \hat{Y}_{p,\sigma}\hat{Z}_{p-1,\sigma}\cdots \hat{Z}_{0,\sigma}.
\end{align}
Similarly, for monomer B we have:
\begin{align}
    \hat{\gamma}_{q\tau,0} = \hat{X}_{q,\tau}\hat{Z}_{q-1,\tau}\cdots \hat{Z}_{0,\tau}, \\
    \hat{\gamma}_{q\tau,1} = \hat{Y}_{q,\tau}\hat{Z}_{q-1,\tau}\cdots \hat{Z}_{0,\tau}.
\end{align}
The $\mathsf{PREPARE}$ and $\mathsf{SELECT}$ operators for these operators are given explicitly in \cite{babbush2018encoding}. Here, we summarize the total $\ell_1$ norm for all of the SAPT operators: 
\begin{align}
    \lambda^{(s)}_V &=  \sum_{q_1q_2} \Big| \sum_p {v}^{pp}_{q_1q_2}\Big| + \sum_{p_1p_2} \Big|  \sum_q v^{p_1p_2}_{qq} \Big|  + \sum_{\substack{p_1p_2\\q_1q_2}} |v^{p_1p_2}_{q_1q_2}|, \\
    \lambda^{(s)}_P &= \tfrac{1}{2}\sum_{q_1q_2} \Big| \sum_p S^p_{q_2}S^p_{q_1}\Big| + \tfrac{1}{2}\sum_{p_1p_2} \Big| \sum_q S^{p_1}_{q}S^{p_2}_{q}\Big|  + \tfrac{1}{2}\Big[\sum_{pq} |S^p_{q}|\Big]^2,  \\ 
    \lambda^{(s)}_{VP} &=  \lambda^{(s)}_{VP_A} + \lambda^{(s)}_{VP_B} + \lambda^{(s)}_{VP_{1\mathrm{m}}} + \lambda^{(s)}_{VP_{1\ell}} + \lambda^{(s)}_{VP_2} + \lambda^{(s)}_{VP_3} +  \lambda^{(s)}_{P}\sum_{\substack{p_1p_2\\q_1q_2}} |v^{p_1p_2}_{q_1q_2}|,
\end{align}
where
\begin{align}
    \lambda^{(s)}_{VP_A} &= \tfrac{1}{2} \sum_{p_1p_2} |\kappa^{(A)}_{p_1p_2}| + \tfrac{1}{2} \sum_{ \substack{p_1>p_2\\p_3>p_4}} |\Lambda^{p_1p_2}_{p_3p_4} - \Lambda^{p_1p_4}_{p_3p_2}| + \tfrac{1}{4}\sum_{ \substack{p_1p_2\\p_3p_4}} |\Lambda^{p_1p_2}_{p_3p_4}|, \\ 
    \lambda^{(s)}_{VP_B} &= \tfrac{1}{2} \sum_{q_1q_2} |\kappa^{B}_{q_1q_2}| +  \tfrac{1}{2} \sum_{ \substack{q_1>q_2\\q_3>q_4}} |\Lambda^{q_1q_2}_{q_3q_4} - \Lambda^{q_1q_4}_{q_3q_2}| + \tfrac{1}{4}\sum_{ \substack{q_1q_2\\q_3q_4}} |\Lambda^{q_1q_2}_{q_3q_4}|, \\
    \lambda^{(s)}_{VP_{1\mathrm{m}}} &= \tfrac{1}{2}\sum_{\substack{p_1p_2\\q_1q_2}} |\Lambda^{p_1p_2}_{q_1q_2}|, \\ 
    \lambda^{(s)}_{VP_{1\ell}} &= \tfrac{1}{2}\sum_{\substack{p_1p_2\\q_1q_2}} |\Lambda^{p_1q_2}_{q_1p_2}|, \\
    \lambda^{(s)}_{VP_2} &= \tfrac{1}{2}\!\!\!\!\sum_{\substack{p_1p_2p_3p_4\\q_1q_2}}\!\!\!\!\! | \Lambda^{p_1p_2}_{q_1p_4}S^{p_3}_{q_2}|, \\    
    \lambda^{(s)}_{VP_3} &= 
    \tfrac{1}{2}\!\!\!\!\sum_{\substack{p_1p_2\\q_1q_2q_3q_4}} \!\!\!| \Lambda^{p_1q_4}_{q_1q_2}S^{p_2}_{q_3}|
\end{align}
Here, the superscript ${(s)}$ is used to denote the sparse representation. The total number $L_F$ of operator terms for each of the SAPT operators $\hat{F}$ will scale as $L_V^{(s)} = \mathcal{O}(N_A^2 N_B^2), L_P^{(s)} = \mathcal{O}(N_A^2 N_B^2)$, and $L_{VP}^{(s)} = \mathcal{O}(N_A^4 N_B^4)$ using a naive block encoding. As we show below, however, it is possible to reduce this scaling with respect to the total number of terms $L_F$ by using factorization circuits especially designed to describe the product of two operators, e.g. $\hat{V}$ and $\hat{P}$. This well help reduce the scaling of certain terms, such as the second, third, and fourth terms: $\widehat{VP}_2$, $\widehat{VP}_3$ and $\widehat{VP}_4$. For instance, the compilation scaling of the fourth term in the electrostatic exchange operator, Eq. \eqref{eq:VP4_majorana}, will reduce to $L_{VP}^{(s)} = \mathcal{O}(N_A^4 + N_B^4)$. It is also worth pointing out that we used the anti-symmetry property of the fermionic operators to reduce the $\ell_1$ norm of the intra-monomer contributions as pointed out in Ref.~\cite{koridon2021orbital}.

\subsection{Factorized tensor representation}
\label{sec:factorization}
While the sparse representation serves to provide a first-pass implementation of symmetry-adapted perturbation theory, it does not scale well with the number of terms $L$, nor does it scale well with the $\ell_1$ norm of the operator. In the following, we present a tensor factorization scheme that is analogous to the low-rank double factorization scheme~\cite{VonBurg2021} used in Hamiltonian simulation. The advantage afforded by this encoding scheme is two-fold. First, it will be shown to drastically reduce the $\ell_1$ norm of each of the three operators. Fundamentally, this is explained by the fact that the Schatten 1-norm is strictly less than or equal to the entry-wise 1-norm of tensors. This property was pointed out by Von Burg et al. and becomes apparent in the numerical benchmark data sets we present in the results section. Second, it reduces the total number of terms $L$ required for the data-loading oracle. It is important to emphasize that without an appropriate factorization scheme, the implementation of SAPT on the quantum computer would be much less scalable. The complete tensor factorization procedure is presented in the main text. Using this procedure, all of the unique four-index tensors in first order SAPT theory may be expanded as:
\begin{align}
    v^{p_1p_2}_{q_1q_2} &= \sum_{tkl} s^{(v)}_t\alpha^{(A_{v})}_{tk}\alpha^{(B_{v})}_{tl} U^{(v)}_{tkp_1} U^{(v)}_{tkp_2} V^{(v)}_{tlq_1} V^{(v)}_{tlq_2}, \\    
    \text{sym}(\Lambda^{p_1p_2}_{p_3p_4}) &= \sum_{tkl} s^{(A_2)}_t\alpha^{(A_2)}_{tk}\alpha^{(A_2)}_{tl} U^{(A_2)}_{tkp_1}U^{(A_2)}_{tkp_2} U^{(A_2)}_{tlp_3} U^{(A_2)}_{tlp_4} \label{A_expansion} ,\\
    \text{sym}(\Lambda^{q_1q_2}_{q_3q_4}) &= \sum_{tkl} s^{(B_2)}_t\alpha^{(B_2)}_{tk}\alpha^{(B_2)}_{tl} V^{(B_2)}_{tkq_1}V^{(B_2)}_{tkq_2} V^{(B_2)}_{tlq_3} V^{(B_2)}_{tlq_4} \label{B_expansion},\\
    \text{sym}(\Lambda^{p_1q_2}_{q_1p_2}) &= \sum_{tkl} s^{(1\ell)}_t\beta^{(1\ell)}_{tk}\beta^{(1\ell)}_{tl}
    U^{(1\ell)}_{tkp_1}V^{(1\ell)}_{tkq_2} U^{(1\ell)}_{tlp_2} V^{(1\ell)}_{tlq_1} ,\\
    \text{sym}(\Lambda^{p_1p_2}_{q_1q_2}) &= \sum_{tkl} s^{(1\mathrm{m})}_t\alpha^{(A_{1\mathrm{m}})}_{tk}\alpha^{(B_{1\mathrm{m}})}_{tl} 
    U^{(1\mathrm{m})}_{tkp_1} U^{(1\mathrm{m})}_{tkp_2} V^{(1\mathrm{m})}_{tlq_1} V^{(1\mathrm{m})}_{tlq_2} ,\\
    \Lambda^{p_1p_2}_{q_1p_4} &= \sum_{tkl} s^{(2)}_t\alpha^{(2)}_{tk}\beta^{(2)}_{tl}
    U^{(2)}_{tkp_1} U^{(2)}_{tkp_2} \tilde{U}^{(2)}_{tlp_4} V^{(2)}_{tlq_1} ,\\
    \Lambda^{p_1q_4}_{q_1q_2} &= \sum_{tkl} s^{(3)}_t\alpha^{(3)}_{tk}\beta^{(3)}_{tl} 
    U^{(3)}_{tlp_1}\tilde{V}^{(3)}_{tkq_4} V^{(3)}_{tlq_1}V^{(3)}_{tlq_2},
\end{align}
which is exact when none of the terms are truncated. In the following subsections, we use this expansion to summarize the tensor-factorized SAPT operators applicable in both the all-electron (full space) and active space pictures. Throughout the following sections, we make use of the orbital-transformed Majorana operators,
\begin{align}
    \tilde{\gamma}_{k\sigma,i} &= \sum_p U_{pk} \hat{\gamma}_{\textsc{p},i} = \hat{G}^{(A)\dagger}_{\sigma} \hat{\gamma}_{k\sigma,i} \hat{G}^{(A)}_{\sigma},    \\
    \tilde{\gamma}_{l\tau,i} &= \sum_q V_{ql} \hat{\gamma}_{l\tau,i} = \hat{G}^{(B)\dagger}_{\tau} \hat{\gamma}_{l\tau,i} \hat{G}^{(B)}_{\tau},
\end{align}
where $i=\{0,1\}$ and the second equality in both expressions is defined with respect to the Givens operator, $\hat{G}^{(A)}_{\sigma} = \exp(\sum_{pk}[\log \mathbf{U}]_{pk} \hat{a}^\dagger_{\textsc{p}} \hat{a}_{k\sigma} )$ and $\hat{G}^{(B)}_{\tau} = \exp(\sum_{ql}[\log \mathbf{V}]_{ql} \hat{b}^\dagger_{q\tau} \hat{b}_{l\tau} )$ respectively.

\subsubsection{Factorization of electrostatic operator: $\hat{V}$}
\label{sec:fact_electrostatics}
Following tensor factorization procedure from the main text, we find the final form of the electrostatic operator,
\begin{align}
    \hat{V} &=  \sum_{pq}v^{pp}_{qq} + \tfrac{i}{2}\sum_{k,\sigma} s_k^{(A)} \tilde{\gamma}_{k\sigma,0}\tilde{\gamma}_{k\sigma,1} 
    +  \tfrac{i}{2}\sum_{l,\sigma} s_l^{(B)} \tilde{\gamma}_{l\sigma,0}\tilde{\gamma}_{l\sigma,1} 
    - \tfrac{1}{4}\sum_{\substack{tkl \\ \sigma\tau}} s^{(v)}_t\alpha^{(A_{v})}_{tk}\alpha^{(B_{v})}_{tl} \tilde{\gamma}_{k\sigma,0}\tilde{\gamma}_{k\sigma,1}
    \tilde{\gamma}_{l\tau,0}\tilde{\gamma}_{l\tau,1}, \\
    &=  \sum_{pq}v^{pp}_{qq} - \tfrac{1}{2}\sum_{k,\sigma} s_k^{(A)} \hat{G}_\varnothing^{(A)\dagger}\hat{Z}_{k\sigma}\hat{G}_\varnothing^{(A)} 
    - \tfrac{1}{2}\sum_{l,\sigma} s_l^{(B)} \hat{G}_\varnothing^{(B)\dagger}\hat{Z}_{l\sigma}  \hat{G}_\varnothing^{(B)} 
     \label{V_tensor_factorization} \\
    &+\tfrac{1}{4} \sum_{\substack{tkl \\ \sigma\tau}} [s^{(v)}_t\alpha^{(A_{v})}_{tk}\alpha^{(B_{v})}_{tl}] \Big(\hat{G}_t^{(A)\dagger}\hat{Z}_{k\sigma}\hat{G}_t^{(A)} \Big)\otimes \Big(\hat{G}_t^{(B)\dagger}\hat{Z}_{l\tau}\hat{G}_t^{(B)}\Big) .
    \nonumber
\end{align}
In the second line, we used the following Jordan-Wigner identity for each monomer, 
\begin{align}
    \hat{\gamma}_{k\sigma,0}\hat{\gamma}_{k\sigma,1} &= i\hat{Z}_{k\sigma}, \\
    \hat{\gamma}_{l\tau,0}\hat{\gamma}_{l\tau,1} &= i\hat{Z}_{l\tau}, 
\end{align}
and we have also defined the Givens operators, $\hat{G}^{(X)}_\varnothing = \hat{G}^{(X)}_{\varnothing,\alpha} \otimes \hat{G}^{(X)}_{\varnothing,\beta}$ and $\hat{G}^{(X)}_t = \hat{G}^{(X)}_{t,\alpha} \otimes \hat{G}^{(X)}_{t,\beta}$ for $X\in \{A,B\}$, where each spin-block is defined as:
\begin{align}
    \hat{G}^{(A)}_{\varnothing,\sigma} &=  \exp\left(\sum_{p>k}[\log \mathbf{U}^{(f)}]_{pk} ( \hat{E}^{\sigma}_{pk} - \hat{E}^{\sigma}_{kp})  \right), \\
    \hat{G}^{(B)}_{\varnothing,\tau} &=  \exp\left(\sum_{q>l}[\log \mathbf{V}^{(f)}_t]_{ql} ( \hat{E}^{\tau}_{ql} - \hat{E}^{\tau}_{lq})  \right), 
\end{align}
and
\begin{align}
    \hat{G}^{(A)}_{t,\sigma} &=  \exp\left(\sum_{p>k}[\log \mathbf{U}^{(v)}_t]_{pk} ( \hat{E}^{\sigma}_{pk} - \hat{E}^{\sigma}_{kp})  \right), \\
    \hat{G}^{(B)}_{t,\tau} &=  \exp\left(\sum_{q>l}[\log \mathbf{V}^{(v)}_t]_{ql} ( \hat{E}^{\tau}_{ql} - \hat{E}^{\tau}_{lq})  \right) .
\end{align}
The corresponding $\ell_1$ norm for the tensor-factorized electrostatic operator is given by,
\begin{equation}
    \lambda_V^{(\mathrm{tf})} = \sum_k |s_k^{(A)}| + \sum_l |s_l^{(B)}| + \sum_{tkl} |s^{(v)}_t\alpha^{(A_{v})}_{tk}\alpha^{(B_{v})}_{tl}|.
\end{equation}

\subsection{Factorization of Exchange Operator: $\hat{P}$}
\label{sec:fact_exchange}
We consider the tensor factorization procedure for the active space exchange operator since it is more general than the all-electron  (full space) version. Substituting the factorization of the intermolecular overlap matrix, we obtain:
\begin{align}
    \hat{P}_\mathrm{active} &= -\tfrac{1}{2}\sum_{pq}S^{p}_{q}S^{p}_{q} - \tfrac{i}{4}\sum_{k,\sigma} s_k^{(P_{A_1})} \tilde{\gamma}_{k\sigma,0}\tilde{\gamma}_{k\sigma,1} 
    - \tfrac{i}{4}\sum_{l,\sigma} s_l^{(P_{B_1})} \tilde{\gamma}_{l\sigma,0}\tilde{\gamma}_{l\sigma,1} \\
    &+ \tfrac{1}{16}\sum_{\substack{kl\\ \sigma  }} s_k s_l (\tilde{\gamma}_{l\sigma,0}\tilde{\gamma}_{k\sigma,1} + \tilde{\gamma}_{k\sigma,0}\tilde{\gamma}_{l\sigma,1} )\otimes (\tilde{\gamma}_{l\sigma,0}\tilde{\gamma}_{k\sigma,1} + \tilde{\gamma}_{k\sigma,0}\tilde{\gamma}_{l\sigma,1} ) \nonumber \\
    &= -\tfrac{1}{2}\sum_{pq}S^{p}_{q}S^{p}_{q} + \tfrac{i}{4}\sum_{k,\sigma} s_k^{(P_{A_1})} \hat{G}^{(P_{A_1})\dagger}_\varnothing \hat{Z}_{k\sigma} \hat{G}^{(P_{A_1})}_\varnothing + \tfrac{1}{4}\sum_{l,\sigma} s_l^{(P_{B_1})} \hat{G}^{(P_{B_1})\dagger}_\varnothing \hat{Z}_{l\sigma}\hat{G}^{(P_{B_1})}_\varnothing  \label{P_JordanWigner} \\
    &- \tfrac{1}{4}\sum_{k,\sigma} s_k^2 \Big(\hat{G}^{(A_S)\dagger}\hat{Z}_{k\sigma}\hat{G}^{(A_S)}\Big)\otimes\Big(\hat{G}^{(B_S)\dagger}\hat{Z}_{k\sigma} \hat{G}^{(B_S)\dagger} \Big)  \nonumber \\ 
    &+\tfrac{1}{16}\sum_{ \substack{ k\neq l \\ \sigma}}
    s_ks_l \Big(\hat{G}^{(A_S)\dagger}[\hat{X}_{k\sigma}\vec{Z}_{kl}^\sigma \hat{X}_{l\sigma} + \hat{Y}_{k\sigma}\vec{Z}_{kl}^\sigma \hat{Y}_{l\sigma}]\hat{G}^{(A_S)}\Big)\otimes \Big(\hat{G}^{(B_S)\dagger}[\hat{X}_{k\sigma}\vec{Z}_{kl}^\sigma \hat{X}_{l\sigma} + \hat{Y}_{k\sigma}\vec{Z}_{kl}^\sigma \hat{Y}_{l\sigma}]\hat{G}^{(B_S)} \Big) , \nonumber
\end{align}
where
\begin{equation}
    \vec{Z}^\sigma_{kl} = 
    \begin{cases}
        \hat{Z}_{k-1,\sigma}\cdots \hat{Z}_{l+1,\sigma} & \text{if}\;\; k>l \\
        \hat{Z}_{l-1,\sigma}\cdots \hat{Z}_{k+1,\sigma} & \text{if}\;\; l>k.
    \end{cases}
\end{equation}
We have also defined the Givens operators, $\hat{G}^{(P_{X_1})}_\varnothing = \hat{G}^{(P_{X_1})}_{\varnothing,\alpha} \otimes \hat{G}^{(P_{X_1})}_{\varnothing,\beta}$ and $\hat{G}^{(X_S)} = \hat{G}^{(X_S)}_{\alpha} \otimes \hat{G}^{(X_S)}_{\beta}$ for $X\in \{A,B\}$, where each spin-block is defined as:
\begin{align}
    \hat{G}^{(P_{A_1})}_{\varnothing,\sigma} &=  \exp\left(\sum_{p>k}[\log \mathbf{U}^{(P_{A_1})}]_{pk} ( \hat{E}^{\sigma}_{pk} - \hat{E}^{\sigma}_{kp})  \right), \\
    \hat{G}^{(P_{B_1})}_{\varnothing,\tau} &=  \exp\left(\sum_{q>l}[\log \mathbf{V}^{(P_{B_1})}_t]_{ql} ( \hat{E}^{\tau}_{ql} - \hat{E}^{\tau}_{lq})  \right), 
\end{align}
and
\begin{align}
    \hat{G}^{(A_S)}_{\sigma} &=  \exp\left(\sum_{p>k}[\log \mathbf{U}^{(s)}]_{pk} ( \hat{E}^{\sigma}_{pk} - \hat{E}^{\sigma}_{kp})  \right), \\
    \hat{G}^{(B_S)}_{\tau} &=  \exp\left(\sum_{q>l}[\log \mathbf{V}^{(s)}]_{ql} ( \hat{E}^{\tau}_{ql} - \hat{E}^{\tau}_{lq})  \right) .
\end{align}

The corresponding $\ell_1$ norm is given by,
\begin{equation}
    \lambda_P^{(\mathrm{tf})} = \tfrac{1}{2}\sum_k |s_k^{(P_{A_1})}| + \tfrac{1}{2}\sum_l |s_l^{(P_{B_1})}| + \tfrac{1}{2}\left(\sum_n s_n \right)^2.
\end{equation}

\subsection{Factorization of electrostatic-exchange operator: $\widehat{VP}$}
\label{sec:fact_electrostatic_exchange}

The electrostatic-exchange factorization is decomposed with respect to six different four-index contributions, which
we outline below. The first two correspond to monomer-only contributions, $\widehat{VP}_A$ and $\widehat{VP}_B$ 
respectively. In the following, we provide the factorization procedure which arrives at the same result
as the standard double factorization procedure derived by Von Burg et al. 

\subsubsection*{$\widehat{VP}_A$/$\widehat{VP}_B$:}
Substituting the factorization of the two-body terms into Eqs.~\eqref{eq:VPA_majorana}-\eqref{eq:VPB_majorana}, we find
\begin{align}
    \widehat{VP}_A &= \sum_k \tilde{s}_k^{(A)}  \tilde{\gamma}_{k\sigma,0}\tilde{\gamma}_{k\sigma,1} 
    - \sum_{tkl} s^{(A_2)}_t\alpha^{(A_2)}_{tk}\alpha^{(A_2)}_{tl} \tilde{\gamma}_{k\sigma,0}\tilde{\gamma}_{k\sigma,1}
    \tilde{\gamma}_{l\tau,0}\tilde{\gamma}_{l\tau,1}, \\
    \widehat{VP}_B &= \sum_k \tilde{s}_k^{(B)} \tilde{\gamma}_{k\sigma,0}\tilde{\gamma}_{k\sigma,1} 
    - \sum_{tkl} s^{(B_2)}_t\alpha^{(B_2)}_{tk}\alpha^{(B_2)}_{tl} \tilde{\gamma}_{k\sigma,0}\tilde{\gamma}_{k\sigma,1}
    \tilde{\gamma}_{l\tau,0}\tilde{\gamma}_{l\tau,1}.
\end{align}
The transformed Majorana operators may be found in Ref.~\cite{VonBurg2021} with the appropriate orbital transformation matrices found in Eqs.~\eqref{A_expansion}-\eqref{B_expansion}. Since the two-body terms for each monomer have a complete-square form, it is possible to use a quantum signal processing technique to load a second order Chebyshev polynomial in order implement the second body term \cite{VonBurg2021,Lee2021}. This effectively reduces the $\ell_1$ norm of the two-body term by a factor of two, leading to the following expressions:
\begin{align}
    \lambda_{VP_A} &= \tfrac{1}{2}\sum_k |\tilde{s}_k^{(A)}| + \tfrac{1}{4} \sum_t |s_t^{(A_2)}| \Big(\sum_k |\alpha^{(A_2)}_{tk}|\Big)^2, \\
    \lambda_{VP_B} &= \tfrac{1}{2}\sum_l |\tilde{s}_l^{(B)}| + \tfrac{1}{4} \sum_t |s_t^{(B_2)}| \Big(\sum_l |\alpha^{(B_2)}_{tl}|\Big)^2 .
\end{align}

\subsubsection*{$\widehat{VP}_{1\mathrm{m}}$:}
The first order mixed-spin term is equivalent to the electrostatic operator apart from one-body contributions. The tensor factorized operator may be written explicitly as,
\begin{align}
    \widehat{VP}_{1\mathrm{m}} &= \tfrac{1}{8}\sum_{\substack{tkl \\ \sigma\tau}} s^{(1\mathrm{m})}_t\alpha_{kt}^{(A_{1\mathrm{m}})}\alpha_{lt}^{(B_{1\mathrm{m}})} \tilde{\gamma}^{(A_{1\mathrm{m}})}_{k\sigma,0}\tilde{\gamma}^{(A_{1\mathrm{m}})}_{k\sigma,1}
    \tilde{\gamma}^{(B_{1\mathrm{m}})}_{l\tau,0}\tilde{\gamma}^{(B_{1\mathrm{m}})}_{l\tau,1}, \\
    &= -\tfrac{1}{8} \sum_{\substack{tkl \\ \sigma\tau}} [s^{(1\mathrm{m})}_t\alpha_{kt}^{(A_{1\mathrm{m}})}\alpha_{lt}^{(B_{1\mathrm{m}})}] \Big(\hat{G}_t^{(A_{1\mathrm{m}})\dagger}\hat{Z}_{k\sigma}\hat{G}_t^{(A_{1\mathrm{m}})} \Big)\otimes \Big(\hat{G}_t^{(B_{1\mathrm{m}})\dagger}\hat{Z}_{l\tau}\hat{G}_t^{(B_{1\mathrm{m}})}\Big) .
\end{align}
where the Givens operators for each spin-block are defined as:
\begin{align}
    \hat{G}^{(A_{1\mathrm{m}})}_{t,\sigma} &=  \exp\left(\sum_{p>k}[\log \mathbf{U}^{(1\mathrm{m})}_t]_{pk} ( \hat{E}^{\sigma}_{pk} - \hat{E}^{\sigma}_{kp})  \right), \\
    \hat{G}^{(B_{1\mathrm{m}})}_{t,\tau} &=  \exp\left(\sum_{q>l}[\log \mathbf{V}^{(1\mathrm{m})}_t]_{ql} ( \hat{E}^{\tau}_{ql} - \hat{E}^{\tau}_{lq})  \right) .
\end{align}
The corresponding $\ell_1$ norm is given by:
\begin{equation}
    \lambda_{VP_{1\mathrm{m}}} = \tfrac{1}{2}\sum_{tkl} |s^{(1\mathrm{m})}_t\alpha_{kt}^{(A_{1\mathrm{m}})}\alpha_{lt}^{(B_{1\mathrm{m}})}|.
\end{equation}

\subsubsection*{$\widehat{VP}_{1\ell}$:}
The first order spin-locked contribution is similar to the exchange operator but requires a four-index factorization procedure outlined in previous terms. The tensor factorized operator is written as:
\begin{align}
    \widehat{VP}_{1\ell} &= \tfrac{1}{4}\sum_{\substack{tkl \\ \sigma}} s^{(1\ell)}_t\beta^{(1\ell)}_{tk}\beta^{(1\ell)}_{tl} \tilde{\gamma}^{A_{1\ell}}_{k\sigma,0}\tilde{\gamma}^{A_{1\ell}}_{l\sigma,1}
    \tilde{\gamma}^{B_{1\ell}}_{k\sigma,0}\tilde{\gamma}^{B_{1\ell}}_{l\sigma,1}, \\
    &= - \tfrac{1}{4}\sum_{\substack{tk \\ \sigma}} s^{(1\ell)}_t\beta^{(1\ell)}_{tk}\beta^{(1\ell)}_{tk} \Big(\hat{G}_t^{(A_{1\ell})\dagger}\hat{Z}_{k\sigma}\hat{G}_t^{(A_{1\ell})} \Big)\otimes \Big(\hat{G}_t^{(B_{1\ell})\dagger}\hat{Z}_{k\sigma}\hat{G}_t^{(B_{1\mathrm{m}})}\Big) 
    \label{eq:VP1l_TF} \\
    &+\tfrac{1}{16}\sum_{ \substack{ t \sigma \\ k\neq l}}
    s^{(1\ell)}_t\beta^{(1\ell)}_{tk}\beta^{(1\ell)}_{tl} \Big(\hat{G}_t^{(A_{1\ell})\dagger}[\hat{X}_{k\sigma}\vec{Z}_{kl}^\sigma \hat{X}_{l\sigma} + \hat{Y}_{k\sigma}\vec{Z}_{kl}^\sigma \hat{Y}_{l\sigma}]\hat{G}_t^{(A_{1\ell})}\Big)\otimes \Big(\hat{G}_t^{(B_{1\ell})\dagger}[\hat{X}_{k\sigma}\vec{Z}_{kl}^\sigma \hat{X}_{l\sigma} + \hat{Y}_{k\sigma}\vec{Z}_{kl}^\sigma \hat{Y}_{l\sigma}]\hat{G}_t^{(B_{1\ell})}\Big) .\nonumber
\end{align}
where we used the permutation symmetries of the four-index tensor to go from the first line to the second line and we have defined the Givens operators for each spin-block as:
\begin{align}
    \hat{G}^{(A_{1\ell})}_{t,\sigma} &=  \exp\left(\sum_{p>k}[\log \mathbf{U}^{(1\ell)}_t]_{pk} ( \hat{E}^{\sigma}_{pk} - \hat{E}^{\sigma}_{kp})  \right), \\
    \hat{G}^{(B_{1\ell})}_{t,\tau} &=  \exp\left(\sum_{q>l}[\log \mathbf{V}^{(1\ell)}_t]_{ql} ( \hat{E}^{\tau}_{ql} - \hat{E}^{\tau}_{lq})  \right) .
\end{align}
The corresponding $\ell_1$ norm given by:
\begin{equation}
    \lambda_{VP_{1\ell}} = \tfrac{1}{2}\sum_{tkl} |s^{(1\ell)}_t\beta^{(1\ell)}_{tk}\beta^{(1\ell)}_{tl}|.
\end{equation}

\subsubsection*{$\widehat{VP}_{2}$/$\widehat{VP}_{3}$:}
The second and third terms have similar structure.  Here, we provide the explicit form for the second term taking into account all of the permutations involved for $\text{sym}(\Lambda^{p_1p_2}_{q_1p_4}S^{p_3}_{q_2})$: 
\begin{align}
    \widehat{VP}_2 &= -\tfrac{i}{8} \sum_{\substack{tkl\\ n \sigma\tau}} [s_n s^{(2)}_t\alpha^{(2)}_{tk}\beta^{(2)}_{tl}] \tfrac{1}{8}\Big( \tilde{\gamma}^{(2_\alpha)}_{k\sigma,0}\tilde{\gamma}^{(2_\alpha)}_{k\sigma,1} \Big[\tilde{\gamma}^{(A_S)}_{n\tau,0}\tilde{\gamma}^{(A_{2\beta})}_{l\tau,1}\otimes \tilde{\gamma}^{(B_{2\beta})}_{l\tau,0}\tilde{\gamma}^{(B_S)}_{n\tau,1} + \tilde{\gamma}^{(A_S)}_{n\tau,0}\tilde{\gamma}^{(A_{2\beta})}_{l\tau,1}\otimes \tilde{\gamma}^{(B_S)}_{n\tau,0}\tilde{\gamma}^{(B_{2\beta})}_{l\tau,1} \\
    &+ \tilde{\gamma}^{(A_{2\beta})}_{l\tau,0}\tilde{\gamma}^{(A_S)}_{n\tau,1}\otimes\tilde{\gamma}^{(B_{2\beta})}_{l\tau,0}\tilde{\gamma}^{(B_S)}_{n\tau,1} + 
    \tilde{\gamma}^{(A_{2\beta})}_{l\tau,0}\tilde{\gamma}^{(A_S)}_{n\tau,1}\otimes \tilde{\gamma}^{(B_S)}_{n\tau,0}\tilde{\gamma}^{(B_{2\beta})}_{l\tau,1}\Big] \nonumber \\
    &+ \Big[\tilde{\gamma}^{(A_S)}_{n\tau,0}\tilde{\gamma}^{(A_{2\beta})}_{l\tau,1}\otimes \tilde{\gamma}^{(B_{2\beta})}_{l\tau,0}\tilde{\gamma}^{(B_S)}_{n\tau,1} + \tilde{\gamma}^{(A_S)}_{n\tau,0}\tilde{\gamma}^{(A_{2\beta})}_{l\tau,1}\otimes \tilde{\gamma}^{(B_S)}_{n\tau,0}\tilde{\gamma}^{(B_{2\beta})}_{l\tau,1} \nonumber\\
    &+ \tilde{\gamma}^{(A_{2\beta})}_{l\tau,0}\tilde{\gamma}^{(A_S)}_{n\tau,1}\otimes \tilde{\gamma}^{(B_{2\beta})}_{l\tau,0}\tilde{\gamma}^{(B_S)}_{n\tau,1} + 
    \tilde{\gamma}^{(A_{2\beta})}_{l\tau,0}\tilde{\gamma}^{(A_S)}_{n\tau,1}\otimes \tilde{\gamma}^{(B_S)}_{n\tau,0}\tilde{\gamma}^{(B_{2\beta})}_{l\tau,1}\Big] \tilde{\gamma}^{(2_\alpha)}_{k\sigma,0}\tilde{\gamma}^{(2_\alpha)}_{k\sigma,1} \Big), \nonumber
\end{align}
where all of the orbital-transformed Majorana operators are explicitly defined as:
\begin{align}
    \tilde{\gamma}_{k\sigma,i}^{(2_\alpha)} &= \sum_p U^{(2)}_{tkp} \hat{\gamma}_{p\sigma,i}, \\
    \tilde{\gamma}_{l\sigma,i}^{(A_{2\beta})} &= \sum_p \tilde{U}^{(2)}_{tlp} \hat{\gamma}_{p\sigma,i},  \\
    \tilde{\gamma}_{l\sigma,i}^{(B_{2\beta})} &= \sum_q V^{(2)}_{tlq} \hat{\gamma}_{q\sigma,i},\\
    \tilde{\gamma}_{n\sigma,i}^{(A_S)} &= \sum_p U^{(s)}_{np} \hat{\gamma}_{p\sigma,i}, \\
    \tilde{\gamma}_{n\sigma,i}^{(B_S)} &= \sum_q V^{(s)}_{nq} \hat{\gamma}_{q\sigma,i}.
\end{align} 
The corresponding $\ell_1$ norm is given by:
\begin{align}
    \lambda_{VP_2} &= \tfrac{1}{2}\lambda_s\sum_{tkl} |s^{(2)}_t\alpha^{(2)}_{tk}\beta^{(2)}_{tl}|.
\end{align}
The $\ell_1$ norm of the third term is also found to be given by,
\begin{align}
    \lambda_{VP_3} &= \tfrac{1}{2}\lambda_s\sum_{tkl} |s^{(3)}_t\alpha^{(3)}_{tk}\beta^{(3)}_{tl}|.
\end{align}

\subsubsection*{$\widehat{VP}_{4}$:}
The fourth term consists of the symmetric product of two operators, $\hat{V}'$ and $\hat{P}'$ defined above.
The corresponding $\ell_1$ norm is simply given by,
\begin{equation}
    \lambda_{VP_4} = \lambda_P^{(\mathrm{tf})}\sum_{tkl} |s^{(v)}_t\alpha^{(A_{v})}_{tk}\alpha^{(B_{v})}_{tl}|.
\end{equation}
To implement this operator on the quantum computer, we have two options which we discuss in more detail in Appendix \ref{subsec:PBEO}. We can block encode the entire operator $\widehat{VP}_4$ which will consist of $L_VL_P$ total terms without truncation. Using the tensor factorization procedures from previous sections, we have found empirically that the number of terms of the electrostatic operator, $L_V$, will scale between $\mathcal{O}( N^2 ) - \mathcal{O}( N^3 )$ depending on whether we increase the number of basis orbitals in the continuum limit (increasing the basis set size) or increase the number of active space orbitals with fixed filling fraction. Since $L_p$ scales as $\mathcal{O}(N^2)$ then the quantum circuit for $\widehat{VP}_4$ will have an overall complexity of $\mathcal{O}(N^5)$ using a naive implementation. For large system sizes, this implies that the compilation overhead for this operator will be substantial. While we have shown in the main manuscript that the eigenstate reflection circuit is the primary bottleneck in the SAPT-EVE algorithm, it is still worth understanding how we can improve this scaling to ensure that the block encoding of $\widehat{VP}_4$ is not the rate-limiting step once improved eigenstate preparation techniques are developed in the future. As a result, we propose the self-inverse product of block encoded operators circuit outlined in more detail in Appendix \ref{subsec:PBEO}. While the total normalization constant (which affects the total run-time of the expectation value estimation algorithm) will remain equal to the product, $\lambda_a\lambda_b$, the compilation cost is additive with respect to the two operators, $H_a$ and $H_b$, scaling asymptotically as $\mathcal{O}(L_a + L_b)$ rather than $\mathcal{O}(L_aL_b)$. 

\section{SAPT block encoding}
\label{sec:SAPT_block_encoding}

block encodings provide a powerful framework for performing non-unitary operations on a quantum computer. In the following, we describe the block encoding of the electrostatic, exchange and dominant terms of the electrostatic-exchange operator. We provide the block encoding of all of the SAPT operators in the active space picture because they contain more terms than the full space picture. The full space operator block encoding represents a special case of the active space operator with some terms excluded. This Appendix is divided into four sections: Sec. \ \ref{subsec:Block_encoding_overview} presents an overview of the block encoding methodology, Sec. \ref{subsec:data_loading_oracle} presents the data-loading oracle, Sec. \ref{subsec:PBEO} presents the product of block encoded operator circuit proposed in this work, and Sec. \ref{subsec:SAPT_block_encoding_details} outlines the full circuits for different SAPT operators.

\subsection{Overview of block encoding framework}
\label{subsec:Block_encoding_overview}
We consider the general block encoding of an operator $\hat{F}$ expressed as a linear combination of unitaries (LCU),
\begin{equation}
    \hat{F} = \sum_n \alpha_n H_n
    \label{Op_F}
\end{equation}
where $\alpha_n$ are real coefficients and $H_n$ are unitary and Hermitian operators that are assumed to be self-inverse, $H_n^2 = \mathbb{1}$. This operator can be prepared on a quantum computer using standard block encoding techniques where the Hermitian operator $\hat{F}$ is embedded within a larger unitary operator, $\mathcal{B}[\hat{F}]$. To perform the appropriate block encoding, we require:
\begin{align}
    \mathsf{PREPARE}_F\ket{0}\ket{\psi} &= \sum_n \sqrt{\frac{\alpha_n}{\lambda}} \ket{n}\!\ket{\psi} \\
    \mathsf{SELECT}_F &= \sum_n \ket{n}\!\!\bra{n} \otimes H_n 
\end{align}
where $\mathsf{PREPARE}_F$ is a unitary circuit that prepares the coefficients in the LCU representation of the operator $\hat{F}$ in Eq. \eqref{Op_F}, and $\mathsf{SELECT}_F$ is a reflection which coherently loads each unitary in the LCU representation of $\hat{F}$ ~\cite{childs2012hamiltonian}. These subroutines satisfy the block encoding equation,
\begin{equation}
    \braket{0|\mathsf{PREPARE}_F^\dagger\, \mathsf{SELECT}_F\, \mathsf{PREPARE}_F|0} = \hat{F}/\lambda
\end{equation}
Here, $\lambda = \sum_n |\alpha_n|$ denotes the $\ell_1$ norm for the vector of coefficients $\{\alpha_n\}$ and is needed to ensure that the operator remains unitary. It is well established in qubitization as well as other energy estimation algorithms with block encodings that the $\ell_1$ norm significantly affects the resource cost of the algorithm. This will also be true for observable estimation algorithms such as the SAPT-EVE algorithm.

\subsection{Data-loading oracle}
\label{subsec:data_loading_oracle}
The efficient implementation of the $\mathsf{PREPARE}$ and $\mathsf{SELECT}$ oracles has been the subject of ongoing work over the past few years. In the following, we briefly review current state-of-the-art implementations which make use of the data-lookup oracle $\mathsf{QROM}$ (quantum read-only memory) that performs the task \cite{babbush2018encoding,childs2018toward},
\begin{equation}
    \mathsf{QROM} \ket{\mathbf{\mathbf{x}}}\ket{0} = \ket{\mathbf{\mathbf{x}}}\ket{a_x}.
\end{equation}
As pointed out in \cite{low2018trading}, by utilizing additional qubits (beyond the standard $\log L$ qubits), it is possible to reduce the overall Toffoli cost. The asymptotic Toffoli count for this type of $\mathsf{QROM}$ is given by,
\begin{equation}
    C_T = \mathcal{O}\Big(\Big\lceil \frac{L}{k} \Big\rceil + b(k-1) \Big),
\end{equation}
where $k$ represents a tunable power-of-two number copies of the first register. Minimizing this quantity with respect to $k$, we obtain:
\begin{equation}
    k = \sqrt{\frac{L}{b}}
\end{equation}
which requires $\mathcal{O}(\sqrt{L})$ additional qubits. This optimization will be exploited throughout in order to minimize the overall runtime of the algorithm. We also take advantage of the efficient quantum adders, coherent alias sampling, and unprepare subroutines discussed in \citet{babbush2018encoding}.

\subsection{Product of block encoded operators}
\label{subsec:PBEO}
\begin{figure}[h!]
        \includegraphics[scale=1]{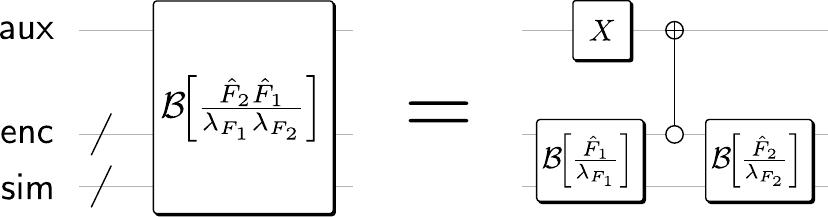}
    \caption{Not self-inverse block encoding of a product of operators. Product of normalized hermitian operators $\hat{F}_2$ and$\hat{F}_1$ (with $\ell_1$ norms $\lambda_{F_2}$, $\lambda_{F_1}$, respectively), that each  act on a register $\mathsf{sim}$. The block encoding of the product features the block encoding of each operator, such that a register $\mathsf{enc}$ of auxiliary qubits is shared. When the number of auxiliary qubits to encode $\hat{F}_j$  is $\log L_j$, then $\mathsf{enc}$ only needs to hold  enough qubits  to encode the operators separately, i{.}e{.}~a total of $\max\lbrace \log L_1, \log L_2\rbrace$ qubits, rather than their combined cost $\log L_1+\log L_2$. This comes at the cost of only one additional qubit labeled $\mathsf{aux}$, as well as a Toffoli gate over the entire $\mathsf{enc}$ register.}
    \label{pbeo_figure}
\end{figure}

\begin{figure}[h!]
    \includegraphics[scale=1]{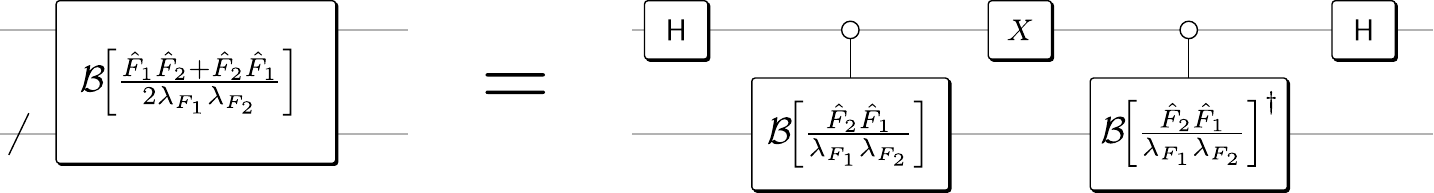}
    \caption{Self-inverse circuit for block encoding of an operator product. Using the non-self-inverse block encodings of a product of normalized hermitian operators $\hat{F}_2$ and $\hat{F}_1$ (with $\ell_1$ norms $\lambda_{F_2}$, $\lambda_{F_1}$, respectively), we can turn said block encoding into a self-inverse encoding of $(\hat{F}_1\hat{F}_2 + \hat{F}_2\hat{F}_1)/2$ using one query to the controlled versions of each -- the original block encoding and its inverse. The cost for this is an additional qubit, which has to be initialized on the Hadamard basis.}
    \label{self_inverse_pbeo}
\end{figure}

Before proceeding with the exposition of the block encoding of the SAPT observables, we first discuss the block encoding of products of operators which is required for the electrostatic-exchange $\widehat{VP}$ operator. For illustration purposes, we consider two independent operators $\hat{F}_1$ and $\hat{F}_2$ given by,
\begin{align}
    \hat{F}_1 &= \sum_n^{L_1} \alpha^{(1)}_n H^{(1)}_n \;\;\text{and}\;\; \lambda_{F_1} = \sum_n |\alpha^{(1)}_n|, \\
    \hat{F}_2 &= \sum_n^{L_2} \alpha^{(2)}_n H^{(2)}_n \;\;\text{and}\;\; \lambda_{F_2} = \sum_n |\alpha^{(2)}_n|.
\end{align}
The block encoding of the product of two operators, $\hat{F}_2\hat{F}_1$, can proceed in different ways. For instance, consider the following product of block encoding representation,
\begin{align}
    \prepare^\dagger_{F_2F_1}\select_{F_2F_1}\prepare_{F_2F_1}.
\end{align}
This block encoding uses two sets of non-contiguous auxiliary qubit registers with sizes $\log L_1$ and $\log L_2$ respectively. For instance, the select oracle is explicitly written as:
\begin{equation}
    \select_{F_2F_1} = \sum_{\substack{n=0 \\ m=0 } }^{\substack{n = L_1-1 \\ m = L_2-1}} \ket{m}\!\bra{m} \otimes \ket{n}\!\bra{n} \otimes H_m^{(2)} H_n^{(1)}.
\end{equation}
On the other hand, the prepare oracle may be expressed as the Kronecker product of each individual prepare oracle, $\prepare_{F_2F_1} = \prepare_{F_2}\otimes\prepare_{F_1}$. Based on these definitions, one can verify that the projection with respect to the zero-state of both auxiliary registers recovers the appropriate product of operators, $\hat{F}_2\hat{F}_1$,
\begin{align}
    \bra{0}_2 \bra{0}_1 \prepare_{F_2}^\dagger\prepare_{F_1}^\dagger \select_{F_2F_1}\prepare_{F_1}\prepare_{F_2} \ket{0}_1 \ket{0}_2 = \hat{F}_2\hat{F}_1/(\lambda_{F_1}\lambda_{F_2}).
\end{align}
This implementation, however, will have a QROM cost scaling as $\mathcal{O}(L_1L_2)$ which becomes prohibitive for large values of $L_1$ and $L_2$. Instead, we consider the product of block encoded operator (PBEO) circuit in Fig. \ref{pbeo_figure}.
This circuit is a simplification of the one first presented in the work by Von Burg \emph{et al.} \cite{VonBurg2021}. To ensure that this operator remains self-inverse, we make a modification by adding a single auxiliary qubit initialized in the Hadamard basis as shown in Fig. \ref{self_inverse_pbeo},
which is manifestly self-inverse. The advantage of this approach comes from the asymptotic cost of block encoding which now scales as $\mathcal{O}(L_1 + L_2)$, providing a substantial reduction Toffoli cost in the large $L$ limit.

\subsection{SAPT operators}
\label{subsec:SAPT_block_encoding_details}

Due to the factorized representations of the various SAPT operators detailed in Appendix \ref{sec:factorization}, the circuit implementations of $\mathsf{PREPARE}$ and $\mathsf{SELECT}$ must be greatly modified from their most naive, canonical forms. Below we present the block encoding circuits for each SAPT term with qualitative descriptions of the reproduced circuit diagrams to provide intuition for the constructions. We highlight the following notational conventions used throughout the circuit diagrams:

\begin{itemize}
    \item We use hexagonal controls to indicate ``multiplexing'', or ``uniformly-controlling'' a routine \cite{gidney2018halving}. 
    \item We distinguish temporary registers and persistent qubit registers whose state must remain coherent throughout the entirety of an algorithm by where the drawn wires begin and end; persistent registers (such as index registers and system qubits) extend the entire length of a diagram, while temporary qubits begin at particular subroutines that allocate them and terminate at subroutines that deallocate them.
    \begin{itemize}
        \item For example, data-loaders allocate a number of so-called clean auxiliary qubits, and the uncomputation of data-loaders deallocate (measure out) those same qubits; in these diagrams, the target registers where data is loaded will never flow into a compute portion of a data-loader or flow out of the uncompute portion of a data-loader, but rather, will emerge and terminate (respectively) from these routines. As such, these are ``temporary" qubit registers.
    \end{itemize}
    \item As an extension of the Gidney elbow notation introduced in Ref.~\cite{gidney2018halving} for a Toffoli gate targeting a newly-allocated clean ancilla qubit, we also depict certain subroutines with emerging elbows to denote when a routine ANDs the result of a computation (via a Toffoli gate) onto a newly-allocated, single clean ancillary qubit.
    \begin{itemize}
        \item For example, a comparator circuit flips the state of a zero qubit depending on the result of a particular inequality check. This ``result'' qubit can be allocated by this comparator, and so we depict this resulting qubit and the Toffoli that targets it as an elbow emerging from a comparator box.
    \end{itemize}
\end{itemize}

\subsection*{block encoding circuit for the factorized electrostatics operator $\hat{V}$}

\begin{figure}[t!]
    \centering
    \includegraphics[scale=.8]{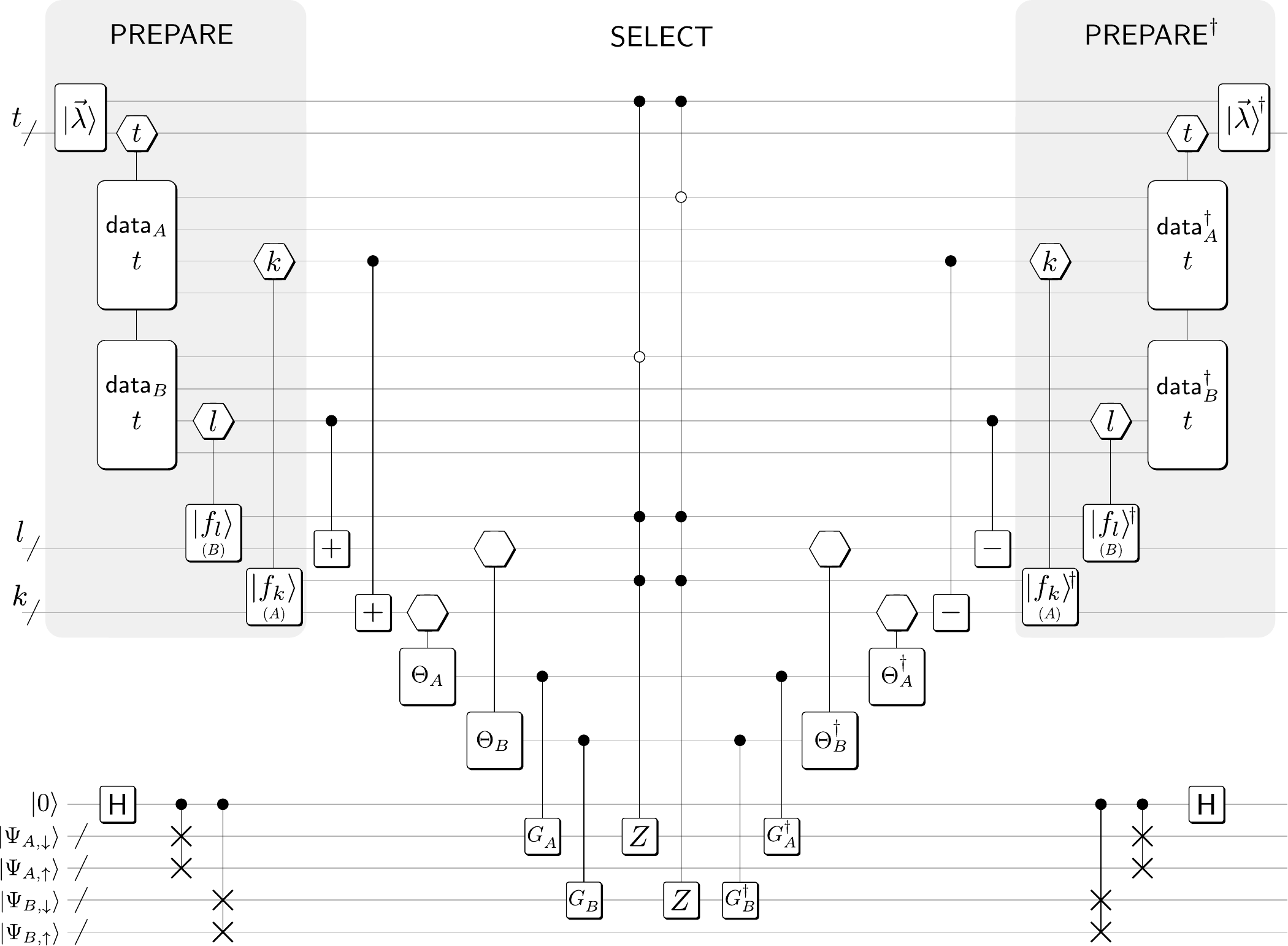}
    \caption{block encoding of the electrostatic operator.  Here, the gates labeled  `$|\vec{\lambda}\rangle$ ', `$|f_k\rangle$' and `$|f_l\rangle$'  implement linear superpositions, while hexagonal gates are data loaders multiplexing over indices such as $t$, $k$ and $l$. Controlled gates labeled `$+$' (`$-$') denote in place additions (subtractions) of the control number to the target number. The labels  $\Theta_A$ and $\Theta_B$ are sets of angles for the Givens rotations $G_A$ and $G_B$. The $S$ gate acts on the computational basis as $S=\mathrm{diag}(1,i)$. We also highlight the $\mathsf{PREPARE}$ and $\mathsf{SELECT}$ sections of this circuit near the top of the figure. $\mathsf{H}$ is the Hadamard gate.}
    \label{electrostatic_block_encoding}
\end{figure}

Just as the factorized form of the electrostatics operator (after a Jordan-Wigner transformation) shown in Eq.~\eqref{V_tensor_factorization} is analogous to the double factorization procedure used for electronic structure Hamiltonians, the block encoding circuit in Figure~\ref{electrostatic_block_encoding} is analogous to the double factorization circuit introduced in ~\citet{VonBurg2021} and modified in~\citet{Lee2021}. In fact, the circuit largely resembles Fig. 16 in Appendix C of \citet{Lee2021}. 

Only a handful of the qubit registers in the circuit are persistent qubit registers; the rest are temporary auxiliary qubits used for loading information, and conditioning subsequent operations on those registers. The persistent registers are: a register that indexes over the first rank (logarithmic in $N_1\leq \text{max}(N_A^2,N_B^2)$), a register that indexes over the second internal rank for monomer A (logarithmic in $N_2^{(A)}\leq N_A$), a register that indexes over the second internal rank for monomer B (logarithmic in $N_2^B\leq N_B$), a single ancilla used to index over spin for each monomer, and a pair of registers representing alpha (up) and beta (down) spin-orbitals of monomers A and B.

In spirit, the circuit proceeds just as the canonical double factorization circuit~\cite{Lee2021}: the $\mathsf{PREPARE}$ portion begins by preparing a superposition over the coefficients indexed by the outer rank on one register, then uses this register to load data necessary for state preparation on a separate register for the inner rank of each leaf in the double factorization, and performs this multiplexed state preparation over the inner rank. The $\mathsf{SELECT}$ portion proceeds by coherently loading angles for basis-transforming Givens rotations indexed by the inner, second rank, and then applies these Givens rotations onto the system qubits before applying single-qubit Pauli Zs.

The apparent differences between the original double factorization circuit~\cite{Lee2021} and the one depicted in this work arise because we are dealing with two monomers rather than a single set of system qubits. For this reason, we must coherently load two sets of data for the inner rank state preparations over each monomer individually, and for rotations that target each monomer's system qubits register. Additionally, rather than only applying single-qubit Pauli Z gates on the system qubits after the Givens rotations, we must select between $Z_A \otimes Z_B$, $Z_A \otimes  \mathbb{1}$, and $ \mathbb{1} \otimes Z_B$. To do this, we introduce two additional auxiliary qubits to act as one-body, single monomer flags to either turn on or turn off the application of a Pauli Z on the opposite monomer (top wires of the $\text{data}_A$ and $\text{data}_B$ subroutines).

The bulk of the cost for this block encoding is in loading the Givens rotation angles for monomers A and B, and then executing these rotations. For more details on subroutine-wise costs, see Ref.~\cite{Lee2021}.

\subsection*{block encoding circuit for the factorized exchange operator $\hat{P}$}

\begin{figure}[t!]
    \centering
    \includegraphics[scale=0.8]{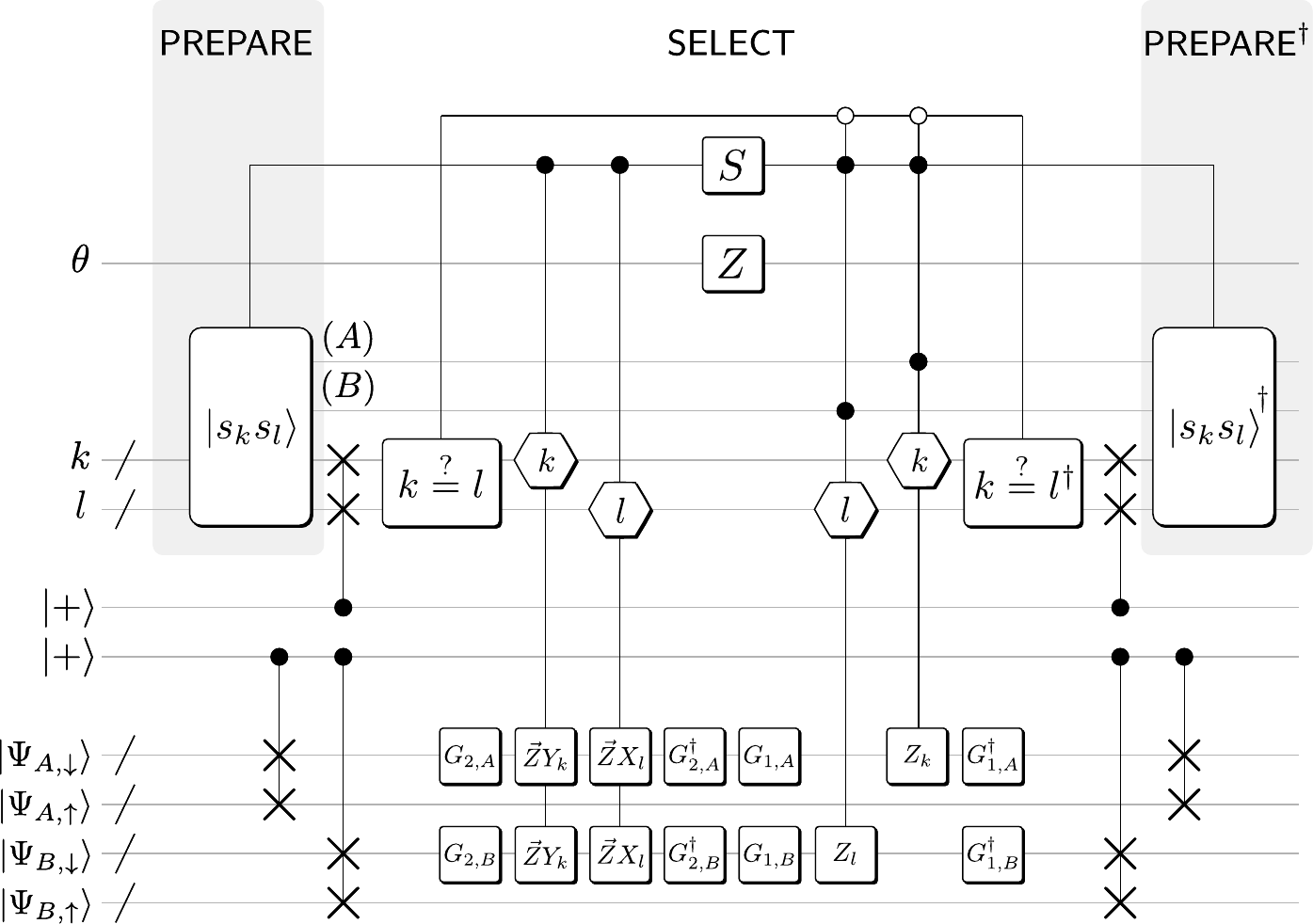}
    \caption{block encoding of exchange operator. Here, the gates labeled `$|s_ks_l\rangle$' implement linear superpositions of data entries $k$ and $l$, while
    hexagonal gates are data loaders multiplexing over said indices. The gates labeled `$k\stackrel{?}{=}l$' compare the numbers in the $k$ and $l$ registers with an adder. Wires born from and ending with rectangular gates denote Gidney elbows, see Ref.~\cite{babbush2018encoding} Figure 4. The gates labeled  `$G_{i,A}$' and `$G_{j,B}$' are Givens rotations of monomers A and B, where $i,j\in \lbrace 1, 2\rbrace$ label whether it corresponds to the intra-monomer or inter-monomer Givens rotations respectively. The $S$ gate acts on the computational basis as $S=\mathrm{diag}(1,i)$. We have also highlighted the $\mathsf{PREPARE}$ and $\mathsf{SELECT}$ sections of this circuit near the top of the figure.}
    \label{exchange_block_encoding}
\end{figure}

The factorization circuit used for the exchange circuit in Fig.~\ref{exchange_block_encoding} is similar in spirit to the single factorization circuit introduced in \citet{Berry_2019} and modified in \citet{Lee2021}. Just as in that case, the $\mathsf{PREPARE}$ circuit prepares a superposition over the coefficients resulting from the factorization (in this case, the two indices obtained from the single-rank decomposition of both monomers), and then the $\mathsf{SELECT}$ portion must choose between a number of different Pauli strings. The differences here again owe to the fact that we are dealing with two monomers rather than a single system register.

It is worth noting that in the full space formulation, the Givens operators for the inter-monomer and intra-monomer terms are equivalent, i.e. $G_{2,X} = G_{1,X}$ where $X \in \{A,B \}$ (see Eq.~\eqref{P_JordanWigner}), while in the active space case, they are different, i.e. $G_{2,X} \neq G_{1,X}$.

Furthermore, the Pauli strings are implemented using a similar control logic to that detailed in Ref.~\cite{Berry_2019}. We have the operations $\vec{Z}Y_k$ and $\vec{Z}X_l$ targeting the same sets of qubits (depicted in Fig. 9 of Ref.~\cite{babbush2018encoding}), which implement $X\vec{Z}X$, $Y\vec{Z}Y$, or $Z$ (up to signs and phases) depending on whether $k < l$, $k > l$, or $k = l$. The signs and phases are cleaned up by the Pauli $Z$ and $S$ gates in the diagram. To distinguish the one-body single-monomer cases ($Z_{k}\otimes  \mathbb{1}$ and $ \mathbb{1} \otimes Z_{l}$), we also introduce additional one-body single-monomer temporary qubits emerging (or terminating) at the joint state preparation (or its conjugate). These auxiliary qubits conditionally control the application of a $k$-indexed $Z_k$ operation or an $l$-indexed $Z_l$ operation, also controlled on $k \neq l$ (zero-controls at the top of the diagram).

Here the only persistent qubit registers are the two index registers which sum over the first rank indices for both monomers, a single ancilla in the plus state to symmetrize over both index registers after joint state preparation, a single ancilla plus state to index over spin, another single ancilla on which we apply a Pauli Z to correct a phase on the resulting Pauli strings, and the up-and-down spin-orbital registers for both monomers. 

\subsection*{block encoding circuit for the factorized electrostatic-exchange operator $\widehat{VP}$}
As detailed in Appendix~\ref{sec:fact_electrostatic_exchange}, the factorization for the complete electrostatic-exchange operator is quite verbose. For illustration purposes, here we present the exact block encoding of only the $\widehat{VP}_{1\ell}$ and $\widehat{VP}_4$ terms. The latter is the asymptotically most expensive part of the block encoding. We then describe how to combine all the resulting terms to realize a block encoding of the entire electrostatic-exchange operator $\widehat{VP}$.

\begin{figure}[t!]
    \centering
    \includegraphics[scale=0.8]{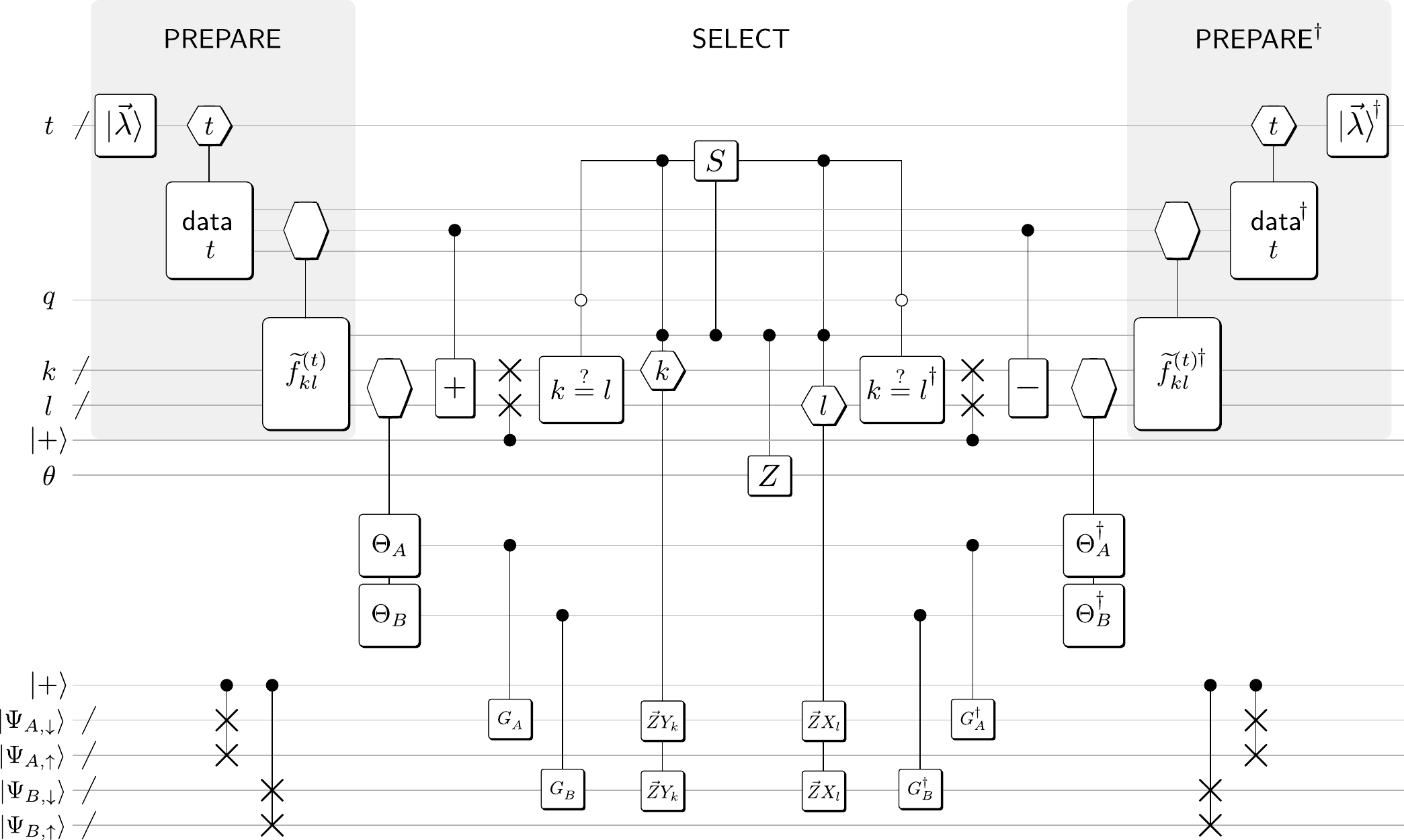}
    \caption{block encoding of the electrostatic-exchange operator, $\widehat{VP}_{1\ell}$. Here, the gates labeled `$|\vec{\lambda}\rangle$'  implement linear superpositions, while hexagonal gates are data loaders multiplexing over indices such as $t$, $k$, and $l$, in the corresponding registers. The gates labeled `$k\stackrel{?}{=}l$' compare the numbers in the $k$ and $l$ registers with an adder. The labels `$\widetilde{f}_{kl}^{(t)}$' denote coefficients, and $\Theta_A$, $\Theta_B$ are sets of angles for the Givens rotations $G_A$ and $G_B$. The $S$ gate acts on the computational basis as $S=\mathrm{diag}(1,i)$. The controlled `$+$' gate is an in-place addition of numbers from the control register into the target register. We have highlighted the $\mathsf{PREPARE}$ and $\mathsf{SELECT}$ sections of this circuit near the top of the figure.}
    \label{fig:VP1_block_encoding}
\end{figure}

\subsubsection{$\widehat{VP_{1\ell}}$}

The complete block encoding circuit for the electrostatic-exchange operator $\widehat{VP}_{1\ell}$ from Eq.~\eqref{eq:VP1l_TF} is shown in Fig.~\ref{fig:VP1_block_encoding}. This operator is conceptually a hybrid between the electrostatic and exchange operators discussed in the previous sections. Compared to the previous operators, $\widehat{VP}_{1\ell}$ does not contain any one-body terms. However, unlike the exchange operator, it requires a double factorization procedure where the first factorization is denoted by the index $t$ and the second factorization(s) are denoted by $k$ and $l$ respectively, where $k$ and $l$ are analogous to the $k$ and $l$ indices in the exchange operator.

The only persistent registers are the ones holding the indices over the various factorization indices ($t$, $k$, and $l$), a plus state to symmetrize over the $k$ and $l$ registers, qubits labeled $q$ and $\theta$ to help implement different Pauli strings in the case distinction (analogous to the case distinction discussed for the exchange operator), a plus state to symmetrize over spin, and the pairs of up and down spin-orbital system qubits for each monomer.

Just as in the exchange operator block encoding, we need control logic to distinguish various Pauli string cases. Here the case distinction is exactly like the one described in Ref.~\cite{Berry_2019}, except we have two monomers. For each monomer, we must apply $X\vec{Z}X + Y\vec{Z}Y$ and $ \mathbb{1} - Z$, which we can do via inequality checks for $k$ and $l$ and using the $q$ wire in a plus state to choose between $Z$ and $ \mathbb{1}$ for the case of $k = l$. The controlled Pauli Z and controlled S are used to correct for the sign and imaginary unit that arise from the product of the appropriate Pauli operators.

\subsubsection{$\widehat{VP_{4}}$}

\begin{figure}[ht!]
    \centering
    \includegraphics[scale=1]{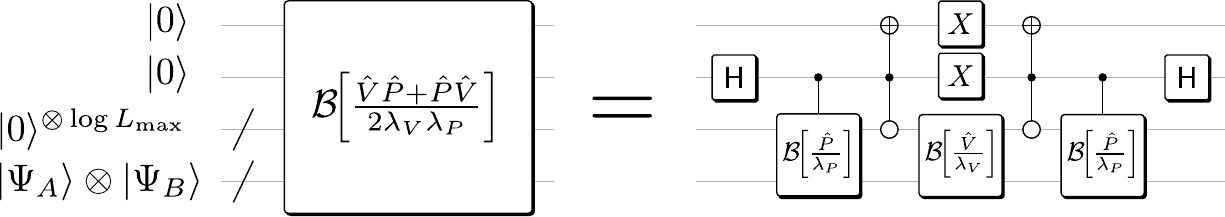}
    \caption{block encoding of the electrostatic-exchange operator, $\widehat{VP}_4$. The symmetric and self-inverse product of $\hat{V}$ and $\hat{P}$ is block encoded on the simulator registers of monomers A and B, with a number $\log L_{\mathrm{max}}+2$ of auxiliary qubits, where $L_{\mathrm{max}}=\max \lbrace L_V, L_P\rbrace$, meaning that the two types of block encodings reuse each other's auxiliary qubits. The price of that is just one extra qubit, plus the qubit to implement the superposition. The circuit calls three instances of the block encoding circuits, where we can make the choice for the more expensive one, here: $\mathcal{B}[\hat{V}/\lambda_V]$,  to be called only once.}
    \label{VP4_block_encoding}
\end{figure}

Finally, we consider the implementation of, $\widehat{VP}_4 = \tfrac{1}{2}(\hat{V}\hat{P} + \hat{P}\hat{V})$, written as the symmetric product of two operators. Using the established block encoding circuit primitives for $\mathcal{B}[\hat{V}]$ and $\mathcal{B}[\hat{P}]$ found above, it is then possible to use the circuits in Figs. \ref{pbeo_figure} and \ref{self_inverse_pbeo} to implement the self-inverse product of block encoding of $\widehat{VP}_4$. Explicitly, the complete circuit for the $\widehat{VP}_4$ term is given in Fig.~\ref{VP4_block_encoding}.

\subsubsection{Combining all terms}

In order to implement the complete electrostatic-exchange operator, we recall that the following terms are required:
\begin{equation}
    \widehat{VP}_\mathrm{s} = \widehat{VP}_A + \widehat{VP}_B + \widehat{VP}_{1\mathrm{m}} + \widehat{VP}_{1\ell} + \widehat{VP}_2 + \widehat{VP}_3 + \widehat{VP}_4. 
\end{equation}
The first two terms correspond to monomer-only terms that are exactly analogous to the standard monomer Hamiltonians which use the standard double factorization procedure for its implementation. The $\widehat{VP}_{1\mathrm{m}}$ term corresponds to a unique term that only arises in the active space picture and is exactly equal to the electrostatic operator without any 1-body contributions. We have also accounted for $\widehat{VP}_{1\ell}$ and $\widehat{VP}_4$ in the previous paragraphs. As discussed in Appendix \ref{sec:fact_electrostatic_exchange}, in the limit of a large number of orbitals, this operator will always correspond to the most dominant in terms of Toffoli cost. To implement the complete electrostatic-exchange operator, we use the following quantum circuit primitive from Ref.~\cite{VonBurg2021}:
\begin{figure}[h!]
    \centering
    \includegraphics[scale=1]{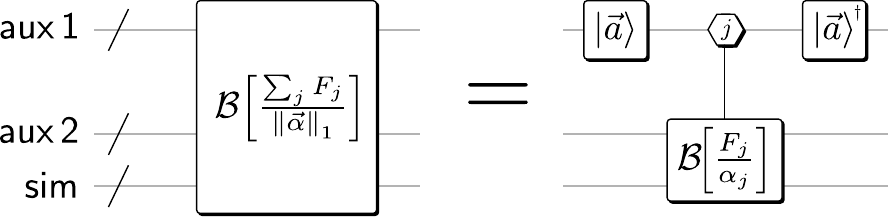}
    \label{LinearComination_Block_Encodings}
    \caption{Linear combination of block encodings. The block encoding of a hermitian operator $\sum_j F_j / ||\vec{\alpha}||_1$ (where $\alpha_j$ is the one-norm of $F_j$, and $F_j$ is hermitian) acting on a register $\mathsf{sim}$, is block encoded by preparing the state $\sum_j a_j |j\rangle$ in one auxiliary qubit register $\mathsf{aux\, 1}$, such that $a_j = |\alpha_j| / ||\vec{\alpha}||_1$, and then multiplexing the block encoding of the  $F_j/\alpha_j$ over the computational basis state $|j\rangle$ in $\mathsf{aux\, 1}$ using the auxiliary register $\mathsf{aux\, 2}$ to encode the operator.}
\end{figure}
\\
Which only requires three additional ancillae in order to implement all seven terms appropriately. 

\section{Benchmark set for small molecules}\label{sec:Appendix_Small_Benchmark}
The benchmark molecules have been chosen according to previous benchmark studies found in the classical SAPT literature \cite{jurevcka2006benchmark}. The $\mathtt{XYZ}$ files for all of the small benchmark cases may be found in in the benchmark energy and geometry database for noncovalent complexes: $\texttt{http://www.begdb.org}$. Resource estimates for the SAPT-EVE algorithm for the small molecule benchmark system as well as the heme-artemisinin system are also presented here for completeness.

\section{Benchmark for drug design: Heme-artemisinin}\label{sec:Appendix_Benchmark_Artemisinin}

\subsection{Computational Details}
The initial pre-complex structure of heme and artemisinin was prepared using the MOE suit employing the Amber10:EHT force field~\cite{moe}. Unrestricted density functional theory calculations (DFT)~\cite{hohenberg1964inhomogeneous, kohn1965self} were performed to obtain the transition state (TS) structure using Gaussian 16, Revision C.01 (G16)~\cite{g16}. The $\omega$B97X-D functional~\cite{wb97xd} and a mixed basis set (Fe: def2-TZVP~\cite{def2basis}, N: def2-SVPD and def2-SVP for C, O, and H) were used with standard G16 settings for SCF convergence, exchange-correlation grid and geometry optimization convergence parameters. The level shift was enabled ($vshift = 1000$) to accelerate the SCF convergence. Relaxed potential-energy surface scans along the Fe-O bond prepared the initial structure of the transition state. The transition state was subsequently located using the G16 standard TS search algorithm. We used ROHF in combination with the cc-pVDZ~\cite{dunning1989a, balabanov2005a, balabanov2006a} basis to generate orbitals and integrals for the active space selection and subsequent DMRG~\cite{chan2002highly, wouters2014density} calculations using the Pyscf~\cite{sun2020recent} and block2~\cite{block2} packages. All important data describing the molecular systems used in this paper can be found on the open access repository~\cite{data_zenodo}.

\begin{figure}[h!]
    \centering
    \includegraphics[page=2, width=1\textwidth]{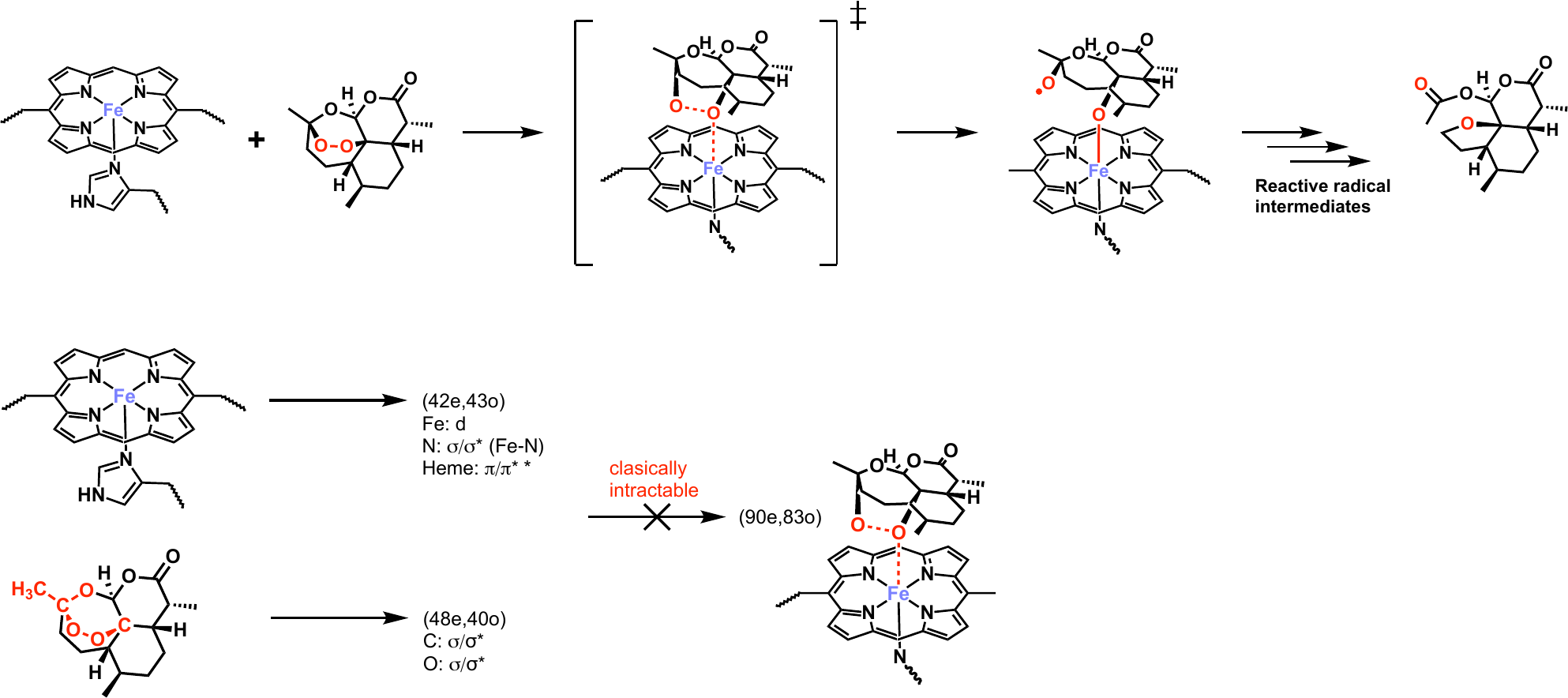}
    \caption{(Top) Proposed mechanism of the decomposition of artemisinin with heme. The key initial reaction step involves the coordination of artemisinin to the Fe center and a homolytic cleavage of the \ce{O-O} bond. (Bottom) Summary of the key orbitals for both artemisinin and heme in order to describe the complete decomposition mechanism, which must be included in the active space (see main text for a detailed justification). }
    \label{fig:art_chemdraw}
\end{figure}

\subsection{Active space selection}

\begin{figure}
    \centering
    \includegraphics[width=0.45\textwidth]{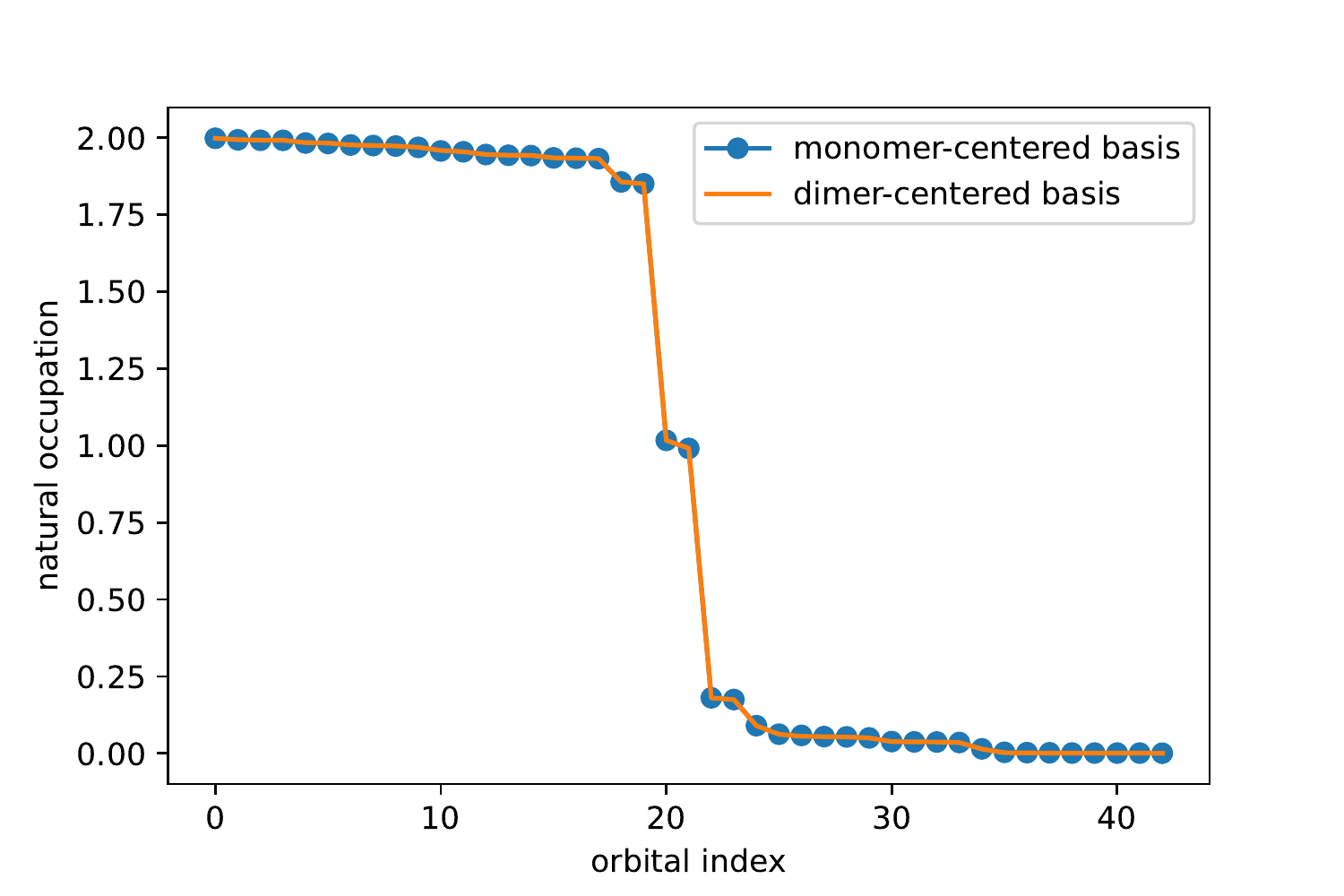}
        \includegraphics[width=0.45\textwidth]{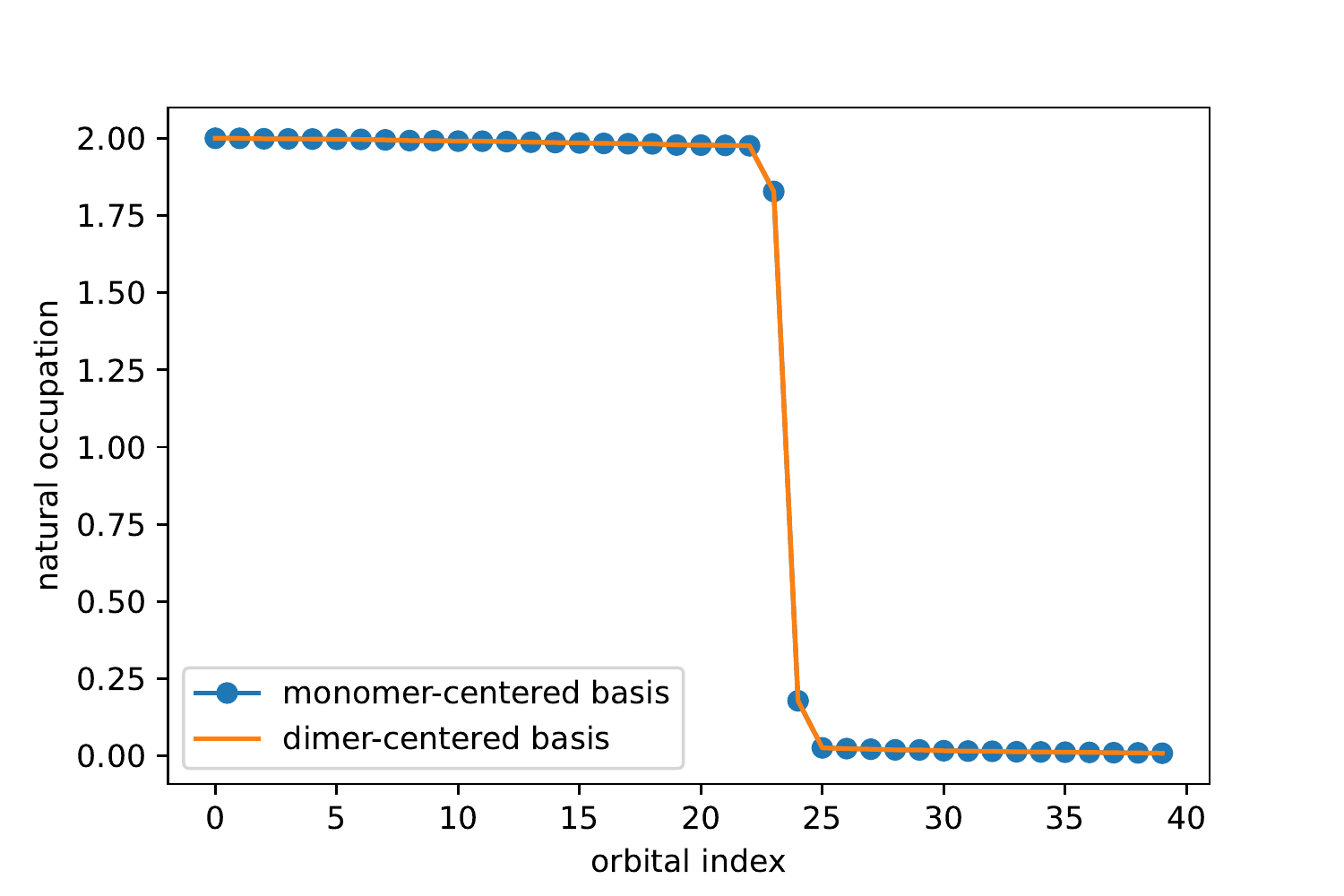}
    \caption{Natural occupation numbers vs{.} the index of the active space orbitals from DMRG calculations; left: the heme monomer (bond dimension $M=2000$); right: artemisinin monomer ($M=1000$). Deviation from integer values ($0,1,2$) indicates a strongly correlated nature that is not well described by single-reference wavefunctions. Both the heme and the artemisinin molecule have orbitals around the Fermi-level that substantially deviate from integer occupation, while the heme natural orbital occupations also show long tails of smaller corrections away from the Fermi-level that would require large active space calculations to describe correctly. }
    \label{fig:noons}
    
\end{figure}

The active space was generated for each monomer separately. For heme, we followed the approach from Ref.~\cite{goings} and included Fe 3d and 4d orbitals, $\pi$ and $\pi^*$ orbitals of nitrogen and carbon in the heme ring, and bonding and antibonding Fe-N orbitals. This resulted in a (42o,42e) active space. We picked the orbitals individually from the Pipek-Mezej localized orbitals based on a high-spin ROHF reference state. The natural orbital occupation from DMRG calculation are plotted in Fig.~\ref{fig:noons}. For the dimer-centered basis, we started with the converged monomer-centered ROHF density matrix to converge the ROHF calculation for which the same procedure was repeated.

For artemisinin, we have used the atomic valence active space (AVAS) method~\cite{sayfutyarova2017automated}. We have also included all $2s$ and $2p$ orbitals of the peroxo moiety (\ce{C-O-O-C}) in the selection prompt. In addition, we included four more atoms close to the moiety, which play an important role during the decomposition (see Fig.~\ref{fig:art_chemdraw}, colored in red). This yields an active of (40o, 48e) for the monomer-centered basis, the natural orbital occupation numbers are plotted in Fig.~\ref{fig:noons}. The same procedure was repeated for the dimer-centered basis.

The active space for the dimer system has been generated by first selecting the artemisinin orbitals through AVAS from a high-spin ROHF reference. The remaining orbitals were localized through the Pipek-Mezej routine and the same set of orbitals from the heme monomer were then selected by hand. This yields a total active space of (82o, 90e) for the supersystem, which is intractable for most quantum chemistry methods. The active space choice is not unique, and FTQC simulations would be required actually to confirm this choice. However, we believe by carefully analyzing each monomer DMRG result, using chemical intuition and established mechanistic insights, that this is a representable active space for studying the initial reaction step for the artemisinin decomposition.

For the resource estimates, we also require the energy gap between the ground state of the Hamiltonian and the first excited state and the overlap with the initial state. We do not have access to the full configuration interaction energies or states but we use accurate DMRG calculations as a proxy. In Table~\ref{tab:gaps_overlaps}, we list the results of our active space calculations for the heme and artemisinin monomers in the monomer-centered and dimer-centered bases. We chose bond dimension $M=2000$ for the heme and $M=1000$ for artemisinin. The ground states and excited states are found in the same spin sector: $S=1$ for the heme and $S=0$ for artemisinin. 

\begin{table}
\caption{Molecular data from DMRG calculations for benchmark heme artemisinin system. $E_0$/$E_1$ are ground and first-excited state energies measured in Hartrees (Ha). $\Delta = E_1 - E_0$ is the energy gap. The last column represents the overlap between the restricted open-shell Hartree Fock wavefunction and the DMRG wavefunction. Data is provided on both monomer-centered basis (MCB) and dimer-centered basis (DCB).}
\label{tab:gaps_overlaps}
\renewcommand{\arraystretch}{1.5}
\begin{tabular}{lc@{\hskip 5mm}rrr@{\hskip 5mm}c@{\hskip 5mm}}
\hline\hline
System & Basis & $E_0$[Ha] & $E_1$[Ha] & $\Delta$[Ha] & $\left|\braket{\Psi_{\rm ROHF}|\Psi_{\rm DMRG}}\right|^2$ \\\hline
Heme      & MCB   & -2469.858136 & -2469.851239 & 0.006897 & 0.068174\\
                & DCB   & -2469.858690 & -2469.851043 & 0.007646 & 0.067537\\\hline
Artemisinin     & MCB   & -916.399835   & -916.278659   & 0.121176 & 0.800254\\
                & DCB   & -916.405542   & -916.284444   & 0.121098 & 0.800654\\
\hline\hline
\end{tabular}
\end{table}

\section{Supermolecular resource data}\label{sec:supermolecular_data}

Supermolecular resource data is provided in Tab.~\ref{tab:supermolecular} for the small molecule benchmark system in the all-electron (full space) picture as well as the active-space-optimized heme-artemisinin dimer system. In all cases, the supermolecular (SM) resource estimation consists of three standard quantum phase estimation runs with the block-encoded double-factorized Hamiltonians for each system (for more details, see Refs.~\cite{VonBurg2021,Lee2021}). Note that we have performed an appropriate error budgeting procedure to reduce the overall Toffoli gate cost of the total supermolecular calculation. 
\begin{table}[t!]
    \resizebox{16cm}{!}{
    \begin{tabular}{llrrrrrr@{\hskip 6mm}rrrr@{\hskip 6mm}rrrr}
\hline\hline
 &  & \multicolumn{6}{c}{System Parameters} & \multicolumn{4}{c}{Gates} & \multicolumn{4}{c}{Qubits} \\
 &  & $N_{AB}$ & $N_A$ & $N_B$ & $\lambda_{H_{AB}}$ & $\lambda_{H_A}$ & $\lambda_{H_B}$ & SM & $\textrm{E}_{AB}$ & $\textrm{E}_{A}$ & $\textrm{E}_{A}$ & SM & $\textrm{E}_{AB}$ & $\textrm{E}_{A}$ & $\textrm{E}_{A}$ \\

\midrule
\multirow[c]{3}{*}{Water} & STO-3G & 14 & 7 & 7 & 142.8 & 53.9 & 53.9 & $2.11\!\times\!10^{9}$ & $1.30\!\times\!10^{9}$ & $4.09\!\times\!10^{8}$ & $4.09\!\times\!10^{8}$ & 709 & 709 & 440 & 440 \\
 & 6-31G & 26 & 13 & 13 & 226.5 & 73.1 & 73.1 & $5.70\!\times\!10^{9}$ & $3.60\!\times\!10^{9}$ & $1.05\!\times\!10^{9}$ & $1.05\!\times\!10^{9}$ & 1154 & 1154 & 675 & 675 \\
 & cc-pVDZ & 48 & 24 & 24 & 1027.7 & 328.1 & 328.1 & $4.98\!\times\!10^{10}$ & $3.16\!\times\!10^{10}$ & $9.07\!\times\!10^{9}$ & $9.06\!\times\!10^{9}$ & 2096 & 2096 & 1144 & 1144 \\
\cline{1-16} \multirow[c]{3}{*}{Ammonia} & STO-3G & 16 & 8 & 8 & 127.3 & 44.6 & 44.6 & $2.06\!\times\!10^{9}$ & $1.28\!\times\!10^{9}$ & $3.87\!\times\!10^{8}$ & $3.87\!\times\!10^{8}$ & 777 & 777 & 474 & 474 \\
 & 6-31G & 30 & 15 & 15 & 278.3 & 84.0 & 84.0 & $7.93\!\times\!10^{9}$ & $5.13\!\times\!10^{9}$ & $1.40\!\times\!10^{9}$ & $1.40\!\times\!10^{9}$ & 1330 & 1330 & 743 & 743 \\
 & cc-pVDZ & 58 & 29 & 29 & 1405.1 & 433.8 & 433.8 & $8.20\!\times\!10^{10}$ & $5.30\!\times\!10^{10}$ & $1.45\!\times\!10^{10}$ & $1.45\!\times\!10^{10}$ & 2535 & 2535 & 1331 & 1331 \\
\cline{1-16} \multirow[c]{3}{*}{Benz.-Water} & STO-3G & 43 & 36 & 7 & 592.6 & 453.6 & 53.9 & $2.97\!\times\!10^{10}$ & $1.67\!\times\!10^{10}$ & $1.21\!\times\!10^{10}$ & $8.38\!\times\!10^{8}$ & 1897 & 1897 & 1594 & 455 \\
 & 6-31G & 79 & 66 & 13 & 1495.6 & 1158.5 & 73.1 & $1.36\!\times\!10^{11}$ & $7.72\!\times\!10^{10}$ & $5.58\!\times\!10^{10}$ & $2.74\!\times\!10^{9}$ & 3440 & 3440 & 2850 & 696 \\
 & cc-pVDZ & 138 & 114 & 24 & 6104.4 & 4563.9 & 328.1 & $1.01\!\times\!10^{12}$ & $5.82\!\times\!10^{11}$ & $4.08\!\times\!10^{11}$ & $2.23\!\times\!10^{10}$ & 6237 & 6237 & 5084 & 1176 \\
\cline{1-16} Heme-Art. & cc-pVDZ[opt] & 83 & 43 & 40 & 942.0 & 232.2 & 361.8 & $7.91\!\times\!10^{10}$ & $5.14\!\times\!10^{10}$ & $1.28\!\times\!10^{10}$ & $1.49\!\times\!10^{10}$ & 3515 & 3515 & 1852 & 1744 \\
\cline{1-16} \hline\hline
\end{tabular} }
    \caption{Resource estimates of the supermolecular approach. $N_X$ denotes the number of spatial orbitals, $\lambda_X$ denotes the $\ell_1$ norm for the double factorized system Hamiltonian ($\hat{H}^{(\mathrm{df})}_X$) in units of Hartree, where $X\in \{A,B,AB\}$. The total cost of the supermolecular (SM) algorithm is measured in terms of Toffoli gates and is decomposed according to each of the three eigenphase estimates ($E_A$, $E_B$, and $E_{AB}$) required to compute the interaction energy in units of Hartree. The same decomposition is performed with respect to the total number of qubits. It is important to note that we have performed an appropriate error budgeting procedure in order to minimize the overall Toffoli  gate cost of the total supermolecular calculation via the Lagrange method, setting the targeted errors of each QPE as $\varepsilon_X\propto \sqrt{\lambda_X}$.}
\label{tab:supermolecular}
\end{table}

\newpage
\section{Comprehensive resource data}\label{sec:data_code}

\begin{figure*}[h!]
    \subfloat[\label{fig:sub11}]{%
  \includegraphics[width=.49\linewidth]{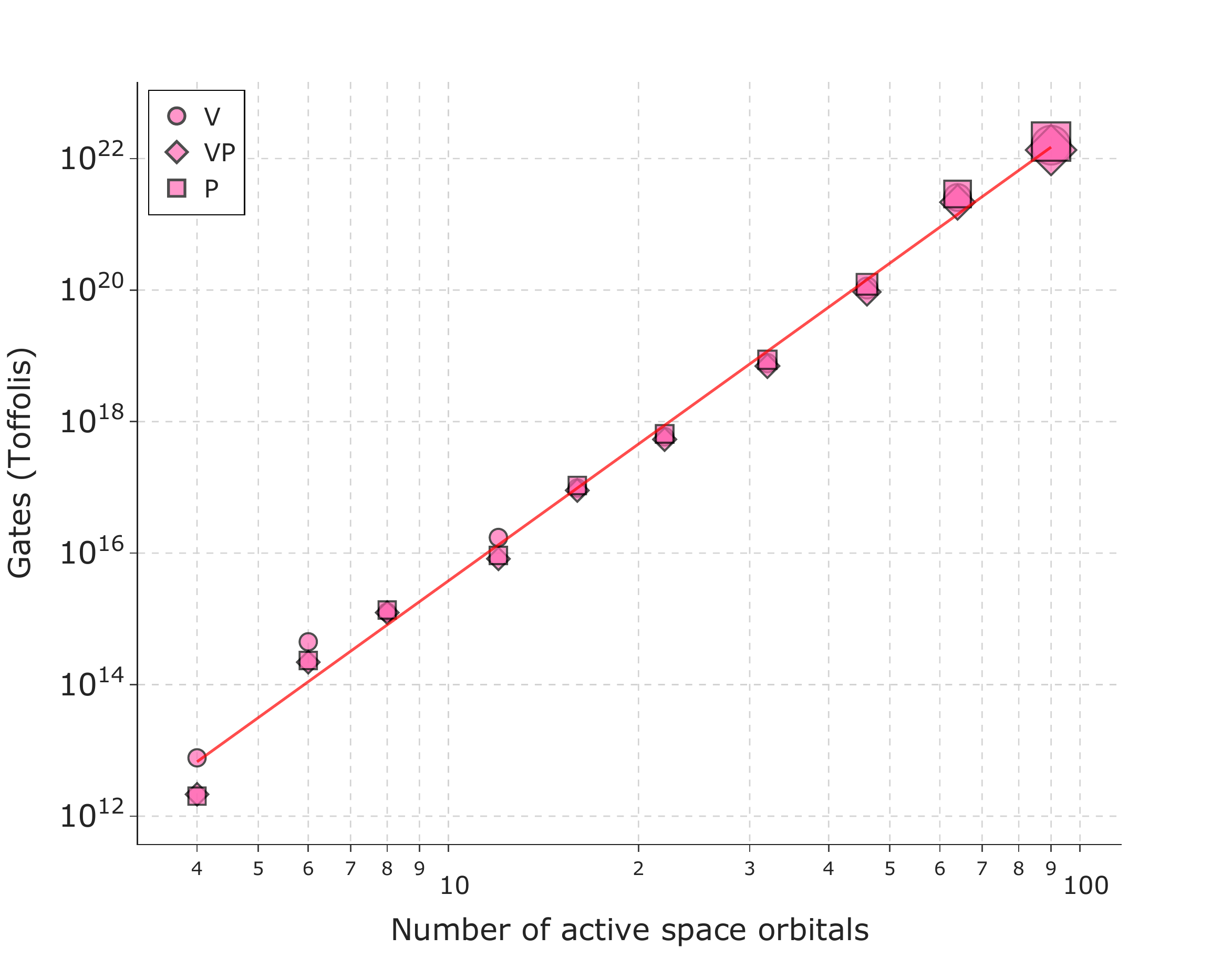}%
}\hfill
\subfloat[\label{fig:sub22}]{%
  \includegraphics[width=.49\linewidth]{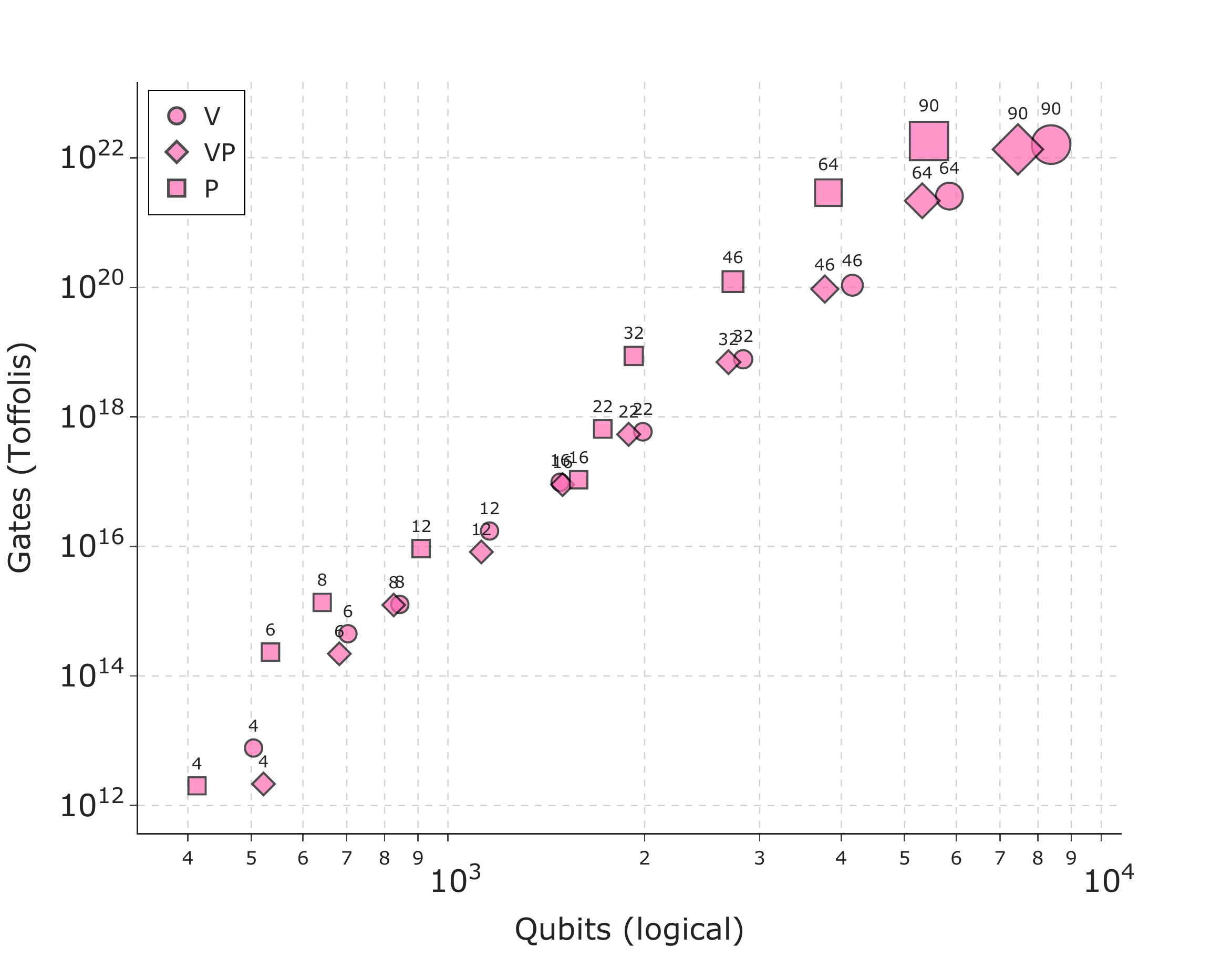}%
}
\caption{Quantum resource estimation in the active space picture. (a) Toffoli gate cost for target heme-artemisinin system as a function of active space spatial orbitals. The active space orbitals are not optimized and are symmetrically chosen around the Fermi level. (b) Toffoli and logical qubit dependence of the SAPT-EVE algorithm for the three SAPT observables, $\hat{F}=\{\hat{V},\hat{P},\widehat{VP}\}$, for the non-optimal active space system on the left. The size of the data points corresponds to the size of the $\ell_1$ norm for the corresponding observable. The numbers on the right figure correspond to the number of active space orbitals.}
    \label{fig:trend_line}
\end{figure*}

\begin{figure}[h!]
    \resizebox{15cm}{!}{
    \includegraphics[trim={5px, 5px, 5px, 5px}, clip, width=7.5cm]{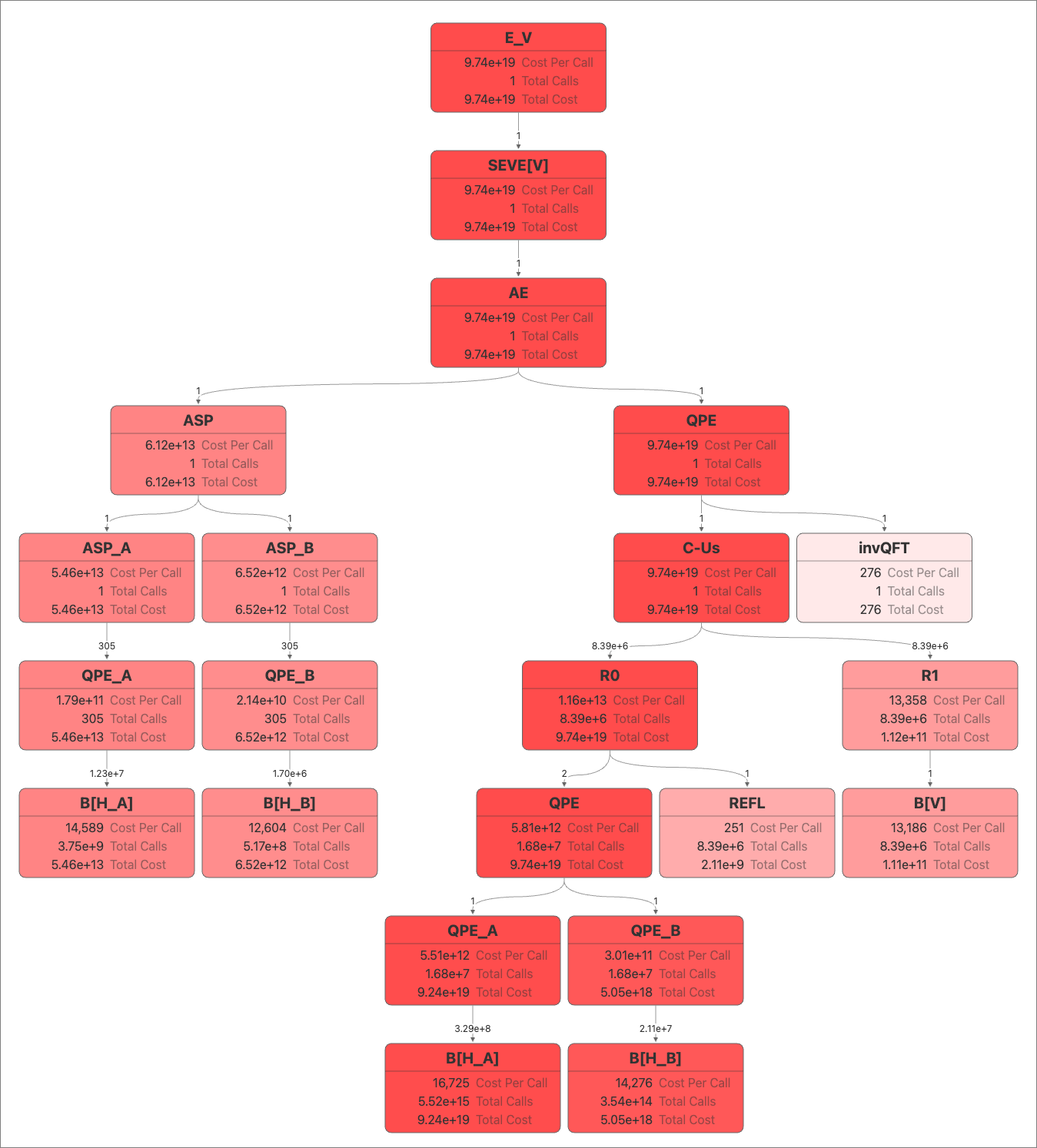}
    }
    \caption{SAPT-EVE algorithm call graph for the electrostatic operator, $\hat{V}$. The total call graph depicts the estimated resources required for computing all of the SAPT operators for heme-artemisinin benchmark system using QSP-EVE. The graph displays the distribution of the costs among various subroutines, and where B[H] is the block encoding of the Hamiltonian and $\mathcal{R}_\tau$ is the controlled block encoding of the observable. All costs are given in terms of Toffoli gate counts, with each subroutine node depicting the per-call cost, the total number of calls and cost when taken over the full algorithm (i.e. overall parent calls). Note that some routines are deliberately omitted in the count, as they contribute very little. Edge numbers define the number of calls of the target routine within a single call of its parent routine. Darker shading indicates a greater total gate cost for the subroutine. The ASP comprises several low-precision phase estimation routines in the repeat-until-success circuit. }
    \label{fig:call_graph1}
\end{figure}
\begin{figure}
    \resizebox{15cm}{!}{
    \includegraphics[trim={5px, 5px, 5px, 5px}, clip, width=7.5cm]{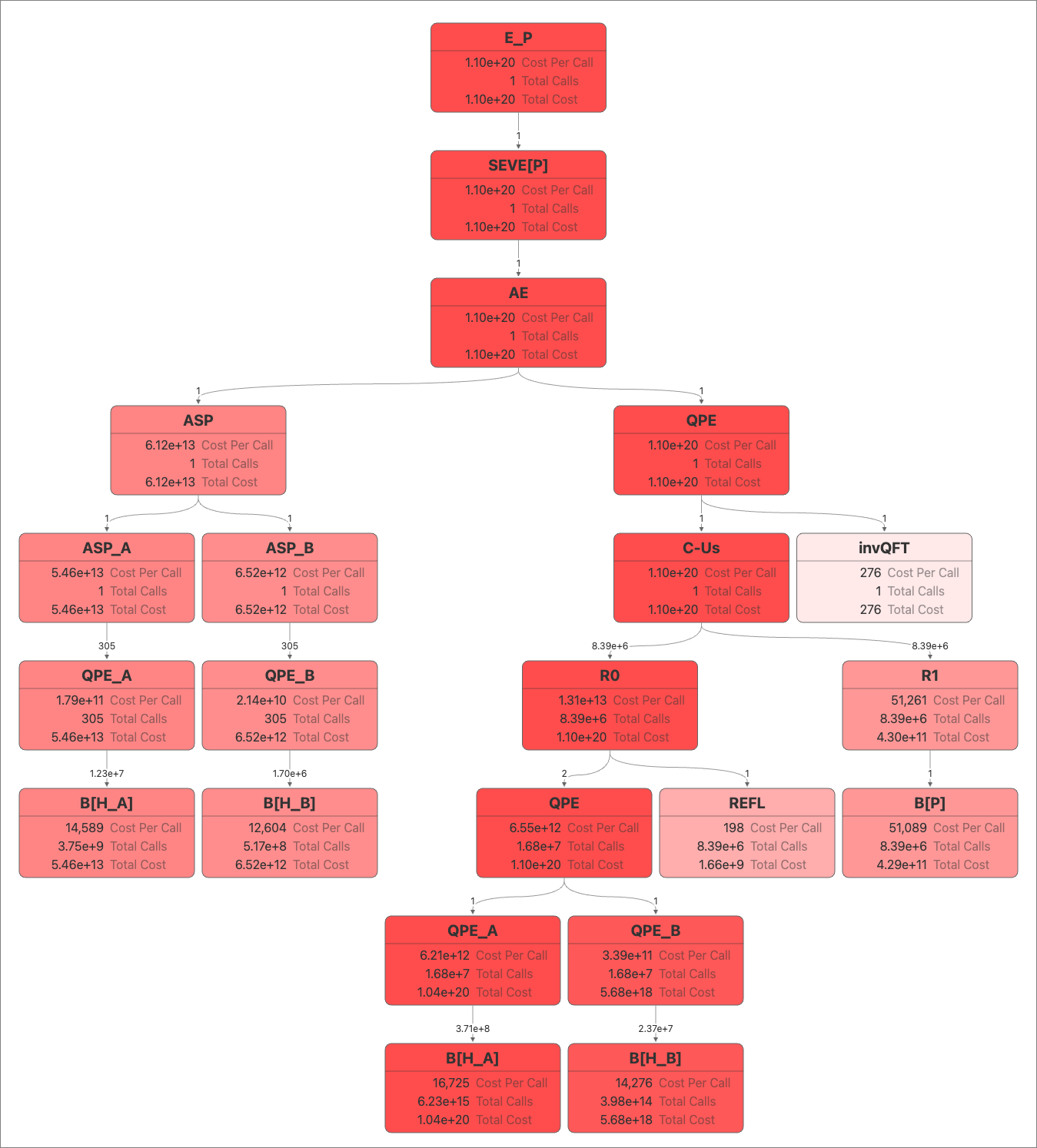}
    }
    \caption{SAPT-EVE algorithm call graph for the exchange operator, $\hat{P}$. The total call graph depicts the estimated resources required for computing all of the SAPT operators for heme-artemisinin benchmark system using QSP-EVE. The graph displays the distribution of the costs among various subroutines, and where B[H] is the block encoding of the Hamiltonian and $\mathcal{R}_\tau$ is the controlled block encoding of the observable. All costs are given in terms of Toffoli gate counts, with each subroutine node depicting the per-call cost, the total number of calls and cost when taken over the full algorithm (i.e. over all parent calls). Note that some routines are deliberately omitted in the count, as they contribute very little. Edge numbers define the number of calls of the target routine within a single call of its parent routine. Darker shading indicates a greater total gate cost for the subroutine. The ASP comprises several low-precision phase estimation routines in the repeat-until-success circuit. }
    \label{fig:call_graph2}
\end{figure}
\begin{sidewaysfigure}
    \resizebox{25cm}{!}{
    \includegraphics[trim={5px, 5px, 5px, 5px}, clip, width=8.5cm]{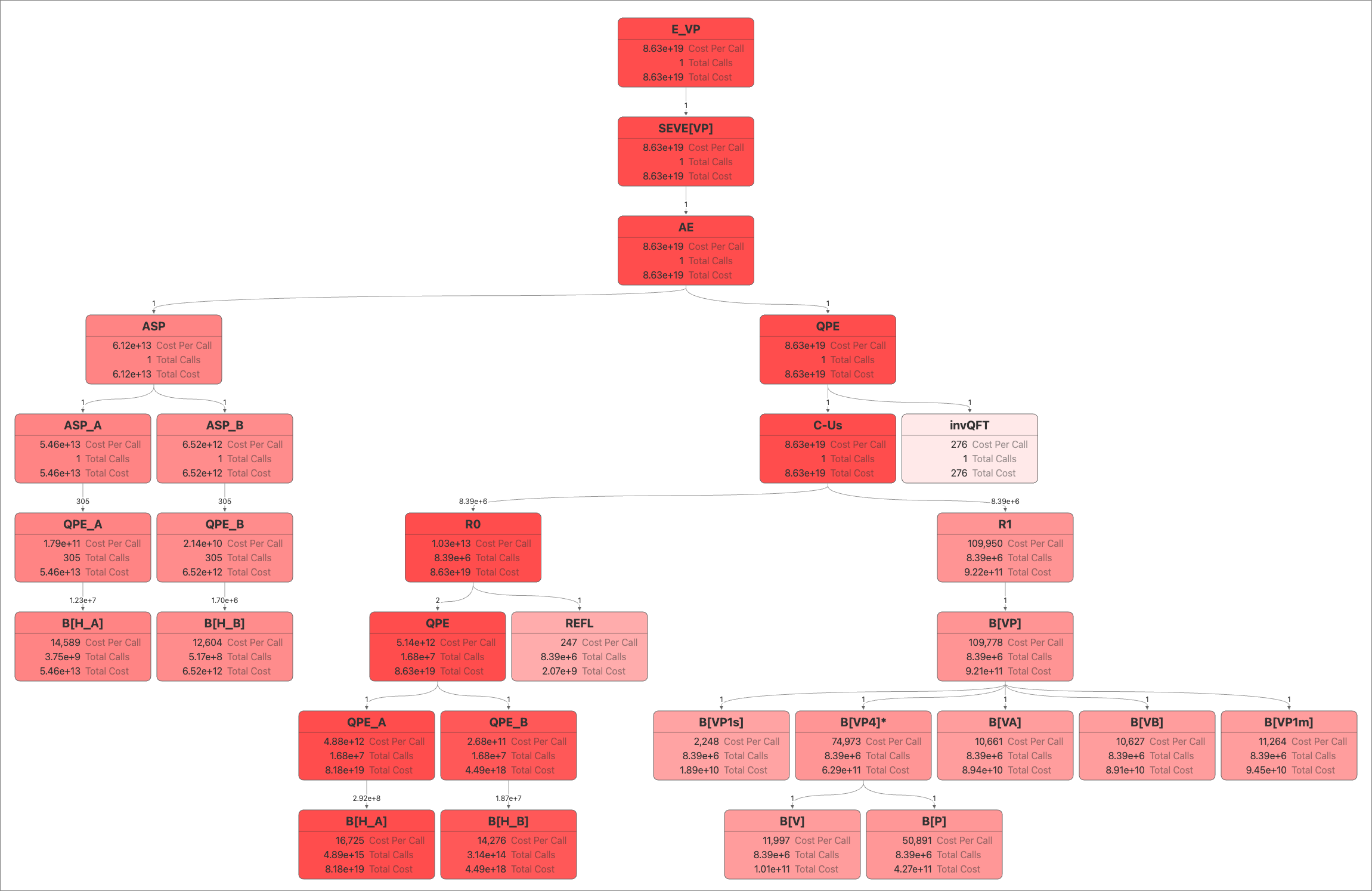}
    }
    \caption{SAPT-EVE algorithm call graph for the electrostatic-exchange operator $\widehat{VP}$. The total call graph depicts the estimated resources required for computing all of the SAPT operators for heme-artemisinin benchmark system using QSP-EVE. The graph displays the distribution of the costs among various subroutines, and where B[H] is the block encoding of the Hamiltonian and $\mathcal{R}_\tau$ is the controlled block encoding of the observable. All costs are given in terms of Toffoli gate counts, with each subroutine node depicting the per-call cost, the total number of calls and cost when taken over the full algorithm (i.e. over all parent calls). Note that some routines are deliberately omitted in the count, as they contribute very little. Edge numbers define the number of calls of the target routine within a single call of its parent routine. Darker shading indicates a greater total gate cost for the subroutine. The ASP comprises several low-precision phase estimation routines in the repeat-until-success circuit. }
    \label{fig:call_graph3}
\end{sidewaysfigure}

\newpage
\begin{sidewaystable}
\caption{Molecular properties (sub-table $a$) and resource estimates for the electrostatic operator $\hat{V}$ (sub-tables $b$ and $c$), electrostatic-exchange operator $\widehat{VP}_\mathrm{s}$ (sub-tables $d$ and $e$), and exchange operator $\hat{P}$ (sub-tables $f$ and $g$). $N_A/N_B$ denote spatial orbitals of monomers A/B. $\lambda_A/\lambda_B$ denote $\ell_1$ norms for Hamiltonians $H_A/H_B$ in units of Hartree. $\Delta_A/\Delta_B$ denote the spectral gaps for monomers $A/B$ in units of Hartree. $\lambda_F$ denotes the $\ell_1$ norm for observable, $\hat{F}$. $\lambda_V/\lambda_{VP}$ are in units of Hartree, while $\lambda_P$ is unitless. $\varepsilon_F$ is the allocated precision for observable  $\hat{F} \in \{\hat{V},\hat{P},\widehat{VP}_\mathrm{s}\}$.  Spectral gaps, $\ell_1$ norms and precisions are displayed in units of Hartree. The final cost of the algorithm measured in total number of Toffoli gates is decomposed in terms of total cost (E$_F$), initial state preparation cost (ASP), outer QPE cost (oQPE), eigenstate reflection cost ($\mathcal{R}_\pi$), inner QPE cost for each monomer (iQPE$_A$ and iQPE$_B$), block encoding cost for each monomer Hamiltonian (B[H$_A$] and B[H$_B$]), second reflection cost ($\mathcal{R}_\tau$). We have also performed the same decomposition for the total number of qubits. Note that the block encoding cost for the electrostatic-exchange operator, $\widehat{VP}_\mathrm{s}$, include the $\widehat{VP}_A$, $\widehat{VP}_B$, $\widehat{VP}_{1\mathrm{m}}$, $\widehat{VP}_{1\ell}$, and $\widehat{VP}_4$ contributions as discussed in the main text.}
\label{fig:Resource_estimate_data}
$(a)$ Properties  \vspace*{.5cm} \\
\begin{tabular}{llrrrrrrrrrrrr}					
\toprule										
System	&	Basis	&	$N_A$	&	$N_B$	&	$\lambda_{A}$	&	$\lambda_{B}$	&	$\Delta_{A}$	&	$\Delta_{B}$	&	$\lambda_V$	&	$\lambda_{VP}$	&	$\lambda_P$	&	$\varepsilon_V$	&	$\varepsilon_{VP}$	&	$\varepsilon_P$		\\
	
\midrule				
\multirow{3}{*}{Water}	&	STO-3G	&	7	&	7	&	53.92	&	53.92	&	0.455	&	0.455	&	0.43	&	0.11	&	0.04	&	$9.04\!\times\!10^{-4}$	&	$4.72\!\times\!10^{-4}$	&	$4.33\!\times\!10^{-4}$		\\
	
	&	6-31G	&	13	&	13	&	73.12	&	73.08	&	0.311	&	0.311	&	3.22	&	7.09	&	0.96	&	$2.96\!\times\!10^{-4}$	&	$6.14\!\times\!10^{-4}$	&	$1.26\!\times\!10^{-4}$		\\
	
	&	cc-pVDZ	&	24	&	24	&	328.11	&	328.11	&	0.302	&	0.302	&	13.05	&	39.5	&	1.53	&	$2.20\!\times\!10^{-4}$	&	$6.09\!\times\!10^{-4}$	&	$3.32\!\times\!10^{-5}$		\\
\midrule										
\multirow{3}{*}{Ammonia}	&	STO-3G	&	8	&	8	&	44.59	&	44.59	&	0.532	&	0.532	&	0.55	&	0.07	&	0.03	&	$1.02\!\times\!10^{-3}$	&	$3.79\!\times\!10^{-4}$	&	$3.09\!\times\!10^{-4}$		\\
	
	&	6-31G	&	15	&	15	&	84.03	&	84.03	&	0.289	&	0.289	&	3.98	&	8.44	&	0.97	&	$2.96\!\times\!10^{-4}$	&	$6.06\!\times\!10^{-4}$	&	$1.03\!\times\!10^{-4}$		\\
	
	&	cc-pVDZ	&	29	&	29	&	433.76	&	433.76	&	0.28	&	0.28	&	16.92	&	54.9	&	1.72	&	$2.04\!\times\!10^{-4}$	&	$6.05\!\times\!10^{-4}$	&	$2.61\!\times\!10^{-5}$		\\
\midrule										
\multirow{3}{*}{Benz.-Water}	&	STO-3G	&	36	&	7	&	453.61	&	53.92	&	0.211	&	0.455	&	1.94	&	0.14	&	0.02	&	$1.12\!\times\!10^{-3}$	&	$3.05\!\times\!10^{-4}$	&	$8.12\!\times\!10^{-5}$		\\
		
	&	6-31G	&	66	&	13	&	1158.48	&	73.1	&	0.198	&	0.311	&	12.6	&	22.23	&	0.93	&	$3.14\!\times\!10^{-4}$	&	$5.78\!\times\!10^{-4}$	&	$3.32\!\times\!10^{-5}$		\\
		
	&	cc-pVDZ	&	114	&	24	&	4563.95	&	328.15	&	0.197	&	0.302	&	47.44	&	146.51	&	1.82	&	$1.99\!\times\!10^{-4}$	&	$5.88\!\times\!10^{-4}$	&	$9.51\!\times\!10^{-6}$		\\
\midrule																												
\multirow{11}{*}{Heme-Art.}	&	cc-pVDZ[4]	&	4	&	4	&	1.38	&	2.03	&	0.007	&	0.121	&	0.37	&	0.05	&	0.05	&	$9.58\!\times\!10^{-4}$	&	$3.59\!\times\!10^{-4}$	&	$6.22\!\times\!10^{-4}$		\\
		
	&	cc-pVDZ[6]	&	6	&	6	&	2.81	&	3.97	&	0.007	&	0.121	&	1.9	&	1.59	&	0.82	&	$3.74\!\times\!10^{-4}$	&	$4.62\!\times\!10^{-4}$	&	$2.40\!\times\!10^{-4}$		\\
		
	&	cc-pVDZ[8]	&	8	&	8	&	4.74	&	6.93	&	0.007	&	0.121	&	2.71	&	2.34	&	0.88	&	$3.60\!\times\!10^{-4}$	&	$4.59\!\times\!10^{-4}$	&	$1.71\!\times\!10^{-4}$		\\
		
	&	cc-pVDZ[12]	&	12	&	12	&	10.51	&	13.61	&	0.007	&	0.121	&	5.39	&	8.2	&	1.48	&	$2.53\!\times\!10^{-4}$	&	$4.91\!\times\!10^{-4}$	&	$8.97\!\times\!10^{-5}$		\\
	
	&	cc-pVDZ[16]	&	16	&	16	&	18.98	&	23.44	&	0.007	&	0.121	&	9.91	&	23.97	&	2.31	&	$1.80\!\times\!10^{-4}$	&	$5.09\!\times\!10^{-4}$	&	$5.01\!\times\!10^{-5}$		\\
	
	&	cc-pVDZ[22]	&	22	&	22	&	36.67	&	44.71	&	0.007	&	0.121	&	19.56	&	68.73	&	3.28	&	$1.34\!\times\!10^{-4}$	&	$5.21\!\times\!10^{-4}$	&	$2.58\!\times\!10^{-5}$		\\
	
	&	cc-pVDZ[32]	&	32	&	32	&	84.56	&	98.26	&	0.007	&	0.121	&	41.28	&	225.01	&	5.02	&	$9.28\!\times\!10^{-5}$	&	$5.31\!\times\!10^{-4}$	&	$1.24\!\times\!10^{-5}$		\\
		
	&	cc-pVDZ[46]	&	46	&	46	&	191.62	&	215.95	&	0.007	&	0.121	&	94.15	&	901.4	&	9.01	&	$5.46\!\times\!10^{-5}$	&	$5.35\!\times\!10^{-4}$	&	$5.51\!\times\!10^{-6}$		\\
	
	&	cc-pVDZ[64]	&	64	&	64	&	386.59	&	455.65	&	0.007	&	0.121	&	184.92	&	2775.08	&	14.47	&	$3.50\!\times\!10^{-5}$	&	$5.34\!\times\!10^{-4}$	&	$2.83\!\times\!10^{-6}$		\\
		
	&	cc-pVDZ[90]	&	90	&	90	&	808.5	&	1068.83	&	0.007	&	0.121	&	347.93	&	8059.98	&	22.22	&	$2.31\!\times\!10^{-5}$	&	$5.37\!\times\!10^{-4}$	&	$1.51\!\times\!10^{-6}$		\\
	
	&	cc-pVDZ[opt]	&	43	&	40	&	232.2	&	361.78	&	0.007	&	0.121	&	65.54	&	537.3	&	6.35	&	$7.29\!\times\!10^{-5}$	&	$5.66\!\times\!10^{-4}$	&	$7.60\!\times\!10^{-6}$		\\
	
\bottomrule	
\end{tabular}
\end{sidewaystable}

\begin{sidewaystable}
{ \footnotesize 
$(b)$ $\hat{V}$: gate counts  \vspace*{.5cm} \\
\begin{tabular}{llrrrrrrrrrrrr}																											
\toprule																		
System	&	Basis	&	$E_{V}$	&	ASP	&	$\textrm{aQPE}_A$	&	$\textrm{aQPE}_B$	&	oQPE	&	$\mathcal{R}_\pi$	&	$\textrm{iQPE}_A$	&	$\textrm{B}[H_A]$	&	$\textrm{iQPE}_B$	&	$\textrm{B}[H_B]$	&	$\mathcal{R}_\tau$	&	B[V]	\\
	
\midrule																											
\multirow{3}{*}{Water}	&	STO-3G	&	$3.67\!\times\!10^{13}$	&	$1.15\!\times\!10^{10}$	&	$1.88\!\times\!10^{7}$	&	$1.88\!\times\!10^{7}$	&	$3.67\!\times\!10^{13}$	&	$8.95\!\times\!10^{9}$	&	$2.24\!\times\!10^{9}$	&	$2.26\!\times\!10^{3}$	&	$2.24\!\times\!10^{9}$	&	$2.26\!\times\!10^{3}$	&	$1.73\!\times\!10^{3}$	&	$1.61\!\times\!10^{3}$	\\
		
	&	6-31G	&	$4.80\!\times\!10^{15}$	&	$5.69\!\times\!10^{10}$	&	$9.34\!\times\!10^{7}$	&	$9.32\!\times\!10^{7}$	&	$4.80\!\times\!10^{15}$	&	$3.66\!\times\!10^{10}$	&	$9.16\!\times\!10^{9}$	&	$4.14\!\times\!10^{3}$	&	$9.15\!\times\!10^{9}$	&	$4.14\!\times\!10^{3}$	&	$3.62\!\times\!10^{3}$	&	$3.48\!\times\!10^{3}$	\\
	
	&	cc-pVDZ	&	$1.72\!\times\!10^{17}$	&	$2.48\!\times\!10^{12}$	&	$4.07\!\times\!10^{9}$	&	$4.07\!\times\!10^{9}$	&	$1.72\!\times\!10^{17}$	&	$3.28\!\times\!10^{11}$	&	$8.20\!\times\!10^{10}$	&	$8.28\!\times\!10^{3}$	&	$8.19\!\times\!10^{10}$	&	$8.27\!\times\!10^{3}$	&	$6.88\!\times\!10^{3}$	&	$6.72\!\times\!10^{3}$	\\
\midrule																											
\multirow{3}{*}{Ammonia}	&	STO-3G	&	$1.97\!\times\!10^{13}$	&	$7.29\!\times\!10^{9}$	&	$1.19\!\times\!10^{7}$	&	$1.19\!\times\!10^{7}$	&	$1.97\!\times\!10^{13}$	&	$4.81\!\times\!10^{9}$	&	$1.20\!\times\!10^{9}$	&	$2.47\!\times\!10^{3}$	&	$1.20\!\times\!10^{9}$	&	$2.47\!\times\!10^{3}$	&	$2.00\!\times\!10^{3}$	&	$1.88\!\times\!10^{3}$	\\
	
	&	6-31G	&	$5.60\!\times\!10^{15}$	&	$9.50\!\times\!10^{10}$	&	$1.56\!\times\!10^{8}$	&	$1.56\!\times\!10^{8}$	&	$5.60\!\times\!10^{15}$	&	$4.27\!\times\!10^{10}$	&	$1.07\!\times\!10^{10}$	&	$4.83\!\times\!10^{3}$	&	$1.07\!\times\!10^{10}$	&	$4.83\!\times\!10^{3}$	&	$4.10\!\times\!10^{3}$	&	$3.96\!\times\!10^{3}$	\\
	
	&	cc-pVDZ	&	$4.28\!\times\!10^{17}$	&	$5.77\!\times\!10^{12}$	&	$9.46\!\times\!10^{9}$	&	$9.46\!\times\!10^{9}$	&	$4.28\!\times\!10^{17}$	&	$4.08\!\times\!10^{11}$	&	$1.02\!\times\!10^{11}$	&	$1.02\!\times\!10^{4}$	&	$1.02\!\times\!10^{11}$	&	$1.02\!\times\!10^{4}$	&	$8.43\!\times\!10^{3}$	&	$8.27\!\times\!10^{3}$	\\
\midrule																											
\multirow{3}{*}{Benz.-Water}	&	STO-3G	&	$7.66\!\times\!10^{15}$	&	$5.24\!\times\!10^{12}$	&	$1.72\!\times\!10^{10}$	&	$1.88\!\times\!10^{7}$	&	$7.66\!\times\!10^{15}$	&	$4.67\!\times\!10^{11}$	&	$2.32\!\times\!10^{11}$	&	$1.27\!\times\!10^{4}$	&	$2.20\!\times\!10^{9}$	&	$2.26\!\times\!10^{3}$	&	$4.99\!\times\!10^{3}$	&	$4.84\!\times\!10^{3}$	\\
	
	&	6-31G	&	$1.10\!\times\!10^{18}$	&	$7.03\!\times\!10^{13}$	&	$2.30\!\times\!10^{11}$	&	$9.33\!\times\!10^{7}$	&	$1.10\!\times\!10^{18}$	&	$2.09\!\times\!10^{12}$	&	$1.04\!\times\!10^{12}$	&	$2.49\!\times\!10^{4}$	&	$9.03\!\times\!10^{9}$	&	$4.14\!\times\!10^{3}$	&	$1.07\!\times\!10^{4}$	&	$1.05\!\times\!10^{4}$	\\
	
	&	cc-pVDZ	&	$3.57\!\times\!10^{19}$	&	$2.09\!\times\!10^{15}$	&	$6.86\!\times\!10^{12}$	&	$4.07\!\times\!10^{9}$	&	$3.57\!\times\!10^{19}$	&	$1.70\!\times\!10^{13}$	&	$8.43\!\times\!10^{12}$	&	$4.59\!\times\!10^{4}$	&	$8.20\!\times\!10^{10}$	&	$8.28\!\times\!10^{3}$	&	$2.00\!\times\!10^{4}$	&	$1.98\!\times\!10^{4}$	\\
\midrule																											
\multirow{11}{*}{Heme-Art.}	&	cc-pVDZ[4]	&	$7.71\!\times\!10^{12}$	&	$1.43\!\times\!10^{8}$	&	$4.22\!\times\!10^{5}$	&	$4.61\!\times\!10^{4}$	&	$7.71\!\times\!10^{12}$	&	$1.88\!\times\!10^{9}$	&	$8.43\!\times\!10^{8}$	&	$1.18\!\times\!10^{3}$	&	$9.90\!\times\!10^{7}$	&	$1.06\!\times\!10^{3}$	&	$1.06\!\times\!10^{3}$	&	$9.70\!\times\!10^{2}$	\\
		
	&	cc-pVDZ[6]	&	$4.48\!\times\!10^{14}$	&	$8.95\!\times\!10^{8}$	&	$2.67\!\times\!10^{6}$	&	$2.68\!\times\!10^{5}$	&	$4.48\!\times\!10^{14}$	&	$6.84\!\times\!10^{9}$	&	$3.07\!\times\!10^{9}$	&	$1.82\!\times\!10^{3}$	&	$3.55\!\times\!10^{8}$	&	$1.65\!\times\!10^{3}$	&	$1.70\!\times\!10^{3}$	&	$1.60\!\times\!10^{3}$	\\
	
	&	cc-pVDZ[8]	&	$1.27\!\times\!10^{15}$	&	$3.53\!\times\!10^{9}$	&	$1.05\!\times\!10^{7}$	&	$1.13\!\times\!10^{6}$	&	$1.27\!\times\!10^{15}$	&	$1.94\!\times\!10^{10}$	&	$8.66\!\times\!10^{9}$	&	$2.47\!\times\!10^{3}$	&	$1.03\!\times\!10^{9}$	&	$2.26\!\times\!10^{3}$	&	$2.24\!\times\!10^{3}$	&	$2.12\!\times\!10^{3}$	\\
		
	&	cc-pVDZ[12]	&	$1.73\!\times\!10^{16}$	&	$2.70\!\times\!10^{10}$	&	$8.15\!\times\!10^{7}$	&	$7.02\!\times\!10^{6}$	&	$1.73\!\times\!10^{16}$	&	$6.59\!\times\!10^{10}$	&	$2.94\!\times\!10^{10}$	&	$3.77\!\times\!10^{3}$	&	$3.52\!\times\!10^{9}$	&	$3.55\!\times\!10^{3}$	&	$3.42\!\times\!10^{3}$	&	$3.29\!\times\!10^{3}$	\\
		
	&	cc-pVDZ[16]	&	$9.77\!\times\!10^{16}$	&	$1.21\!\times\!10^{11}$	&	$3.67\!\times\!10^{8}$	&	$2.89\!\times\!10^{7}$	&	$9.77\!\times\!10^{16}$	&	$1.86\!\times\!10^{11}$	&	$8.81\!\times\!10^{10}$	&	$5.30\!\times\!10^{3}$	&	$5.02\!\times\!10^{9}$	&	$4.75\!\times\!10^{3}$	&	$4.69\!\times\!10^{3}$	&	$4.55\!\times\!10^{3}$	\\
		
	&	cc-pVDZ[22]	&	$5.89\!\times\!10^{17}$	&	$6.59\!\times\!10^{11}$	&	$2.01\!\times\!10^{9}$	&	$1.51\!\times\!10^{8}$	&	$5.89\!\times\!10^{17}$	&	$5.61\!\times\!10^{11}$	&	$2.65\!\times\!10^{11}$	&	$7.50\!\times\!10^{3}$	&	$1.52\!\times\!10^{10}$	&	$6.76\!\times\!10^{3}$	&	$6.67\!\times\!10^{3}$	&	$6.52\!\times\!10^{3}$	\\
		
	&	cc-pVDZ[32]	&	$7.75\!\times\!10^{18}$	&	$5.26\!\times\!10^{12}$	&	$1.61\!\times\!10^{10}$	&	$1.14\!\times\!10^{9}$	&	$7.75\!\times\!10^{18}$	&	$1.85\!\times\!10^{12}$	&	$8.73\!\times\!10^{11}$	&	$1.15\!\times\!10^{4}$	&	$5.03\!\times\!10^{10}$	&	$1.04\!\times\!10^{4}$	&	$9.98\!\times\!10^{3}$	&	$9.82\!\times\!10^{3}$	\\
	
	&	cc-pVDZ[46]	&	$1.07\!\times\!10^{20}$	&	$4.12\!\times\!10^{13}$	&	$1.27\!\times\!10^{11}$	&	$8.44\!\times\!10^{9}$	&	$1.07\!\times\!10^{20}$	&	$6.41\!\times\!10^{12}$	&	$2.87\!\times\!10^{12}$	&	$1.74\!\times\!10^{4}$	&	$3.39\!\times\!10^{11}$	&	$1.62\!\times\!10^{4}$	&	$1.51\!\times\!10^{4}$	&	$1.50\!\times\!10^{4}$	\\
	
	&	cc-pVDZ[64]	&	$2.56\!\times\!10^{21}$	&	$2.48\!\times\!10^{14}$	&	$7.58\!\times\!10^{11}$	&	$5.41\!\times\!10^{10}$	&	$2.56\!\times\!10^{21}$	&	$3.81\!\times\!10^{13}$	&	$1.80\!\times\!10^{13}$	&	$2.54\!\times\!10^{4}$	&	$1.03\!\times\!10^{12}$	&	$2.33\!\times\!10^{4}$	&	$2.16\!\times\!10^{4}$	&	$2.14\!\times\!10^{4}$	\\
	
	&	cc-pVDZ[90]	&	$1.60\!\times\!10^{22}$	&	$1.60\!\times\!10^{15}$	&	$4.81\!\times\!10^{12}$	&	$4.44\!\times\!10^{11}$	&	$1.60\!\times\!10^{22}$	&	$1.19\!\times\!10^{14}$	&	$5.62\!\times\!10^{13}$	&	$3.74\!\times\!10^{4}$	&	$3.26\!\times\!10^{12}$	&	$3.44\!\times\!10^{4}$	&	$3.13\!\times\!10^{4}$	&	$3.11\!\times\!10^{4}$	\\
	
	&	cc-pVDZ[opt]	&	$9.74\!\times\!10^{19}$	&	$6.12\!\times\!10^{13}$	&	$1.79\!\times\!10^{11}$	&	$2.14\!\times\!10^{10}$	&	$9.74\!\times\!10^{19}$	&	$1.16\!\times\!10^{13}$	&	$5.51\!\times\!10^{12}$	&	$1.67\!\times\!10^{4}$	&	$3.01\!\times\!10^{11}$	&	$1.43\!\times\!10^{4}$	&	$1.34\!\times\!10^{4}$	&	$1.32\!\times\!10^{4}$	\\
	
\bottomrule																											
\end{tabular}	} \\				
{\footnotesize  

\, \vspace*{.5cm}\\
$(c)$ $\hat{V}$: qubit counts \vspace*{.5cm} \\ 
\begin{tabular}{llrrrrrrrrrrrr}			
\toprule						
System	&	Basis	&	$E_{V}$	&	ASP	&	$\textrm{aQPE}_A$	&	$\textrm{aQPE}_B$	&	oQPE	&	$\mathcal{R}_\pi$	&	$\textrm{iQPE}_A$	&	$\textrm{B}[H_A]$	&	$\textrm{iQPE}_B$	&	$\textrm{B}[H_B]$	&	$\mathcal{R}_\tau$	&	B[V]	\\
		
\midrule							
\multirow{3}{*}{Water}	&	STO-3G	&	706	&	457	&	380	&	380	&	674	&	597	&	449	&	75	&	449	&	75	&	635	&	513	\\
	
	&	6-31G	&	1221	&	686	&	591	&	591	&	1187	&	887	&	700	&	94	&	700	&	94	&	1141	&	1005	\\
	
	&	cc-pVDZ	&	2081	&	1246	&	1112	&	1112	&	2042	&	1479	&	1249	&	126	&	1249	&	126	&	1990	&	1834	\\
\midrule							
\multirow{3}{*}{Ammonia}	&	STO-3G	&	768	&	471	&	394	&	394	&	737	&	618	&	469	&	76	&	469	&	76	&	700	&	580	\\
	
	&	6-31G	&	1342	&	776	&	674	&	674	&	1307	&	979	&	788	&	98	&	788	&	98	&	1261	&	1125	\\
	
	&	cc-pVDZ	&	2480	&	1479	&	1331	&	1331	&	2440	&	1730	&	1486	&	137	&	1486	&	137	&	2387	&	2229	\\
\midrule																											
\multirow{3}{*}{Benz.-Water}	&	STO-3G	&	2041	&	1708	&	1631	&	380	&	2000	&	1985	&	1828	&	154	&	449	&	75	&	1538	&	1395	\\
	
	&	6-31G	&	3663	&	3093	&	2998	&	591	&	3618	&	3598	&	3402	&	220	&	700	&	94	&	3056	&	2900	\\
	
	&	cc-pVDZ	&	6452	&	5706	&	5572	&	1112	&	6403	&	6381	&	6143	&	325	&	1249	&	126	&	5507	&	5332	\\
\midrule																											
\multirow{11}{*}{Heme-Art.}	&	cc-pVDZ[4]	&	504	&	275	&	230	&	194	&	478	&	409	&	291	&	57	&	255	&	51	&	442	&	350	\\
		
	&	cc-pVDZ[6]	&	703	&	367	&	310	&	268	&	674	&	529	&	385	&	67	&	343	&	61	&	632	&	528	\\
	
	&	cc-pVDZ[8]	&	844	&	462	&	394	&	346	&	813	&	633	&	478	&	76	&	430	&	70	&	769	&	655	\\
	
	&	cc-pVDZ[12]	&	1158	&	645	&	561	&	501	&	1125	&	832	&	653	&	90	&	606	&	84	&	1077	&	951	\\
	
	&	cc-pVDZ[16]	&	1485	&	827	&	731	&	652	&	1449	&	1051	&	857	&	103	&	761	&	95	&	1399	&	1265	\\
	
	&	cc-pVDZ[22]	&	1988	&	1125	&	1010	&	890	&	1950	&	1363	&	1146	&	120	&	1026	&	112	&	1897	&	1753	\\
		
	&	cc-pVDZ[32]	&	2831	&	1589	&	1444	&	1321	&	2790	&	1911	&	1660	&	146	&	1500	&	138	&	2733	&	2577	\\
	
	&	cc-pVDZ[46]	&	4160	&	2306	&	2125	&	1959	&	4116	&	2715	&	2418	&	179	&	2256	&	173	&	4054	&	3886	\\
		
	&	cc-pVDZ[64]	&	5856	&	3283	&	3059	&	2771	&	5809	&	3807	&	3461	&	222	&	3173	&	214	&	5742	&	5562	\\
	
	&	cc-pVDZ[90]	&	8380	&	4660	&	4377	&	4076	&	8330	&	5432	&	5023	&	279	&	4631	&	271	&	8260	&	8070	\\
	
	&	cc-pVDZ[opt]	&	3724	&	2224	&	2052	&	1785	&	3679	&	2615	&	2331	&	175	&	2002	&	163	&	3617	&	3445	\\
	
\bottomrule							
\end{tabular}}									
\end{sidewaystable}
\begin{sidewaystable}
    {\footnotesize
$(d)$ $\widehat{VP}_\mathrm{s}$: gate counts \vspace*{.5cm} \\ 
    \begin{tabular}{llrrrrrrrrrrrr}
\toprule							
System	&	Basis	&	$E_{VP}$	&	ASP	&	$\textrm{aQPE}_A$	&	$\textrm{aQPE}_B$	&	oQPE	&	$\mathcal{R}_\pi$	&	$\textrm{iQPE}_A$	&	$\textrm{B}[H_A]$	&	$\textrm{iQPE}_B$	&	$\textrm{B}[H_B]$	&	$\mathcal{R}_\tau$	&	B[VP]	\\
	
\midrule						
\multirow{3}{*}{Water}	&	STO-3G	&	$1.92\!\times\!10^{13}$	&	$1.15\!\times\!10^{10}$	&	$1.88\!\times\!10^{7}$	&	$1.88\!\times\!10^{7}$	&	$1.92\!\times\!10^{13}$	&	$9.39\!\times\!10^{9}$	&	$2.35\!\times\!10^{9}$	&	$2.26\!\times\!10^{3}$	&	$2.35\!\times\!10^{9}$	&	$2.26\!\times\!10^{3}$	&	$7.66\!\times\!10^{3}$	&	$7.53\!\times\!10^{3}$	\\
		
	&	6-31G	&	$4.52\!\times\!10^{15}$	&	$5.69\!\times\!10^{10}$	&	$9.34\!\times\!10^{7}$	&	$9.32\!\times\!10^{7}$	&	$4.52\!\times\!10^{15}$	&	$3.45\!\times\!10^{10}$	&	$8.62\!\times\!10^{9}$	&	$4.14\!\times\!10^{3}$	&	$8.61\!\times\!10^{9}$	&	$4.14\!\times\!10^{3}$	&	$1.70\!\times\!10^{4}$	&	$1.68\!\times\!10^{4}$	\\
	
	&	cc-pVDZ	&	$1.60\!\times\!10^{17}$	&	$2.48\!\times\!10^{12}$	&	$4.07\!\times\!10^{9}$	&	$4.07\!\times\!10^{9}$	&	$1.60\!\times\!10^{17}$	&	$3.06\!\times\!10^{11}$	&	$7.65\!\times\!10^{10}$	&	$8.28\!\times\!10^{3}$	&	$7.65\!\times\!10^{10}$	&	$8.27\!\times\!10^{3}$	&	$3.83\!\times\!10^{4}$	&	$3.82\!\times\!10^{4}$	\\
\midrule																											
\multirow{3}{*}{Ammonia}	&	STO-3G	&	$1.05\!\times\!10^{13}$	&	$7.29\!\times\!10^{9}$	&	$1.19\!\times\!10^{7}$	&	$1.19\!\times\!10^{7}$	&	$1.05\!\times\!10^{13}$	&	$5.13\!\times\!10^{9}$	&	$1.28\!\times\!10^{9}$	&	$2.47\!\times\!10^{3}$	&	$1.28\!\times\!10^{9}$	&	$2.47\!\times\!10^{3}$	&	$9.14\!\times\!10^{3}$	&	$9.02\!\times\!10^{3}$	\\
		
	&	6-31G	&	$5.35\!\times\!10^{15}$	&	$9.50\!\times\!10^{10}$	&	$1.56\!\times\!10^{8}$	&	$1.56\!\times\!10^{8}$	&	$5.35\!\times\!10^{15}$	&	$4.08\!\times\!10^{10}$	&	$1.02\!\times\!10^{10}$	&	$4.83\!\times\!10^{3}$	&	$1.02\!\times\!10^{10}$	&	$4.83\!\times\!10^{3}$	&	$2.17\!\times\!10^{4}$	&	$2.15\!\times\!10^{4}$	\\
	
	&	cc-pVDZ	&	$4.00\!\times\!10^{17}$	&	$5.77\!\times\!10^{12}$	&	$9.46\!\times\!10^{9}$	&	$9.46\!\times\!10^{9}$	&	$4.00\!\times\!10^{17}$	&	$3.81\!\times\!10^{11}$	&	$9.53\!\times\!10^{10}$	&	$1.02\!\times\!10^{4}$	&	$9.53\!\times\!10^{10}$	&	$1.02\!\times\!10^{4}$	&	$5.24\!\times\!10^{4}$	&	$5.22\!\times\!10^{4}$	\\
\midrule																											
\multirow{3}{*}{Benz.-Water}	&	STO-3G	&	$2.09\!\times\!10^{15}$	&	$5.24\!\times\!10^{12}$	&	$1.72\!\times\!10^{10}$	&	$1.88\!\times\!10^{7}$	&	$2.08\!\times\!10^{15}$	&	$5.08\!\times\!10^{11}$	&	$2.51\!\times\!10^{11}$	&	$1.27\!\times\!10^{4}$	&	$2.42\!\times\!10^{9}$	&	$2.26\!\times\!10^{3}$	&	$2.46\!\times\!10^{4}$	&	$2.44\!\times\!10^{4}$	\\
	
	&	6-31G	&	$5.27\!\times\!10^{17}$	&	$7.03\!\times\!10^{13}$	&	$2.30\!\times\!10^{11}$	&	$9.33\!\times\!10^{7}$	&	$5.27\!\times\!10^{17}$	&	$2.01\!\times\!10^{12}$	&	$9.97\!\times\!10^{11}$	&	$2.49\!\times\!10^{4}$	&	$8.75\!\times\!10^{9}$	&	$4.14\!\times\!10^{3}$	&	$5.59\!\times\!10^{4}$	&	$5.57\!\times\!10^{4}$	\\
	
	&	cc-pVDZ	&	$3.36\!\times\!10^{19}$	&	$2.09\!\times\!10^{15}$	&	$6.86\!\times\!10^{12}$	&	$4.07\!\times\!10^{9}$	&	$3.36\!\times\!10^{19}$	&	$1.60\!\times\!10^{13}$	&	$7.94\!\times\!10^{12}$	&	$4.59\!\times\!10^{4}$	&	$7.65\!\times\!10^{10}$	&	$8.28\!\times\!10^{3}$	&	$1.15\!\times\!10^{5}$	&	$1.14\!\times\!10^{5}$	\\
\midrule																											
\multirow{11}{*}{Heme-Art.}	&	cc-pVDZ[4]	&	$2.15\!\times\!10^{12}$	&	$1.43\!\times\!10^{8}$	&	$4.22\!\times\!10^{5}$	&	$4.61\!\times\!10^{4}$	&	$2.15\!\times\!10^{12}$	&	$2.10\!\times\!10^{9}$	&	$9.39\!\times\!10^{8}$	&	$1.18\!\times\!10^{3}$	&	$1.10\!\times\!10^{8}$	&	$1.06\!\times\!10^{3}$	&	$4.48\!\times\!10^{3}$	&	$4.39\!\times\!10^{3}$	\\
	
	&	cc-pVDZ[6]	&	$2.20\!\times\!10^{14}$	&	$8.95\!\times\!10^{8}$	&	$2.67\!\times\!10^{6}$	&	$2.68\!\times\!10^{5}$	&	$2.20\!\times\!10^{14}$	&	$6.71\!\times\!10^{9}$	&	$3.01\!\times\!10^{9}$	&	$1.82\!\times\!10^{3}$	&	$3.48\!\times\!10^{8}$	&	$1.65\!\times\!10^{3}$	&	$7.68\!\times\!10^{3}$	&	$7.57\!\times\!10^{3}$	\\
		
	&	cc-pVDZ[8]	&	$1.25\!\times\!10^{15}$	&	$3.53\!\times\!10^{9}$	&	$1.05\!\times\!10^{7}$	&	$1.13\!\times\!10^{6}$	&	$1.25\!\times\!10^{15}$	&	$1.90\!\times\!10^{10}$	&	$8.50\!\times\!10^{9}$	&	$2.47\!\times\!10^{3}$	&	$1.01\!\times\!10^{9}$	&	$2.26\!\times\!10^{3}$	&	$1.03\!\times\!10^{4}$	&	$1.02\!\times\!10^{4}$	\\
		
	&	cc-pVDZ[12]	&	$8.20\!\times\!10^{15}$	&	$2.70\!\times\!10^{10}$	&	$8.15\!\times\!10^{7}$	&	$7.02\!\times\!10^{6}$	&	$8.20\!\times\!10^{15}$	&	$6.26\!\times\!10^{10}$	&	$2.79\!\times\!10^{10}$	&	$3.77\!\times\!10^{3}$	&	$3.35\!\times\!10^{9}$	&	$3.55\!\times\!10^{3}$	&	$1.70\!\times\!10^{4}$	&	$1.69\!\times\!10^{4}$	\\
	
	&	cc-pVDZ[16]	&	$9.00\!\times\!10^{16}$	&	$1.21\!\times\!10^{11}$	&	$3.67\!\times\!10^{8}$	&	$2.89\!\times\!10^{7}$	&	$9.00\!\times\!10^{16}$	&	$1.72\!\times\!10^{11}$	&	$8.12\!\times\!10^{10}$	&	$5.30\!\times\!10^{3}$	&	$4.63\!\times\!10^{9}$	&	$4.75\!\times\!10^{3}$	&	$2.42\!\times\!10^{4}$	&	$2.40\!\times\!10^{4}$	\\
	
	&	cc-pVDZ[22]	&	$5.36\!\times\!10^{17}$	&	$6.59\!\times\!10^{11}$	&	$2.01\!\times\!10^{9}$	&	$1.51\!\times\!10^{8}$	&	$5.36\!\times\!10^{17}$	&	$5.11\!\times\!10^{11}$	&	$2.42\!\times\!10^{11}$	&	$7.50\!\times\!10^{3}$	&	$1.38\!\times\!10^{10}$	&	$6.76\!\times\!10^{3}$	&	$3.70\!\times\!10^{4}$	&	$3.68\!\times\!10^{4}$	\\
	
	&	cc-pVDZ[32]	&	$6.99\!\times\!10^{18}$	&	$5.26\!\times\!10^{12}$	&	$1.61\!\times\!10^{10}$	&	$1.14\!\times\!10^{9}$	&	$6.99\!\times\!10^{18}$	&	$1.67\!\times\!10^{12}$	&	$7.89\!\times\!10^{11}$	&	$1.15\!\times\!10^{4}$	&	$4.48\!\times\!10^{10}$	&	$1.04\!\times\!10^{4}$	&	$6.65\!\times\!10^{4}$	&	$6.64\!\times\!10^{4}$	\\
		
	&	cc-pVDZ[46]	&	$9.38\!\times\!10^{19}$	&	$4.12\!\times\!10^{13}$	&	$1.27\!\times\!10^{11}$	&	$8.44\!\times\!10^{9}$	&	$9.38\!\times\!10^{19}$	&	$5.59\!\times\!10^{12}$	&	$2.50\!\times\!10^{12}$	&	$1.74\!\times\!10^{4}$	&	$2.96\!\times\!10^{11}$	&	$1.62\!\times\!10^{4}$	&	$1.46\!\times\!10^{5}$	&	$1.46\!\times\!10^{5}$	\\
	
	&	cc-pVDZ[64]	&	$2.17\!\times\!10^{21}$	&	$2.48\!\times\!10^{14}$	&	$7.58\!\times\!10^{11}$	&	$5.41\!\times\!10^{10}$	&	$2.17\!\times\!10^{21}$	&	$3.23\!\times\!10^{13}$	&	$1.52\!\times\!10^{13}$	&	$2.54\!\times\!10^{4}$	&	$8.84\!\times\!10^{11}$	&	$2.33\!\times\!10^{4}$	&	$3.74\!\times\!10^{5}$	&	$3.74\!\times\!10^{5}$	\\
	
	&	cc-pVDZ[90]	&	$1.35\!\times\!10^{22}$	&	$1.60\!\times\!10^{15}$	&	$4.81\!\times\!10^{12}$	&	$4.44\!\times\!10^{11}$	&	$1.35\!\times\!10^{22}$	&	$1.00\!\times\!10^{14}$	&	$4.74\!\times\!10^{13}$	&	$3.74\!\times\!10^{4}$	&	$2.76\!\times\!10^{12}$	&	$3.44\!\times\!10^{4}$	&	$1.20\!\times\!10^{6}$	&	$1.20\!\times\!10^{6}$	\\
	
	&	cc-pVDZ[opt]	&	$8.63\!\times\!10^{19}$	&	$6.12\!\times\!10^{13}$	&	$1.79\!\times\!10^{11}$	&	$2.14\!\times\!10^{10}$	&	$8.63\!\times\!10^{19}$	&	$1.03\!\times\!10^{13}$	&	$4.88\!\times\!10^{12}$	&	$1.67\!\times\!10^{4}$	&	$2.68\!\times\!10^{11}$	&	$1.43\!\times\!10^{4}$	&	$1.10\!\times\!10^{5}$	&	$1.10\!\times\!10^{5}$	\\
		
\bottomrule																											
\end{tabular}																											

    }

{ \footnotesize
\, \vspace*{.5cm}\\
$(e)$ $\widehat{VP}_\mathrm{s}$: qubit counts \vspace*{.5cm} \\ 

\begin{tabular}{llrrrrrrrrrrrr}																											
\toprule							
System	&	Basis	&	$E_{VP}$	&	ASP	&	$\textrm{aQPE}_A$	&	$\textrm{aQPE}_B$	&	oQPE	&	$\mathcal{R}_\pi$	&	$\textrm{iQPE}_A$	&	$\textrm{B}[H_A]$	&	$\textrm{iQPE}_B$	&	$\textrm{B}[H_B]$	&	$\mathcal{R}_\tau$	&	B[VP]	\\
	
\midrule								
\multirow{3}{*}{Water}	&	STO-3G	&	731	&	457	&	380	&	380	&	699	&	605	&	449	&	75	&	449	&	75	&	661	&	539	\\
	
	&	6-31G	&	1160	&	686	&	591	&	591	&	1126	&	886	&	700	&	94	&	700	&	94	&	1080	&	944	\\
	
	&	cc-pVDZ	&	1976	&	1246	&	1112	&	1112	&	1937	&	1478	&	1249	&	126	&	1249	&	126	&	1885	&	1729	\\
\midrule																											
\multirow{3}{*}{Ammonia}	&	STO-3G	&	801	&	471	&	394	&	394	&	770	&	629	&	469	&	76	&	469	&	76	&	734	&	614	\\
	
	&	6-31G	&	1452	&	776	&	674	&	674	&	1417	&	981	&	788	&	98	&	788	&	98	&	1371	&	1235	\\
	
	&	cc-pVDZ	&	2355	&	1479	&	1331	&	1331	&	2315	&	1729	&	1486	&	137	&	1486	&	137	&	2262	&	2104	\\
\midrule										
\multirow{3}{*}{Benz.-Water}	&	STO-3G	&	2050	&	1708	&	1631	&	380	&	2009	&	1996	&	1828	&	154	&	449	&	75	&	1643	&	1500	\\
	
	&	6-31G	&	3664	&	3093	&	2998	&	591	&	3619	&	3600	&	3402	&	220	&	700	&	94	&	2975	&	2819	\\
	
	&	cc-pVDZ	&	6451	&	5706	&	5572	&	1112	&	6402	&	6380	&	6143	&	325	&	1249	&	126	&	5222	&	5047	\\
\midrule																											
\multirow{11}{*}{Heme-Art.}	&	cc-pVDZ[4]	&	522	&	275	&	230	&	194	&	496	&	417	&	291	&	57	&	255	&	51	&	462	&	370	\\
	
	&	cc-pVDZ[6]	&	682	&	367	&	310	&	268	&	653	&	528	&	385	&	67	&	343	&	61	&	612	&	508	\\
	
	&	cc-pVDZ[8]	&	826	&	462	&	394	&	346	&	795	&	635	&	478	&	76	&	430	&	70	&	751	&	637	\\
	
	&	cc-pVDZ[12]	&	1125	&	645	&	561	&	501	&	1092	&	831	&	653	&	90	&	606	&	84	&	1045	&	919	\\
	
	&	cc-pVDZ[16]	&	1498	&	827	&	731	&	652	&	1462	&	1050	&	857	&	103	&	761	&	95	&	1412	&	1278	\\
	
	&	cc-pVDZ[22]	&	1891	&	1125	&	1010	&	890	&	1853	&	1362	&	1146	&	120	&	1026	&	112	&	1800	&	1656	\\
	
	&	cc-pVDZ[32]	&	2688	&	1589	&	1444	&	1321	&	2647	&	1907	&	1660	&	146	&	1500	&	138	&	2590	&	2434	\\
	
	&	cc-pVDZ[46]	&	3775	&	2306	&	2125	&	1959	&	3731	&	2711	&	2418	&	179	&	2256	&	173	&	3669	&	3501	\\
		
	&	cc-pVDZ[64]	&	5321	&	3283	&	3059	&	2771	&	5274	&	3800	&	3461	&	222	&	3173	&	214	&	5207	&	5027	\\
	
	&	cc-pVDZ[90]	&	7456	&	4660	&	4377	&	4076	&	7406	&	5425	&	5023	&	279	&	4631	&	271	&	7336	&	7146	\\
	
	&	cc-pVDZ[opt]	&	3449	&	2224	&	2052	&	1785	&	3404	&	2611	&	2331	&	175	&	2002	&	163	&	3342	&	3170	\\
	
\bottomrule											
\end{tabular}									
}
\end{sidewaystable}

\begin{sidewaystable}
    {\footnotesize
$(f)$ $\hat{P}$: gate counts \vspace*{.5cm}\\
    \begin{tabular}{llrrrrrrrrrrrr}				
\toprule											
System	&	Basis	&	$E_{P}$	&	ASP	&	$\textrm{aQPE}_A$	&	$\textrm{aQPE}_B$	&	oQPE	&	$\mathcal{R}_\pi$	&	$\textrm{iQPE}_A$	&	$\textrm{B}[H_A]$	&	$\textrm{iQPE}_B$	&	$\textrm{B}[H_B]$	&	$\mathcal{R}_\tau$	&	B[P]	\\
	
\midrule										
\multirow{3}{*}{Water}	&	STO-3G	&	$9.78\!\times\!10^{12}$	&	$1.15\!\times\!10^{10}$	&	$1.88\!\times\!10^{7}$	&	$1.88\!\times\!10^{7}$	&	$9.77\!\times\!10^{12}$	&	$9.54\!\times\!10^{9}$	&	$2.39\!\times\!10^{9}$	&	$2.26\!\times\!10^{3}$	&	$2.39\!\times\!10^{9}$	&	$2.26\!\times\!10^{3}$	&	$8.70\!\times\!10^{2}$	&	$7.48\!\times\!10^{2}$	\\
	
	&	6-31G	&	$2.51\!\times\!10^{15}$	&	$5.69\!\times\!10^{10}$	&	$9.34\!\times\!10^{7}$	&	$9.32\!\times\!10^{7}$	&	$2.51\!\times\!10^{15}$	&	$3.83\!\times\!10^{10}$	&	$9.57\!\times\!10^{9}$	&	$4.14\!\times\!10^{3}$	&	$9.56\!\times\!10^{9}$	&	$4.14\!\times\!10^{3}$	&	$2.18\!\times\!10^{3}$	&	$2.04\!\times\!10^{3}$	\\
	
	&	cc-pVDZ	&	$1.90\!\times\!10^{17}$	&	$2.48\!\times\!10^{12}$	&	$4.07\!\times\!10^{9}$	&	$4.07\!\times\!10^{9}$	&	$1.90\!\times\!10^{17}$	&	$3.62\!\times\!10^{11}$	&	$9.06\!\times\!10^{10}$	&	$8.28\!\times\!10^{3}$	&	$9.06\!\times\!10^{10}$	&	$8.27\!\times\!10^{3}$	&	$7.47\!\times\!10^{3}$	&	$7.31\!\times\!10^{3}$	\\
\midrule						
\multirow{3}{*}{Ammonia}	&	STO-3G	&	$5.35\!\times\!10^{12}$	&	$7.29\!\times\!10^{9}$	&	$1.19\!\times\!10^{7}$	&	$1.19\!\times\!10^{7}$	&	$5.34\!\times\!10^{12}$	&	$5.21\!\times\!10^{9}$	&	$1.30\!\times\!10^{9}$	&	$2.47\!\times\!10^{3}$	&	$1.30\!\times\!10^{9}$	&	$2.47\!\times\!10^{3}$	&	$1.22\!\times\!10^{3}$	&	$1.10\!\times\!10^{3}$	\\
		
	&	6-31G	&	$3.01\!\times\!10^{15}$	&	$9.50\!\times\!10^{10}$	&	$1.56\!\times\!10^{8}$	&	$1.56\!\times\!10^{8}$	&	$3.01\!\times\!10^{15}$	&	$4.59\!\times\!10^{10}$	&	$1.15\!\times\!10^{10}$	&	$4.83\!\times\!10^{3}$	&	$1.15\!\times\!10^{10}$	&	$4.83\!\times\!10^{3}$	&	$3.76\!\times\!10^{3}$	&	$3.63\!\times\!10^{3}$	\\
	
	&	cc-pVDZ	&	$2.39\!\times\!10^{17}$	&	$5.77\!\times\!10^{12}$	&	$9.46\!\times\!10^{9}$	&	$9.46\!\times\!10^{9}$	&	$2.39\!\times\!10^{17}$	&	$4.56\!\times\!10^{11}$	&	$1.14\!\times\!10^{11}$	&	$1.02\!\times\!10^{4}$	&	$1.14\!\times\!10^{11}$	&	$1.02\!\times\!10^{4}$	&	$1.45\!\times\!10^{4}$	&	$1.43\!\times\!10^{4}$	\\
\midrule																											
\multirow{3}{*}{Benz.-Water}	&	STO-3G	&	$1.13\!\times\!10^{15}$	&	$5.24\!\times\!10^{12}$	&	$1.72\!\times\!10^{10}$	&	$1.88\!\times\!10^{7}$	&	$1.12\!\times\!10^{15}$	&	$5.48\!\times\!10^{11}$	&	$2.71\!\times\!10^{11}$	&	$1.27\!\times\!10^{4}$	&	$2.64\!\times\!10^{9}$	&	$2.26\!\times\!10^{3}$	&	$1.17\!\times\!10^{3}$	&	$1.03\!\times\!10^{3}$	\\
	
	&	6-31G	&	$6.17\!\times\!10^{17}$	&	$7.03\!\times\!10^{13}$	&	$2.30\!\times\!10^{11}$	&	$9.33\!\times\!10^{7}$	&	$6.17\!\times\!10^{17}$	&	$2.35\!\times\!10^{12}$	&	$1.17\!\times\!10^{12}$	&	$2.49\!\times\!10^{4}$	&	$1.04\!\times\!10^{10}$	&	$4.14\!\times\!10^{3}$	&	$3.22\!\times\!10^{3}$	&	$3.06\!\times\!10^{3}$	\\
	
	&	cc-pVDZ	&	$4.10\!\times\!10^{19}$	&	$2.09\!\times\!10^{15}$	&	$6.86\!\times\!10^{12}$	&	$4.07\!\times\!10^{9}$	&	$4.10\!\times\!10^{19}$	&	$1.96\!\times\!10^{13}$	&	$9.68\!\times\!10^{12}$	&	$4.59\!\times\!10^{4}$	&	$9.72\!\times\!10^{10}$	&	$8.28\!\times\!10^{3}$	&	$1.09\!\times\!10^{4}$	&	$1.07\!\times\!10^{4}$	\\
\midrule																											
\multirow{11}{*}{Heme-Art.}	&	cc-pVDZ[4]	&	$2.02\!\times\!10^{12}$	&	$1.43\!\times\!10^{8}$	&	$4.22\!\times\!10^{5}$	&	$4.61\!\times\!10^{4}$	&	$2.02\!\times\!10^{12}$	&	$1.97\!\times\!10^{9}$	&	$8.81\!\times\!10^{8}$	&	$1.18\!\times\!10^{3}$	&	$1.03\!\times\!10^{8}$	&	$1.06\!\times\!10^{3}$	&	$4.67\!\times\!10^{2}$	&	$3.75\!\times\!10^{2}$	\\
	
	&	cc-pVDZ[6]	&	$2.33\!\times\!10^{14}$	&	$8.95\!\times\!10^{8}$	&	$2.67\!\times\!10^{6}$	&	$2.68\!\times\!10^{5}$	&	$2.33\!\times\!10^{14}$	&	$7.11\!\times\!10^{9}$	&	$3.19\!\times\!10^{9}$	&	$1.82\!\times\!10^{3}$	&	$3.68\!\times\!10^{8}$	&	$1.65\!\times\!10^{3}$	&	$8.69\!\times\!10^{2}$	&	$7.65\!\times\!10^{2}$	\\
																											
	&	cc-pVDZ[8]	&	$1.36\!\times\!10^{15}$	&	$3.53\!\times\!10^{9}$	&	$1.05\!\times\!10^{7}$	&	$1.13\!\times\!10^{6}$	&	$1.36\!\times\!10^{15}$	&	$2.08\!\times\!10^{10}$	&	$9.31\!\times\!10^{9}$	&	$2.47\!\times\!10^{3}$	&	$1.10\!\times\!10^{9}$	&	$2.26\!\times\!10^{3}$	&	$1.27\!\times\!10^{3}$	&	$1.16\!\times\!10^{3}$	\\
	
	&	cc-pVDZ[12]	&	$9.22\!\times\!10^{15}$	&	$2.70\!\times\!10^{10}$	&	$8.15\!\times\!10^{7}$	&	$7.02\!\times\!10^{6}$	&	$9.22\!\times\!10^{15}$	&	$7.03\!\times\!10^{10}$	&	$3.14\!\times\!10^{10}$	&	$3.77\!\times\!10^{3}$	&	$3.76\!\times\!10^{9}$	&	$3.55\!\times\!10^{3}$	&	$2.57\!\times\!10^{3}$	&	$2.45\!\times\!10^{3}$	\\
	
	&	cc-pVDZ[16]	&	$1.07\!\times\!10^{17}$	&	$1.21\!\times\!10^{11}$	&	$3.67\!\times\!10^{8}$	&	$2.89\!\times\!10^{7}$	&	$1.07\!\times\!10^{17}$	&	$2.04\!\times\!10^{11}$	&	$9.65\!\times\!10^{10}$	&	$5.30\!\times\!10^{3}$	&	$5.41\!\times\!10^{9}$	&	$4.75\!\times\!10^{3}$	&	$4.25\!\times\!10^{3}$	&	$4.12\!\times\!10^{3}$	\\
	
	&	cc-pVDZ[22]	&	$6.50\!\times\!10^{17}$	&	$6.59\!\times\!10^{11}$	&	$2.01\!\times\!10^{9}$	&	$1.51\!\times\!10^{8}$	&	$6.50\!\times\!10^{17}$	&	$6.20\!\times\!10^{11}$	&	$2.93\!\times\!10^{11}$	&	$7.50\!\times\!10^{3}$	&	$1.67\!\times\!10^{10}$	&	$6.76\!\times\!10^{3}$	&	$8.34\!\times\!10^{3}$	&	$8.20\!\times\!10^{3}$	\\
	
	&	cc-pVDZ[32]	&	$8.71\!\times\!10^{18}$	&	$5.26\!\times\!10^{12}$	&	$1.61\!\times\!10^{10}$	&	$1.14\!\times\!10^{9}$	&	$8.71\!\times\!10^{18}$	&	$2.08\!\times\!10^{12}$	&	$9.82\!\times\!10^{11}$	&	$1.15\!\times\!10^{4}$	&	$5.64\!\times\!10^{10}$	&	$1.04\!\times\!10^{4}$	&	$2.43\!\times\!10^{4}$	&	$2.42\!\times\!10^{4}$	\\
	
	&	cc-pVDZ[46]	&	$1.22\!\times\!10^{20}$	&	$4.12\!\times\!10^{13}$	&	$1.27\!\times\!10^{11}$	&	$8.44\!\times\!10^{9}$	&	$1.22\!\times\!10^{20}$	&	$7.30\!\times\!10^{12}$	&	$3.27\!\times\!10^{12}$	&	$1.74\!\times\!10^{4}$	&	$3.81\!\times\!10^{11}$	&	$1.62\!\times\!10^{4}$	&	$8.44\!\times\!10^{4}$	&	$8.42\!\times\!10^{4}$	\\
	
	&	cc-pVDZ[64]	&	$2.89\!\times\!10^{21}$	&	$2.48\!\times\!10^{14}$	&	$7.58\!\times\!10^{11}$	&	$5.41\!\times\!10^{10}$	&	$2.89\!\times\!10^{21}$	&	$4.31\!\times\!10^{13}$	&	$2.04\!\times\!10^{13}$	&	$2.54\!\times\!10^{4}$	&	$1.16\!\times\!10^{12}$	&	$2.33\!\times\!10^{4}$	&	$2.88\!\times\!10^{5}$	&	$2.88\!\times\!10^{5}$	\\
	
	&	cc-pVDZ[90]	&	$1.81\!\times\!10^{22}$	&	$1.60\!\times\!10^{15}$	&	$4.81\!\times\!10^{12}$	&	$4.44\!\times\!10^{11}$	&	$1.81\!\times\!10^{22}$	&	$1.35\!\times\!10^{14}$	&	$6.37\!\times\!10^{13}$	&	$3.74\!\times\!10^{4}$	&	$3.69\!\times\!10^{12}$	&	$3.44\!\times\!10^{4}$	&	$1.07\!\times\!10^{6}$	&	$1.07\!\times\!10^{6}$	\\
	
	&	cc-pVDZ[opt]	&	$1.10\!\times\!10^{20}$	&	$6.12\!\times\!10^{13}$	&	$1.79\!\times\!10^{11}$	&	$2.14\!\times\!10^{10}$	&	$1.10\!\times\!10^{20}$	&	$1.31\!\times\!10^{13}$	&	$6.21\!\times\!10^{12}$	&	$1.67\!\times\!10^{4}$	&	$3.39\!\times\!10^{11}$	&	$1.43\!\times\!10^{4}$	&	$5.13\!\times\!10^{4}$	&	$5.11\!\times\!10^{4}$	\\
	
\bottomrule										
\end{tabular}																											

    }
    {     \footnotesize
    \, \vspace*{.5cm}\\
$(g)$ $\hat{P}$: qubit counts \vspace*{.5cm} \\ 

    \begin{tabular}{llrrrrrrrrrrrr}				
\toprule											
System	&	Basis	&	$E_{P}$	&	ASP	&	$\textrm{aQPE}_A$	&	$\textrm{aQPE}_B$	&	oQPE	&	$\mathcal{R}_\pi$	&	$\textrm{iQPE}_A$	&	$\textrm{B}[H_A]$	&	$\textrm{iQPE}_B$	&	$\textrm{B}[H_B]$	&	$\mathcal{R}_\tau$	&	B[P]	\\
	
\midrule										
\multirow{3}{*}{Water}	&	STO-3G	&	608	&	457	&	380	&	380	&	576	&	565	&	449	&	75	&	449	&	75	&	307	&	185	\\
	
	&	6-31G	&	895	&	686	&	591	&	591	&	861	&	844	&	700	&	94	&	700	&	94	&	790	&	654	\\
	
	&	cc-pVDZ	&	1712	&	1246	&	1112	&	1112	&	1673	&	1434	&	1249	&	126	&	1249	&	126	&	1621	&	1465	\\
\midrule																											
\multirow{3}{*}{Ammonia}	&	STO-3G	&	627	&	471	&	394	&	394	&	596	&	585	&	469	&	76	&	469	&	76	&	411	&	291	\\
	
	&	6-31G	&	1507	&	776	&	674	&	674	&	1472	&	939	&	788	&	98	&	788	&	98	&	1427	&	1291	\\
	
	&	cc-pVDZ	&	1804	&	1479	&	1331	&	1331	&	1764	&	1684	&	1486	&	137	&	1486	&	137	&	1712	&	1554	\\
\midrule																											
\multirow{3}{*}{Benz.-Water}	&	STO-3G	&	1998	&	1708	&	1631	&	380	&	1957	&	1945	&	1828	&	154	&	449	&	75	&	497	&	354	\\
	
	&	6-31G	&	3614	&	3093	&	2998	&	591	&	3569	&	3550	&	3402	&	220	&	700	&	94	&	988	&	832	\\
	
	&	cc-pVDZ	&	6403	&	5706	&	5572	&	1112	&	6354	&	6332	&	6143	&	325	&	1249	&	126	&	1955	&	1780	\\
\midrule																											
\multirow{11}{*}{Heme-Art.}	&	cc-pVDZ[4]	&	413	&	275	&	230	&	194	&	387	&	376	&	291	&	57	&	255	&	51	&	201	&	109	\\
	
	&	cc-pVDZ[6]	&	535	&	367	&	310	&	268	&	506	&	490	&	385	&	67	&	343	&	61	&	315	&	211	\\
		
	&	cc-pVDZ[8]	&	642	&	462	&	394	&	346	&	611	&	594	&	478	&	76	&	430	&	70	&	465	&	351	\\
	
	&	cc-pVDZ[12]	&	910	&	645	&	561	&	501	&	877	&	789	&	653	&	90	&	606	&	84	&	830	&	704	\\
	
	&	cc-pVDZ[16]	&	1585	&	827	&	731	&	652	&	1549	&	1006	&	857	&	103	&	761	&	95	&	1499	&	1365	\\
	
	&	cc-pVDZ[22]	&	1726	&	1125	&	1010	&	890	&	1688	&	1316	&	1146	&	120	&	1026	&	112	&	1635	&	1491	\\
	
	&	cc-pVDZ[32]	&	1925	&	1589	&	1444	&	1321	&	1884	&	1861	&	1660	&	146	&	1500	&	138	&	1825	&	1669	\\
	
	&	cc-pVDZ[46]	&	2731	&	2306	&	2125	&	1959	&	2687	&	2662	&	2418	&	179	&	2256	&	173	&	2096	&	1928	\\
	
	&	cc-pVDZ[64]	&	3823	&	3283	&	3059	&	2771	&	3776	&	3749	&	3461	&	222	&	3173	&	214	&	2251	&	2071	\\
	
	&	cc-pVDZ[90]	&	5450	&	4660	&	4377	&	4076	&	5400	&	5372	&	5023	&	279	&	4631	&	271	&	2467	&	2277	\\
		
	&	cc-pVDZ[opt]	&	2631	&	2224	&	2052	&	1785	&	2586	&	2562	&	2331	&	175	&	2002	&	163	&	1981	&	1809	\\
	
\bottomrule																											
\end{tabular}																											

    }
\end{sidewaystable}

\end{document}